\documentclass[english]{article}
\usepackage[T1]{fontenc}
\usepackage[latin9]{inputenc}
\usepackage{geometry}
\usepackage{fancyhdr}
\usepackage{color}
\usepackage{textcomp}
\usepackage{amsmath}
\usepackage{amssymb}
\usepackage{graphicx}
\usepackage{rotfloat}
\usepackage{setspace}
\usepackage{wasysym}
\usepackage[authoryear]{natbib}

\makeatletter
\newcommand{\lyxaddress}[1]{
	\par {\raggedright #1
	\vspace{1.4em}
	\noindent\par}
}

\usepackage{floatpag,mwe}
\usepackage{url}

\makeatother

\usepackage{babel}
\begin{document}
\noindent \begin{flushleft}
\vspace*{-0.3in}
\hspace*{-0.4in}%
\begin{minipage}[t]{1.2\columnwidth}%
\begin{doublespace}
\noindent \begin{flushleft}
\textbf{\textcolor{blue}{\large{}\thispagestyle{plain}}}\textbf{\textcolor{black}{\LARGE{}An
Electromagnetic Calculation of Ionospheric Conductance that}}\\
\textbf{\textcolor{black}{\LARGE{}seems to Override the Field Line
Integrated Conductivity}}\textbf{\textcolor{blue}{\large{}\bigskip{}
}}
\par\end{flushleft}
\end{doublespace}
\end{minipage}\vspace{0.3in}
\\
\hspace*{-0.4in}%
\begin{minipage}[t]{0.8\columnwidth}%
\textbf{Russell B. Cosgrove}\\
November 19, 2022\medskip{}

\lyxaddress{\noindent Center for Geospace Studies, SRI International, Menlo Park,
CA, USA\\
Florida Space Institute, University of Central Florida, Orlando, FL,
USA\\
russell.cosgrove@me.com}
\noindent \begin{flushleft}
\textbf{\textcolor{black}{\large{}Abstract}}\textcolor{black}{{} }We
derive a steady-state, electromagnetic solution for collisional plasma
and apply it to computing the total conductance for a vertically stratified
ionosphere interrogated by a 3D signal with finite transverse wavelength,
which we compare to the field-line-integrated conductivity, finding
significant differences on all scales investigated. The approximate
solution is derived from an exact linearization of the electromagnetic
5-moment fluid equations, and writing down the integral form of the
driven steady-state solution, which is expressed as a sum over modal
contributions associated with the eigenvectors/eigenvalues of the
set of equations. We find the eigenvectors/eigenvalues numerically
and use them to argue that only two of the modes are capable of transmitting
energy through the ionosphere. Approximating their integrands with
analytic functions yields a solution valid for homogeneous plasma,
expressed in terms of the eigenvectors/eigenvalues for the two modes.
The solution may be understood using wavepacket concepts. Transmission
line theory is then used to extend the homogeneous-plasma solution
to a solution for the vertically-stratified ionosphere. We review
how transmission line theory contains electrostatic theory as a special
case, and show that our solution reproduces electrostatic theory exactly
when the eigenvectors/eigenvalues have the right properties, which
requires that they be artificially modified. Otherwise we find short-wavelength,
mode-mixing, and wave-admittance effects with broad relevance to ionospheric
science, and to global modeling of the Sun-Earth system. Since resonant
systems can be very sensitive, quantitative accuracy may require tuning,
but the physical conclusions seem unavoidable and to affect all transverse
scales.
\par\end{flushleft}

\section{\textcolor{black}{Introduction}\label{sec:Introduction}}

Electrostatic theory is an extremely convenient, steady-state simplification
of electromagnetic theory used extensively in electrical engineering
(e.g., resistors, capacitors, and inductors), and also in the physics
of collisional plasmas, such as the ionosphere. Transmission line
(TL) theory is also a steady-state theory, which applies to physical
systems that cross the boundaries of scale where electrostatic theory
becomes inapplicable {[}e.g., \citealp{collin-microwave-1966}; \citealp{malherbe-1979}{]}.
Thus, TL theory allows for testing electrostatic theory which, when
applicable, arises from the long-wavelength limit of TL theory. However,
TL theory is generally applied to engineered physical systems where
only one mode of propagation is supported and the transverse dimension
is suppressed. Specifically, applications of TL theory to the ionosphere
have been very limited {[}\citealp{mallinckrodt+carlson-1978}; \citealp{bhattacharyya+burke-2000}{]},
and as a result, no rigorous criteria\linebreak{}
\end{minipage}~\hspace*{0.1in}%
\begin{minipage}[t]{0.35\columnwidth}%
\noindent \begin{flushleft}
\vspace{2in}
\begin{minipage}[t]{0.9\columnwidth}%
\noindent \begin{flushleft}
{\footnotesize{}\hspace*{0.2in}}\textbf{\footnotesize{}Key Points\vspace{-0.1in}
}{\footnotesize\par}
\par\end{flushleft}
\begin{itemize}
\item \begin{flushleft}
{\footnotesize{}An electromagnetic calculation of the ionospheric
conductance finds something quite different from the field line integrated
conductivity}\textbf{\footnotesize{}\vspace{-0.1in}
}{\footnotesize\par}
\par\end{flushleft}
\item \begin{flushleft}
{\footnotesize{}The electric field does not map through the ionosphere,
and the penetration depth is highly dependent on scale and density}\textbf{\footnotesize{}\vspace{-0.1in}
}{\footnotesize\par}
\par\end{flushleft}
\item \begin{flushleft}
{\footnotesize{}There are effects from short parallel wavelengths
and multiple modes, such that wave-like and resonant behavior are
supported}{\footnotesize\par}
\par\end{flushleft}
\end{itemize}
\end{minipage}\\
\vspace*{0.5in}
\begin{minipage}[t]{0.9\columnwidth}%
\begin{spacing}{0.85}
\noindent \begin{flushleft}
{\footnotesize{}\hspace*{0.19in}}\textbf{\footnotesize{}10 Peer Reviews\vspace*{-0.35in}
}{\footnotesize\par}
\par\end{flushleft}
\end{spacing}
\begin{itemize}
\item \begin{flushleft}
\textcolor{black}{\footnotesize{}Supplementary Information includes
10 anonymous peer reviews with detailed replies, accumulated from
earlier versions}{\footnotesize\par}
\par\end{flushleft}
\end{itemize}
\end{minipage}\vspace*{0.55in}
\hspace*{0.25in}%
\noindent\begin{minipage}[t]{1\columnwidth}%
\textbf{\footnotesize{}© 2022, CC BY SA license}%
\end{minipage}
\par\end{flushleft}%
\end{minipage}
\par\end{flushleft}

\vspace*{-0.08in}
\lhead{}

\newpage{}

\noindent have been developed for use of electrostatic theory in the
ionosphere {[}\citealp{vasyliunas-2012}; \citealp{cosgrove-2016}{]}. In
this work we examine the validity of electrostatic theory in the ionosphere
using a bi-modal form of TL theory for collisional plasma, and derive
a steady-state electromagnetic solution for the vertically-inhomogeneous
ionosphere interrogated by a plane-wave signal with finite transverse
wavelength (i.e., the Fourier components of a 3D signal). We show
how this solution would reproduce electrostatic theory if the collisional
waves in the ionosphere had the right properties. However, we find
that the ionospheric waves do not generally have the right properties.
In fact, there is a wide range of parameter space where the TL theory
predicts profound wave-like effects that should have important systemic
consequences. For example, the ionospheric conductance is different
from the field line integrated conductivity.

Hence, we find that an electromagnetic theory of the ionosphere is
required, and we put forth our TL theory as the most natural generalization
of the current electrostatic baseline. Although needing to be expanded
in terms of the parameter space that it covers, this TL-theory calculation
amounts to a rigorous, electromagnetic theory of the ionosphere that
can replace electrostatic theory in many applications, and which can
be used to validate the numerical simulations being pursued by others.
By virtue of its connection to circuit theory, the TL theory also
provides a physically intuitive description of the electrical steady-state
of the ionosphere, which appears to be much more interesting than
previously expected. (It should be noted that the author is a former
microwave filter engineer who formed a deep appreciation for TL theory
through practical application.)

\textbf{Disambiguation: }This article is addressed primarily to ionospheric
physics, which traditionally employs a version of electrostatic theory
that is very close to the textbook theory {[}e.g., \citealp{jackson-1975}{]},
where all time derivatives are set to zero and what remains of the
equations of motion are linearized and applied to a boundary value
problem {[}\citealp{farley-1959}; \citealp{farley-1960}; \citealp{spreiter+briggs-1961};
\citealp{kelley-2009}{]}. By ``electrostatic theory,'' or ``full
electrostatic theory,'' for emphasis, we refer to the practice of
assuming that this solution is an approximation for the linear, steady-state
solution found using the full equations of motion. One of the major
electrostatic concepts we will test is the ``Farley factor'' {[}e.g.,
\citealp{kelley+gelinas-2000}; \citealp{kagan+etal-2000}{]}, which provides
a way to assess the transverse-scale-size dependence of the distance
that the influence of a disturbance will extend along geomagnetic
field lines, for a uniform plasma. Specifically, this ``mapping distance''
is estimated as the transverse scale-size multiplied by the Farley
factor, which is often referred to as the Farley mapping distance,
$d_{F}=\sqrt{\sigma_{0}/\sigma_{P0}}/k$ (where $k=\left|\vec{k}\right|$
is the transverse wavenumber, and $\sigma_{0}$ and $\sigma_{P0}$
are the field-aligned and zero-frequency Pedersen conductivities,
respectively). This concept is used to estimate the degree of coupling
between different regions of the ionosphere, and detailed electric
field distributions may be calculated by solving an associated boundary
value problem for the actual inhomogeneous plasma {[}e.g., \citealp{kelley-2009}
and references therein{]}. The same concept also gives rise to the
idea that the ionospheric conductance can be calculated by integrating
the local conductivity along geomagnetic field lines, which is essentially
an assumption that the electric field maps unchanged through the whole
of the ionospheric thickness. By solving the associated boundary value
problem, it has been determined that this should be true if the transverse-scale-size
of the imposing signal is greater than about 10~km {[}e.g., \citealp{farley-1960};
\citealp{spreiter+briggs-1961}; \citealp{hysell+etal-2002b}{]}. Similar
mapping assumptions and boundary-value calculations underly much of
ionospheric physics, including analysis of auroral arcs {[}e.g., \citealp{marclund+etal-1982};
\citealp{aikio+etal-2002}; \citealp{juusola+etal-2016}{]}, analysis of
equatorial plasma bubbles {[}e.g., \citealp{tsunoda+etal-1982}; \citealp{aviero+hysell-2012};
\citealp{Tsunoda-2015}{]}, analysis of the pre-reversal enhancement
{[}e.g., \citealp{farley+etal-1986}; \citealp{eccles+etal-2015}{]}, analysis
of sporadic E layers {[}e.g., \citealp{haldoupis+etal-1996b}; \citealp{cosgrove+tsunoda-2002b};
\citealp{hysell+etal-2004}{]}, and analysis of data from incoherent
scatter radars {[}e.g., \citealp{heinselman+nicolls-2008}; \citealp{nicolls+etal-2014}{]},
to name a few.

A weaker form of electrostatic assumption involves substitution of
electrostatic waves for electromagnetic waves, and assuming they are
essentially equivalent. In this case the time-derivative is dropped
in Faraday's law (so that $\vec{\nabla}\times\vec{E}=0$, and Maxwell's
equations become the Poisson equation), but retained in at least some
of the remaining equations, so that waves are still supported. In
ionospheric physics this ``electrostatic-wave theory'' is used to
analyze plasma instabilities, and to compute the spectrum of backscatter
for incoherent scatter radar. The validity of electrostatic-wave theory
is a distinct question from that of the full electrostatic theory,
which does not include wavelike effects. We also analyze this weaker
assumption, by substituting the Poisson equation for the Maxwell equations.
We find that electrostatic waves may be good substitutes below about
100~m in transverse wavelength, but above 100~m they have different
properties that would produce very different results in the TL theory
calculation. In the case of magnetohydrodynamic (MHD) waves, which
are non-collisional, a referee has stated that this is a known result,
which follows from a more general result that Alfvén waves become
electrostatic when the transverse scale becomes less than the electron
inertial length {[}however see \citealp{Saleem-2001}{]}. But we are
not aware of equivalent results for the fully collisional ionosphere,
and it is often suggested that the high collision frequencies preclude
electromagnetic effects.

\textbf{Focus on Ionospheric Conductance:} Although the TL method
includes calculation of the local field quantities like electric and
magnetic field, we focus discussion on the transverse-scale-size dependence
of the ``ionospheric conductance'' {[}e.g., \citealp{fuller-rowell+evans-1987};
\citealp{Lysak-MI-1990}{]}, which is a linearized and non-local quantity
the non-locality of which is normally handled with electrostatic theory,
through the electric field mapping assumption. As such, the ionospheric
conductance provides an appropriate differentiator for testing electrostatic
theory, such as the scale-size-dependent electric-field mapping it
entails. In addition, MHD and other models of the magnetosphere generally
employ an inner boundary condition that is established by considering
the ionosphere as an electrical load, characterized by this ionospheric
conductance, which is meant to represent the input admittance seen
from above the ionosphere {[}e.g., \citealp{Raeder+etal-1998}; \citealp{gombosi+etal-2001};
\citealp{toffoletto+etal-2003}; \citealp{lyon+etal-2004}; \citealp{janhunen+etal-2012};
\citealp{lotko+etal-2014}{]}. This input admittance is therefore an
important quantity for large-scale modeling of the Sun-Earth interaction.
Under electrostatic theory it is derived by integrating the (zero
frequency) ionospheric conductivity along the geomagnetic field, based
on the mapping assumption. Dropping the assumption that it is purely
real, and dropping the assumption that it can be derived from electrostatic
theory, we arrive at an important problem that is well suited to TL
theory, that of deriving the ionospheric input admittance.

In doing this we want to emphasize that we are doing it as a function
of the transverse wavelength of the imposing signal, in order to make
comparisons with electrostatic theory as it is commonly used in the
ionosphere. The most comparable previous results of which we are aware
are those found by \citeauthor{knudsen+etal-1992}~{[}1992{]}, which
do support the calculation of conductance as the field line integrated
conductivity, for sufficiently low-frequencies. However, for reasons
that are discussed in Section~\ref{sec:Background}, and more fully
in \citeauthor{cosgrove-2016}~{[}2016{]}, we believe that these
results must be subordinated to the results presented herein. There
are also a number of time domain simulations that might seem applicable
{[}e.g., \citealp{seyler-1990}; \citealp{Birk+Otto-1996}; \citealp{zhu+etal-2001};
\citealp{otto+zhu-2003}; \citealp{dao+etal-2013}; \citealp{lysak+etal-2013};
\citealp{Tu+Song-2016}; \citealp{tu+song-2019}{]}. But we have not been
able to find where any of these actually compute the ionospheric conductance
as a function of the transverse wavelength. Some likely reasons for
this are discussed in the background section~(\ref{sec:Background}),
although with humility as we are not in a position to properly critique
the full collection of complex and sophisticated modeling efforts.
The summary is that these time-domain simulations are addressed primarily
to complex transient phenomena on global scales, and are not optimized
for calculating the wavelength dependent conductance, which is only
accessible to them through implementation of potentially impractical
boundary conditions, and then by running the model for a long time,
until everything stops changing (steady state).

What appears to be true is that the calculation provided in the present
article represents the most rigorous attempt yet to calculate the
ionospheric conductance as a function of the transverse wavelength.
We give an exact linearization of the electromagnetic 5-moment fluid
equations without making any of the usual ``resistive MHD'' or other
simplifications. Some such simplifications are made in every other
work we have found, and we do not believe there is any way to know
if they might be affecting the conductance found (or which would be
found) for the highly collisional \emph{E} region ionosphere, except
by removing the simplifications as we have done in this work. The
approximations that we do make are all ones that are clearly beneficial
to the prospects for recovering electrostatic theory at long transverse
wavelengths (or low frequencies), with the only possible exception
being use of the usual understanding of wavepacket propagation, which
we are mostly able to validate. We also note that our method involves
a low level of numerical error, allows for continuous vertical resolution,
is sanctioned by the principle of causality, and results in a much
more analytical and physically intuitive description of electromagnetic
ionospheric physics.

An additional motivation for this work comes from recent observational
findings that the magnetosphere-ionosphere (MI) coupling interaction
may actually occur over scales significantly shorter than previously
expected. For example, using data from the Swarm satellites, \citeauthor{pakhotin+etal-2021}~{[}2021{]}
found that half the Poynting flux may be associated with events having
transverse-scale-size less than 250~km. If this is true then having
an accurate transverse-wavelength-dependent characterization of ionospheric
conductance is in fact very important for large scale modeling, and
so it is important to either validate or replace the current electrostatic
paradigm for conductance.

\textbf{Contents and Barrier to Entry:} Due to the many years that
electrostatic theory has been used in ionospheric physics, and the
many papers that have been written on waves in the magnetosphere,
mostly using other techniques, we anticipate the need to prove the
veracity of our calculation. At the same time, we anticipate the need
for a relatively simple example that illustrates the breakdown of
electrostatic theory and motivates the calculation. 

To meet the first need, in Sections~\ref{sec:Solution}~and~\ref{sec:modleIonosphereAdmittance}
we derive our generalized form of TL theory from an exact solution
to the full equations of motion for the linearized problem, while
describing every approximation in detail. We also review why linearization
is appropriate. Then in Section~\ref{sec:Wave-Modes-Relevant-to-Ionospheric-Science}
we use the results of Section~\ref{sec:Solution} to give a (brief)
quantitative presentation of the properties of the wave modes that
are relevant to ionospheric physics, and to compare the electrostatic
and electromagnetic waves with respect to the critical modeling parameters
of parallel wavelength and admittance.

To meet the need for a simple example, in Section~\ref{sec:Gedanken-Experiment}
we reframe selected results from \citeauthor{cosgrove-2016}~{[}2016{]}
in the form of a gedanken experiment that illustrates the breakdown
of electrostatic theory and introduces TL theory. This discussion
is continued in Section~\ref{sec:ElectricFieldMapping} where the
quantitative results of Section~\ref{sec:Wave-Modes-Relevant-to-Ionospheric-Science}
are added, in order to discuss the likely physical behavior of the
ionosphere, and to justify retaining only two modes. Together these
sections provide a framework for understanding the major results,
and we refer back to them often in the final presentation of results
in Section~\ref{sec:Model-Results}.

The following background section is for readers seeking a more thorough
introduction. Other readers may wish to skip to Section~\ref{sec:Gedanken-Experiment}.

\section{Background\label{sec:Background}}

\textbf{TL Theory:} TL theory uses the Laplace transform to derive
the response of a uniform transmission line to a harmonic source at
its input that turns on, and then continues operating indefinitely
{[}e.g., \citealp{nilsson-1984}; \citealp{miano+maffucci-2001}{]}. The
Fourier transform is applied in the spatial domain so that the steady-state
part of the linearized response can be characterized by quantities
such as wave-number and wave-admittance, where usually only one wave
mode is retained. Then, finite length sections of transmission line
having different characteristics (e.g., different admittances) are
pieced together, such that each contains a superposition of oppositely-directed
pieces of its unique steady-state response, and suitable boundary
conditions are enforced across the junctions. The oppositely directed
pieces can be thought of as incident and reflected waves, that is,
for the case when one of the ends is not connected to a source. This
formalism allows for deriving the steady-state response of an assemblage
of these lines, which can include branches, loop backs, or whatever.
An amazing variety of real, three-dimensional circuits have been approximated
in this way {[}e.g., \citealp{Matthaei-1980}{]}, using specialized software
packages such as Touchstone {[}\citeauthor{Touchstone}{]}.

Note that although the term transmission ``line'' is used, the formalism
can be applied to a slab-like geometry to produce a 3D model for a
signal incident on a vertically inhomogeneous ionosphere, where the
transverse wavevector arises as an independent parameter. Note also
that although the formalism sounds inherently approximate, we can
derive it from an exact solution, and the approximations can be well
understood. Thus, the idea of extending these methods to the ionosphere
is attractive, and one could imagine that software similar to Touchstone
could be made available for ionospheric modeling.

\textbf{Difficulties for Temporal/Spatial-Domain Models:} TL theory
has strengths and weaknesses that are complementary to those of the
nonlinear models that are more commonly used in the ionosphere and
magnetosphere, which resolve the system in time and space. To begin
with, note that circuit theory in electrical engineering is essentially
a theory of linear systems in steady state, where these expedients
allow for a source-independent characterization of the system. If
we wish to deploy the concepts of admittance and conductance to ionospheric
science then we must adopt a linear and steady-state description.
There are two major aspects of temporal/spacial-domain models that
make this a difficult proposition: first, the source itself must be
unvarying; and second, the model must be run for a long time, until
everything stops changing. We consider these two difficulties in turn.

What is meant by the the source being unvarying is that the source
should produce an unchanging superposition of waves incident on the
system. Ideally there should be only a single wave, which can be varied
so as to fill out the Fourier spectrum. A single wave incident on
the system is characterized by its wave vector, polarization vector,
frequency, phase, and sufficiently-small amplitude. This creates a
difficulty for spatial domain models because in such models the source
usually takes the form of a boundary condition imposed at the input
to the system. The system reflects some of the incident waves, such
that the needed boundary condition is the sum of the incident and
reflected waves. But since the goal is to characterize the system,
it follows that the system has not yet been characterized, and so
there is no way to predict the reflected waves from the incident waves.
There is no way to know, a priori, the correct boundary condition
corresponding to a particular, chosen, incident wave. And an arbitrarily
chosen boundary condition may not reflect any such case, it likely
reflects a source producing a \emph{changing }superposition of incident
waves, which defeats the steady-state characterization.

One way around this problem is to extend the modeled domain so that
the source is a long way from the region of interest, so that the
system can be characterized before the reflected wave returns to corrupt
the response {[}e.g., \citealp{streltsov+lotko-2003b}{]}. However, to
find the steady-state response from a time-domain model it is necessary
to run the model for a long time, until everything stops changing.
Hence the system must come to steady-state before the reflected wave
returns, which does not seem very practical.

In addition, if the model contains nonlinearities then the steady-state
will depend on the amplitude of the excitation, and it may well be
that it never reaches a steady state. In either of these cases the
model is not applicable to calculating the admittance, unless it is
modified by removing the non-linearities. And what is the point of
a numerical simulation of linear wave propagation when an analytical
solution is available? A numerical simulation involves errors that
build up over time, and it is very difficult to analyze these errors,
especially in the limit as time goes to infinity.

\textbf{Advantages of TL Theory:} By contrast, TL theory does not
encounter any of these difficulties, because it utilizes the analytical,
steady-state solution that is available once the system is expressed
in the Laplace/Fourier domain, and linearized. And this has the additional
advantage of providing a result that is easy to interpret physically:
the solution is expressed as a sum over steady-state wave-modes in
analytical form, plus a background state. Thus TL theory is both intuitive
and natively addressed to calculating circuit quantities, using the
assumptions of linearity and steady-state that are crucial elements
in their definition.

TL theory also has the advantage that the scale-size-dependence of
the ionospheric conductance can be calculated using a model that is
effectively one-dimensional; that is, the background state can be
one dimensional, while the incident signal is represented in the Fourier
domain (so that it is three-dimensional). This is in contrast to the
situation for a nonlinear model where one-dimensional means that all
transverse derivatives vanish, including those for the incident signal,
so that there is no field-aligned current {[}e.g., \citealp{tu+etal-2014}{]}.
In order to address current-closure through the ionosphere, and the
scale size dependence of electric field mapping, non-linear models
must be at least two-dimensional, and this creates a much heavier
numerical burden.

Another advantage of TL theory is the ability to explicitly remove
the parasitic contributions from high frequency modes without any
distortion of the retained modes, which arises from working in the
Laplace/Fourier domain. In the time-domain, finite-difference approach
it is necessary either to use a time-step short enough to resolve
the radio-frequency waves (which is generally impractical), or to
introduce low-frequency approximations to the governing equations
that kill these modes. The low-frequency approximations have some
effect on the retained modes, and may cause distortion in the $E$-region
ionosphere, where the collision frequency is very high. Since the
$E$-region contains most of the ionospheric conductivity, any compromise
with respect to collision frequency introduces uncertainty into calculation
of the ionospheric conductance.

Another advantage of TL theory over spatial-domain models is the ability
to achieve continuous spatial-resolution. Spatial domain models cannot
find Fourier components above the Nyquist frequency for the spatial
sampling, although below this frequency there is continuous resolution
of the Fourier components. This situation is reversed for Fourier-domain
models: Fourier domain models cannot model locations farther away
than a certain distance from the center of the spatial domain, but
within this region there is continuous spatial resolution. This dichotomy
means that while spatial-domain models may be more useful for large-scale
modeling, they will have quite a bit of difficulty resolving the sharp
features that we find below are critical to the ionospheric conductance.
Hence it is important that both kinds of models are developed and
used in tandem.

\textbf{Boundary Value Methods:} There is also another kind of solution
for the equations of motion that we should mention, which is often
used to study ELF/VLF waves in the ionosphere. Instead of the Laplace
transform, the Fourier transform is applied to the time axis and either
the ionospheric Ohms law or cold-plasma dielectric tensor are incorporated
into the Maxwell equations, in lieu of the plasma momentum equations
{[}e.g., \citealp{hughes-1974}; \citealp{knudsen+etal-1992}; \citealp{kuzichev+etal-2018}{]}.
Application of the Fourier transform to the time axis was discussed
in Section~7 of \citeauthor{cosgrove-2016}~{[}2016{]}, where it
was explained that it produces a boundary value problem. This approach
does not apply the equations of motion to the Cauchy problem for which
they are sanctioned by causality. The dispersion relation for the
equations of motion is solved for complex $\vec{k}$ as a function
of a real-valued $\omega$, whereas for the Cauchy problem the (same)
dispersion relation is solved for complex $\omega$ as a function
of real $\vec{k}$. These are two different operating points on the
dispersion surface and so almost-certainly the solutions will be different.
If different, only one of them can be physical.

This relationship was studied by \citeauthor{cosgrove-2016}~{[}2016{]},
and a number of concrete differences were found. We mention two of
them that are particularly simple to understand. First, the parallel
wavelengths corresponding to the same frequency (real part of $\omega$)
do not agree with each other. And second, the conductivity equations
are evaluated either with, or without an imaginary component in $\omega$
(and we note that the conductivity equations do not depend on $\vec{k}$).
In Figure~16 of \citeauthor{cosgrove-2016}~{[}2016{]} the Pedersen
conductivities evaluated in these two ways were compared, and at least
for the near-DC frequencies of interest in this work, it was found
that the imaginary part of $\omega$ is very important. And, the imaginary
part of $\omega$ also depends on the mode of propagation.

Thus, in using the Ohms-law/dielectric-tensor in Maxwell's equations
without the imaginary part of frequency there is an effect of ``squeezing
together'' the modes under a single, inaccurate conductivity, in
order to accommodate the smaller equation set (i.e., momentum equations
omitted). Coupling this with the parallel wavelengths being different,
it appears that the boundary-value methods will produce a solution
very different from the Cauchy-problem methods used below. In the
discussion of equations~(17)~and~(18) in \citeauthor{cosgrove-2016}~{[}2016{]},
it was shown that the boundary value method might lead to a solution
very close to electrostatic equilibrium. But based on the fundamental
principle of causality in physics, the solution of the Cauchy problem
must be considered as the physical one. In what follows we use TL
theory to solve the Cauchy problem for the steady-state solution.

\textbf{Generalizing the TL Theory: }The advantages explained in the
previous paragraphs allow this work to achieve its goal of an electromagnetic
calculation of the scale-size dependence of the ionospheric conductance,
and of the associated electric field mapping and current closure through
the ionosphere, etc. Temporal/Spatial-domain models are not really
appropriate for this purpose, for all the reasons expressed above,
and this may explain why such a calculation has never been done, to
our knowledge.

However, TL models in electrical engineering normally involve a single
wave mode, and circuits are generally designed to ensure that only
one mode is supported. In the ionosphere we do not have this luxury,
and so we will generalize the usual TL theory to one that includes
multiple modes. Doing this means that the scalar admittance, which
relates the electric and magnetic fields, must be replaced by a tensor
such that there is an ``admittance vector'' or ``polarization vector''
for each mode. A problem arises in that the boundary conditions connecting
different sections must be expanded, and we are not aware of any universal
rules that can be applied. However, we find that as long as consideration
is limited to scales above a few tens of meters, the ionosphere supports
only two relevant modes, which can be roughly associated with the
usual, physically-defined Alfvén and Whistler waves. In this case
there are sufficient criteria available for setting the boundary conditions,
and we are able to proceed with a bi-modal form of TL theory.

Another added complexity arises from the equations of motion, which
must describe the evolution of a collisional plasma, and therefore
must be extended beyond Maxwell's equations. In earlier work {[}\citealp{cosgrove-2016}{]}
we discussed waves in the ionosphere (a collisional plasma), and showed
how they can be derived from the combined Maxwell and plasma momentum
equations without simplifications. The methods employed were actually
quite general, and in this work we extend them to the 5-moment fluid
equations {[}\citealp{schunk+nagy-2009}{]}, which will form the equations
of motion for this work. The waves that inhabit our transmission lines
are the waves of the electromagnetic 5-moment fluid equations, formally
defined through eigenvectors and eigenvalues, and validated by numerical
evaluation of the exact integral expressions. We have also tested
the model using the smaller equation-set from \citeauthor{cosgrove-2016}~{[}2016{]},
and the results appear to be exactly the same. Using the larger equation-set
allows for evaluating possible contributions from lower-frequency
waves, and for possibly including some of these waves in later editions. 

\section{Gedankenexperiment and Breakdown of Electrostatic Theory\label{sec:Gedanken-Experiment}}

\begin{figure}[p]
\thisfloatpagestyle{plain}\includegraphics{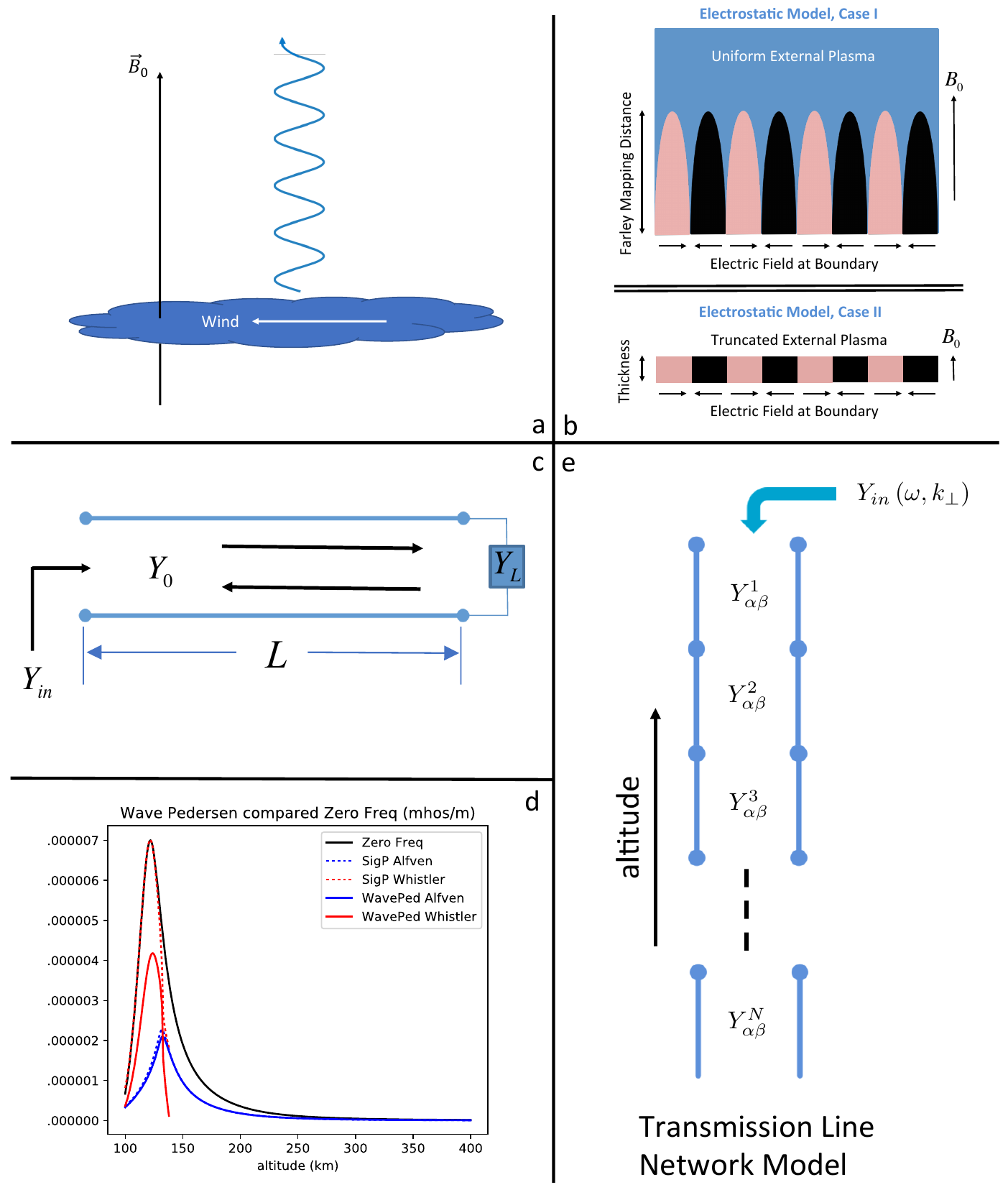}

\caption{Electrostatic versus TL theory: panel~a, schematic for the gedanken
experiment; panel~b, electrostatic approach; panel~c, TL terminated
by load admittance $Y_{L}$; panel~d, wave-Pedersen conductivity
compared with the standard zero-frequency Pedersen conductivity, and
the full Pedersen conductivity, for the Alfvén and Whistler modes
($\lambda_{\perp}=1000\,\mathrm{km,}\:n_{e}=4.7\times10^{9}\mathrm{m}^{-3}$);
panel~e, ionosphere modeled as a cascade of short transmission line
sections. Each section represents a thin slab of homogeneous plasma.\label{fig:Electrostatic-versus-TL}}

\end{figure}

To motivate the construction of an electromagnetic model it is useful
to analyze a toy example that illustrates the limitations of electrostatic
theory as it pertains to the ionosphere. Consider the case of the
horizontally extended, vertically thin \emph{E}-region dynamo shown
in panel~a of Figure~\ref{fig:Electrostatic-versus-TL}, which is
driven by a horizontal neutral wind, and embedded at the bottom a
semi-infinite homogeneous plasma with vertical geomagnetic field (and
no wind). Such a vertically localized source excites the characteristic
modes at the surface of the surrounding plasma, at the frequency and
wavelength of the source, which we imagine to be a pure tone. Those
that are able to propagate at that frequency and wavelength carry
energy away from the source, while gradually dissipating {[}\citealp{bernstein-1958}{]}.
Those that cannot propagate will produce a near field that we will
ignore, or, what amounts to the same thing, we will consider the upper
boundary of the source to be beyond the near field. As time goes to
infinity a steady state condition arises where the plasma oscillates
harmonically with the source. The time scale for reaching steady-state
is generally thought very short compared to other relevant ionospheric
time scales, and this is a key assumption of electrostatic theory
{[}\citealp{farley-1959}{]}, which also applies to TL theory. Nonlinear
interactions of the waves may direct the evolution away from any kind
of steady state, and also introduce a non-trivial dependence on the
amplitude of the source. Thus this kind of a description assumes signal
amplitudes are small enough that nonlinearities can be ignored, and
applies to the far zone (aka, radiation zone) of the source after
the system has reached steady state. 

In order to apply the standard TL theory, consider the hypothetical
case where there is only one propagating mode, and calculate the input
admittance seen by the dynamo of panel~a, Figure~\ref{fig:Electrostatic-versus-TL},
as it looks up at the uniform plasma above it. This is the admittance
that loads the dynamo, affecting the net electric field and current
(and etc.) that are produced, in accordance with the dynamo's own
internal admittance. We can calculate the input admittance using both
electrostatic theory and TL theory, which in this case is a single
mode version of electromagnetic theory. Most of this material was
covered by {[}\citealp{cosgrove-2016}{]}, although here we use standard
transmission line formulas instead of deriving the results directly,
and organize the material in a way that might serve to make it more
accessible.

\textbf{Electrostatic Calculation:} For the electrostatic calculation,
take the electric field at the surface of the dynamo to have the form
$\vec{E}=\left(\hat{y}\tilde{E}_{y}+\hat{z}\tilde{E}_{z}\right)\mathrm{e}^{i(\omega t+k_{z}y+k_{z}z)}$,
where $\hat{y}$ is transverse to the geomagnetic field (tangent to
the boundary), and $\hat{z}$ is along the geomagnetic field. There
is no $\hat{x}$ component because the curl of the electric field
must be zero under electrostatic theory, and we assume the convention
that the wavevector is perpendicular to $\hat{x}$. Under electrostatic
theory, as illustrated in the panel~b of Figure~\ref{fig:Electrostatic-versus-TL}
(Case~I), the transverse electric field at the surface of the dynamo
$\left(\vec{E}_{\perp}=\hat{y}\tilde{E}_{y}\mathrm{e}^{i(\omega t+k_{y}y)}\right)$
maps a distance $d_{F}$ along $\vec{B}_{0}$ into the homogeneous
plasma, and therefore drives a field-line-integrated transverse current,
$\vec{J}_{\perp}=\vec{E}_{\perp}\sigma_{P0}d_{F}-\vec{E}_{\perp}\times\hat{z}\sigma_{H0}d_{F}$,
where $\sigma_{P0}$ and $\sigma_{H0}$ are the zero-frequency Pedersen
and Hall conductivities, respectively. The distance $d_{F}$ is the
Farley mapping distance, which is the distance an electric field is
expected to map along the geomagnetic field under electrostatic theory
(see Section~\ref{sec:Introduction}). Beyond the distance $d_{F}$
there is no electric field, and thus no current. Therefore, by current
continuity, the region of homogeneous plasma returns a current density
obeying $j_{\parallel}=\vec{\nabla}\cdot\vec{J}_{\perp}=\sigma_{P0}d_{F}\vec{\nabla}\cdot\vec{E}_{\perp}$,
where $j_{\parallel}$ is the current density into the adjoining plasma
(parallel to the geomagnetic field). That is, under electrostatic
theory, the adjacent homogeneous plasma presents a load to the dynamo-source
that can be characterized by the input admittance,
\begin{equation}
Y_{in\mathrm{Matched}}=\frac{j_{\parallel}}{\vec{\nabla}\cdot\vec{E}_{\perp}}=\sigma_{P0}d_{F}.\label{eq:load_under_el_eq_matched}
\end{equation}

If instead the region of homogeneous plasma ends a short distance
from the dynamo layer, such that the conductivity beyond is zero,
then Case~I of panel~b is replaced by Case~II. Under electrostatic
theory, the electric field maps across the thin region of uniform
plasma, and beyond there is no current. Thus, the thickness $L$ of
the region of homogeneous plasma replaces $d_{F}$ in the expressions
of the previous paragraph, and the input admittance becomes,
\begin{equation}
Y_{in\mathrm{OC}}=\frac{j_{\parallel}}{\vec{\nabla}\cdot\vec{E}_{\perp}}=\sigma_{P0}L,\label{eq:load_under_el_eq_open_circuit}
\end{equation}
where ``OC'' stands for ``open circuit.''

\textbf{Electromagnetic Calculation:} Now consider calculating the
input admittance using the (single-mode) TL theory. When the dynamo
is bounded by a semi-infinite homogeneous plasma, as in panel~a,
there can be no back reflection, and so there is only a single wave
propagating away from the dynamo. So to obtain $Y_{in\mathrm{Matched}}$
under TL theory we need to compute the quantity $j_{\parallel}/\vec{\nabla}\cdot\vec{E}_{\perp}$
for a single propagating wave, which we can do from the results in
\citeauthor{cosgrove-2016}~{[}2016{]} (or using our results below).
However, we do not actually need specific values to make our major
point. 

The input admittance associated with a single wave propagating away
is commonly referred to as the characteristic admittance of the wave,
$Y_{0}$. It is this quantity that enters into the well-known TL expression
for the input admittance to a transmission line terminated by a load
admittance $Y_{L}$ {[}e.g., \citealp{collin-microwave-1966}{]},
\begin{equation}
Y_{in}=Y_{0}\frac{iY_{0}\tan\phi+Y_{L}}{iY_{L}\tan\phi+Y_{0}},\label{eq:TL-InputAdmittance}
\end{equation}
where the electrical length $\phi=k_{z}L$ is the phase rotation along
the transmission line, $k_{z}$ is the parallel wavevector, and $L$
is the line length. More precisely, the formula~(\ref{eq:TL-InputAdmittance})
actually allows $\phi$ to be complex, and is thus slightly generalized
from what can be found in textbooks like \citeauthor{collin-microwave-1966}~{[}1966{]},
where dissipation is not included. Panel~c of Figure~\ref{fig:Electrostatic-versus-TL}
illustrates the application for the formula. The arrows in the figure
illustrate how the formula is derived, as the superposition of two
waves traveling in opposite directions, related by the reflection
coefficient at the load, $\left(Y_{0}-Y_{L}\right)/\left(Y_{0}+Y_{L}\right)$.
Note the assumption that the waves have finished traversing the line
and come to steady state. Note, also, the assumption that the waves
have small enough amplitudes that non-linear interactions can be ignored,
so that linear superposition is valid. These assumptions underlie
the whole of circuit theory in electrical engineering, and the definition
of admittance depends on them.

Using equation~(\ref{eq:TL-InputAdmittance}), the result for the
semi-infinite case is found by matching the load to the characteristic
admittance, so there is no back-reflected wave, that is, $Y_{L}=Y_{0}$.
Substituting into equation~(\ref{eq:TL-InputAdmittance}) gives,
\begin{equation}
Y_{in\mathrm{Matched}}=Y_{0},\label{eq:input_admittance_matched_electromagnetic}
\end{equation}
which, of course, just shows that the formula is working as intended;
we have defined $Y_{0}$ to be the input admittance when the load
is matched to the line, that is, when there is only a single wave
propagating away. Note that this result recalls the formula $j_{\parallel}/\vec{\nabla}\cdot\vec{E}_{\perp}=(\mu_{0}V_{A})^{-1}$
that is used in classical magnetosphere-ionosphere coupling papers,
such as {[}\citealp{mallinckrodt+carlson-1978}, in Gaussian units{]},
where $(\mu_{0}V_{A})^{-1}$ is the characteristic admittance of the
(collisionless) Alfvén wave, and $V_{A}$ is the (collisionless) Alfvén
velocity. Note also that if the displacement current is included in
$j_{\parallel}$, our definition for characteristic admittance encompasses
the definition in terms of electric and magnetic fields that is commonly
used in free space (the displacement current is thought negligible
in the ionosphere {[}\citealp{song+lysak-2006}{]}).

Now consider the second case, when the region of homogenous plasma
is bounded, instead of being semi-infinite. In this case there will
be a back-reflected wave, and we actually need equation~(\ref{eq:TL-InputAdmittance}).
When the homogeneous plasma has an upper boundary the current above
the boundary must be zero, that is, $Y_{L}=0$. Making this substitution
into equation~(\ref{eq:TL-InputAdmittance}) gives the electromagnetic
prediction for the input admittance,
\begin{equation}
Y_{in\mathrm{OC}}=iY_{0}\tan\phi.\label{eq:input_admittance_OC_electromagnetic}
\end{equation}

\textbf{Comparison of Results:} Assuming only that the waves are relatively
low loss, so that $\phi$ is mostly real, we can see that the electrostatic
results are not consistent with the electromagnetic (TL theory) results.
The electrostatic expressions~(\ref{eq:load_under_el_eq_matched})~and~(\ref{eq:load_under_el_eq_open_circuit})
are both purely real-valued, whereas one of the electromagnetic expressions
has an $i$ in front. This dichotomy is well known in TL theory, where
a matched line generally looks like a resistor, while a short open-circuited
line generally looks like a capacitor. The electrostatic results do
not capture this dichotomy, which is solidly established in TL theory.
This is what we mean by the ``breakdown'' of electrostatic theory.

However, in the usual case for TL-theory $Y_{0}$ is a real number,
which makes $Y_{in\mathrm{OC}}$ imaginary (a capacitor). For the
ionospheric case we have found, both in this work (panels~g~and~h
of Figure~\ref{fig:wave_analysis_and_phase_rotation}) and in \citeauthor{cosgrove-2016}~{[}2016{]},
that $Y_{0}$ is strongly imaginary, which makes $Y_{in\mathrm{OC}}$
a real number. This makes sense because it allows the electrostatic
and electromagnetic results to agree (with respect to real/imaginary,
anyway) when the system size is small. This is what we expect; if
electrostatic theory is to apply, it should apply when the system
size is small compared to a wavelength. When we go to calculate the
input admittance to the ionosphere, we will essentially be calculating
$Y_{in\mathrm{OC}}$, where $L$ is the thickness of the ionosphere.
$Y_{in\mathrm{OC}}$ being a real number allows for the possibility
of salvaging the usual view that the input admittance to the ionosphere
should be the field line integrated conductivity, a real number.

So how well do the electrostatic and electromagnetic results agree
for $Y_{in\mathrm{OC}}$? Equations~(\ref{eq:input_admittance_OC_electromagnetic})~and~(\ref{eq:load_under_el_eq_open_circuit})
have a different functional dependence on $L$, and cannot agree in
general. The tangent factor in equation~(\ref{eq:input_admittance_OC_electromagnetic})
describes well-known wave effects on admittance: when moving along
an open-circuited transmission line (or as the line gets longer) the
input admittance goes through infinities that alternate in sign, which
occur where the electric field of the reflected wave cancels that
of the incident wave. Since $Y_{in\mathrm{OC}}$ is usually imaginary,
these are infinite reactances that alternate between being capacitive
and being inductive. But for the ionosphere $Y_{in\mathrm{OC}}$ is
a real number. So these would be infinities in the ionospheric conductance,
which could become negative if the electrical thickness ($\phi$)
of the ionosphere becomes greater than $90^{\circ}$. This result
is a harbinger of results from the model developed below. Note that
an infinity in the admittance simply corresponds to a zero in the
electric field, and so is not a cause for alarm. 

The electrostatic and electromagnetic expressions for $Y_{in\mathrm{OC}}$
can be compared in the limiting case where the thickness $L$ of the
plasma slab is much less than a wavelength. Substituting for $\phi=k_{z}L$
in equation~(\ref{eq:input_admittance_OC_electromagnetic}) and letting
the thickness $L$ become small, we obtain $Y_{in}=iY_{0}k_{z}L$.
This expression has the same $L$-dependence as equation~(\ref{eq:load_under_el_eq_open_circuit}),
and so we can compare the $L$-coeficients, $iY_{0}k_{z}$~and~$\sigma_{P0}$.
Making this same comparison, \citeauthor{cosgrove-2016}~{[}2016{]}
suggested defining a wave-Pedersen conductivity (their equation~(14)),
which in our current analysis takes the form,
\begin{equation}
\sigma_{Pwj}=iY_{0j}k_{jz},\label{eq:wave_pedersen_conductivity}
\end{equation}
and we have added subscripts $j$ to denote different wave modes.
In this gedanken experiment we are considering only one mode, but
we find below that there are in fact two important modes. The wave-Pedersen
conductivity is distinct from the local Pedersen conductivity, which,
also as explained in \citeauthor{cosgrove-2016}~{[}2016{]}, can
either be calculated under the assumption that both the real and imaginary
parts of frequency $(\omega)$ are zero ($\sigma_{P0},$ the usual
electrostatic Pedersen conductivity), or using the full complex $\omega$
(what we call the ``full Pedersen conductivity''), where the imaginary
part of $\omega$ was shown to be very important. 

\citeauthor{cosgrove-2016}~{[}2016{]} explained that $\sigma_{Pwj}$
and $\sigma_{P0}$ cannot agree in general, because $\sigma_{Pwj}$
depends on wavelength and $\sigma_{P0}$ does not. The full Pedersen
conductivity does have a wavelength dependence, which arises through
its dependence on $\omega$. It agrees much better with $\sigma_{Pwj}$,
and this shows the importance of the imaginary part of $\omega$ in
the conductivity equations (see Figure~16\footnote{Errata: In the caption for Figure~16 of \citeauthor{cosgrove-2016}~{[}2016{]},
``equation~(1)'' should be ``equation~(A9).''} from \citealp{cosgrove-2016}). Panel~d of Figure~\ref{fig:Electrostatic-versus-TL}
shows the results from an example calculation of $\sigma_{Pwj}$ for
the Alfvén and Whistler waves, using a 1000~km transverse wavelength
and plasma density of $4.7\times10^{9}\mathrm{m}^{-3}$. For comparison,
both the zero-frequency and full Pedersen conductivities are shown,
where in the latter case there is one for each wave type. In the case
of the Alfvén wave, the plot shows fairly good agreement between $\sigma_{Pwj}$
and the full Pedersen conductivity. In the case of the Whistler wave,
the plot shows agreement between $\sigma_{P0}$ and the full Pedersen
conductivity in the low altitude region. This is because the Whistler
wave has a long dissipation-time-scale in the low altitude region
(i.e., imaginary part of frequency is small). But $\sigma_{Pwj}$
does not agree with $\sigma_{P0}$ for either wave, and this is the
result that is relevant to our discussion.

The real/imaginary inconsistency noted above can be avoided if the
electrical length $\phi$ is strongly imaginary. In the usual case
for TL theory the dissipation scale length is much longer than the
wavelength, and is ignored. But if it were not ignorable, the electrical
length would acquire an imaginary component, $\phi=k_{z}L=2\pi L/\lambda_{z}-iL/l_{dz}$,
where $l_{dz}$ is the dissipation scale length and $\lambda_{z}$
is the wavelength. Note that including this loss term ``cures''
the infinities noted earlier, because the reflected wave can never
completely cancel the electric field of the incident wave. If it were
to be the case that both $l_{dz}$ is much less than $\lambda_{z}$,
and that the characteristic admittance is strongly real, then both
$Y_{in\mathrm{OC}}$ and $Y_{in\mathrm{Matched}}$ would be strongly
real, as predicted by electrostatic theory. Physically, this would
be the case where the ``waves'' are so lossy that they do not really
behave like waves. This special situation essentially allows the system
size requirements to be removed.

\textbf{Criteria for Electrostatic Theory:} In summary, we have shown
that when electromagnetic theory can be approximated using a single
wave mode, then electrostatic theory is also applicable when the following
three conditions hold: (1) the wave-Pedersen conductivity ($\sigma_{Pwj}$)
agrees with the zero-frequency Pedersen conductivity ($\sigma_{P0}$);
(2) the wavelength is long compared to the system size; and (3) the
dissipation scale length is long compared to the system size. There
is also a special case where if the dissipation scale length is much
less than the wavelength, and is also equal to the Farley mapping
distance, then requirements~(2)~and~(3) may be replaced with the
requirement that the dissipation scale length is short compared to
the system size (which essentially removes the system size requirements). 

However, both from the results below and from the results in \citeauthor{cosgrove-2016}~{[}2016{]},
we find that these conditions do not hold; the wave-Pedersen conductivity
does not equal the zero-frequency Pedersen conductivity, the dissipation
scale length is much longer than the wavelength (for both modes),
and the wavelength is not sufficiently long compared to the ionospheric
thickness. And when multiple modes are present there are additional
unlikely requirements that must be added, which we touch on briefly
in Section~\ref{subsec:Artificial-Examples}. Fortunately, the characteristic
admittance is strongly imaginary, and so at least the ionospheric
input admittance is strongly real.

In Section~S.3
of the Supplementary Information
we provide some additional commentary on energy conservation in this
unusual situation. In sorting all this out it is important to observe
that wave dissipation is not a critical or necessary pathway for energy
absorbed by the system. Rather, the critical pathway is energy carried
away by waves, converted to/from some other form (like kinetic energy
{[}\citealp{vasyliunas+song-2005}; \citealp{cosgrove-2016}{]}), or dissipated
by nonlinear effects. If the wave amplitudes are sufficiently small,
the nonlinear effects may be ignored in their computation. As long
as the wave amplitudes are accurate, the Poynting flux computed from
the superposed waves is valid, and the heating may also be calculated.
And when there \emph{is} significant wave dissipation the signal is
absorbed before reaching the bottom of the ionosphere, which means,
among other things, that the electric field does not map through the
ionosphere.

The discussion above is essentially a restatement of results from
\citeauthor{cosgrove-2016}~{[}2016{]}, which although largely denying
electrostatic theory in the ionosphere, did not provide a practical
replacement. In this work we put the ideas into action and perform
an electromagnetic calculation that forms the basis for a deployable
electromagnetic model, which can replace electrostatic theory in many
cases, and which provides for a significant amount of physical intuition. 

\section{Solution of the Equations of Motion, and Model Building Blocks\label{sec:Solution}}

We adopt the electromagnetic 5-moment fluid equations {[}e.g., \citealp{schunk+nagy-2009}{]}
for description of collisional plasma, such as the ionosphere. These
well known equations of motion are presented in Appendix~A~(\ref{eq:generalIono}),
along with the derivation of the matrix $H_{5}$~(Figure~\ref{fig:H_Matrix})
used below, which contains their linear parts. In addition to the
Maxwell equations, the equations of motion~(\ref{eq:generalIono})
include the continuity, momentum, and energy equations for electrons
and one species of ion. Sources such as wind, background electric
field, photo-electrons, and gravity are set to zero or omitted, except
in the case of the explicit external source we add to derive the driven
steady-state solution. Hence, the equations of motion expand on those
used in \citeauthor{cosgrove-2016}~{[}2016{]} by adding the continuity
and energy equations. They are appropriate for exploring the relationship
between electrostatic and electromagnetic theories of the ionosphere
for scale-sizes greater than a few tens of meters, but they do not
include kinetic effects, the dynamics of the neutral gas, or interactions
between different species of ion (although the latter two categories
could, in principle, be added to the analysis with little change in
method). Additional details can be found in Appendix A. Here we focus
on deriving the linearized, driven steady-state solution using an
abstract matrix notation. The methods employed are standard textbook
methods, and our contribution is only to apply them to this problem.

Taking the Fourier transform in space, the equations of motion~(\ref{eq:generalIono})
can be written,
\begin{equation}
\frac{\partial\vec{X}}{\partial t}-iH_{5}\vec{X}=\vec{F}(t),\label{eq:matrixEqnWithNonlin}
\end{equation}
where $\vec{X}$ is a vector containing the deviations from thermal
equilibrium of the 16 dependent variables, $H_{5}$ is a time-independent
$16\times16$ matrix containing the linear parts of the equations,
and $\vec{F}(t)$ is a time-dependent length-16 vector containing
the nonlinear terms. The variables in $\vec{X}$ are defined in Appendix
A, and consist of scaled versions of the densities, velocities, pressures,
and electric and magnetic fields. The matrix $H_{5}$ is given in
Figure~\ref{fig:H_Matrix}, and to be clear, is derived without making
approximations (details are in Appendix A). The non-linear term $\vec{F}$
is not given, since we do not use it here.

\textbf{The Intuitive Derivation of the Steady-State Solution:} Before
giving a (more) rigorous derivation of the model equations, we first
motivate the results with a simple intuitive derivation based on the
initial value solution, which makes contact with the discussion in
\citeauthor{cosgrove-2016}~{[}2016{]}. We also remind the reader
of the reasoning that makes it appropriate to ignore nonlinear terms.

To better understand the role of the nonlinear terms, consider diagonalizing
the matrix $H_{5}$ as $\Omega=U^{-1}H_{5}U$, where $U$ is the matrix
having the eigenvectors of $H_{5}$ as columns, and $\Omega$ is diagonal.
By applying $U^{-1}$ from the left hand side, the equations of motion~(\ref{eq:matrixEqnWithNonlin})
breakup into individual scalar equations for each eigenmode, coupled
only by nonlinear ``source terms'' on the right hand side:
\begin{equation}
\frac{\partial X_{j}^{\prime}}{\partial t}-i\omega_{j}X_{j}^{\prime}=F_{j}^{\prime}(t),\label{eq:diagonalMatrixEqnWithNonlin}
\end{equation}
where $\vec{X}^{\prime}=U^{-1}\vec{X}$, $\vec{F}^{\prime}=U^{-1}\vec{F}$,
$\omega_{j}=\Omega_{jj}$, and $j$ indexes the eigenmodes. Because
the electromagnetic 5-moment fluid equations comprise 16 scalar equations,
there are 16 eigenmodes, and equation~(\ref{eq:diagonalMatrixEqnWithNonlin})
shows that the nonlinear terms have the effect of coupling energy
from one mode to another (nonlinear leakage). As the size of the disturbance
is reduced, there is (at least) a linear reduction of $F_{j}^{\prime}$
relative to $\frac{\partial X_{j}^{\prime}}{\partial t}$ and $\omega_{j}X_{j}^{\prime}$,
and the rate of nonlinear leakage between modes is reduced. So for
a sufficiently small disturbance the rate of nonlinear leakage is
negligible, the eigenmodes evolve independently, and the general solution
can be obtained. (As further described in Appendix A, we mean a small
disturbance with respect to thermal equilibrium, and electrostatic
theory is also formulated in this way.)

Therefore, consider the general homogeneous solution (in the language
of differential equations) to equation~(\ref{eq:matrixEqnWithNonlin}),
that is, the solution when the nonlinear ``source'' term, $\vec{F}$,
is neglected, which is given by,\footnote{The general homogeneous solution (i.e., with $\vec{F}=0$) to equation~\ref{eq:matrixEqnWithNonlin}
is given by the matrix exponential, $\mathrm{e}^{iH_{5}t}$ {[}e.g.,
\citealp{Artin-1991}{]}. However, as a practical matter, we find that
$H_{5}$ is diagonalizable ``almost everywhere,'' and this allows
us to write the solution in a more explicit form, by taking advantage
of the invertibility of the matrix of eigenvectors, $U$.}
\begin{eqnarray}
\vec{X}\left(t,\vec{k}\right) & = & \sum_{j=1}^{16}a_{0j}(\vec{k})\,\vec{h}_{j}(\vec{k})\,\mathrm{e}^{i\omega_{j}(\vec{k})t},\;\mathrm{or,\,taking\,the\,inverse\,Fourier\,transform,}\nonumber \\
\vec{X}\left(t,\vec{r}\right) & = & \sum_{j=1}^{16}\int\mathrm{d}^{3}k\:a_{0j}(\vec{k})\,\vec{h}_{j}(\vec{k})\,\mathrm{e}^{i\left(\omega_{j}(\vec{k})t+\vec{k}\cdot\vec{r}\right)},\label{eq:solutionHomogenious}
\end{eqnarray}
where \textcolor{black}{$\left\{ \vec{h}_{j}\right\} $ and $\left\{ \omega_{j}\right\} $
are the eigenvectors and eigenvalues of $H_{5}$, respectively, $\vec{k}$
is the Fourier transform variable (the wavevector), and the $\left\{ a_{0j}\right\} $
are 16 arbitrary coefficients for each $\vec{k}$. The solution is
easily verified by direct substitution. }Since $H_{5}$ is diagonalizable
its eigenvectors are linearly independent, and so the solution~(\ref{eq:solutionHomogenious})
is the complete initial-value solution\textcolor{black}{; an arbitrary
initial state can be chosen by choosing the 16 $a_{0j}(\vec{k})$,
and then evolution consists of 16 independently evolving eigenmodes.}

Equation~(\ref{eq:solutionHomogenious}) shows that the eigenvalues
of $H_{5}$ (i.e., the $\omega_{j}$) play the role of complex frequency.
Since equation~(\ref{eq:matrixEqnWithNonlin}) is the Fourier transform
of real-valued equations of motion, the eigenvalues of $iH_{5}$ come
in complex conjugate pairs. This means that when the real part of
frequency $\left(\omega_{jr}=\mathrm{real}(\omega_{j})\right)$ is
not zero there are paired two oppositely propagating eigenmodes (the
complex conjugate negates $\omega_{jr}$, reversing the direction
of propagation), and hence there are 8 potential types of propagating
eigenmode. For the smaller equation-set used in \citeauthor{cosgrove-2016}~{[}2016{]}
there were 6 potential types of propagating eigenmode. However, it
is possible for $\omega_{jr}$ to be zero, for some $j$ and range
of $\vec{k}$, and in this range the eigenmode is evanescent; it is
unpaired and does not effectively transmit energy, since the real
part of its group velocity is zero. 

To envision the source-free initial-value solution~(\ref{eq:solutionHomogenious})
we might imagine a narrow-band antenna with transmitter that turns
on, transmits for a few cycles, and then turns off. To the extent
the stimulus matches the natural energy-carrying modes of the plasma,
energy will be transferred into the plasma, and carried away. To the
extent it does not energy will not be able to transition into the
plasma, only non-propagating modes will be excited, and the transmitter
will feedback on itself in an effect analogous to a radio-frequency
antenna that is not well matched to the free space impedance. By ``non-propagating
modes'' we mean terms in equation~(\ref{eq:solutionHomogenious})
where the frequency and transverse wavelength of the antenna cannot
be matched by the dispersion relation $\omega_{j}(\vec{k})$, and
it is intended to imagine an unusually-large antenna that extends
many wavelengths in the transverse direction, so that it will be transverse-wavelength
selective.

After the transmitter turns off, and at any significant distance,
there will remain a localized disturbance consisting mainly of the
modes that are able to propagate at the transmitter's frequency and
transverse wavelength. Finding the \textcolor{black}{$\left\{ a_{0j}\right\} $
that fits }the initial-value-solution to the disturbance, the integrals
in~(\ref{eq:solutionHomogenious}) should produce a superposition
of propagating wave-packets centered on wavevectors $\left\{ \vec{k}_{0j}\right\} $
such that $\mathrm{real}\left(\omega_{j}(\vec{k}_{0j})\right)$ matches
the transmitter frequency, $\omega_{0}$, where the transverse part
of each $\vec{k}_{0j}$ matches the antenna wavelength. 

Thus, using the standard interpretation of wave-packet propagation,
we might imagine the subsequent evolution as a superposition of a
few independently evolving wave packets centered on wavevectors $\vec{k}_{0j}$,
propagating with group velocities $\vec{v}_{gjr}=-\left.\mathrm{real}\left(\vec{\nabla}_{k}\omega_{j}\right)\right|_{\vec{k}_{0j}}$,
dissipating with time scales $\omega_{ji}^{-1}=1/\mathrm{imag}\left(\omega_{j}(\vec{k}_{0j})\right)$,
oscillating with frequencies $\mathrm{real}\left(\omega_{j}(\vec{k}_{0j})\right)=\omega_{0}$,
and disturbing the plasma with polarization vectors $\vec{h}_{j}(\vec{k}_{0j})$\textcolor{black}{.
If the transmitter did this repeatedly, that is, if it did not turn
off, there would result a steady-state disturbance around the source
that decreases with distance according to the dissipation scale length
$\vec{v}_{gjr}/\omega_{ji}$ (for each mode). This behavior amounts
to, essentially, the steady-state solutions~(\ref{eq:drivenSteadyStateSolution})~and}~(\ref{eq:solutionFromQuadratic})\textcolor{black}{{}
that we now derive more rigorously, where the second of these applies
to eigenmodes that become cutoff below some frequency (i.e., become
non-propagating, i.e., not able to satisfy }$\mathrm{real}\left(\omega_{j}(\vec{k}_{0j})\right)=\omega_{0}$\textcolor{black}{).}

\textbf{Validating the Steady-State Solution:} The TL theory is based
on the driven steady-state solution associated with a distant source
transmitting into an infinite (and homogeneous) medium. Although the
intuitive description just given seems almost adequate, there are
in fact a number of assumptions that go into the wave packet interpretation,
and so we would like to test it with a more rigorous derivation. We
do this now, and for the most part find that the usual wave-packet
interpretation appears to be a good one for our purposes. However,
we do find some uncertainty with respect to the lower $E$-region
ionosphere ($\sim100$~km altitude), where our method of validation
is not fully applicable. This difficult problem is deferred to future
work, and hence we regard our model as an important baseline associated
with the usual wave packet interpretation, which is almost but not
completely validated down to 100~km in altitude. In Section~\ref{sec:Model-Results}
we describe some sensitivity tests that show little effect from changes
near the bottom of the modeling domain, and these reinforce the efforts
at validation just below.

Therefore, consider the source,
\begin{equation}
\vec{f}_{A}\left(t,\vec{r}\right)=\vec{A}u(t)\mathrm{e}^{i\omega_{0}t}\delta(z)\mathrm{e}^{ik_{0y}y},\label{eq:source}
\end{equation}
which turns on at $t=0$ ($u$ is the unit step function) and is localized
in the $z$ dimension, being non-zero only at $z=0$ ($\delta$ is
the Dirac delta function, and we assume that the geomagnetic field
is along~$z$). Taking the Laplace transform (with respect to time)
of the left hand side of the general equation~(\ref{eq:matrixEqnWithNonlin}),
and setting it equal to the Laplace and Fourier Transforms of the
source~(\ref{eq:source}), gives the following equation for the plasma
response,
\begin{equation}
s\vec{X}\left(s,\vec{k}\right)-iH_{5}\vec{X}\left(s,\vec{k}\right)=\vec{X}(t=0^{-})+4\pi^{2}\frac{\vec{A}}{s-i\omega_{0}}\delta\left(k_{x}\right)\delta\left(k_{y}-k_{0y}\right),\label{eq:LaplaceTransformEqn}
\end{equation}
where $s$ is the Laplace transform variable. This equation is solved
for $\vec{X}\left(s,\vec{k}\right)$ by inverting the matrix $sI-iH_{5}=U\left(sI-i\Omega\right)U^{-1}$,
where $sI-i\Omega$ is diagonal. Then taking the inverse Laplace and
Fourier transforms, and assuming zero for the initial condition $\left(\vec{X}(t=0^{-})=0\right)$,
we find the integral-form solution,
\begin{eqnarray}
\vec{X}\left(t,\vec{r}\right) & = & \mathrm{e}^{i(\omega_{0}t+k_{0y}y)}\frac{i}{2\pi}\sum_{j=1}^{16}\varint_{-\infty}^{\infty}\mathrm{d}k_{z}\frac{\vec{h}_{j}(0,k_{0y},k_{z})a_{j}(0,k_{0y},k_{z})}{\omega_{j}(0,k_{0y},k_{z})-\omega_{0}}\mathrm{e}^{ik_{z}z}\nonumber \\
 & - & \mathrm{e}^{ik_{0y}y}\frac{i}{2\pi}\sum_{j=1}^{16}\varint_{-\infty}^{\infty}\mathrm{d}k_{z}\frac{\vec{h}_{j}(0,k_{0y},k_{z})a_{j}(0,k_{0y},k_{z})}{\omega_{j}(0,k_{0y},k_{z})-\omega_{0}}\mathrm{e}^{i\left(\omega_{j}(0,k_{0y},k_{z})t+k_{z}z\right)},\label{eq:laplaceIntegralFormSolution}
\end{eqnarray}
where $\vec{a}=U^{-1}\vec{A}$. Because $\mathrm{imag}\left(\omega_{j}\right)>0$,
the second sum of integrals represents the transient response, which
goes to zero as $t\rightarrow\infty$. Therefore, the driven steady-state
solution is a sum over separate contributions from each eigenmode,
according to the source vector $\vec{A},$ which are found by evaluating
the integrals in the first sum of equation~(\ref{eq:laplaceIntegralFormSolution}).
We consider first an approximate analytical solution, and then test
it by numerical evaluation of the integrals for a few select cases.
We will consider the case with $\omega_{0}>0$, which, comparing~(\ref{eq:laplaceIntegralFormSolution})
with the exact initial value solution~(\ref{eq:solutionHomogenious}),
means that there can be no contribution from modes with $\mathrm{real}(\omega_{j})\leqq0$,
and so we have at most 8 modes to consider.

The integrals in~(\ref{eq:laplaceIntegralFormSolution}) involve
eigenvalues, $\omega_{j}$, and eigenvectors, $\vec{h}_{j}$, and
it is well known that such quantities are not analytic over the entire
complex plane. For example, eigenvalues are found as roots of a polynomial,
and so expressions for them contain square and higher roots that are
not analytic over the entire complex plane; analyticity fails where
the argument of a square root vanishes, which is where two eigenvalues
coincide {[}e.g., \citealp{Tsing-et-al-1994}{]}. So while it is tempting
to consider analytic continuation off of the real-$k_{z}$ axis, and
use the residue theorem for the integrals, it will not do to simply
take a residue where $\omega_{j}=\omega_{0}$. The dependence of $\omega_{j}$
on $k_{z}$ is not analytic, and so the residue theorem does not apply
(for example, see Figure~8 from \citeauthor{cosgrove-2016}~{[}2016{]},
which demonstrates non-analytic behavior). However, if we approximate
the integrand along the real axis by a function that \emph{is} analytic
over the entire complex plane, then the residue theorem can be made
to apply.

Because the integrand has denominator $\omega_{j}(0,k_{0y},k_{z})-\omega_{0}$,
it should be peaked around the $k_{z}$ having $\mathrm{real}\left(\omega_{j}(0,k_{0y},k_{z})\right)=\omega_{0}$.
Furthermore, the initial value solution~(\ref{eq:solutionHomogenious})
suggests that a plasma disturbance centered on wavenumber $k_{z}$
should oscillate with frequency $\mathrm{real}\left(\omega_{j}(0,k_{0y},k_{z})\right)$,
and the driven steady-state solution~(\ref{eq:laplaceIntegralFormSolution})
oscillates purely at $\omega_{0}$. Therefore, consider approximating
the integrand by expanding around $k_{z}=k_{0jz}$ such that $\mathrm{real}\left(\omega_{j}(0,k_{0y},k_{0jz})\right)=\omega_{0}$.
If we were to expand the denominator in a Taylor series and retain
only the linear term, it would produce a simple pole at $k_{z}=k_{0jz}+i\omega_{ji}/v_{gj}$,
where $\omega_{ji}=\left.\mathrm{imag}(\omega_{j})\right|_{k_{0z}}$,
and $v_{gj}=-\left.\partial\omega_{j}/\partial k_{z}\right|_{k_{0jz}}$.
Therefore, consider approximating the integrand $\Gamma$ by writing
$\Gamma=\Psi\mathrm{e}^{ik_{z}z}/(k_{z}-k_{0jz}-i\omega_{ji}/v_{gj})$
and seeking a linear approximation for $\Psi$ in the vicinity of
$k_{0jz}$. The factor $(k_{z}-k_{0jz}-i\omega_{ji}/v_{gj})/(\omega_{j}(0,k_{0y},k_{z})-\omega_{0})$
is stationary at $k_{z}=k_{0jz}$ (with value $-1/v_{gj}$), and so
the only first order variation of $\Psi$ comes from the factor $\vec{h}_{j}(0,k_{0y},k_{z})a_{j}(0,k_{0y},k_{z})$.
Approximating $\vec{h}_{j}(0,k_{0y},k_{z})a_{j}(0,k_{0y},k_{z})$
by an analytic function gives the following approximation for the
$j$th-eigenmode integral,
\begin{equation}
\varint_{-\infty}^{\infty}\mathrm{d}k_{z}\frac{\vec{h}_{j}^{\prime}(k_{z}-k_{0jz})a_{j}^{\prime}(k_{z}-k_{0jz})}{i\omega_{ji}-(k_{z}-k_{0jz})v_{gj}}\mathrm{e}^{ik_{z}z},\label{eq:integralWithLinearizations}
\end{equation}
where $\vec{h}_{j}^{\prime}\left(k_{z}-k_{0jz}\right)=\Theta\left(k_{z}-k_{0jz}\right)\vec{h}_{j}(0,k_{0y},k_{0jz})$,
$\vec{a}^{\prime}\left(k_{z}-k_{0jz}\right)=U^{-1}\left(k_{0jz}\right)\Theta^{-1}\left(k_{z}-k_{0jz}\right)\vec{A}$,
and $\Theta\left(k_{z}-k_{0jz}\right)$ is a parameterization of the
16-dimensional special unitary transformations $\left(SU(16)\right)$
about the identity, with the $k_{z}$ dependence of the angle linearized,
which is described more fully in Section~S.4
of the Supplementary Information.
The introduction of $\Theta$ is really just a formality since the
validations below provide that we need never actually evaluate $\Theta$.
What matters here is just that we can approximate $\vec{h}_{j}(0,k_{0y},k_{z})a_{j}(0,k_{0y},k_{z})$
by some analytic function, and so we can apply contour integration,
where the issue of the boundary term is discussed in Section~S.4
of the Supplementary Information.
Therefore, evaluating the residue at $k_{z}=k_{0jz}+i\omega_{ji}/v_{gj}$
and closing the contour in the lower-half-plane for $z<0$, and in
the upper-half-plane for $z>0$, gives the following result for the
contribution of the $j$th eigenmode to the driven steady-state solution,
\begin{equation}
\vec{X}_{j}(t,\vec{r})=\pm\frac{1}{v_{gj}}\vec{h}_{j}^{\prime}\left(i\omega_{ji}/v_{gj}\right)a_{j}^{\prime}\left(i\omega_{ji}/v_{gj}\right)\mathrm{e}^{i\left[\omega_{0}t+k_{0y}y+k_{0jz}z+iz\omega_{ji}/v_{gj}\right]},\label{eq:drivenSteadyStateSolution}
\end{equation}
where the plus sign applies when $z>0$, the minus sign applies when
$z<0$, $\vec{k}_{0j}=\left[0,k_{0y},k_{0jz}\right]$, and we note
that $\vec{h}_{j}^{\prime}(0)=\left.\vec{h}_{j}\right|_{\vec{k}=\vec{k}_{0j}}$
and $\vec{a}^{\prime}(0)=\left.a_{j}\right|_{\vec{k}=\vec{k}_{0j}}$,
since $\Theta(0)=I$. For $z>0$, $k_{0jz}$ is the negative value
of $k_{z}$ such that $\omega_{jr}(k_{z})=\omega_{0}$, and for $z<0$,
$k_{0jz}$ is the positive value of $k_{z}$ such that $\omega_{jr}(k_{z})=\omega_{0}$.
Note that the real part of the group velocity $(v_{gjr})$ changes
sign with $k_{z}$, and so there is exponential decay in both the
positive and negative directions. Hence the solution~(\ref{eq:drivenSteadyStateSolution})
includes both negative-going and positive-going waves, which can be
separately extracted for use in the TL model.

Using the term ``mode'' to refer to the contribution from a particular
eigenmode, that is, to one of the 16 terms in~(\ref{eq:laplaceIntegralFormSolution}),
the linear approximation provides that each propagating mode is approximated
by a single decaying wave (which may propagate in either direction),
such as for the wave-packet interpretation. It would also be possible
to use a higher-order polynomial approximation, in which case the
denominator would have multiple zeros, and the contour integration
would involve multiple residues. In this case each mode would be approximated
by a sum of complementary waves. Although there do not appear to be
any critical obstacles to a higher order implementation, using the
single-wave approximations makes the bi-modal model much easier to
interpret, and less complex to implement. An increase in order requires
calculation of higher order derivatives, which is a more difficult
numerical problem. Therefore, in this work we will limit our consideration
to the linear approximation, that is, to single-wave representations
of the modes. The adequacy of the linear approximation is the major
criteria that must be satisfied for the wave-packet interpretation
to be valid. 

In the case where the linear approximation is known to be accurate
there are also some additional criteria. The displacement from $k_{0jz}$
to the location for the residue contains real and imaginary parts,

\begin{eqnarray}
k_{jz} & = & k_{0jz}+i\omega_{ji}/v_{gj}\nonumber \\
 & = & k_{0jz}\pm v_{gji}/(l_{djz}v_{gjr})\pm i/l_{djz},\quad\mathrm{where},\nonumber \\
l_{djz} & = & \left|v_{gj}\right|^{2}/\left|\omega_{ji}v_{gjr}\right|,\label{eq:dissipation_scale_length}
\end{eqnarray}
and the subscripts $r$ and $i$ denote the real and imaginary parts,
respectively. The imaginary part of the displacement, $1/l_{djz}$,
provides the lowest order correction to infinite dissipation scale
length. But the real part of the displacement, $v_{gji}/(l_{djz}v_{gjr})$,
is a correction to $k_{0jz}$, and if it is not negligible the $z$-directed
wavelength $(\lambda_{jz})$ in the solution~(\ref{eq:drivenSteadyStateSolution})
could be different from $2\pi/k_{0jz}$. Although this still provides
for a single-wave representation, it constitutes a correction to the
usual wavepacket interpretation. In addition, $l_{djz}$ in equation~(\ref{eq:dissipation_scale_length})
does not match the expression from the usual wavepacket interpretation,
$\left|v_{gjr}/\omega_{ji}\right|$, unless $v_{gjr}^{2}\gg v_{gji}^{2}$.
And finally, there are corrections to the polarization vectors such
that they are not exactly the eigenvectors. However, since we cannot
be certain of the linear approximation, we really need an independent
evaluation both of it and of these potential corrections to the simple
wave packet interpretation. 

What we actually want to know is whether an expression of the form~(\ref{eq:drivenSteadyStateSolution})
provides a reasonably accurate representation of the steady state
solution, for some wavelength $(\lambda_{jz})$, some dissipation
scale length $(l_{djz})$, and some polarization vector $(\vec{h}_{j}^{\prime})$.
And, if so, we want values for these three parameters. The only real
way to solve this problem is to evaluate the exact integral-form solution
(\ref{eq:laplaceIntegralFormSolution}) numerically. Hence, we undertake
to evaluate the steady state solution exactly for a few select cases,
using numerical integration.

The first step in doing this is to discuss the issue of the near zone
of the source~(\ref{eq:source}), and how it relates to TL theory.
The source~(\ref{eq:source}), being an impulse-function in $z$,
excites all wavelengths equally. Hence it has a near zone consisting
of contributions to the integrals away from $k_{z}$ such that $\mathrm{real}\left(\omega_{j}(0,k_{0y},k_{z})\right)=\omega_{0}$,
and this near-zone would be different if the source had a different
$z$-dependence. This brings up an important distinction between the
meaning of the term ``source'' in circuit theory (including TL theory),
and the application of that term to the ``source''~(\ref{eq:source}).
TL theory does not actually seek to characterize the response of the
system to the particular source~(\ref{eq:source}). Rather, TL theory
seeks to characterize the generic behavior of the system when it is
placed in an arbitrary circuit involving other generic circuit elements,
for example a generic family of sources. While the near-zone will
depend very specifically on the exact source excitation, the radiation-zone
should depend only on the frequency $(\omega_{0})$, transverse wavelength
$(k_{0y})$, amplitude $(\vec{A})$, and propagation medium in which
the waves from the source must propagate (and in which any reflected
waves must propagate). Thus TL theory limits consideration to the
radiation zone and thereby achieves a generic or ``universal'' characterization
of the system.

The expedient of assuming the excitation is sufficiently far away
may be placed in the same category as linearization, in that they
both allow for condensing-out the universal aspects of the system
behavior, which will apply under many different circumstances. This
is the basic formula for circuit theory, which has been found to have
great practical value; it is very often true that the signal amplitude
is sufficiently small and the excitation sufficiently far away. Taking
the circuit-theory concepts of admittance and conductance into ionospheric
science means eschewing the near-zone in order to achieve a universal
description of the ionospheric response. Thus, after carrying out
each of the numerical integrations in~(\ref{eq:laplaceIntegralFormSolution}),
we must identify the radiation-zone of the source.

To evaluate the integrals in the exact solution we simply discretize
the kernel of the inverse Fourier transform in~(\ref{eq:laplaceIntegralFormSolution})
and apply the inverse discrete Fourier transform. We take this simple
approach because it makes it easy to understand the parameters that
govern accuracy. We require that the spectrum be sampled finely enough
that the spatial-domain response remains valid beyond the near zone
and beyond the expected dissipation scale length, to avoid aliasing.
We also require that the spectrum is sampled widely enough that there
is no corruption from windowing, where the inverse Fourier transform
of the window is convolved with the desired spatial-domain response.
As part of the evaluation we test our sampling on the Fourier transform
of the approximate solution~(\ref{eq:drivenSteadyStateSolution}),
which is one of the standard forms having an analytical solution.
We consider the ability to reproduce the form~(\ref{eq:drivenSteadyStateSolution})
as a minimum requirement for the sampling, and then we look for any
other evidence that the criteria just mentioned may not be satisfied.
When satisfied with the result we identify the radiation zone visually
and fit the functional form~(\ref{eq:drivenSteadyStateSolution})
to a section a couple of wavelengths long. The fit region is usually
about 10 to 20 wavelengths from the source, except in the lower \emph{E}-region
where it must be much closer, due to the dissipation scale length
being (possibly) less than the wavelength in some cases. 

An extended discussion of the validation results is provided in Section~S.5
of the Supplementary Information,
where three figures are presented. Here we wish just to assert that
we feel the usual wave-packet interpretation has been validated, at
least in the sense of providing for an important baseline model that
may be obtained from the eigenmodes, and which should provide at least
a qualitatively accurate description of the ionospheric behavior.
The wavelength comparisons all show very good agreement, and with
the exception of the Alfvén mode in the lower \emph{E} region, the
polarization vectors also agree very well with the eigenvectors. The
dissipation scale lengths do not match so well, but they are never
found to be shortened to such a degree as to have an impact (recall
from Section~\ref{sec:Gedanken-Experiment} that once the dissipation
scale is sufficiently long, making it longer does not produce any
change). In fact, the dissipation scale found from the group velocity
can itself become so short as to have an effect on the Alfvén mode
at the bottom of the \emph{E} region, and the numerical integrations
suggest that perhaps they are in fact longer than this, so as to eliminate
the effect. However, all in all the numerical integrations serve to
reinforce confidence in the usual wave-packet interpretation, with
meaningful differences arising only in cases where there are good
reasons not to trust the results.

The differences that do arise arise almost exclusively in the lower
\emph{E} region, and involve one of two things: either the sampling
required for the Whistler mode is impractical using the computing
power available, or the dissipation scale for the Alfvén mode becomes
very short, and may not extend past the near-zone of the source. While
the first problem is solvable in the future, the second problem creates
an ambiguous situation that may be unresolvable; we do not have a
recipe for removing near-zone effects. If the source produces a near
zone that extends past the radiation zone, then it is not useful for
testing our hypothesis. But this does not mean that the eigenmodes
do not well characterize the plasma response, under the wave-packet
interpretation. Rather, it seems likely that this is just a difficulty
arising from the impulse-function feature of the source~(\ref{eq:source}),
which excites all wavelengths equally, and therefore is not realistic.
We generally expect that radiation-zone waves will preferentially
excite other radiation-zone waves, due to their inherent similarity.
Thus, it seems likely that our method of validation is simply not
applicable to the lower \emph{E} region, and that the eigenmodes are
still the appropriate description of the plasma response, just as
they are at higher altitudes.

Fortunately, testing shows that corrections made in the lower \emph{E}-region
do not produce a substantial difference in the results for a signal
incident from above. Not very much energy makes it to the lower \emph{E}
region, as will be seen below. Thus we feel it most appropriate to
implement the eigenmode-based model in it's pure form, using the usual
wave-packet interpretation. This model is a baseline electromagnetic
model that is the most natural extension of the current electrostatic
baseline. 

\textbf{Handling the Cutoff Modes:} The solution~(\ref{eq:drivenSteadyStateSolution})
does not apply for eigenmodes where there does not exist a $k_{z}$
such that $\mathrm{real}\left(\omega_{j}(0,k_{0y},k_{z})\right)=\omega_{0}$.
In the case of ionospheric science, where $\omega_{0}$ is generally
associated either with a wind-driven dynamo or a magnetospheric source
moving near the convection velocity, the so-called ``light wave''
or ``radio frequency'' modes are much too high in frequency to satisfy
this matching criterion. Inability to satisfy the matching criterion
means that these modes do not propagate at the source's frequency
and transverse wavelength, and so should not be effective at transmitting
energy away from the source. This is why these modes, known to be
electromagnetic, are always excluded from analysis of ionospheric
and magnetospheric dynamics through some form of low frequency approximation
{[}e.g., \citealp{maltsev+etal-1977}; \citealp{mallinckrodt+carlson-1978};
\citealp{goertz+boswell-1979}; \citealp{seyler-1990}; \citealp{streltsov+lotko-2003b};
\citealp{otto+zhu-2003}; \citealp{dao+etal-2013}; \citealp{lysak+etal-2013};
\citealp{tu+song-2019}; and many others{]}. Removing the three radio-frequency
modes leaves us with five modes to consider for ionospheric applications.

The eigen-decomposition of $H_{5}$ also includes two purely-evanescent
eigenmodes having $\mathrm{real}\left(\omega_{j}\right)=0$ (to numerical
precision). Since these eigenmodes cannot satisfy the matching criterion
and in fact have group velocity of zero, they also cannot be associated
with energy transmission. Hence we are now down to four modes that
we might possibly need to consider for ionospheric science. 

Although the radio-frequency modes are cutoff and so should have only
a parasitic effect, there is one important higher-frequency mode that
can propagate in the lower altitude regions of the ionosphere, but
which becomes cutoff above a certain altitude. This is the mode that
we refer to as the Whistler mode. We need a mathematical treatment
for the cutoff of the Whistler mode so that it can be included in
the model. Therefore, in Appendix~B we derive an additional, quadratic
integral approximation applicable to the higher frequency modes, which
models both the propagation and low-frequency cutoff of these modes.
The resulting solution~(\ref{eq:solutionFromQuadratic}) supports
the assessment that cutoff modes will not contribute significantly
to the transmission of energy; the dissipation scale length decreases
dramatically across the $k_{0y}$ threshold where $\mathrm{real}\left(\omega_{j}(0,k_{0y},k_{z})\right)=\omega_{0}$
cannot be satisfied for any $k_{z}$. Away from cutoff, the solution~(\ref{eq:solutionFromQuadratic})
reduces to the solution~(\ref{eq:drivenSteadyStateSolution}), and
so the validation just described applies to both of these approximate
solutions. 

One other of the four ionospheric-frequency modes can also become
cutoff, except with the cutoff occurring at high frequencies, instead
of low. However, even when this mode does propagate (i.e., satisfy
$\mathrm{real}\left(\omega_{j}(0,k_{0y},k_{z})\right)=\omega_{0}$),
it has a very short dissipation scale length (see Sections~\ref{subsec:Basic-Properties-of-modes}~and~\ref{sec:ElectricFieldMapping}).
So this mode is also not relevant to the transmission of energy through
the ionosphere. Also, it disappears when the energy equations are
omitted from the analysis (e.g., the analysis in \citeauthor{cosgrove-2016}~{[}2016{]}),
and the energy equations should not be required to derive electrostatic
theory. So although a solution very similar to~(\ref{eq:solutionFromQuadratic})
should be applicable, we have not yet derived a solution for this
mode, which we call the Thermal mode.

The remaining two ionospheric-frequency modes do not seem to be susceptible
to cutoff, and so our original solution~(\ref{eq:drivenSteadyStateSolution})
is applicable. We call these modes the Alfvén and Ion modes. (See
Section~\ref{sec:Wave-Modes-Relevant-to-Ionospheric-Science} for
a discussion of the modes and their naming.)

What is important for this work is that there do exist relatively
low-loss modes for which one of the solutions~(\ref{eq:drivenSteadyStateSolution})~or~(\ref{eq:solutionFromQuadratic})
should be a valid approximation, and these modes are found (below)
to dominate energy transmission through the ionosphere. In Section~\ref{subsec:Basic-Properties-of-modes}
we show that electrostatic waves cannot be used as approximations
for these electromagnetic modes, by means of a quantitative evaluation
of the parallel wavelength, dissipation scale length, and characteristic
admittance as a function of transverse wavelength and altitude. In
Section~\ref{sec:ElectricFieldMapping} we make the argument for
retaining only two modes, and perform a ``back of the envelope''
calculation that predicts wavelike effects. And in Section~(\ref{sec:Model-Results})
we present first results from our model calculation, which development
is completed in the following Section~(\ref{sec:modleIonosphereAdmittance}).

\section{Electromagnetic Modeling for the Ionosphere\label{sec:modleIonosphereAdmittance}}

The steady-state solutions found in Section~\ref{sec:Solution} apply
for small deviations from thermal equilibrium, which means that they
apply for small deviations from a homogeneous plasma configuration.
We would like to extend these solutions to ones that apply to small
deviations from an inhomogeneous plasma configuration like the ionosphere.
This situation is similar to the one that exists for electrostatic
theory: although technically electrostatic theory applies to small
deviations about thermal equilibrium (see Appendix~A), it is often
applied to other background states that are just assumed to satisfy
the equations of motion, even though they may not actually do so.
The idea is that if we observe this state in nature, it must be some
kind of quasi-equilibrium of the actual physical system, which may
not be adequately described by the equations of motion. This gives
a heuristic justification for just pretending that some observed state
satisfies the equations of motion~(\ref{eq:generalIono}), and using
it as a background. Hence we will adopt this same justification and
use TL theory to extend our driven steady-state solutions~(\ref{eq:laplaceIntegralFormSolution})~and~(\ref{eq:solutionFromQuadratic})
to ones that apply to small perturbations about a background state
with an arbitrary inhomogeneity in one dimension (the vertical dimension). 

A wave propagating along a waveguide with varying characteristic admittance
can be modeled in steady-state by cascading together a number of short
sections of transmission line having constant characteristic admittance,
and imposing boundary conditions at each interface to determine the
transmitted and reflected components. Similarly, a wave propagating
in a three-dimensional medium that is inhomogeneous in one direction
only can be modeled in steady-state by stacking a number of thin homogeneous
slabs, and imposing boundary conditions at each interface to determine
the transmitted and reflected components. Hence, adopting the language
and imagery of TL theory and labeling the thin slabs as TL sections,
we propose to estimate the ionospheric admittance and model the ionosphere
using a cascade of short transmission line sections stacked in the
vertical direction, as represented in panel~e of Figure~\ref{fig:Electrostatic-versus-TL}.
The model applies to the far zone for sources of the form~(\ref{eq:source}),
where we can regard $\lambda_{y}=2\pi/k_{0y}$ as the relevant transverse
scale, while eventually planning to form transverse wavepackets by
regarding each solution as a Fourier component, and superposing the
solutions from a spectrum of $\lambda_{y}$. Although this geometry
assumes a vertically stratified ionosphere, the geomagnetic field
can be tilted. At a higher level of approximation we can consider
geometries adapted to curved geomagnetic field lines and wave packets
guided along them, in order to treat ionospheric phenomena at low
and equatorial latitudes, such as the pre-reversal enhancement, and
equatorial spread-\emph{F} (see Section~\ref{sec:Introduction}). 

Solving the cascade of line sections (panel~e of Figure~\ref{fig:Electrostatic-versus-TL})
involves solving for the modal coefficients in each line section,
such that the boundary conditions are satisfied across each interface.
By ``modal coefficients'' we mean the complex constants that are
applied either to the solution~(\ref{eq:drivenSteadyStateSolution})
or the solution~(\ref{eq:solutionFromQuadratic}), as appropriate
for the mode, in order to adjust the amplitude and phase. The positive-going
or negative-going halves of each solution will be separately employed,
as determined by causality. The complete solution for the potentially
large number of boundaries can be obtained by bootstrapping from the
last section of line: the input admittance to the last section of
line is obtained, and then used as the load terminating the second
to last section to obtain its input admittance, which is used as the
load for the third to last section, etc., until the input to the cascade
is reached. 

However, the inclusion of multiple modes means that the input admittance
can no longer be represented by a single complex number. In general,
the admittance is a two index tensor, where one index runs over the
modes to be retained and the two directions, and the other index runs
over the components of $\vec{X}$ for which boundary conditions will
be enforced (the tensor must be square). We will use the term ``polarization
vector'' to refer to the part of the admittance tensor associated
with a particular mode and direction. To establish an equation for
the boundary condition joining the last two sections of transmission
line in panel~e of Figure~\ref{fig:Electrostatic-versus-TL}, consider
a multi-modal component $\vec{E}_{N}$ incident on the $N$th boundary,
and originating at the $(N-1)$th boundary. ($\vec{E}_{N}$ contains
only the components of $\vec{X}$ for which boundary conditions will
be enforced.) Using the solution~(\ref{eq:drivenSteadyStateSolution})
(for brevity, both integral approximations will be used in combination),
and dropping the temporal and transverse dependence (which are the
same for all modes), this incident component takes the form,
\begin{eqnarray}
\vec{E}_{N}(z) & = & \sum_{\stackrel{j(l)}{l\in n_{N-1,+}}}\frac{1}{v_{gj}}\vec{h}_{j}^{\prime\prime}\varepsilon_{j}\mathrm{e}^{(z-z_{N-1})(ik_{0zj}-\omega_{ji}/v_{gjr})_{\vec{k}=\vec{k}_{0j}}}\nonumber \\
 & = & E_{N-1}(z)\vec{\varepsilon}_{N}\label{eq:EN}
\end{eqnarray}
where $z_{N-1}$ is the location of the $(N-1)$th boundary, $n_{N-1,+}$
is the set of retained modes within the $(N-1)$th line section that
were generated at the $(N-1)$th boundary ($+$ indicates that these
are positive-going waves), $E_{N-1}(z)$ is a matrix, $\vec{\varepsilon}_{N}$
is the vector with elements $\left\{ \varepsilon_{j}\right\} $, and
the double-prime on the eigenvector $\vec{h}_{j}$ indicates removing
the elements for which boundary conditions will not be enforced. Equation~(\ref{eq:EN})
defines the notation $\left(\vec{E}_{N}(z),E_{N-1}(z),\vec{\varepsilon}_{N}\right)$
for describing the wave incident on the $N$th boundary, where the
matrix carries the subscripts $N-1$ to indicate that the wave is
traveling in the $(N-1)$th line section, and the coefficient $\vec{\varepsilon}$
and wave $\vec{E}$ carry the subscripts $N$ to indicate that the
wave participates in the boundary condition at the $N$th boundary.
In the same way we define the notation $\left(\vec{B}_{N}(z),B_{N-1}(z),\vec{b}_{N}\right)$
and $\left(\vec{C}_{N}(z),C_{N}(z),\vec{c}_{N}\right)$ for the reflected
and transmitted components associated with the $N$th boundary condition,
respectively. The reflected component employs the negative-going waves,
in this case $n_{N-1,-}$, by the notation above. Note that this notation
can be employed for any of the boundaries, and regardless of which
solution, (\ref{eq:drivenSteadyStateSolution})~or~(\ref{eq:solutionFromQuadratic}),
has been used for the various modes.

At the boundary between the $N$th and $(N-1)$th sections of line
(the $N$th boundary), the component $\vec{E}_{N}$ produces the transmitted
component $\vec{C}_{N}$ in the $N$th section, and the reflected
component $\vec{B}_{N}$ in the $(N-1)$th section. The component
$\vec{E}_{N}$ is, in turn, produced by the component $\vec{E}_{N-1}$
incident on the $(N-1$)th boundary, and so on. This association provides
that there is a constraint, $\vec{c}_{\alpha-1}=\vec{\varepsilon}_{\alpha}$.
We are now in a position to describe the bootstrapping procedure.

For illustration, assume that the boundary conditions are simply those
of continuity (which is our actual case). The boundary condition at
the $N$th boundary can be written $\left.\vec{E}_{N}\right|_{z=z_{N}}+\left.\vec{B}_{N}\right|_{z=z_{N}}=\left.\vec{C}_{N}\right|_{z=z_{N}}$,
where $\vec{B}_{N}$ represents the modes in the $(N-1)$th section
that were generated at the $N$th boundary, and $\vec{C}_{N}$ represents
the modes in the $N$th section and generated at this same boundary.
Since there are no boundaries beyond the $N$th boundary (panel~e~of~Figure~\ref{fig:Electrostatic-versus-TL}),
the $N$th boundary condition is unique in that there is no need to
consider reflections of $\vec{C}_{N}$. Using the notation just developed,
the boundary condition at the $N$th boundary allows for expressing
$\vec{E}_{N}$ in terms of the modal coefficients for the transmitted
and reflected components,
\begin{eqnarray}
\left.\vec{E}_{N}\right|_{z=z_{N}} & = & C_{N}(z_{N})\vec{c}_{N}-B_{N-1}(z_{N})\vec{b}_{N}\nonumber \\
 & = & \left[C_{N}(z_{N})\cup B_{N-1}(z_{N})\right]\left[\begin{array}{c}
\vec{c}_{N}\\
\vec{b}_{N}
\end{array}\right]\label{eq:boundaryCondition_2}
\end{eqnarray}
where the symbol $\cup$ indicates appending the columns of one matrix
to another, with a negative sign applied to the second. Note that
the presence of evanescent modes would render the matrix $\left[C_{N}\cup B_{N-1}\right]$
non-invertible, because the evanescent modes would be present in both
$C_{N}$ and $B_{N-1}$. Hence, this formulation requires that only
propagating modes are retained.

A bootstrapping procedure analogous to that for the single mode case
can be used to solve the transmission line cascade of Figure~\ref{fig:Electrostatic-versus-TL},~panel~e.
Beginning with the last ($N$th) boundary, combine equations~(\ref{eq:EN})~and~(\ref{eq:boundaryCondition_2})
and solve for the modal coefficients $\left(\vec{c}_{N},\,\vec{b}_{N}\right)$
at the $N\mathrm{th}$ boundary, in terms of the modal coefficients
$\vec{c}_{N-1}$ at the $(N-1)\mathrm{th}$ boundary,

\begin{eqnarray}
\left[\begin{array}{c}
\vec{c}_{N}\\
\vec{b}_{N}
\end{array}\right] & = & \left[C_{N}(z_{N})\cup B_{N-1}(z_{N})\right]^{-1}\left.\vec{E}_{N}\right|_{z=z_{N}}\nonumber \\
 & = & \left[C_{N}(z_{N})\cup B_{N-1}(z_{N})\right]^{-1}E_{N-1}(z_{N})\vec{\varepsilon}_{N}\\
 & = & \left[C_{N}(z_{N})\cup B_{N-1}(z_{N})\right]^{-1}E_{N-1}(z_{N})\vec{c}_{N-1},\label{eq:bootstrapppingEqnWithModalCoeficients}
\end{eqnarray}
where we have used the constraint $\vec{\varepsilon}_{N}=\vec{c}_{N-1}$
noted above. 

Now moving to the $(N-1)$th boundary, the component $\vec{E}_{N-1}$
produces the component $\vec{E}_{N}$, but associated with $\vec{E}_{N}$
is the reflected component $\vec{B}_{N}$, so that effectively $\vec{E}_{N-1}$
produces a ``transmitted'' component $\vec{C}_{N-1}=\vec{E}_{N}+\vec{B}_{N}$
in the $(N-1)$th line section. To write the equation for the $(N-1)$th
boundary in a form similar to~(\ref{eq:boundaryCondition_2}), we
need an equation for $\vec{C}_{N-1}$ in terms of $\vec{c}_{N-1}$
alone, which we can obtain using the result~(\ref{eq:bootstrapppingEqnWithModalCoeficients})
from the $N$th boundary,
\begin{eqnarray}
\vec{C}_{N-1}(z) & = & \vec{E}_{N}(z)+\vec{B}_{N}(z)\nonumber \\
 & = & \left(E_{N-1}(z)+B_{N-1}(z)\overline{\left[C_{N}(z_{N})\cup B_{N-1}(z_{N})\right]^{-1}}E_{N-1}(z_{N})\right)\vec{c}_{N-1}\nonumber \\
 & = & C_{N-1}\vec{c}_{N-1},\label{eq:effectiveTransmittedWave-2}
\end{eqnarray}
where the over-bar indicates that the top half of the rows have been
removed, so that only $\vec{b}_{N}$ is present in the output. 

The columns of $C_{N-1}$ are a set of effective steady-state modes
for transmission into the $(N-1)$th line section. Thus we can write
the $(N-1)$th boundary-condition equation, $\left.\vec{E}_{N-1}\right|_{z=z_{N-1}}=\left.\vec{C}_{N-1}\right|_{z=z_{N-1}}-\left.\vec{B}_{N-1}\right|_{z=z_{N-1}}$,
in the form of equation~(\ref{eq:boundaryCondition_2}), and obtain
the invertible matrix $\left[C_{N-1}(z_{N-1})\cup B_{N-2}(z_{N-1})\right]$
for the next iteration: $\left[\vec{c}_{N-1};\,\vec{b}_{N-1}\right]=\left[C_{N-1}(z_{N-1})\cup B_{N-2}(z_{N-1})\right]^{-1}E_{N-2}(z_{N-1})\vec{c}_{N-2}$.

Continuing to iterate to the input of the cascade we finally obtain,
\[
\left[\begin{array}{c}
\vec{c}_{1}\\
\vec{b}_{1}
\end{array}\right]=\left[C_{1}(z_{1})\cup B_{0}(z_{1})\right]^{-1}\vec{E}_{1}(z_{1}),
\]
which allows for finding the reflected and transmitted components
associated with a component $\vec{E}_{1}$ incident on the modeled
region. The coefficient $\vec{b}_{1}$ determines the reflected component
through the relation similar to~(\ref{eq:EN}), and the coefficient
$\vec{c}_{1}$ determines the transmitted component through the relation
similar to~(\ref{eq:effectiveTransmittedWave-2}). The input admittance
to the modeled region follows directly from this result, as do all
the field quantities at all altitudes. 

If there were no evanescent modes then, in principle, the above would
provide a general procedure for deriving the complete solution for
the cascade of line sections. Otherwise, the evanescent modes could
be removed and there should be only a very small effect, since we
do not expect them to be excited by incoming waves and they cannot
transmit energy. However, there is one critical piece missing: implementation
requires knowledge of the physical boundary conditions for all the
retained components of $\vec{X}$. In fact, we only know of three
unambiguous boundary conditions: continuity of the two components
of electric field parallel to the boundary, and continuity of the
single component of magnetic field perpendicular to the boundary.
To include even two modes we need one additional boundary condition.
For this fourth boundary condition we will use continuity of the component
of current density perpendicular to the boundary, which provides that
the closure of the field aligned current through transverse currents
will happen only through the plasma within the line sections, and
will not happen through boundary terms. An electrostatic treatment
would not involve any such boundary terms, and so, among other justifications,
we will use this boundary condition in order to allow for the possibility
of reproducing electrostatic theory.

In what follows we argue that there are, in fact, only two modes that
are effective in transmitting energy through the ionosphere, and that
only these two modes need be included to produce a suitable model.
These two modes can be modeled by the solutions~(\ref{eq:drivenSteadyStateSolution})~and~(\ref{eq:solutionFromQuadratic}).
In the next section we find that they are roughly analogous to the
Alfvén and Whistler waves, respectively, although our formal definition
does not map one-to-one to the usual physically-defined waves.

Examples of the ionospheric input admittance computed using one and
two mode versions of the model are presented in Section~\ref{sec:Model-Results},
along with some selected field quantities resolved in the vertical
direction. Note that in practice we follow the matrix inversions with
optimization steps, to ensure that the boundary conditions are enforced
to the full numerical precision available. 

\section{Wave Modes Relevant to Ionospheric Science\label{sec:Wave-Modes-Relevant-to-Ionospheric-Science}}

The solution~(\ref{eq:solutionHomogenious}) to the initial value
problem allows us to understand the ionospheric plasma motion as a
sum of characteristic modes, which will begin to couple as the excitation
becomes larger and nonlinear terms become important, but which will
evolve independently for small excitations of a homogeneous background.
A source excites the characteristic modes and those that are able
to propagate will carry energy away, through the ionosphere, while
gradually dissipating. As time goes to infinity a steady state condition
arises where the plasma oscillates harmonically with the source, and
also matches the source in the transverse spatial dimensions (from
equation~(\ref{eq:laplaceIntegralFormSolution})). At any appreciable
distance the only remaining modes are those that can propagate at
the source-frequency and transverse wavelength. In this context we
can determine which modes are relevant to ionospheric physics, and
which can be safely ignored, by considering typical ionospheric scale
sizes and drift velocities.

\subsection{Basic Properties of the Modes and Comparison to Electrostatic Waves\label{subsec:Basic-Properties-of-modes}}

Applying numpy.linalg.eig from the Numpy Python library we find the
eigenvalues and eigenvectors of the matrix $H_{5}$, which was introduced
in Section~\ref{sec:Solution}. Searching the transverse wavelength
range from 100~km to 1000~km, with a 40~m/s transverse phase velocity,
we find that the electromagnetic five-moment fluid equations support
four modes that can propagate at the determined frequencies $\left(\mathrm{i.e.,\:}\frac{40\,\mathrm{m/s}}{\lambda_{\perp}}\right)$,
in the plasma density range $4.7\times10^{9}\,\mathrm{m}^{-3}$ to
$1.0\times10^{11}\,\mathrm{m}^{-3}$, and altitude range 100~km to
400~km. The three radio-frequency modes cannot propagate, and neither
can the potential lowest frequency mode, which decouples into two
evanescent (zero frequency) eigenmodes over this range (to numerical
precision). Thus all 8 of the modes that could be supported by the
set of 16 equations have been accounted for. 

The 40~m/s phase velocity is easily small enough that electrostatic
theory would generally be considered applicable, and we will restrict
our considerations to this phase velocity. Smaller wavelengths and
larger phase velocities were investigated by \citeauthor{cosgrove-2016}~{[}2016{]},
who found only two propagating modes, although using a smaller set
of equations. In fact the two additional propagating modes that we
find are enabled by the larger equation set (see Section~\ref{sec:Solution}),
and could not have been found in the \citeauthor{cosgrove-2016}~{[}2016{]}
study. In decreasing order of real-frequency, we refer to the four
propagating modes as the Whistler, Alfvén, Ion, and Thermal modes.
Note that these are only names for eigenvectors, and so should not
be equated directly to physically defined waves, such as the MHD waves. 

In the first-order integral approximation the modes take the form
of single decaying waves (Section~\ref{sec:Solution}), but we will
continue referring to them as modes since they are actually steady-state
structures, and also to minimize any confusion that may arise in comparisons
to physically defined waves. There may be more physical wave-types
than there are eigenvectors, and so there does not exist a one-to-one
correspondence between the two. Our names refer to eigenvectors that
have been identified over a range of collision frequencies (range
of altitudes), and so may actually include more than one physical
type, in some cases. For example, in \citeauthor{cosgrove-2016}~{[}2016{]}
the usual dispersion relations for the whistler and fast-magnetosonic
waves were found to produce frequencies within an order of magnitude
of each other for the studied conditions, and so seemed to correspond
to the same eigenvector. The whistler wave was a better fit, and so
that name was adopted; but it may not be a better fit at all altitudes
in the ionosphere. There is also an issue in that different subfields
of space physics may consider different naming schemes to be appropriate.
For example we derive the name Ion mode from what ionospheric physicists
call the ion-acoustic wave, but magnetospheric physicists may wish
to call this mode the slow-magnetosonic wave. Since the goal of this
work is to derive the ionospheric conductance, we focus on the \emph{E}-region
manifestations and use the terminology that we think is most comfortable
for ionospheric physics.

\begin{figure}[p]
\vspace*{-0.25in}
\thisfloatpagestyle{plain}\includegraphics{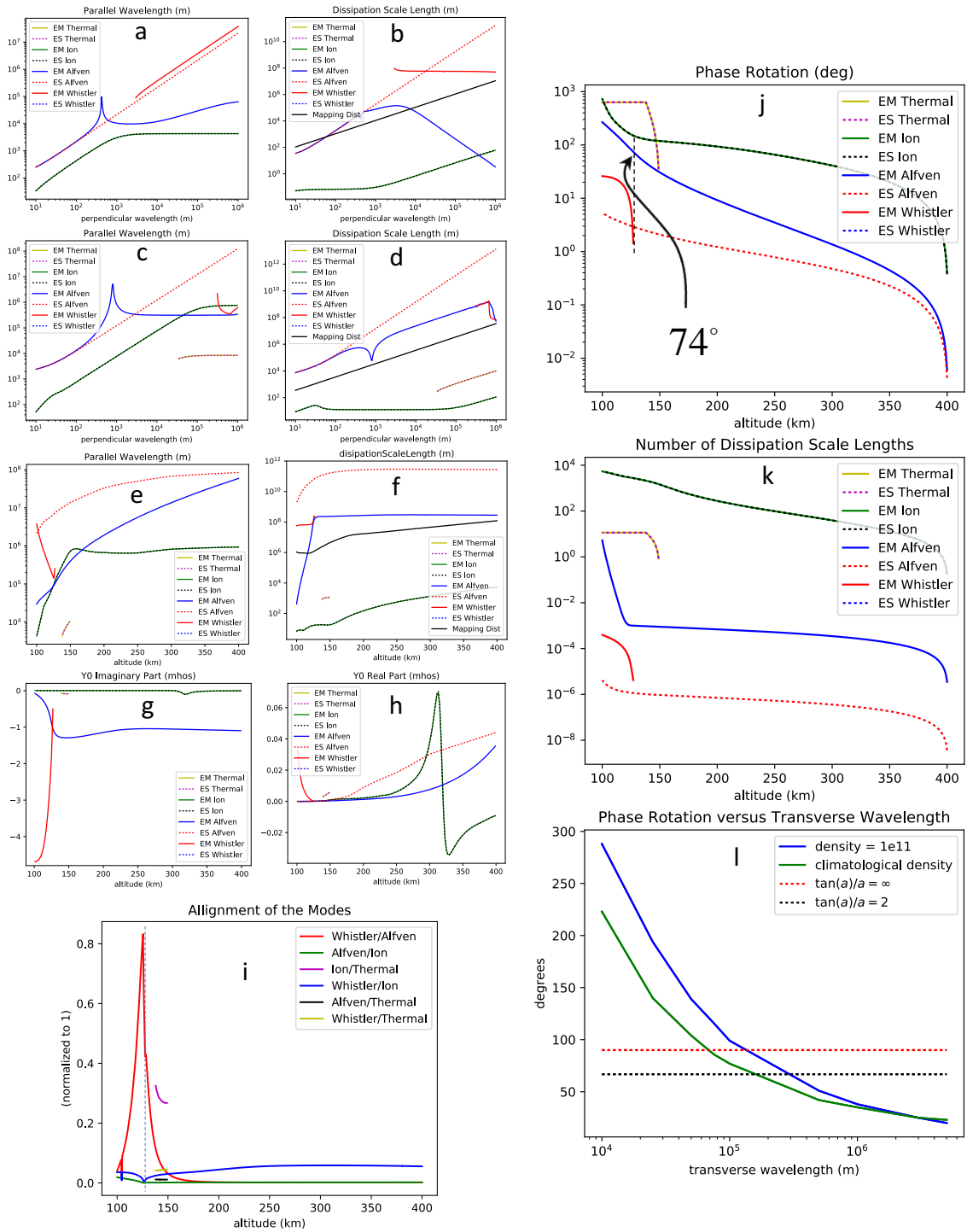}\caption{Wave properties for the relevant modes. For panels~a-i, rows 1 and
2 are for altitudes of 100~km and 145~km, respectively, and rows
3, 4, and 5 are for a 100~km transverse wavelength. Panels~j-l demonstrate
a rough analysis of the phase rotation expected for a signal traversing
the ionosphere, also for a 100~km transverse wavelength. The amount
of phase rotation (panel~j) and number of dissipation scale lengths
(panel~k) are integrated from 400~km in altitude. Panel~l combines
the phase rotations of the Alfvén and Whistler modes to get an estimate
for the electrical thickness of the ionosphere, versus transverse
wavelength: the dashed red line indicates $90^{\circ}$; and the dashed
black line indicates $66.8^{\circ}$, where the electrical length
is just sufficient to double the input conductance. The plasma density
is $10^{11}$~$\mathrm{m}^{-3}$ in all the panels and at all altitudes,
excepting the green curve in panel~l, where the phenomenological
density profile from \citeauthor{cosgrove-2016}~{[}2016{]} is used.
\label{fig:wave_analysis_and_phase_rotation}}
\end{figure}

Figure~\ref{fig:wave_analysis_and_phase_rotation} shows results
using a background plasma density of $10^{11}$~$\mathrm{m}^{-3}$
(at all altitudes), with the remaining parameters determined by the
same altitude profiles used in \citeauthor{cosgrove-2016}~{[}2016{]}
(see their Appendix~B, and included for reference in Section~S.2
of the Supplementary Information). Although
the legend includes all four wave modes, sometimes a particular figure
will not have a corresponding line, or will only have a partial line,
indicating that the frequency matching condition could not be satisfied,
and so the mode does not propagate. For example, the Whistler mode
was not able to satisfy the matching condition in the \emph{F} region,
and demonstrated a short wavelength cutoff that increased with altitude.
Solid lines indicate the full electromagnetic results (i.e., using
$H_{5}$), and dashed lines indicate the most comparable electrostatic
modes found using the matrix $H_{5ES}$, which is equivalent to $H_{5}$
except using the assumption that the curl of the electric field is
zero (which amounts to substituting the Poisson equation for the Maxwell
equations, as explained in detail in Appendix~A). In the case of
the electrostatic modes, absence of a line may also mean simply that
it is outside of the axis limits. The Farley mapping distance is also
shown in the panels for dissipation scale length, as it represents
the comparable electrostatic quantity.

\begin{flushleft}
Panels~a~and~c of Figure~\ref{fig:wave_analysis_and_phase_rotation}
show the wavelength in the direction along the geomagnetic field ($\lambda_{z}$)
for altitudes of 100~km and 145~km, respectively, plotted versus
transverse wavelength. With the exception of the Whistler mode, the
$\lambda_{z}$s are generally quite short, especially at the lower
altitude, and for the Thermal and Ion modes, with 10~km being a representative
number. Where the electromagnetic and electrostatic modes agree the
dashed lines should lie on top of the solid lines. Examining the figure,
the Ion and Thermal modes appear to be well described by the corresponding
electrostatic modes. However, the Alfvén mode appears to be electrostatic
only up to about 100~m in transverse wavelength, and then radically
diverges, with the associated electrostatic mode generally having
a much longer $\lambda_{z}$. And the Whistler mode appears to have
no electrostatic counterpart at all. Similar results are known in
the context of collisionless MHD waves, but we are not aware of equivalent
results for the collisional ionospheric modes.
\par\end{flushleft}

Panels~b~and~d of Figure~\ref{fig:wave_analysis_and_phase_rotation}
show the dissipation scale length ($l_{dz}$) for altitudes of 100~km
and 145~km, respectively, plotted versus transverse wavelength. (Note
that the group velocity is calculated as a finite difference, and
so $l_{dz}$ is subject to more numerical error than $\lambda_{z}$.)
The Ion and Thermal modes are seen to have very short dissipation
scale lengths, 1~km or less, and at the lower of the two altitudes
the Alfvén mode is also quite short, peaking at 50~km. For comparison,
the black line in the panel shows the Farley mapping distance, which
is the commonly accepted estimate for the electric field mapping distance
(derived from electrostatic theory); clearly the agreement is marginal
at best. Comparing the electromagnetic and electrostatic modes, $l_{dz}$
and $\lambda_{z}$ give a similar threshold for the transition to
electrostatic behavior for the Alfvén mode, i.e., below about 100~m.

Looking at panel~b, at the lower altitude (100~km) the Whistler
mode appears to takeover for the Alfvén mode as the transverse wavelength
becomes longer, and the dissipation scale length for the Alfvén mode
is seen to drop below 1~km. Also, comparing panels~a~and~b with
panels~c~and~d, respectively, shows that the Whistler mode requires
a longer transverse wavelength at higher altitudes, in order to be
a propagating mode. And finally, at the higher altitude a kink has
appeared in the curves for the Whistler mode. We later find that this
kink is associated with a change in its physical character. It turns
out that these features are associated with major factors that influence
the wavelength dependence of the input admittance studied below. 

Panels~e,~f,~g,~and~h of Figure~\ref{fig:wave_analysis_and_phase_rotation}
show $\lambda_{z}$, $l_{dz}$, and the real and imaginary parts of
the characteristic wave admittance, respectively, plotted versus altitude,
for a 100~km transverse wavelength. Comparing the solid and dashed
lines in all the panels shows that the Alfvén and Whistler modes cannot
be approximated by electrostatic waves for this transverse wavelength
(100~km), regardless of altitude. The wavelength for the electrostatic
mode is an order of magnitude too long, and the characteristic admittance
is very different (it is outside the axis limits). And there still
does not appear to be any electrostatic counterpart for the Whistler
mode, that is, there are more propagating electromagnetic modes than
there are propagating electrostatic modes. The Thermal mode does seem
to be well approximated by its electrostatic counterpart, but only
arises over a narrow altitude range, around 150~km. The Ion mode
also seems to be well approximated by its electrostatic counterpart,
but has a very short dissipation scale length. Comparing panels~c~and~d
shows that for the Alfvén and Whistler modes, the imaginary parts
of the characteristic admittance are much larger than the real parts,
verifying the assertion made in Section~\ref{sec:Gedanken-Experiment}.
And they are also negative, which provides that the wave-Pedersen
conductivity be positive and (mostly) real, as expected (equation~(\ref{eq:wave_pedersen_conductivity})).

\subsection{Scenario for Energy Transfer Through the Ionosphere, and the Bi-Modal
Model\label{sec:ElectricFieldMapping}}

For energy (and thus electric field, etc.) to be transmitted from,
say, 400~km down to lower altitudes, waves must propagate downward
while encountering varying conditions that may cause them to reflect
and/or to be coupled into other modes at the same frequency and transverse
wavelength. Some of this coupling may be into modes that cannot propagate
under these condition, but such modes cannot transport energy, and
so are essentially loss terms. When conditions are uniform, coupling
between different modes can occur only through the nonlinear term,
$\vec{F}^{\prime}$. However, when conditions change with altitude
(e.g., collision frequency changing) the wave modes also change with
altitude, and so have potential to couple to one-another through this
effect (unlike that due to $\vec{F}^{\prime}$, this form of coupling
does not decrease with wave amplitude). In the TL theory calculation,
this form of coupling occurs at the boundaries between different TL
sections. The coupling potential is increased for eigenmodes that
are not strongly orthogonal, since in this case small variations in
the background conditions have a much greater affect on the modal
allotment. Note that in this subsection we will eschew our practice
of avoiding the term ``wave,'' because of the intuitive benefits
associated with that term for this particular discussion. There is
some additional discussion of the language ambiguities that we face
at the beginning of Section~\ref{sec:Model-Results}.

Focusing on the example of a 100~km transverse wavelength, Figure~\ref{fig:wave_analysis_and_phase_rotation}
(panels~e-h) shows that there are only two propagating waves between
400~km and 150~km in altitude, the Alfvén wave and the Ion wave.
Panel~f shows that the dissipation scale length for the Ion wave
is only about 10~km at 400~km in altitude, and decreases further
with decreasing altitude. Although propagating, the Ion wave cannot
transmit energy over the considered distances, and so any signal arriving
from the magnetosphere must arrive in the form of an Alfvén wave.
Any energy coupled from the Alfvén wave into the Ion wave will be
dissipated without further transmission. This also applies to the
Whistler wave; although non-propagating in this altitude range, some
energy could be coupled into it, and that energy would be rapidly
dissipated. So how much coupling will there be? 

Panel~i of Figure~\ref{fig:wave_analysis_and_phase_rotation} shows
a representation of the degree of alignment of the different modes,
specifically, the complex-conjugate dot-product between the eigenvectors
(which are normalized to unit-length). Modes that are already partially
aligned can be expected to couple more strongly across the TL section
boundaries, whereas modes that are orthogonal should have minimal
coupling. The figure shows that the Alfvén wave is nearly orthogonal
to the Ion wave, and in the region above 150~km is also orthogonal
to the Whistler wave. The vertical dashed line in the figure marks
the altitude where the Whistler begins propagating, whereas the other
waves can propagate everywhere they are shown. (Recall from Section~\ref{sec:Solution}
that we have developed a solution for the Whistler wave that includes
its non-propagating regions, which we are using here.) There is some
alignment between the Whistler and Ion waves, but since the Whistler
wave does not begin to align with the Alfvén wave until below 150~km,
and does not begin propagating until below 127.5~km, it should not
hold much energy until below these altitudes, where its alignment
with the Ion wave is very slight. So there is no clear pathway for
delivering energy to the Ion wave. 

The thermal wave comes briefly available between about 140~km and
150~km in altitude, as can be seen from any of panels~e-i in Figure~\ref{fig:wave_analysis_and_phase_rotation}.
This mode also will not transmit energy, since it also has a very
short dissipation scale length (panel~f of Figure~\ref{fig:wave_analysis_and_phase_rotation}).
However, the Thermal wave has dissipation time-scale longer than any
other mode (not shown). Therefore, it seems possible that the Thermal
wave will fill up with energy. Since the Thermal wave has a very short
parallel wavelength of around 10~km, this could create a significant
anomaly. However, panel~i of Figure~\ref{fig:wave_analysis_and_phase_rotation}
shows that the Thermal wave does not have much alignment with either
the Alfvén or Whistler waves. Also, the Thermal wave does not propagate
if the phenomenological density profile of Figure~S.2 (in Supplementary Information),
which has a lower density in this region, is substituted for the constant
density profile used in Figure~\ref{fig:wave_analysis_and_phase_rotation}.
Also, the Thermal wave disappears when the energy equations are omitted
(not shown). Therefore, it seems that the Thermal wave should only
produce a minor anomaly, except possibly under some particular conditions
that favor it. 

At $127.5$~km in altitude the Whistler wave begins to propagate.
It is right at this altitude that the dissipation scale length for
the Alfvén wave begins to decrease dramatically, while the wavelength
continues to shorten. So if the majority of the incident energy is
to make it down to 100~km, there must be very efficient coupling
from the Alfvén wave to the Whistler wave, such that most of the energy
is transferred to the Whistler wave. From panel~i of Figure~\ref{fig:wave_analysis_and_phase_rotation},
the Whistler and Alfvén waves begin to align strongly below 127.5~km
in altitude, with peak alignment occurring at 125.5~km. There appears
to be a degeneracy or near-degeneracy of the two modes at this altitude,
and so there is potential for strong coupling into the Whistler wave.
However, it seems likely that this condition would result in a splitting
of the energy between the continuing Whistler and Alfvén waves, along
with partial back-reflection of the Alfvén wave due to the impedance
mismatch caused by the new mode suddenly becoming available. 

So in order to evaluate this effect we need to implement the TL theory
calculation of Section~\ref{sec:modleIonosphereAdmittance}, and
both the Whistler and Alfvén waves must be included. However, we have
made the case that the Ion and Thermal waves have only parasitic effects,
and that these effects will only further any deviations from electrostatic
theory that the model may otherwise find. Therefore, in order to accommodate
the limited number of boundary conditions that can be clearly determined,
we will assume that no energy is lost to either the Ion or Thermal
waves, and omit those waves. 

Before studying the results of the full TL theory calculation it is
useful to make a simple, best-case analysis of the electrical length
(phase rotation) associated with energy transfer through the ionosphere.
So consider the case where the Alfvén wave energy is completely transferred
to a downward Whistler wave at 127.5~km, the Ion wave does not siphon
much energy, the Thermal wave does not siphon much energy, and back-reflection
is insubstantial until the wave meets with the bottom of the \emph{E}
region. We also ignore energy coupled into non-propagating modes,
since this effect would only serve to hinder energy transfer. To analyze
this case we integrate the inverse of the parallel wavelength (panel~e
of Figure~\ref{fig:wave_analysis_and_phase_rotation}) over altitude,
from 400~km downward, and display the amount of phase rotation as
a function of altitude in panel~j of Figure~\ref{fig:wave_analysis_and_phase_rotation}
for each mode. When the Whistler wave first begins propagating at
127.5~km in altitude, the Alfvén wave has already rotated $74^{\circ}$.
Here the energy is assumed transferred into the Whistler wave, where,
again from panel~j, there is an additional $25^{\circ}$ of phase
rotation on the way down to 100~km in altitude. So the transmission
line from 400~km down to 100~km has an electrical length of $74^{\circ}+25^{\circ}=99^{\circ}$,
which we might call the electrical thickness of the ionosphere for
a 100~km transverse wavelength and density~of~$10^{11}\:\mathrm{m}^{-3}$.
If instead we use the phenomenological density profile from \citeauthor{cosgrove-2016}~{[}2016{]},
which has a low-density valley region separating the \emph{F} region
from an \emph{E }region arc, we find a reduced electrical thickness
of $77^{\circ}$. 

Panel~k of Figure~\ref{fig:wave_analysis_and_phase_rotation} shows
the results of doing the same integration for the dissipation scale
length. Again using the assumption that the energy is completely transferred
to the Whistler wave at 127.5~km, the amount of dissipation is seen
to be negligible. The Alfvén wave does become significantly dissipative,
but only at the very bottom of the ionosphere, where the Whistler
wave may be dominant. Thus it appears that the ionosphere can be described
by waves that are essentially lossless, which is the usual case for
TL theory, and we verify this in Section~\ref{sec:Model-Results},
where the loss is included. However, it appears that the electrical
thickness of the ionosphere is not negligible (for the 100~km transverse
wavelength and density~of~$10^{11}\:\mathrm{m}^{-3}$).

A $90^{\circ}$ phase rotation means that a maxima at 400~km in altitude
gives rise to a zero at 100~km in altitude. So a $90^{\circ}$ electrical
thickness can be considered as the threshold where the electric field
completely fails to map through the ionosphere. As we have seen in
Section~\ref{sec:Gedanken-Experiment}, equation~(\ref{eq:input_admittance_OC_electromagnetic}),
the input admittance to a section of transmission line terminated
in an open circuit is proportional to the tangent of the electrical
length. Therefore, a $90^{\circ}$ electrical thickness gives rise
to a short circuit condition at 400~km in altitude, meaning that
the wave reflected back from the bottom of the ionosphere completely
cancels the electric field of the incident wave, at 400~km. In this
case the field line integrated conductivity is completely inapplicable
as a measure of the ionospheric admittance. (Note, although wave dissipation
is slight, it will nevertheless produce a small imaginary component
in the electrical thickness that we are ignoring in this discussion,
which will cause the cancelation of electric field to be imperfect.
This lossy effect is, however, included in the model.) 

Repeating the analysis for other wavelengths and one other density
profile, the electrical thickness increases with decreasing transverse
wavelength, and increases with density. Two plots of electrical thickness
versus transverse wavelength are shown in panel~l of Figure~\ref{fig:wave_analysis_and_phase_rotation},
one for the case under discussion (density~$10^{11}\,\mathrm{m}^{-3}$
at all altitudes, blue line), and one for the phenomenological density
profile from \citeauthor{cosgrove-2016}~{[}2016{]} (green line).
For example, for the case under discussion, decreasing the transverse
wavelength to 50~km adds $40^{\circ}$ of electrical length, and
decreasing the transverse wavelength to 25~km adds another $55^{\circ}$,
that is, it adds $95^{\circ}$. So beginning at around 100~km in
transverse wavelength (actually somewhat larger), the electric field
may not map through to the bottom of the E region, and there may even
be positive-feedback effects similar to the ionospheric feedback instability
{[}\citealp{sato-1978}; \citealp{trakhtengertz+feldstein-1984}; \citealp{lysak-1991};
\citealp{pokhotelov+etal-2001}; \citealp{streltsov+lotko-2003b}; \citealp{cosgrove+doe-2010};
\citealp{akbari+etal-2022}{]}. 

For longer transverse wavelengths the electrical thickness decreases,
and becomes negligible. However, other potentially important effects
need to be considered. It seems not unlikely that the dividing of
energy between the Alfvén and Whistler waves, and/or back-reflection
of the Alfvén wave, could have a significant effect. Therefore, to
include these effects, we have implemented the TL theory calculation
described in Section~\ref{sec:modleIonosphereAdmittance}, using
the Alfvén and Whistler waves.

\section{Model Results\label{sec:Model-Results}}

The model of Section~\ref{sec:modleIonosphereAdmittance} has been
implemented using two modes, which in the first-order approximation
take the form of decaying waves, as described in Section~\ref{sec:Solution}.
By ``decaying waves,'' we mean that these are steady-state structures,
not dynamical waves, and so we are modeling the steady state of the
system. There is some difficulty in language in that it is sometimes
convenient to refer, instead, to the dynamical structures that evolve
toward these steady states, and also to relate these to the physically-defined
waves that arise in other contexts. Hence we sometimes refer to the
two modes as the Alfvén and Whistler waves, although we emphasize
that there does not exist a one-to-one mapping to the usual physically
defined waves. Besides the distinction of being steady-state structures,
there is also the distinction that the ``waves'' in our model are
defined over the full altitude extent of the ionosphere, and thus
change their physical character with altitude (Section~\ref{subsec:Basic-Properties-of-modes}).
Although changes in plasma density generally occur with altitude,
in this case we are referring to physical changes caused by the changing
collision frequency, not by changes in plasma density. In order to
simplify the various effects that must be unraveled we keep plasma
density constant with altitude in all the examples below. 

Hoping that the reader can maneuver these ambiguities of language,
we now relax somewhat and describe the model as accounting for the
excitation, transmission, reflection, and coupling of the Alfvén and
Whistler waves, as they carry energy through the ionosphere. Mode-mixing
occurs because the waves change their nature with altitude, because
of the changing collision frequency. And also the waves may be dissipative,
and drop-out as they dissipate. Besides the parasitic modes (Sections~\ref{sec:Solution}~and~\ref{sec:ElectricFieldMapping}),
what is not included in the model is the nonlinear coupling between
waves that happens within each uniform section of line, which becomes
negligible for small amplitude excitations, and which is not part
of the definition for admittance (a quantity from linear circuit theory).
We note that the same situation applies for electrostatic theory,
which drops the same nonlinear terms. 

For pedagogical purposes we have also implemented the model using
one mode at a time. This requires dropping two of the physical boundary
conditions. Using the convention that the $x$-component of the wavevector
is zero ($k_{x}=0$, which we use throughout), and placing the $z$-axis
along the geomagnetic field (assumed vertical for brevity), the omitted
boundary conditions are continuity of the $x$-component of electric
field ($E_{x}$), and continuity of the $z$-component of magnetic
field ($B_{z}$). To the extent that the bi-modal model predicts mode
mixing, the one-mode models are unphysical because they do not respect
these necessary boundary conditions. However, comparison of the one
and two mode models helps us to understand which of the effects are
caused by mode mixing, and which are caused simply by the collision-frequency-dependent
nature of the Alfvén wave.

For all the examples to follow the transverse wavelength is related
to the frequency by assumption of a transverse velocity of 40~m/s,
and (repeating) the plasma density will always be kept constant with
altitude. The remaining ionospheric parameters are the same as used
in \citeauthor{cosgrove-2016}~{[}2016{]} (see their Appendix~B,
and included for reference in Section~S.2
of the Supplementary Information). The
examples all employ one-kilometer-long TL sections, with the exception
of the summary of results shown in Figure~\ref{fig:final_results},
which are described in Section~\ref{subsec:Modeling-Limitations}.
An open-circuit boundary condition is employed at the bottom of the
ionosphere, although we find below that the results are not very sensitive
to this condition.

\subsection{Artificial Examples that Test the Model, and Reproduction of Electrostatic
Theory\label{subsec:Artificial-Examples}}

\begin{figure}[p]
\thisfloatpagestyle{empty}\vspace*{-0.6in}
\includegraphics[viewport=0.8415in 1.721739in 7.07625in 10.33043in,clip]{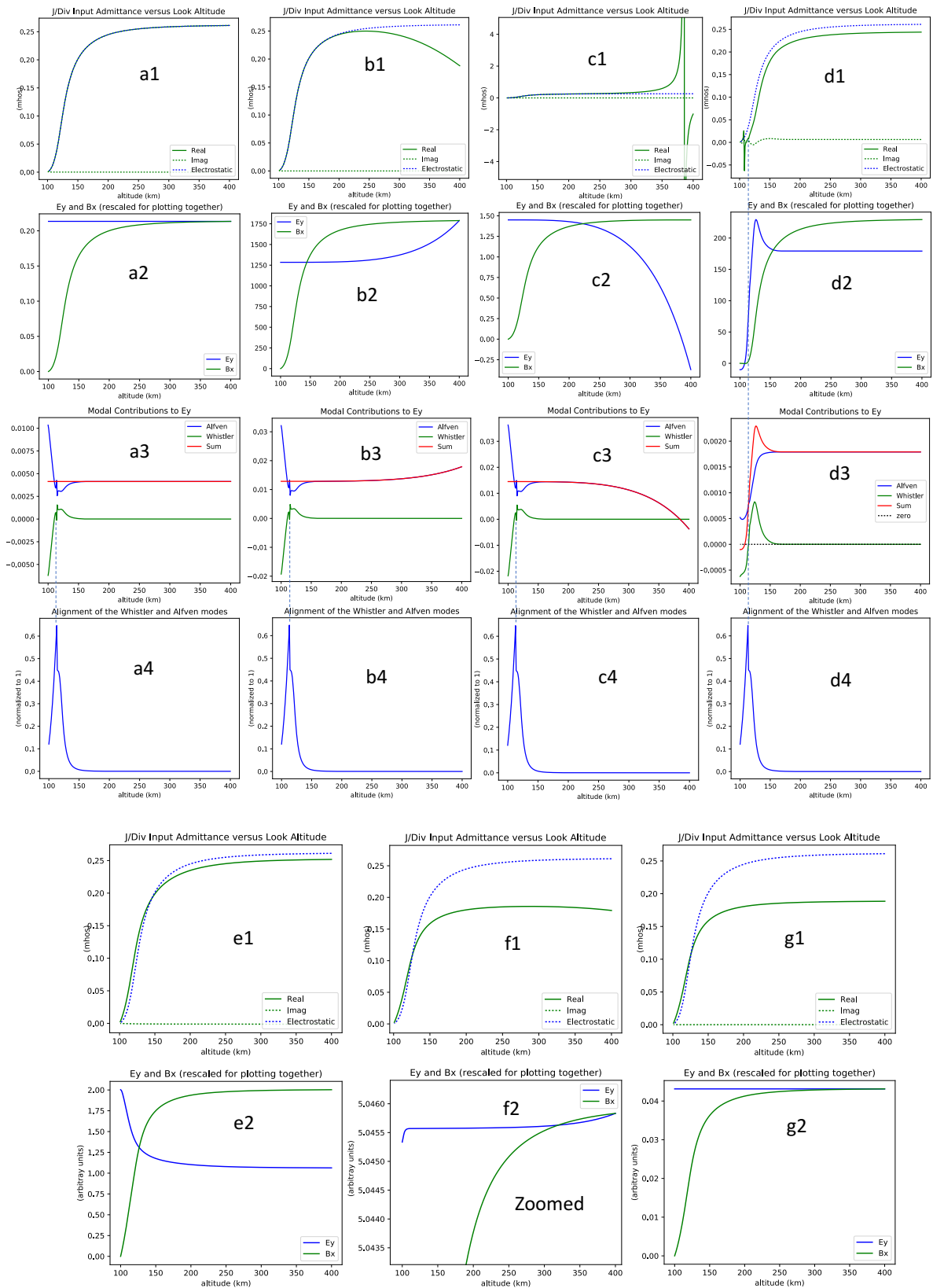}\vspace*{-0.1in}
\caption{Model results using artificial waves that reproduce electrostatic
theory, and show how deviations arise ($\lambda_{\perp}=100\,\mathrm{km,}\:n_{e}=4.7\times10^{9}\mathrm{m}^{-3}$).
Column~a shows exact agreement with electrostatic theory. Column~b
shows what happens when dissipation is added to the case of column~a.
Column~c shows what happens when the wavelengths of both modes are
shortened from the case of column~a. Column~d shows what happens
when the wavelength of the Alfvén mode only is increased from the
case of column~a, so that the wavelengths become unequal. Column~e
shows results using the real wave properties, but with only the Alfvén
mode included in the model. Column~f shows what happens when the
wavelength is artificially lengthened from the case of column~e,
with $E_{y}$ rescaled to maintain the same wave-Pedersen conductivity.
Column~g shows what happens when dissipation is (also) omitted from
the case of column~f. \label{fig:ModelResultsUsingArtificialWaves}}
\end{figure}

We begin with an artificial example that shows how the model would
reproduce electrostatic theory, if the waves had the right characteristics.
In Section~\ref{sec:Gedanken-Experiment} we explained that a single-wave
theory will reproduce electrostatic theory if three conditions hold:
(1) the wave-Pedersen conductivity is equal to the zero-frequency
Pedersen conductivity; (2) the wavelength is long compared to the
system size; and (3) the dissipation scale length is long compared
to the system size. For our two-wave theory we also have the additional
requirements that the waves have the same wavelength, and same dissipation-scale
length, which we show below. Hence, Figure~\ref{fig:ModelResultsUsingArtificialWaves},
column~a shows results where the $z$-directed wavelengths ($\lambda_{z}$)
for both modes have been reset to $5\times10^{8}$~m, the $z$-directed
dissipation scale lengths ($l_{dz}$) have been reset to $\infty$,
and the $y$-components of electric field ($E_{y}$) have been rescaled
to make the wave-Pedersen conductivity equal the zero-frequency Pedersen
conductivity. The starting point for these modifications is the modes
found for a 100~km transverse wavelength and plasma density of $4.7\times10^{9}\:\mathrm{m^{-3}}$
(at all altitudes), and the remaining modal properties are not modified. 

The green curves of Panel~a1, Figure~\ref{fig:ModelResultsUsingArtificialWaves},
show the real (solid) and imaginary (dashed) parts of the downward-looking
input admittance ($\Sigma(z)$) found by the model using the modified
modes, plotted versus altitude, and the electrostatic result is shown
in the dashed blue curve (it is purely real). It is seen that the
model exactly reproduces the electrostatic conductance when the waves
have these modified properties, and panel~a2 shows that $E_{y}$
maps unchanged through the ionosphere. Note that these results apply
regardless of the starting point for the artificial modifications,
and this has been tested for all four of the examples described in
Section~\ref{subsec:Real-Results}.

Panel~a3 shows the modal contributions to $E_{y}$, and it is seen
that the electric field is mostly carried by the Alfvén wave, except
in the low-altitude region where the waves (polarization vectors)
come into near alignment. The degree of alignment is shown in panel~a4,
in terms of the complex-conjugate dot-product between the polarization
vectors for the waves. The resultant $E_{y}$ remains smooth and unchanged,
even though there is substantial mode-mixing and a kinky feature in
the modal allotment associated with the polarization vectors rapidly
coming into alignment.

Column~b of Figure~\ref{fig:ModelResultsUsingArtificialWaves} shows
the results when a significant amount of dissipation is added to the
case of column~a, with the wave-Pedersen conductivity maintained
at the electrostatic value. The conductance now matches the electrostatic
results only when seen from low altitudes, and decreases as the look-altitude
moves away from the conducting region. The conducting region is becoming
disconnected from the altitudes above, as the electric field becomes
shielded.

\noindent Column~c of Figure~\ref{fig:ModelResultsUsingArtificialWaves}
shows the results when the wavelengths for both modes are shortened
to $7\times10^{6}$~m, with the wave-Pedersen conductivities maintained
at the electrostatic value, and without dissipation. There is now
a resonance with a tangent function signature analogous to that discussed
in Section~\ref{sec:Gedanken-Experiment}, equation~(\ref{eq:input_admittance_OC_electromagnetic}),
for a uniform transmission line. We will find a similar effect for
the real case discussed in Section~\ref{subsec:Real-Results}. However,
the effect seen in column~c is very extreme, because the dissipation-scale-length
has been set to $\infty$.
In the real case the slight dissipation rounds-off the peaks of the
tangent function, so that the effect looks much less extreme.

Column~d of Figure~\ref{fig:ModelResultsUsingArtificialWaves} shows
the results when the wavelength of the Alfvén wave only is increased
by a factor of 2.4 from the case of column~a (to $12\times10^{8}$~m),
with the wave-Pedersen conductivity maintained at the electrostatic
value by rescaling $E_{y}$. Otherwise the parameters are the same
as in column~(a). Although the single-wave formulation of Section~\ref{sec:Gedanken-Experiment}
finds that increasing the wavelength should reduce the wavelike effects,
we find something very different in the bi-modal model. Making the
wavelengths unequal has introduced what looks like a low-altitude
resonance, even though the wavelengths are both too long to explain
it, as demonstrated by the results in column~a. There is also a little
kink in the admittance curve, at a slightly higher altitude, and it
looks like the electric field is cutoff right at the kink. Both the
kink and the cutoff of the electric field seem to line-up with the
sharp increase in alignment of the modes seen in panel~d4. A vertical
dashed line has been added to demonstrate the alignment. It looks
like the matrix $H_{5}$ is becoming degenerate or nearly-degenerate
right at his altitude. The conductance seen from above the altitude
of the kink seems to reflect only the portion of the ionosphere that
is above the kink. These results portend the real bi-modal model results
presented below.

To understand what is causing these effects consider replacing the
multi-section TL model with a coarse, two section model, where the
boundary between the sections is right where the modal alignment rapidly
increases (row~4 of Figure~\ref{fig:ModelResultsUsingArtificialWaves}).
Unlike in the upper section, in the lower section the polarization
vectors of the two waves are very similar. The near lack of linear
independence will create a situation where meeting the boundary conditions
requires large and oppositely signed modal coefficients in the lower
section, such that the two large waves are subtracted from each other
to obtain a much smaller resultant. Hence, the signal propagating
in the lower section will have the form $u_{i}=\left((N+\epsilon)\left\{ E_{1x},E_{1y},B_{1x},B_{1z}\right\} \mathrm{e}^{ik_{1}z}-N\left\{ E_{2x},E_{2y},B_{2x},B_{2z}\right\} \mathrm{e}^{ik_{2}z}\right)$,
where for example $\left|N\right|\gg1$ and $\left|\epsilon\right|\sim1$
(although we do not use these relations). There will also be an oppositely
directed signal with the same form other than having negative signs
in the exponential and in front of $B_{1x}$ and $B_{2x}$, but for
simplicity we will ignore this component. We would like to compare
the $z$-derivatives of the fields with those that would be obtained
in the case of a single propagation mode with the same $E_{y}$, such
as a signal of the form $u_{i}^{\prime}=C\left\{ E_{1x},E_{1y},B_{1x},B_{1z}\right\} \mathrm{e}^{ik_{1}z}$.
We now solve for the $\epsilon$ that makes the bi-modal case have
the same $E_{y}$ as the single-mode case (i.e., $E_{y}=E_{y}^{\prime}$
at location $z$) and evaluate the $z$-derivatives. After some algebra
we obtain,
\begin{eqnarray}
\frac{\mathrm{d}E_{y}/\mathrm{d}z}{\mathrm{d}E_{y}^{\prime}/\mathrm{d}z} & = & 1+\frac{NE_{2y}}{E_{y}}\mathrm{e}^{ik_{2}z}\left(1-\frac{k_{2}}{k_{1}}\right)\simeq1+\frac{NE_{2y}}{E_{y}}\left(1-\frac{k_{2}}{k_{1}}\right),\nonumber \\
\frac{\mathrm{d}B_{x}/\mathrm{d}z}{\mathrm{d}B_{x}^{\prime}/\mathrm{d}z} & = & 1+\frac{NB_{2x}}{B_{x}^{\prime}}\mathrm{e}^{ik_{2}z}\left(1-\frac{k_{2}}{k_{1}}\right)\simeq1+\frac{NB_{2x}}{B_{x}^{\prime}}\left(1-\frac{k_{2}}{k_{1}}\right),\nonumber \\
\frac{B_{x}}{B_{x}^{\prime}} & = & 1+\frac{NE_{2y}}{E_{y}}\mathrm{e}^{ik_{2}z}\left(1-\frac{Y_{2}}{Y_{1}}\right)\simeq1+\frac{NE_{2y}}{E_{y}}\left(1-\frac{Y_{2}}{Y_{1}}\right),\label{eq:bi-modal_same_as_single-mode.}
\end{eqnarray}
where $Y_{1}$ and $Y_{2}$ are the wave admittances for the first
and second modes, respectively, and we have also shown results for
the case where $k_{2}z$ is small. Because $N$ may be a very large
(complex) number, these results show that the $z$-derivatives may
be much larger for the bi-modal case. Therefore, the bi-modal case
may exhibit variations over a much shorter scale, and may exhibit
wave-effects over this short scale that are not possible for the single
mode case. This will happen when the modes are combined with large
amplitudes, which we expect in the low-altitude region where they
are strongly aligned.

On the other hand, again from equation~(\ref{eq:bi-modal_same_as_single-mode.}),
when the two modes have the same $k$ (same wavelength and dissipation
scale) the $z$-derivatives become identical to the single mode case,
and if they also have the same wave-admittances then $B_{x}=B_{x}^{\prime}$.
So in this case the bi-modal model should give exactly the same results
as the single-mode model with respect to $E_{y}$ and $B_{x}$, and
hence the same results for the input admittance (which ignoring the
displacement current is $B_{x}/(\mu_{0}E_{y})$, since $k_{x}=0$).
The difference is that the bi-modal model will properly enforce the
boundary conditions for $E_{x}$ and $B_{z}$, and so these quantities
may come out very different from those obtained using the single mode
model. And there may be quite a lot of mode mixing required to do
this, such as is seen in row~3 of Figure~\ref{fig:ModelResultsUsingArtificialWaves},
where there are also kinks that all line up with the alignment-peaks
seen in row~4. In fact, running the model for either of the (similarly-modified)
single modes reproduces exactly the results seen in panels~a1,~a2,~b1,~b2,~c1,~and~c2
of Figure~\ref{fig:ModelResultsUsingArtificialWaves}, and this is
strong evidence that the numerical accuracy in the bi-modal model-evaluation
is sufficient to correctly determine the input admittance that is
our main goal in this work. 

Although there is a lot going on in the bi-modal model, this is a
simple reflection of the fact that it contains a lot of physics. Hopefully,
the above synthetic examples will be enough to convince the reader
that the model is functioning as intended, with sufficient numerical
accuracy, and that it would reproduce electrostatic theory if the
waves had the right characteristics. That the waves do not have the
right characteristics is a simple result of finding the eigenvectors
and eigenvalues of the electromagnetic five-moment fluid equations.
In this regard the wave-Pedersen conductivity is the single most descriptive
parameter, which governs the resultant input conductance when wave
effects are negligible. And when wave effects are not negligible the
electrostatic theory should not have any hope of working.

\subsection{Real Results at the Four Corners of the Modeling Domain\label{subsec:Real-Results}}

We now begin presenting results using the real wave characteristics,
starting with a single-mode example that provides our best match to
electrostatic theory. Hence, consider the case where only the Alfvén
mode is used in the model, and use the real wave characteristics for
a 100~km transverse wavelength, with the rather low plasma density
of $4.7\times10^{9}\:\mathrm{m^{-3}}$ at all altitudes. The green
curves in panel~e1 of Figure~\ref{fig:ModelResultsUsingArtificialWaves}
show $\Sigma(z)$, and the dashed blue curve shows the electrostatic
result. The agreement is within 10\%, which is confusing because in
Section~\ref{sec:Gedanken-Experiment} we said that the Alfvén wave-Pedersen
conductivity is too small to reproduce electrostatic theory. However,
the agreement starts to breakdown when we look at panel~e2, which
shows $E_{y}$ and\textbf{ $B_{x}$}, and reveals that $E_{y}$ does
not map unchanged through the ionosphere. In fact, $E_{y}$ increases
with decreasing altitude. To understand this we try resetting $\lambda_{z}$
to a larger value, while rescaling $E_{y}$ in the polarization vector
so that the wave-Pedersen conductivity is not changed. The results
are shown in panels~f1~and~f2, which show that $\Sigma(z)$ has
been greatly reduced, and that $E_{y}$ now maps almost unchanged
through the ionosphere (note the much finer scale). If we also set
$l_{dz}$ to $\infty$ the results in panels~g1~and~g2 are obtained,
where $E_{y}$ now maps perfectly through the ionosphere, and the
slight droop is removed from $\Sigma(z)$ ($E_{y}$ is again rescaled
to maintain the same wave-Pedersen conductivity). The results now
have the character expected from electrostatic theory, but $\Sigma(z)$
does not quantitatively agree with the electrostatic result. This
shows that the near agreement seen in panel~e1 is something of a
coincidence, which would not occur if it were not for wavelike effects.
Generally the wavelike effects create a much more interesting behavior,
as we see now in exploring the rest of the results. 

\begin{figure}[p]
\thisfloatpagestyle{plain}\includegraphics{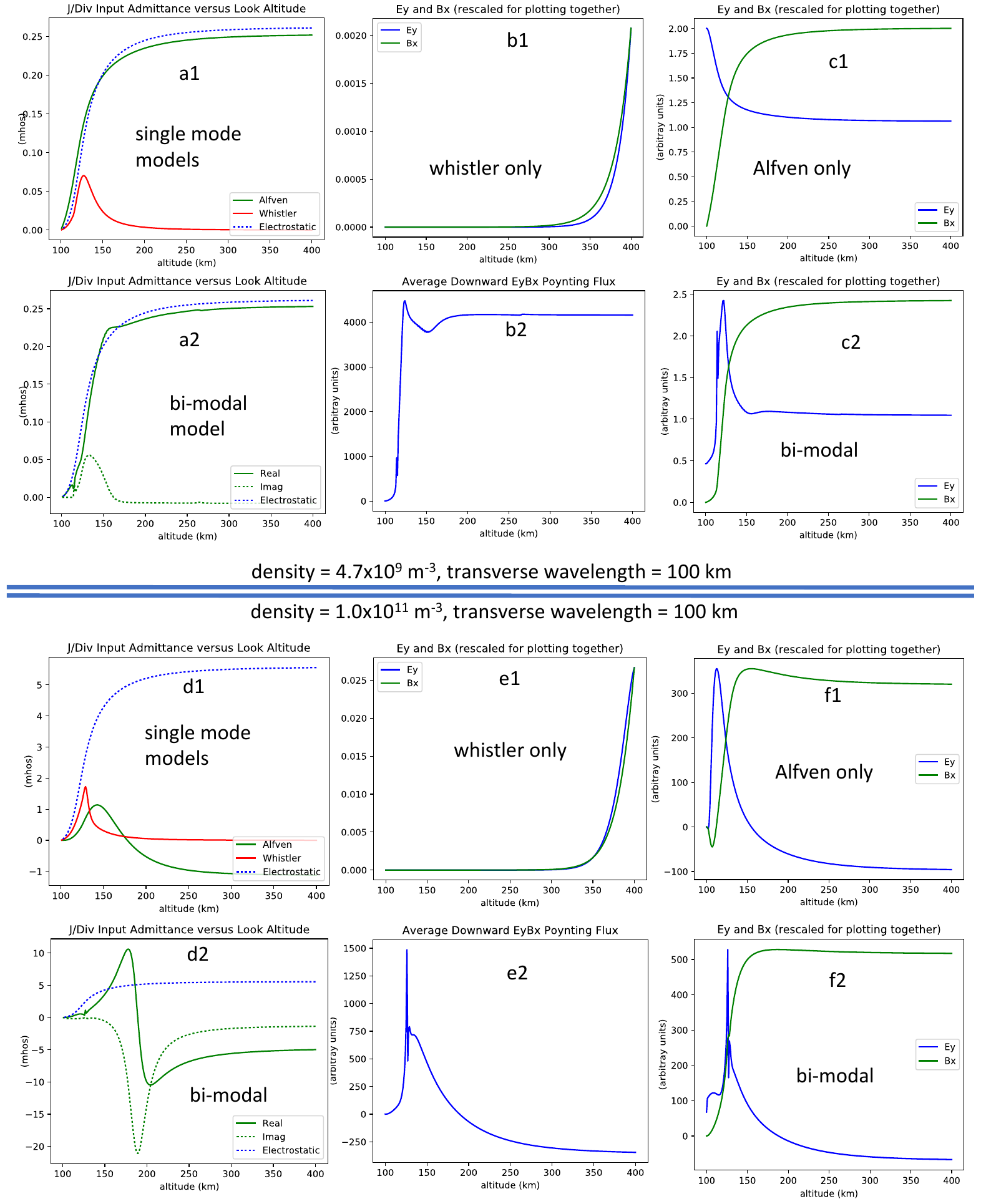}\caption{Single and bi-modal model results for a 100~km transverse wavelength,
and for plasma densities of $4.7\times10^{9}\,\mathrm{m^{-3}}$ (top)
and $1.0\times10^{11}\,\mathrm{m^{-3}}$ (bottom).\label{fig:results_at_the_corners_100km}}
\end{figure}

We now present the results from the one- and two-mode models covering
a transverse scale size from~100~km to~1000~km, and covering a
plasma density range from $4.7\times10^{9}\:\mathrm{m^{-3}}$ to $1.0\times10^{11}\:\mathrm{m^{-3}}$.
Of course we expect the two-mode model results to be the physical
results, but it is useful to compare the single-mode results in order
to evaluate the effects of mode mixing. Figure~\ref{fig:results_at_the_corners_100km}
shows results for a 100~km transverse wavelength, where the upper
six panels are for plasma density $4.7\times10^{9}\:\mathrm{m^{-3}}$,
and the lower six panels are for plasma density $1.0\times10^{11}\:\mathrm{m^{-3}}$.
Panels~a1~and~d1 compare $\Sigma(z)$ (real part only) from the
single mode models with the electrostatic results, for the respective
densities. Because the Whistler mode is cutoff at higher altitudes,
by itself it only provides for a non-zero conductance at look-altitudes
that are quite low. Panels~b1~and~e1 show that the electric field
would not map into the ionosphere if only the Whistler mode existed.
Comparing panels~a1~and~d1 shows that the higher plasma density
shortens the effective wavelength of the Alfvén mode so that it resonates,
and that this resonance has nothing to do with mode-mixing (since
these are single-mode models). This is the kind of tangent-function
resonance that was discussed in Section~\ref{sec:Gedanken-Experiment},
exemplified in column~c of Figure~\ref{fig:ModelResultsUsingArtificialWaves},
and predicted by the simple phase-rotation estimate of Figure~\ref{fig:wave_analysis_and_phase_rotation},
panel~j. However, the lossiness that we noted in Section~\ref{sec:ElectricFieldMapping}
for the Alfvén mode at the very bottom of the \emph{E }region appears
to almost completely eliminate the lossless-case infinities of the
tangent function, for this single-mode model. 

Panels~a2~and~d2 of Figure~\ref{fig:results_at_the_corners_100km}
compare $\Sigma(z)$ from the bi-modal model with the electrostatic
results, for the respective densities $4.7\times10^{9}\:\mathrm{m^{-3}}$
and $1.0\times10^{11}\:\mathrm{m^{-3}}$. Besides the low-altitude
spiky or kinky features that we will discuss later (mostly in Section~\ref{subsec:Modeling-Limitations}),
the bi-modal model gives results that are qualitatively similar to
the model using only the Alfvén wave (compare panels~a1~and~d1,
green curves). For the low-density case the quantitative results are
also quite similar. However, for the higher density case the bi-modal
model has a more distinct resonance effect, with larger overall swings
that are more reminiscent of the expected tangent function behavior.
This suggests that the presence of the Whistler wave has reduced the
amount of energy in the Alfvén wave, in the very low altitude region
where the latter becomes lossy. Finally, similar to the effects seen
above in column~d of Figure~\ref{fig:ModelResultsUsingArtificialWaves}
(for the artificial waves with different wavelengths), panels~c2~and~f2
show that $E_{y}$ experiences a sharp low-altitude peak, and then
is rapidly cutoff before reaching the bottom of the ionosphere. Comparing
panels~c1~and~f1 show that there is a similar effect when only
the Alfvén wave is used in the model, but that the sharp cutoff is
missing or not as sharp.

\begin{figure}[p]
\thisfloatpagestyle{plain}\includegraphics{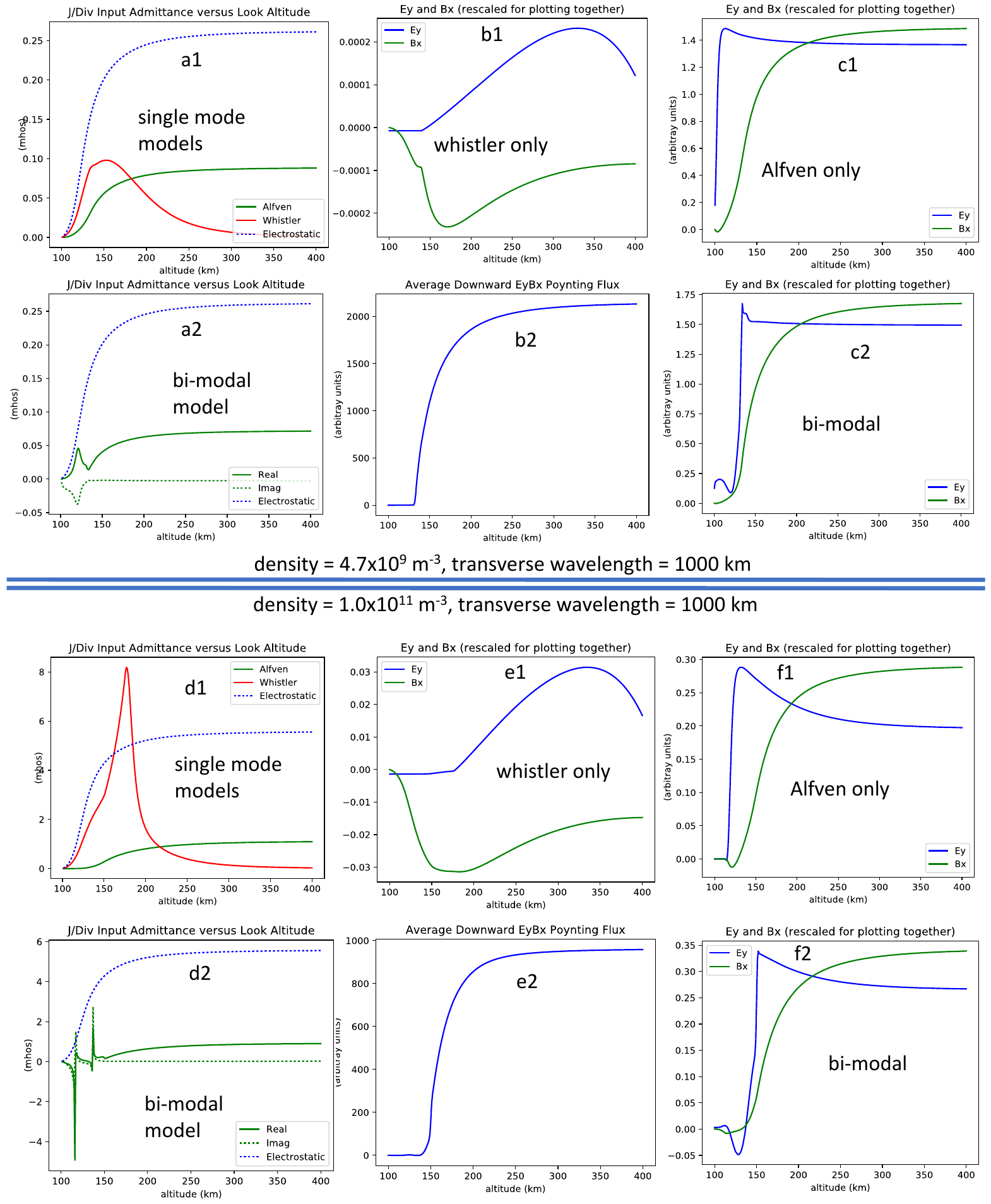}\caption{Single and bi-modal model results for a 1000~km transverse wavelength,
and for plasma densities of $4.7\times10^{9}\,\mathrm{m^{-3}}$ (top)
and $1.0\times10^{11}\,\mathrm{m^{-3}}$ (bottom).\label{fig:results_at_the_corners_1000km}}
\end{figure}

Turning now to the case of the 1000~km transverse wavelength shown
in Figure~\ref{fig:results_at_the_corners_1000km}, we find that
the single-mode-resonance (tangent function of equation~\ref{eq:input_admittance_OC_electromagnetic})
is no longer present. However, $E_{y}$ still does not map to the
bottom of the ionosphere for any of the cases, and the bi-modal conductances
at the top of the ionosphere are only 16\% and 27\% of the electrostatic
predictions, for densities of $1\times10^{11}$ and $4.7\times10^{9}$,
respectively. For the Alfvén-mode-only cases (panels~c1~and~f1),
although we had postulated in the pre-modeling analysis of Section~\ref{sec:ElectricFieldMapping}
that dissipation might keep the electric field from mapping, artificially
increasing the dissipation-scale-length does not make a significant
difference. (This is an example of the lack of low-altitude sensitivity
that has been referenced above, in support of the validation.) Hence
the electric field cutoff in the Alfvén-mode-only cases must be due
mostly to a mismatch caused by rapid changes in the Alfvén mode admittance
quantities, in the lower \emph{E} region (e.g., panel~d of Figure~\ref{fig:scattering_interpretation},
discussed below). It was also postulated in the pre-modeling analysis
that the Whistler wave might prevent this effect and allow the electric
field to map, but in fact the bi-modal results in panels~c2~and~f2
show that instead, $E_{y}$ is cutoff at an even higher altitude.
This is also likely due to a mismatch of some kind, and Section~\ref{subsec:Analysis-of-Sharp}
is dedicated to analyzing this effect, with additional relevant discussion
in Section~\ref{subsec:Modeling-Limitations}. 

Panels~b2~and~e2 in both Figures~\ref{fig:results_at_the_corners_100km}~and~\ref{fig:results_at_the_corners_1000km}
show the $E_{y}\delta B_{x}$ component of the time-averaged downward
Poynting flux for the bi-modal case, which is the component associated
with the field aligned current and $\Sigma(z)$. (Note that continuity
of $\delta B_{y}$ is not enforced and so we will avoid trying to
analyze the $E_{x}\delta B_{y}$ component, which anyway is not directly
related to the field aligned current or $\Sigma(z)$, since $k_{x}=0$
($j_{\parallel}=\hat{z}\cdot\vec{k}\times\delta\vec{B}/\mu_{0}=-k_{y}\delta B_{x}/\mu_{0}$).)
The time-averaged Poynting flux is the real part of the Poynting vector,
and so is not exactly the same as the product of the instantaneous
$E_{y}$ and $\delta B_{x}$ that are shown in the other figures (which
were taken at the instant when the downward Poynting flux maximizes
at 400~km). In all four cases the Poynting flux experiences a sharp
cutoff, such that the lower altitude portion of the ionosphere does
not participate in closing the field aligned current, and does not
contribute to the conductance seen at the top. The cutoff altitude
increases with both density and wavelength, which produces a dependence
that is very different from that expected under electrostatic theory.
This same behavior exists for the Alfvén-mode-only models, with little
change when the dissipation scale length is artificially increased.
These findings all suggest that modeling errors at the bottom of the
ionosphere will have little effect, and thus lend support to the validation
given in Section~\ref{sec:Solution}. Hence, we find that the conductance
seen at the top of the ionosphere is generally much less than expected
under electrostatic theory. The reduction is very significant for
both the bi-modal and Alfvén-only models, but is more pronounced in
the bi-modal case. 

\subsection{Analysis of Sharp Cutoff of Electric Field\label{subsec:Analysis-of-Sharp}}

\begin{figure}[p]
\vspace*{-0.75in}
\includegraphics{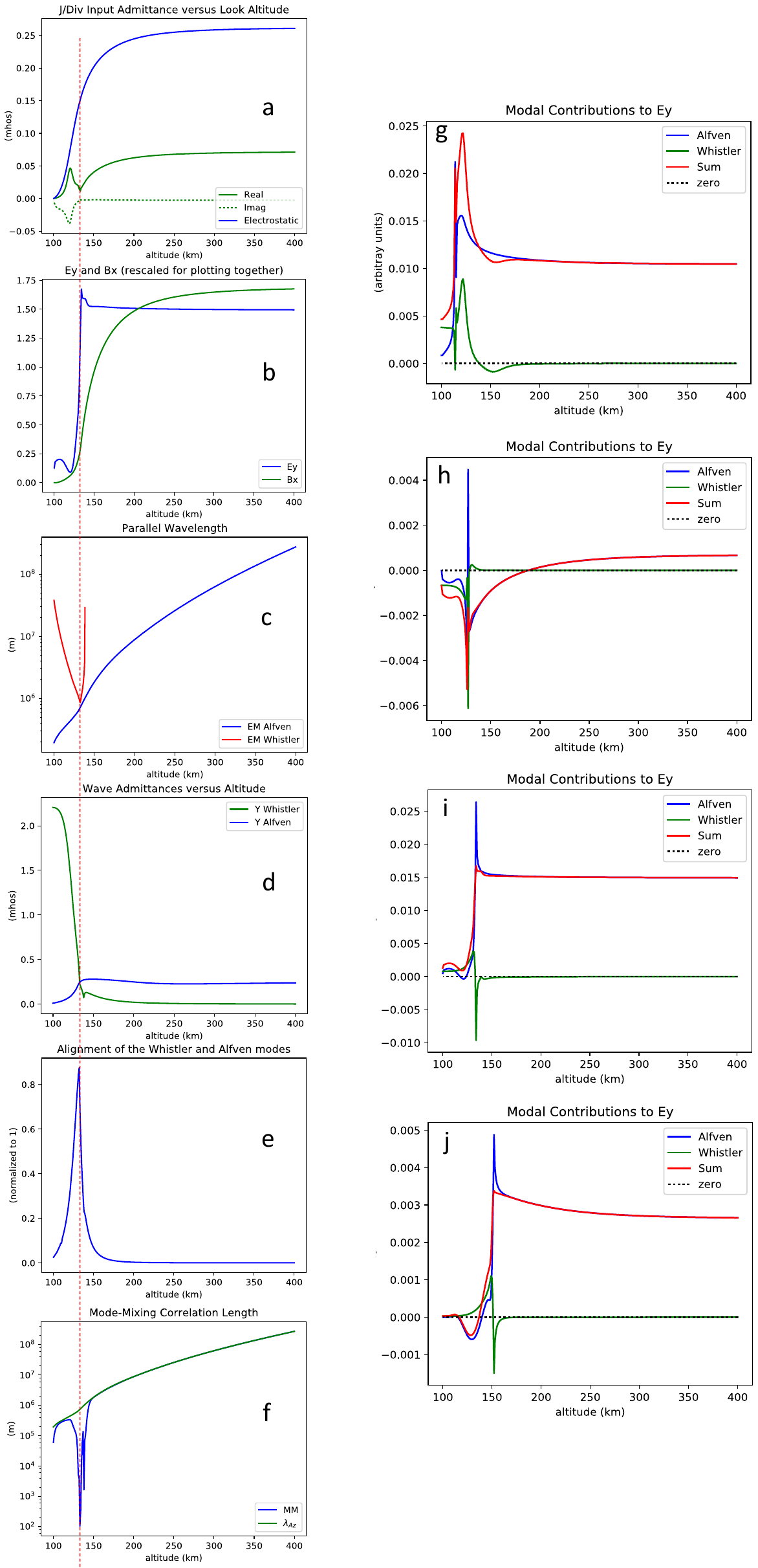}

\vspace*{-0.2in}
\caption{Left Column: Examination of electric field cutoff in the bi-modal
model. Right Column: Modal contributions to $E_{y}$ {[}panel~g,
($\lambda_{\perp}=100\,\mathrm{km,}\:n_{e}=4.7\times10^{9}\mathrm{m}^{-3}$);
panel~h, ($\lambda_{\perp}=100\,\mathrm{km,}\:n_{e}=1.0\times10^{11}\mathrm{m}^{-3}$);
panel~i, ($\lambda_{\perp}=1000\,\mathrm{km,}\:n_{e}=4.7\times10^{9}\mathrm{m}^{-3}$);
panel~j, ($\lambda_{\perp}=1000\,\mathrm{km,}\:n_{e}=1.0\times10^{11}\mathrm{m}^{-3}$){]}.
\label{fig:scattering_interpretation}}
\end{figure}

We have found that there is a blockage of the signal penetration that
arises in both the Alfvén-only and bi-modal models, but which is more
pronounced in the bi-modal case. For the Alfvén-only case we attribute
this effect to wave-reflection caused by a sharp gradient in the Alfvén
admittance, which gradient can be seen in panel~d of Figure~\ref{fig:scattering_interpretation},
for the case of a 1000~km transverse wavelength and density of $4.7\times10^{9}\,\mathrm{m^{-3}}$.
In the pre-modeling analysis of Section~\ref{sec:ElectricFieldMapping}
we had postulated that dissipation might prevent the signal from penetrating,
but since artificially increasing the dissipation-scale-length does
not significantly alleviate the blocking effect, it must instead be
due to wave-reflection. While this is fairly easy to understand, what
is harder to understand is the extremely sharp cutoff of the electric
field that happens in the bi-modal cases, and beyond which the downward
Poynting flux is essentially zero. These locations also seem to be
associated with spiky or kinky features, which we will discuss separately
in Section~\ref{subsec:Modeling-Limitations}, and also in the last
paragraph of this section.

To analyze the cutoff we will focus on the case shown in panels~a2,~b2,~and~c2
of Figure~\ref{fig:results_at_the_corners_1000km}, where the effect
is both clear and relatively simple in form. A closer examination
of this case is shown in Figure~\ref{fig:scattering_interpretation},
panels~a-f, which show from top to bottom the input admittance, $E_{y}$
and $B_{x}$, the parallel wavelengths, the magnitude of the wave
admittances, the alignment of the modes, and a quantity we call the
``mode-mixing coherence length'' to be described below. All these
quantities are plotted versus altitude, and a red-dashed vertical
line is drawn to indicate the altitude of electric field cutoff. We
will attempt to interpret the cutoff as a wave reflection phenomenon
that is enhanced by the presence of more than one wave mode.

Panel~d of Figure~\ref{fig:scattering_interpretation} shows that
over most of the ionosphere the Alfvén admittance is relatively constant,
and so the signal essentially propagates in a uniform transmission
line all the way down to about 134~km in altitude. However, at 134~km
the Whistler mode admittance becomes equal to the Alfvén admittance
and the two modes very-suddenly become strongly aligned. This location
also corresponds to a break in the parallel wavelength of the Whistler
mode, right where that wavelength becomes nearly equal to that of
the Alfvén mode. It appears that the Whistler mode changes its physical
wave-type at that altitude, and that this change is associated with
a degeneracy or near-degeneracy of the matrix $H_{5}$, as was already
mentioned in Section~\ref{sec:ElectricFieldMapping}. The rapidly
changing alignment and crossing wave-admittances suggest that this
could be a point of strong reflection for the Alfvén mode, which might
prevent energy from propagating further downward.

However, the reflection appears to be very sharp, and this is something
of a puzzle considering the wavelengths are not all that short for
either mode. The wavelength plays a role something like a correlation
length, which limits the scale over which the signal can change. When
two scatterers are located much closer together than a wavelength
there cannot be much change between them, and so they are effectively
colocated. Yet in Figure~\ref{fig:scattering_interpretation} the
electric field changes dramatically between the bottom of the ionosphere
and the apparent scattering location at 134~km. These two scattering
locations appear to be behaving independently, as though the wavelength
were shorter than what is shown in panel~c of Figure~\ref{fig:scattering_interpretation}.
Below 134~km the Alfvén wavelength shortens dramatically, and so
this conundrum does not arise for the Alfvén-mode-only case where
the scattering occurs at a lower altitude and is less sharp. But how
can we have such sharp scattering at 134~km? 

To resolve this doubt we refer back to equation~(\ref{eq:bi-modal_same_as_single-mode.})
and the surrounding discussion. Since the $z$-derivative of $E^{\prime}(z)$
is $ik_{1}E^{\prime}(z)$ and we have set $E(z)=E^{\prime}(z)$, we
can define something like an effective $k$ for the mode-mixed case
by,
\[
\frac{\mathrm{d}E_{y}/\mathrm{d}z}{\mathrm{d}E_{y}^{\prime}/\mathrm{d}z}=\frac{k_{MM}}{k_{1}}.
\]
Since we do not actually expect oscillation with wavenumber $k_{MM}$,
we regard this more as a way to estimate a correlation length for
the mixed-mode case. Hence in panel~f we use equation~(\ref{eq:bi-modal_same_as_single-mode.})
to plot $2\pi/k_{MM}$, which we consider as a correlation length
for use in estimating the degree to which the scatterer at 134~km
can be considered as independent from the scattering off the bottom
of the ionosphere. (We note also that panel~d prompted us to try
an alternative, short-circuit boundary condition for the Whistler
wave at the bottom of the ionosphere, and that this had almost no
effect on the results.) Panel~f shows that this correlation length
becomes very short in the neighborhood of 134~km, less than 1~km
at the location of the cutoff, and this can viewed as an explanation
for why the scatter located there produces such sharp and apparently
independent scattering. The two modes together are acting like a wave
with much shorter wavelength, and this can be understood from equation~(\ref{eq:bi-modal_same_as_single-mode.}).

In all the cases the alignment of the cutoff seems to be consistent
with this interpretation. And it is also consistent that for the artificial
bi-modal cases in Figure~\ref{fig:ModelResultsUsingArtificialWaves},
there is no hint of a cutoff until the wavelengths are made unequal
(column~d). With respect to the kinky features, columns~a-d of Figure~\ref{fig:ModelResultsUsingArtificialWaves}
all show kinks in the modal allotment, but there is no hint of an
electric-field cutoff until the wavelengths are made unequal (column~d).

However, at altitudes below the cutoff there sometimes occurs another
feature that looks more like a resonance, that is, a place where the
electric field is nearly canceled. There are two such prominent features
in the admittance function for the case of a 1000~km transverse wavelength
with density $1.0\times10^{11}\,\mathrm{m}^{-3}$, seen in panel~d2
of Figure~\ref{fig:results_at_the_corners_1000km}. There is also
an example in the artificial case shown in column~d of Figure~\ref{fig:ModelResultsUsingArtificialWaves}.
Although dramatic in appearance, it seems that these features are
incidental to the scattering that occurs at the positions of the cutoffs.
In panels~g-j of Figure~\ref{fig:scattering_interpretation} are
shown the separate contributions to $E_{y}$ from the two modes, where
from top to bottom appear the cases with transverse-wavelengths and
densities of $\left(100\,\mathrm{km},4.7\times10^{9}\,\mathrm{m}^{-3}\right)$,
$\left(100\,\mathrm{km},1.0\times10^{11}\,\mathrm{m}^{-3}\right)$,
$\left(1000\,\mathrm{km},4.7\times10^{9}\,\mathrm{m}^{-3}\right)$,
and $\left(1000\,\mathrm{km},1.0\times10^{11}\,\mathrm{m}^{-3}\right)$.
Only the bottom-most of these examples exhibits the resonant-like
effects. However, the bottom three panels all show very similar behavior
for the Alfvén- and Whistler-mode contributions; the wiggles and curves
appear to be driven by the same physics. The only real difference
to be associated with the bottom panel is that the vertical positioning
of the wiggles and curves is such that the resultant $E_{y}$ crosses
the $z$-axis (i.e., passes through zero). There is no dramatic effect
to be seen in $E_{y}$, yet because $E_{y}$ is in the denominator
of the admittance function the latter does display a dramatic effect.
The upper panel (panel~g) is the exception in that it does not display
the same wiggles and curves, but it has the lowest altitude cutoff
of the four, and so there may simply not be room for those phenomena
to occur before meeting with the bottom of the ionosphere. It is the
well known feature of admittance functions that they may include zeros
in the denominator, such as in the case of the tangent function in
equation~(\ref{eq:input_admittance_OC_electromagnetic}). Hence,
there is no anomaly in these findings.

\subsection{Modeling Limitations and Quantitative Results for Wavelength and
Density Dependence\label{subsec:Modeling-Limitations}}

\begin{figure}[p]
\thisfloatpagestyle{plain}\includegraphics{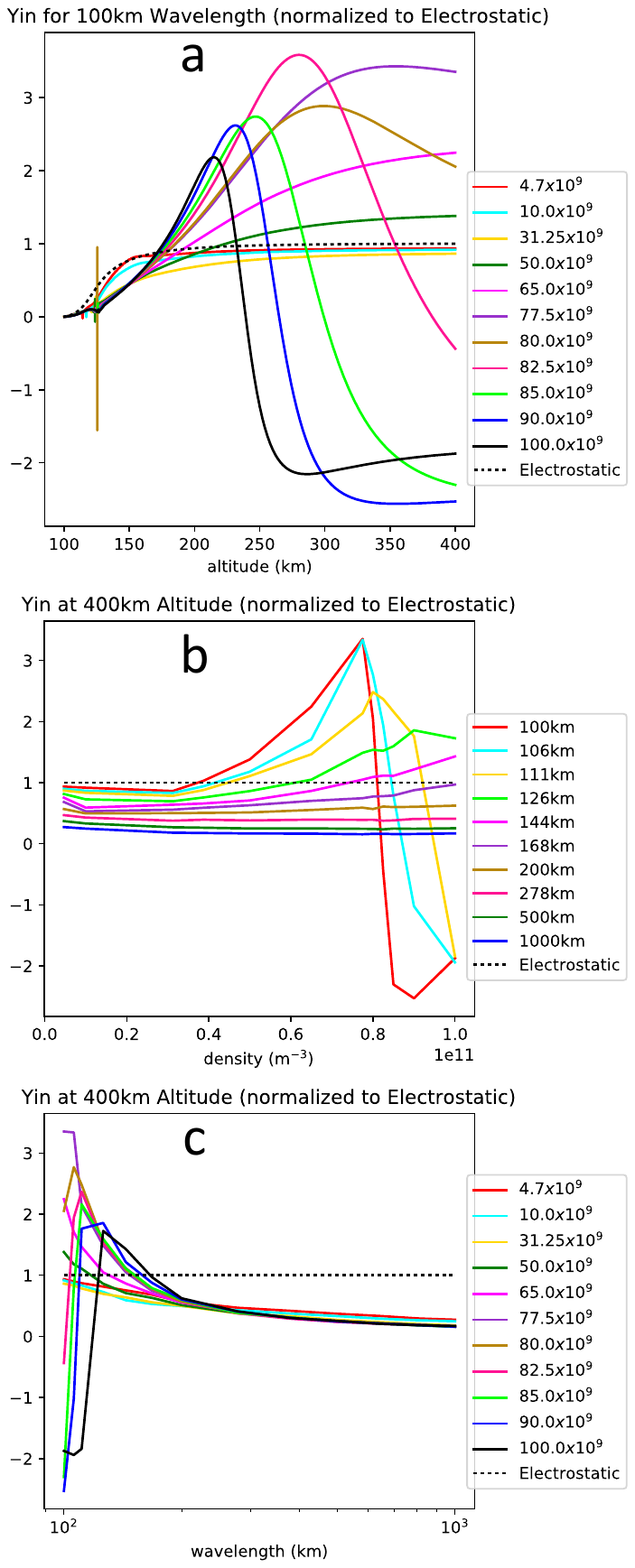}

\caption{Input admittance versus altitude (panel~a), density (panel~b), and
wavelength (panel~c), normalized to the field-line-integrated conductivity
(dashed black lines). The transverse wavelength for panel~a is 100~km.
The legends for panels~a~and~c show the density in m$^{-3}$ (constant
with altitude). The legend for panel~b shows the transverse wavelength.
The curves in panels~b~and~c are made up of end-points of curves
like the ones in panel~a, and as such have a higher level of model
uncertainty.\label{fig:final_results}}
\end{figure}

In this section we provide a summary of the results over the modeled
region, which is bounded by the four corners discussed in Section~\ref{subsec:Real-Results}.
However, before doing this there is one caveat to report concerning
the modal resolution in the vicinity of the degeneracy, which affects
the quantitative results. Panel~a of Figure~\ref{fig:final_results}
shows the real part of the input admittance (the conductance) plotted
versus look altitude for a 100~km transverse wavelength and 11 different
plasma densities between $4.7\times10^{9}\,\mathrm{m}^{-3}$ and $1.0\times10^{11}\,\mathrm{m}^{-3}$.
The curves are normalized to the electrostatic results for the respective
density, which are thus shown by the single dashed black line. The
gradual ascent to the tangent function signature is apparent in panel~a.
However, also seen in panel~a is a very sharp feature colocated with
the cutoff of the electric field and the degeneracy of the matrix
$H_{5}$.

\textbf{Resolution Limitations: }Although the cutoff of the electric
field certainly seems to be a real prediction of the model (Section~\ref{subsec:Analysis-of-Sharp}),
there are sometimes other kinds of sharp features that also occur
right at the location of the degeneracy, and these are more difficult
to assess. (The sharp features occurring \emph{below} the degeneracy
were discussed at the end of Section~\ref{subsec:Analysis-of-Sharp}.)
The sharp feature in the admittance function seen in panel~a of Figure~\ref{fig:final_results}
is associated with a downward spike in the electric field (not shown),
and similar electric-field spikes are seen in all the bi-modal cases
with a 100~km transverse wavelength (e.g., panels~c2~and~f2 of
Figure~\ref{fig:results_at_the_corners_100km}, where these plots
may not reveal the full depth of the spike, since the points are spaced
farther apart {[}1~km{]} than in Figure~\ref{fig:final_results}).
It appears that these electric-field spikes can sometimes pass very
close to zero in the complex plain, giving rise to even sharper spikes
in the admittance function, since the electric field is the denominator
of admittance. Exactly how close to zero would seem to be a highly
sensitive thing to predict, and so the spikes in the admittance function
are probably not accurately calculated. But as for the prediction
of the downward spikes in the electric field, it seems likely that
they are real predictions. However, our ability to probe very close
to the degeneracy is limited by resolution, and not only does this
affect assessment of these electric-field spikes but it also sometimes
affects the quantitative results for the admittance seen at the top
of the ionosphere. Hence, we need to discuss the resolution limitations
for the calculation.

The calculation resolves in the Fourier domain, and so resolution
in this case means resolution in $\vec{k}$-space. As discussed in
Section~\ref{sec:Background}, for a Fourier domain model there is
continuous spatial resolution, in this case by the analytic functions~(\ref{eq:drivenSteadyStateSolution})~and~(\ref{eq:solutionFromQuadratic}).
So since we are solving for one Fourier component at a time, the resolution
would seem to be a question that only affects future applications,
when we try to assemble many Fourier components to create a 3D spatial-domain
representation. Except, we do have the issue that we are stacking
TL sections vertically, and so there does seem to be some kind of
spatial resolution issue. How should we understand this?

In an idealized case where the wave properties are constant with altitude,
we could represent the whole ionosphere by a single TL section. This
provides continuous vertical resolution, even though the ``sampling''
is very coarse (i.e., a single TL section). And similarly, when there
\emph{is} variation in the wave properties (like admittance variations
or wavelength variations) it should be sufficient to refine the sampling
(i.e., use shorter TL sections) only to the extent that the variation
in wave properties is well resolved; beyond this, using shorter TL
sections should have no effect. That is, we do not need to resolve
the electric field variation with the sampling, since that variation
occurs within the TL sections, not across the boundaries as would
be the case for a spatial domain model.

However, for the cases where the resonance is in evidence (e.g., Figure~\ref{fig:results_at_the_corners_100km},
bottom panels) making the sampling finer does make some difference
in the admittance seen at the top of the ionosphere, and it does not
seem possible to achieve a completely stable limit. The resonance
certainly does not go away, but the zero crossing moves a little,
and the conductance swing changes in magnitude. This was at first
confusing because provided that the sampling is sufficient to resolve
the variations in the wave properties, making the sampling finer should
not have an effect.

The explanation seems to be that when the modes are nearly degenerate
(i.e., nearly the same), the modal resolution is not sufficient to
properly differentiate them. We have tested the accuracy of the eigenvectors/eigenvalues
by multiplying the eigenvector by $H_{5}$, and comparing the result
to the eigenvector multiplied by the eigenvalue. Doing this generally
shows agreement to 14 or 15 decimal places, which should be plenty
sufficient. But there is another issue, which is that we find the
$k_{z}$ for each TL section by adjusting it until the real part of
$\omega$ matches the source frequency $(\omega_{0})$ within tolerances,
which are currently set at 1\%. So there is actually error associated
with this tolerance. The error is not associated with finding the
eigenvector/eigenvalue, rather it is associated with finding the \emph{right}
eigenvector/eigenvalue, that is, the ones associated with the frequency
of the source. While the 1\% tolerance seems to be sufficient for
resolving the vertical variation in the wave properties, it may not
be sufficient to properly differentiate the Whistler and Alfvén modes
when they become nearly degenerate.

To understand the effect of this, go back to our discussion of equation~(\ref{eq:bi-modal_same_as_single-mode.}),
which shows that the vertical derivative (derivative with respect
to altitude) increases with the amplitude $N$ of the modes. As explained,
the amplitude will generally increase when the modes come closer into
alignment, and will actually tend to infinity if the modes approach
perfect alignment (become degenerate). So when the modes are nearly
aligned, the error associated with the 1\% tolerance could produce
a large error in their amplitudes, and thus in the vertical derivative.
In refining the altitude sampling we explore closer to the degeneracy
where the modes are nearly aligned, and at some point the modal resolution
will become comparable to the separation between the modes, which
will introduce numerical error into the vertical derivative. If the
sampling is greatly over-refined the error could dominate and produce
artificial spiky structures over very small scales.

However, in exploring different altitude samplings, we find a remarkable
degree of robustness in both the cutoff of the electric field and
the electric-field spikes seen in panels~c2~and~f2 of Figure~\ref{fig:results_at_the_corners_100km}.
In the particular example of a 100~km transverse wavelength with
density $4.7\times10^{9}\,\mathrm{m}^{-3}$, when the sampling is
increased from a few tenths of a kilometer to as much as 4~km, both
the cutoff and the downward electric-field spike are found to persist
with only modest change. Hence it would seem that both the cutoff
and the spike are real predictions of the model. In particular, although
over-refinement of the altitude sampling could produce artificial
spikiness on small scales, the downward electric-field spikes seen
in panels~c2~and~f2 of Figure~\ref{fig:results_at_the_corners_100km}
are not caused by this and appear to be real predictions of the model. 

On the other hand, the spikes seen in the admittance function in panel~a
of Figure~\ref{fig:final_results} are much less likely to be real.
These admittance function spikes are produced when the before-mentioned
electric-field spikes pass very close to zero in the complex plane,
and so could be very sensitive to small changes in the electric-field
spikes. If over-refinement of the sampling has created artificial,
very-small-scale spiky-structures riding on top of the (apparently)
real electric field spike, these could produce very sharp looking
spikes in the admittance function that are also artificial. And it
is certainly reasonable to worry that there could be other, more important
effects. However, with the exception of the cases when the ionosphere
is in resonance (like in the bottom panels of Figure~\ref{fig:results_at_the_corners_100km}),
we find below that the conductance curves, and in particular the conductance
seen at the top of the ionosphere, are not very sensitive to changes
in the altitude sampling, as long as it is sufficiently dense. 

When the sampling is increased to 8~km in the example just discussed,
the cutoff of the electric field is lost and the trace reverts nearly
to what is seen in panel~c1 of Figure~\ref{fig:results_at_the_corners_100km},
which shows the case where only the Alfvén wave is used in the model.
This tells us something about the other extreme, when the sampling
is insufficient. Apparently, insufficient altitude sampling causes
the effect of coupling into the Whistler wave to be lost. So while
we are still learning about what level of sampling is ``sufficiently
dense,'' it is clearly important not to skip over the degeneracy
entirely.

With regard to the spatial domain, finite difference models, there
are a couple of related issues to mention here. First, it seems that
these models will require very-fine altitude-sampling to resolve the
effect of coupling into the Whistler wave. This is because the coarsest
usable sampling of 4~km (in the example just discussed) actually
represents something much finer than 4~km altitude \emph{resolution},
since there is continuous resolution within each 4~km TL section.
The second issue is that these temporal/spatial domain models make
low-frequency approximations (see Section~\ref{sec:Background})
that could affect the degeneracy itself. Even if the coupling effect
were deemed sacrificable, the degeneracy seems to be responsible for
the rapid variation in the Alfvén admittance that reflects the Alfvén
mode, and generally leads to cutoff of the electric field even when
the Whistler mode is not included in the calculation (Figures~\ref{fig:results_at_the_corners_100km}~and~\ref{fig:results_at_the_corners_1000km}).
So in order to replicate the electric field cutoff in any form using
these temporal/spatial domain models, it will likely be important
to make sure that the degeneracy is properly represented in the dispersion
relation. 

\textbf{Mitigation Measures: }For the cases where the tangent function
behavior is in evidence, we have found that the end points of the
conductance curves at 400~km suffer variations according to the particular
sampling of the region around the degeneracy. There is certainly no
doubt about the emergence of the tangent function behavior, since
we found that behavior also in the Alfvén-mode-only model where there
is no issue of modal resolution, and also no kinky or spiky features
(see panel~d1 of Figure~\ref{fig:results_at_the_corners_100km}),
and also since in panel~a of Figure~\ref{fig:final_results} the
behavior is seen to arise gradually and consistently with increasing
density. But the current modal resolution does not seem to be sufficient
to achieve quantitative accuracy for the tangent function signature. 

Hence, in order to provide the quantitative results shown in panels~a,~b,~and~c
of Figure~\ref{fig:final_results}, we have run the model for an
ensemble of different altitude samplings and chosen the median result.
In doing this we ignore the sharp features and choose the curve that
is the median over the greatest fraction of the altitude region. In
some cases the median curve has very sharp features in the admittance
function, at the location of the degeneracy, and so it seems that
these features do not represent any serious corruption of the calculation. 

The ensemble curves are normally very tightly bunched, except when
the electrical thickness becomes large and the tangent function signature
begins to emerge. Since there are now plenty of reasons to feel certain
about the emergence of the tangent function, and since the curves
in panels~f~and~g change continuously and consistently, we view
this as a strictly quantitative limitation, which does not affect
the qualitative conclusions. Thus, it may be that ionospheric behavior
is sometimes very difficult to predict with quantitative accuracy.
However, this does not mean that we cannot understand it. And, a better
understanding of the degeneracy may be possible with additional research.

\textbf{Summary of Quantitative Results: }The conclusions are summarized
in panels~b,~and~c of Figure~\ref{fig:final_results}, which show
the input conductance at 400~km plotted versus density and wavelength,
respectively. The curves are again normalized to the electrostatic
result for each density, which is therefore represented by the horizontal
dashed black line at one. The panels represent two different ways
of presenting the same results, and from either panel we can see that
the (normalized) ionospheric conductance decreases with increasing
wavelength, and has only a weak dependence on density when the transverse
wavelength exceeds about 200~km (meaning that the un-normalized conductance
has a nearly linear dependence on density). This long-wavelength conductance
is much less than the electrostatic prediction (the field line integrated
conductivity). At the other extreme, as the transverse wavelength
decreases the electrical thickness increases, and it is necessary
to refer to the end-points of the curves in panel~a to understand
the behavior. The electrostatic predictions appear to be pretty good
when the density is low and the wavelength short, but this represents
a quite limited region of parameter space, and we have also found
above (see discussion of columns~f~and~g of Figure~\ref{fig:ModelResultsUsingArtificialWaves})
that it is somewhat fortuitous. Outside of this region the electromagnetic
and electrostatic predictions are very different.

\textbf{\textcolor{black}{Implementation Challenges: }}The investigation
has been limited to transverse scales between 100~km and 1000~km
partly because there were difficulties evaluating the model outside
of this range. Consistently sorting the modes is one of the most problematic
aspects of model implementation, as it is difficult to devise an algorithm
that works in all cases. The calculation optimizes the parallel wavelength
$(k_{z})$ to the target operating frequency $(\omega_{0})$, and
a single failure of the mode sorting can cause the optimization to
fail. These failures are easy to detect, but not so easy to fix. Our
current sorting algorithm has been made to work over the range investigated,
but occasionally fails outside of this range. 

And there is also the difficulty with modal resolution that was already
discussed, where $\omega_{0}$ cannot always be matched with sufficient
accuracy. In the wavelength regime greater than 1000~km it was sometimes
not even possible to achieve the relatively-loose 1\% tolerance employed
above. The difficulty is that the real part of $\omega$ $(\omega_{r})$
turns out to be very sensitive to $k_{z}$. The achievable accuracy
appears to be limited by the precision of the representation of $k_{z}$,
and by the precision of the arithmetic for finding $\omega_{r}$ from
$k_{z}$, which is currently double precision. Thus in order to improve
the modal resolution we need to use higher precision arithmetic, such
as quadruple precision arithmetic. And also we need to understand
whether the modes are actually becoming fully degenerate, in which
case increasing the precision cannot fully solve the problem. In this
case a theoretical solution would be needed to fully understand the
signal propagation through the degeneracy. It will also help to use
a higher frequency for $\omega_{0}$ (i.e., higher transverse velocity),
in situations where it is physically appropriate.

The good news is, it is really only necessary to do a one-time evaluation
of the eigenmodes for a sufficiently fine sampling of transverse wavelength,
density, transverse velocity, and altitude (i.e., collision frequency),
and thereafter the model can be run almost instantly for an arbitrary
altitude profile and wavelength. Matching the boundary conditions
does not appear to require any additional numerical precision. Hence,
the model could eventually be made widely available for use on any
computer.

\section{Conclusions}

In Section~\ref{sec:Gedanken-Experiment} we reframe selected results
from \citeauthor{cosgrove-2016}~{[}2016{]} using transmission line
theory. When electromagnetic theory can be approximated using a single
wave mode, then electrostatic theory (see Section~\ref{sec:Introduction}
for disambiguation) is also applicable if the following three conditions
hold: (1) the wave-Pedersen conductivity agrees with the zero-frequency
Pedersen conductivity; (2) the wavelength is long compared to the
system size; and (3) the dissipation scale length is long compared
to the system size. With the exception of condition~(3), it was found
in \citeauthor{cosgrove-2016}~{[}2016{]} that these conditions do
not generally hold for ionospheric science. Moreover in Section~\ref{subsec:Artificial-Examples}
we find that when the system supports two wave modes that couple,
then additional conditions should be added that are also not satisfied. 

Therefore, in this work we develop an electromagnetic calculation
of the ionospheric input-admittance based on transmission line theory.
The real part of the input-admittance would normally be called the
ionospheric conductance, which has been previously estimated as the
field line integrated conductivity, based on electrostatic theory
(such as the assumption that the electric field maps unchanged through
the ionosphere). The electromagnetic calculation, which also provides
the field quantities resolved over altitude, finds that the ionospheric
response can be approximated as the superposition of two wave modes,
which are quantified by the eigenvectors and eigenvalues of equations
of motion derived from fluid theory. The major assumption used in
the calculation is that highly lossy modes do not couple significantly
into the dynamics, and so can be omitted. Especially, the mode often
called the ion-acoustic wave, or alternatively the slow-mode MHD wave,
is assumed not to couple into 100~km-1000~km transverse wavelength
signals arriving from the magnetosphere (the range of scales analyzed
herein). The ion-acoustic mode is found to have a dissipation scale
less, and generally much less than 10~km at these scales. If this
mode were to couple strongly into the dynamics it would prevent the
signal (for example the electric field) from penetrating deeply into
the ionosphere, and thus omitting this mode is required if there is
to be any chance of reproducing electrostatic theory. A more legitimate
reason for omitting this mode is found in the fact that the corresponding
eigenvector is nearly orthogonal to those of the two modes chosen
for inclusion, which we have called the Alfvén and Whistler modes,
in analogy to the physically defined waves bearing those names.

Regarding the naming of the modes, it should be understood that we
are equating eigenmodes (i.e., eigenvectors and eigenvalues) over
the full altitude extent of the ionosphere (actually from 100~km
to 400~km). Because the collision frequencies for ions and electrons
change dramatically over this altitude range, the physical nature
of our modes is not fixed (Section~\ref{sec:Wave-Modes-Relevant-to-Ionospheric-Science}).
The more common approach to defining wave modes is based on physical
categorization, where simplifying assumptions are made in order to
obtain a relatively simple dispersion relation (for example, the high-frequency
assumption that ions are immobile). Hence there is not a one-to-one
mapping between the modes that we define and the waves that are commonly
considered in space physics. For example, in \citeauthor{cosgrove-2016}~{[}2016{]}
we described the reason for choosing the name ``Whistler mode''
and explained that it actually encompasses both the traditional Whistler
wave and the fast-mode MHD wave, which are normally formed from high-
and low-frequency approximations, respectively {[}e.g., \citealp{stix-1992}{]}.
Our methods are not naturally conducive to a physical categorization
of waves, and such a thing would be outside the scope of this work.
Our focus here is on creating an electromagnetic calculation of the
ionospheric conductance that can validate or replace the current electrostatic
baseline (the field line integrated conductivity).

As an initial, validation step we show that the calculation would
reproduce the electrostatic results exactly if the waves found from
the fluid equations satisfied the conditions identified in the first
paragraph of this section. In addition to reproduction of the field
line integrated conductivity, the electric field is also seen to map
unchanged through the ionosphere. These confirmations are made by
artificially modifying the real waves, such that the conditions are
satisfied. After this we move on to employing the actual wave properties
determined from the electromagnetic five-moment fluid equations, and
the results have also been cross-checked using the smaller equation-set
from \citeauthor{cosgrove-2016}~{[}2016{]}. 

The results found using the actual wave properties do not, for the
most part, support the calculation of ionospheric conductance as the
field line integrated conductivity. For a low density ionosphere ($4.7\times10^{9}\,\mathrm{m}^{-3}$)
with 100~km transverse wavelength we find a modest 7\% reduction
from the field line integrated conductivity. But increasing the density
to $1.0\times10^{11}\,\mathrm{m}^{-3}$ produces a resonance associated
with short parallel wavelength, such that the conductance becomes
negative (energetically, this implies that non-electromagnetic forms
of energy are important {[}\citealp{vasyliunas+song-2005}; \citealp{cosgrove-2016}{]}).
The resonance goes away when the transverse wavelength is increased
to 1000~km, but in this case a severe mismatch arises in the \emph{E}
region that prevents the signal from penetrating deeply. There is
a sharp cutoff of the electric field that would be very hard to resolve
using a spatial-domain method. The mismatch is associated with a degeneracy
or near-degeneracy of the equations of motion, and thus might also
be sensitive to the approximations made in resistive MHD simulations.
Together with the fact that the wave-Pedersen conductivity is less
than the usual (zero-frequency) Pedersen conductivity, there is a
result that the conductance is reduced from the field-line-integrated
conductivity by $73\%$ and $84\%$, for densities of $4.7\times10^{9}\,\mathrm{m}^{-3}$
and $1.0\times10^{11}\,\mathrm{m}^{-3}$, respectively. We have also
verified that similar results are obtained when only the Alfvén wave
is used in the calculation (which requires abandoning some of the
physical boundary conditions), including the resonance and its associated
negative conductance. A quantitative summary of the results is given
in Figure~\ref{fig:final_results}, although it may be difficult
to understand this figure without reading (at least) the whole of
Section~\ref{sec:Model-Results}.

We also evaluate the practice of using electrostatic waves as approximations
for electromagnetic waves in the collisional ionospheric plasma. By
electrostatic waves we mean the practice of using $\vec{\nabla}\times\vec{E}=0$
in the Maxwell equations, but otherwise retaining the time dependence
in the equations of motion, so that wave behavior is supported (see
Section~\ref{sec:Introduction} for disambiguation). This simplification
leads to replacing the Maxwell equations with the Poisson equation
(Appendix~A). We find that electrostatic waves may be good approximations
for electromagnetic waves when the transverse wavelength is less than
about 100~m. But above this scale we find significant differences,
such as parallel wavelengths that are too long by orders of magnitude,
and wave-admittances that are very different. And we could find no
electrostatic counterpart for the Whistler wave, which we find to
be an important contributor in the $E$-region ionosphere, where the
wavelength and dissipation scale length for the Alfvén wave become
very short. Hence, our overall finding is not just that a wave description
is required, but that an electromagnetic wave description is required
{[}\citealp{vasyliunas-2012}; \citealp{cosgrove-2016}{]}.

We regard our form of calculation as the natural electromagnetic generalization
of electrostatic theory, just as transmission line theory is used
in electrical engineering to generalize electrostatic theory for circuits
with dimensions comparable to the wavelength. Our method provides
a direct, intuitive, and nearly analytical calculation of the ionospheric
input admittance, which being a circuit quantity is defined only for
a linear system in steady state. Steady state does not mean setting
the time derivatives to zero in the equations of motion, rather it
means allowing evolution to proceed until everything stops changing.
As explained in Section~\ref{sec:Background}, temporal/spatial-domain
simulations are not well suited to this purpose. Hence, our method
of calculation is complimentary to the temporal/spatial-domain approach,
and is especially well suited to our goals, which are as much theoretical
as they are modeling. Our results are of a rigorous, theoretical nature,
while at the same time providing physical understanding. Thus they
allow for evaluating whether complex numerical simulations based on
``resistive MHD'' are respecting the physical fundamentals.

If possible, it should be evaluated whether temporal/spatial-domain
simulations are able to reproduce our results, and if not to understand
where the difference is arising. However, there appear to be some
serious obstacles, which have been discussed briefly in Sections~\ref{sec:Introduction}~and~\ref{sec:Background}.
One of these is the implementation of the source as a boundary condition,
which does not account for the back-reflected wave. Another is the
importance of the degeneracy in the equations of motion, which requires
very high vertical resolution, and may be sensitive to the simplifications
that are usually made to eliminate the radio-frequency modes. To verify
that the dispersion relation is adequate the equations of motion can
be linearized and used in the TL model, to see if there is any difference
from the results using the full dispersion relation. Although, we
note that the TL method also has limitations with respect to the degeneracy,
which arise from the limited modal resolution (Section~\ref{subsec:Modeling-Limitations}).
We hope that our results will stimulate others to study the problem,
as it is really by comparing different methods that the most definitive
results can be obtained.

The results are important in at least two different ways. The results
at high densities and short wavelengths {[}i.e. $(\sim10^{11}\,\mathrm{m}^{-3},\,\sim100\,\mathrm{km})${]}
suggest unexpected behavior on the scale of auroral arcs, which scale
also includes many other ionospheric phenomena, such as medium-scale
traveling ionospheric disturbances {[}e.g., \citealp{garcia+etal-2000};
\citealp{cosgrove+tsunoda-2004a}{]}, \emph{E }region dynamos associated
with sporadic \emph{E} {[}e.g., \citealp{haldoupis+etal-1996b}; \citealp{cosgrove+tsunoda-2002b};
\citealp{hysell+etal-2004}{]}, and equatorial spread \emph{F} {[}e.g.,
\citealp{tsunoda+etal-1982}; \citealp{aviero+hysell-2012}; \citealp{Tsunoda-2015}{]}.
The results at longer wavelengths affect global modeling of the magnetosphere,
which generally uses the ionospheric conductance to establish an inner
boundary condition {[}e.g., \citealp{Raeder+etal-1998}; \citealp{gombosi+etal-2001};
\citealp{toffoletto+etal-2003}; \citealp{lyon+etal-2004}; \citealp{janhunen+etal-2012};
\citealp{lotko+etal-2014}{]}. Our results suggest that it may be important
to, at a minimum, establish an effective scale for the magnetosphere-ionosphere
interaction, and to use the conductance computed for this scale. Research
on Poynting flux events could be used for this purpose {[}e.g., \citealp{knipp+etal-2011};
\citealp{cosgrove+etal-2014}; \citealp{pakhotin+etal-2020}; \citealp{pakhotin+etal-2021}{]}.
The recent paper by \citeauthor{pakhotin+etal-2021}~{[}2021{]} finds
that half the Poynting flux is associated with transverse scales less
than 250~km, and if this relatively short scale characterizes the
magnetosphere-ionosphere interaction the results presented here are
very relevant. And even if the effective scale is longer than our
current upper-scale of 1000~km, the physical understanding derived
herein does not give any reason to expect that electrostatic theory
would become any more relevant.

This work has focused on developing modeling methodology and only
a few initial findings could be communicated. There is a lot more
to do, both in developing the calculation into a model and in investigating
its behavior. Three-dimensional disturbances can be modeled in the
vertically stratified ionosphere by forming wavepackets. The range
of transverse scales covered by the model should be extended both
downward and upward, so that such horizontally localized wave packets
can be constructed. This would allow the ``mapping'' of electric-field
(and the other field-quantities in the equation-set) over altitude
to be modeled for more realistic disturbances, for possible experimental
validation. The heating of the ionosphere should be calculated and
compared to Poynting flux, to see if other forms of energy are playing
a role. The nature of the altitude transition in wave properties should
be studied, as there appears to be a degeneracy or near degeneracy
having a profound effect. Additional boundary conditions should be
developed so that the effects of the Thermal and Ion waves can be
studied. And finally, the end goal is to make a model that is sufficiently
robust that it can be deployed for public use, such as through the
Community Coordinated Modeling Center (CCMC), the Integrated Geoscience
Observatory (InGeo) {[}\citealp{bhatt+etal-2020}{]}, or other systems
for sharing methodology. 

It is difficult to know where the model stands with respect to observational
studies, since it is really necessary to evaluate each observational
campaign in light of the model expectations for the particular event
or statistical sampling of events. The problem is not easy. It is
difficult to arrange for simultaneous and colocated (along $\vec{B_{0}}$)
measurements of electric field in the \emph{E} and \emph{F} regions.
And even though this is accomplished, it is difficult to sufficiently
define the state of the background ionosphere, and the shape of the
incoming signal. Resonant systems are notoriously sensitive and hard
to model quantitatively. Like musical instruments, interferometers
and microwave filters always require an accommodation for tuning.
Thus, since the TL theory is predicting wavelike effects for the ionosphere,
observational validation will require close attention to the sensitivity
in characterization, along with the full cadre of observational capabilities.

TL theory is very useful for designing electrical circuits because
it allows for understanding how they work. In bringing this theory
to the ionosphere, it is important to distinguish between the detailed
results that are presented herein as examples, and the physical conclusions
that are far more robust. The finding that the electrical thickness
of the ionosphere is significant (panel~l of Figure~\ref{fig:wave_analysis_and_phase_rotation})
involves very little theoretical or computational uncertainty; it
does not rely on the full complexity of the transmission line model;
it is just a finding about the parallel wavelengths of the eigenmodes
and how they depend on collision frequency. The finding for electrical
thickness is an indicator for wavelike effects tied closely to the
electromagnetic 5-moment fluid equations themselves, with an accordant
level of physical relevance. The electrical thickness is the amount
of phase rotation for a signal traversing the ionosphere, and we find
that it often exceeds $90^{\circ}$, meaning a complete failure of
electric field mapping, even for transverse wavelengths as long as
100~km. And the findings that the wave-admittances are not compatible
with electrostatic theory (Section~\ref{sec:Gedanken-Experiment}),
and that there are two interacting modes to consider (Section~\ref{sec:Wave-Modes-Relevant-to-Ionospheric-Science}),
and that these modes become (nearly?) degenerate (Section~\ref{sec:Model-Results}),
are similarly robust.

\section{Appendix A: Matrix Form of the Electromagnetic Fluid Equations\label{subsec:The-Fluid-Equations}}

We assume that electrodynamics in the ionosphere can be described
by the 5-moment fluid equations {[}e.g., \citealp{schunk+nagy-2009}
(eqns. 5.22 in the Second Edition){]} for electrons and one species
of ion, plus the dynamical Maxwell's equations, shown together in~(\ref{eq:generalIono}),
where the subscripts denote either ions ($i$), electrons ($e$),
or neutrals ($n$), $m_{\alpha}$ are the masses, $\vec{v}_{\alpha}$
are velocities, $p_{\alpha}$ are the pressures, $\nu_{\alpha\beta}$
is the collision frequency between species~$\alpha$ and species~$\beta$,
$T_{n}$ is the temperature of the neutral atmosphere (assumed independent),
$e$ is the absolute value of charge for an electron, $\epsilon_{0}$
is the permittivity of free space, $\mu_{0}$ is the permeability
of free space, $\xi$ is the recombination coefficient, $Q$ is the
background ionization rate, $\vec{E}$ is the electric field, and
$\delta\vec{B}$ is the perturbation magnetic field (i.e., $\vec{B}=\vec{B}_{0}+\delta\vec{B}$,
where $\vec{B}_{0}$ is a background magnetic field, and it is assumed
that $\nabla\times\vec{B}_{0}=0$):\renewcommand{\theequation}{A.\arabic{equation}}\setcounter{equation}{0}
\begin{eqnarray}
\frac{\partial n_{e}}{\partial t} & = & -\vec{\nabla}\cdot\left(n_{e}\vec{v}_{e}\right)+Q-\xi\left(\frac{n_{e}+n_{i}}{2}\right)^{2}\nonumber \\
\frac{\partial n_{i}}{\partial t} & = & -\vec{\nabla}\cdot\left(n_{i}\vec{v}_{i}\right)+Q-\xi\left(\frac{n_{e}+n_{i}}{2}\right)^{2}\nonumber \\
\frac{\partial\vec{v}_{e}}{\partial t} & = & -\frac{e}{m_{e}}\left(\vec{E}+\vec{v}_{e}\times\vec{B}\right)-\nu_{en}\left(\vec{v}_{e}-\vec{v}_{n}\right)-\nu_{ei}\left(\vec{v}_{e}-\vec{v}_{i}\right)-\left(\vec{v}_{e}\cdot\vec{\nabla}\right)\vec{v}_{e}-\frac{\vec{\nabla}p_{e}}{m_{e}n_{e}}\nonumber \\
\frac{\partial\vec{v}_{i}}{\partial t} & = & \frac{e}{m_{i}}\left(\vec{E}+\vec{v}_{i}\times\vec{B}\right)-\nu_{in}\left(\vec{v}_{i}-\vec{v}_{n}\right)-\nu_{ie}\left(\vec{v}_{i}-\vec{v}_{e}\right)-\left(\vec{v}_{i}\cdot\vec{\nabla}\right)\vec{v}_{i}-\frac{\vec{\nabla}p_{i}}{m_{i}n_{i}}\nonumber \\
\frac{\partial p_{e}}{\partial t} & = & -\left(\vec{v}_{e}\cdot\vec{\nabla}\right)p_{e}-\frac{5}{3}p_{e}\left(\vec{\nabla}\cdot\vec{v}_{e}\right)-2\frac{m_{e}\nu_{en}}{m_{e}+m_{n}}\left(p_{e}-n_{e}k_{B}T_{n}\right)-2\frac{m_{e}\nu_{ei}}{m_{e}+m_{i}}\left(p_{e}-\frac{n_{e}}{n_{i}}p_{i}\right)\nonumber \\
 & + & \frac{2}{3}\frac{n_{e}m_{e}m_{n}\nu_{en}}{m_{e}+m_{n}}\left|\vec{v}_{e}-\vec{v}_{n}\right|^{2}+\frac{2}{3}\frac{n_{e}m_{e}m_{i}\nu_{ei}}{m_{e}+m_{i}}\left|\vec{v}_{e}-\vec{v}_{i}\right|^{2}\nonumber \\
\frac{\partial p_{i}}{\partial t} & = & -\left(\vec{v}_{i}\cdot\vec{\nabla}\right)p_{i}-\frac{5}{3}p_{i}\left(\vec{\nabla}\cdot\vec{v}_{i}\right)-2\frac{m_{i}\nu_{in}}{m_{i}+m_{n}}\left(p_{i}-n_{i}k_{B}T_{n}\right)-2\frac{m_{i}\nu_{ie}}{m_{i}+m_{e}}\left(p_{i}-\frac{n_{i}}{n_{e}}p_{e}\right)\nonumber \\
 & + & \frac{2}{3}\frac{n_{i}m_{i}m_{n}\nu_{in}}{m_{i}+m_{n}}\left|\vec{v}_{i}-\vec{v}_{n}\right|^{2}+\frac{2}{3}\frac{n_{i}m_{i}m_{e}\nu_{ie}}{m_{i}+m_{e}}\left|\vec{v}_{i}-\vec{v}_{e}\right|^{2}\nonumber \\
\epsilon_{0}\frac{\partial\vec{E}}{\partial t} & = & -e\left(n_{i}\vec{v}_{i}-n_{e}\vec{v}_{e}\right)+\mu_{0}^{-1}\vec{\nabla}\times\delta\vec{B}\nonumber \\
\frac{\partial\delta\vec{B}}{\partial t} & = & -\vec{\nabla}\times\vec{E}.\label{eq:generalIono}
\end{eqnarray}
Sources such as gravity and photoelectrons are omitted from equation~(\ref{eq:generalIono}),
and the neutral wind velocity $(\vec{v}_{n})$ will be set to zero.

We follow the approach in \citeauthor{cosgrove-2016}~{[}2016{]}
to analyze the equation-set~(\ref{eq:generalIono}). The equations
are Fourier transformed in space and linearized. Linearization is
justified by the idea that we are exploring the dynamical evolution
of small perturbations (i.e., second order terms negligible) about
some background state. The background state should be an exact solution
to the full, non-linear equation-set~(\ref{eq:generalIono}), so
that the equations will be satisfied in zeroth order, with only the
higher order equations remaining to be solved. Therefore, we will
use the only known exact solution to equation set~(\ref{eq:generalIono}),
which is thermal equilibrium, that is, the zero-velocity homogeneous
plasma with $n_{e}=n_{i}=\bar{n}_{0}$ , temperatures equal to the
neutral temperature ($T_{n}$), pressures equal to $k_{B}T_{n}\bar{n}_{0}$,
$\vec{E}=0$, $\delta\vec{B}=0$, and $Q$ set to balance the recombination
term, so that all the time derivatives are zero.

Thermal equilibrium is technically a special case of electrostatic
equilibrium, in the sense that all the time derivatives are zero.
However, the non-trivial \textcolor{black}{solutions of} \textcolor{black}{electrostatic
theory are derived by linearizing the equation-set~(}\ref{eq:generalIono})
{[}\citealp{farley-1959}{]} (more commonly a reduced set of equations
is used), and so a background state that satisfies the equations is
also needed to derive \textcolor{black}{electrostatic equilibrium
(otherwise, the nonlinear terms cannot be made small). Here, again,
thermal equilibrium is the only rigorous choice, being the only known
exact solution. So both electrostatic equilibrium and the electromagnetic
solution we develop here are, canonically, solutions that describe
small perturbations about thermal equilibrium. (The generalization
to small perturbations about an inhomogeneous background state is
discussed in Section~\ref{sec:modleIonosphereAdmittance}, and is
heuristic in nature.) Hence, it is appropriate to compare our electromagnetic
solution with electrostatic equilibrium, and the former can serve
as a test for the latter.}

In thermal equilibrium the electron and ion velocities and the electric
field are all zero. Hence, we write the dynamical variables as the
sum of zeroth and first order parts as follows,
\begin{align}
n_{e} & =\bar{n}_{0}+\delta n_{e},\nonumber \\
n_{i} & =\bar{n}_{0}+\delta n_{i},\nonumber \\
\vec{v}_{e} & =\delta\vec{v}_{e},\nonumber \\
\vec{v}_{i} & =\delta\vec{v}_{i},\nonumber \\
p_{e} & =k_{B}T_{n}\bar{n}_{0}+\delta p_{e},\nonumber \\
p_{i} & =k_{B}T_{n}\bar{n}_{0}+\delta p_{i,}\nonumber \\
\vec{E} & =\delta\vec{E},\nonumber \\
\vec{B} & =\vec{B}_{0}+\delta\vec{B},\label{eq:perturbationEqns}
\end{align}
where $T_{n}$ is the neutral temperature, $k_{B}$ is Boltzmann's
constant, $\vec{B}_{0}$ is the geomagnetic field, and $\bar{n}_{0}$
is the background plasma density.

Substituting the forms~(\ref{eq:perturbationEqns}) into the equations
of motion~(\ref{eq:generalIono}) and taking the Fourier transform
in space gives, after arranging into matrix notation, the general
equation~(\ref{eq:matrixEqnWithNonlin}) from Section~(\ref{sec:Solution}),
where
\begin{eqnarray*}
\vec{X} & = & \left(V_{ex},V_{ey},V_{ez},V_{ix},V_{iy},V_{iz},E_{x},E_{y},E_{z},c\delta B_{x},c\delta B_{y},c\delta B_{z},\delta N_{e},\delta N_{i},\delta P_{e},\delta P_{i}\right),\\
V_{\alpha\beta} & = & \frac{m_{\alpha}}{e}\omega_{\alpha}v_{\alpha\beta},\\
\delta N_{\alpha} & = & \frac{m_{e}}{m_{i}}\frac{e}{\epsilon_{0}\left|\vec{k}\right|}\delta n_{\alpha},\\
\delta P_{\alpha} & = & \frac{m_{e}}{m_{i}}\frac{e}{\epsilon_{0}\left|\vec{k}\right|k_{B}T_{n}}\delta p_{\alpha},
\end{eqnarray*}
$\omega_{\alpha}=\sqrt{\frac{\bar{n}_{0}e^{2}}{m_{\alpha}\epsilon_{0}}}$
are the electron and ion plasma frequencies, $\vec{k}$ is the wavevector
for the Fourier transform, and the matrix $H_{5}$ is shown in detail
in Figure~\ref{fig:H_Matrix}. In this figure, $\Omega_{i0}=eB_{0}/m_{i}$
is the ion gyro-frequency, $\Omega_{e0}=eB_{0}/m_{e}$ is the electron
gyro-frequency, $\hat{B}=\vec{B}_{0}/B_{0}$, $\vec{V}_{e0}$ and
$\vec{V}_{i0}$ are the (not rescaled) background electron and ion
velocities (which we set to zero, for thermal equilibrium), and $P_{e0}$
and $P_{i0}$ are the background electron and ion pressures (which
we set to $k_{B}T_{n}\bar{n}_{0}$, for thermal equilibrium). The
rescalings provide that the dynamical variables all have units of
electric field, and that the matrix elements of $H_{5}$ all have
units of frequency. To be clear, $H_{5}$ is formed without dropping
any terms or making any approximations.\renewcommand{\thefigure}{A.\arabic{figure}}\setcounter{figure}{0}
\begin{sidewaysfigure}
{\tiny{}{} 
\begin{align*}
H_{5}= & i\left[\begin{array}{cccccc}
\nu_{en}+\nu_{ei}-i\vec{k}\cdot\vec{V}_{e0}+\gamma_{x}V_{e0x} & \Omega_{e0}\hat{B}_{z}+\gamma_{y}V_{e0x} & -\Omega_{e0}\hat{B}_{y}+\gamma_{z}V_{e0x} & -\nu_{ei}\sqrt{\frac{m_{e}}{m_{i}}} & 0 & 0\\
-\Omega_{e0}\hat{B}_{z}+\gamma_{x}V_{e0y} & \nu_{en}+\nu_{ei}-i\vec{k}\cdot\vec{V}_{e0}+\gamma_{y}V_{e0y} & \Omega_{e0}\hat{B}_{x}+\gamma_{z}V_{e0y} & 0 & -\nu_{ei}\sqrt{\frac{m_{e}}{m_{i}}} & 0\\
\Omega_{e0}\hat{B}_{y}+\gamma_{x}V_{e0z} & -\Omega_{e0}\hat{B}_{x}+\gamma_{y}V_{e0z} & \nu_{en}+\nu_{ei}-i\vec{k}\cdot\vec{V}_{e0}+\gamma_{z}V_{e0z} & 0 & 0 & -\nu_{ei}\sqrt{\frac{m_{e}}{m_{i}}}\\
-\nu_{ie}\sqrt{\frac{m_{i}}{m_{e}}} & 0 & 0 & \nu_{in}+\nu_{ie}-i\vec{k}\cdot\vec{V}_{i0}+\gamma_{x}V_{i0x} & -\Omega_{i0}\hat{B}_{z}+\gamma_{y}V_{i0x} & \Omega_{i0}\hat{B}_{y}+\gamma_{z}V_{i0x}\\
0 & -\nu_{ie}\sqrt{\frac{m_{i}}{m_{e}}} & 0 & \Omega_{i0}\hat{B}_{z}+\gamma_{x}V_{i0y} & \nu_{in}+\nu_{ie}-i\vec{k}\cdot\vec{V}_{i0}+\gamma_{y}V_{i0y} & -\Omega_{i0}\hat{B}_{x}+\gamma_{z}V_{i0y}\\
0 & 0 & -\nu_{ie}\sqrt{\frac{m_{i}}{m_{e}}} & -\Omega_{i0}\hat{B}_{y}+\gamma_{x}V_{i0z} & \Omega_{i0}\hat{B}_{x}+\gamma_{y}V_{i0z} & \nu_{in}+\nu_{ie}-i\vec{k}\cdot\vec{V}_{i0}+\gamma_{z}V_{i0z}\\
-\omega_{e} & 0 & 0 & \omega_{i} & 0 & 0\\
0 & -\omega_{e} & 0 & 0 & \omega_{i} & 0\\
0 & 0 & -\omega_{e} & 0 & 0 & \omega_{i}\\
0 & 0 & 0 & 0 & 0 & 0\\
0 & 0 & 0 & 0 & 0 & 0\\
0 & 0 & 0 & 0 & 0 & 0\\
\frac{m_{e}}{m_{i}}\frac{\omega_{e}}{|\vec{k}|}\left(\gamma_{x}-ik_{x}\right) & \frac{m_{e}}{m_{i}}\frac{\omega_{e}}{|\vec{k}|}\left(\gamma_{y}-ik_{y}\right) & \frac{m_{e}}{m_{i}}\frac{\omega_{e}}{|\vec{k}|}\left(\gamma_{z}-ik_{z}\right) & 0 & 0 & 0\\
0 & 0 & 0 & \frac{m_{e}}{m_{i}}\frac{\omega_{i}}{|\vec{k}|}\left(\gamma_{x}-ik_{x}\right) & \frac{m_{e}}{m_{i}}\frac{\omega_{i}}{|\vec{k}|}\left(\gamma_{y}-ik_{y}\right) & \frac{m_{e}}{m_{i}}\frac{\omega_{i}}{|\vec{k}|}\left(\gamma_{z}-ik_{z}\right)\\
\frac{m_{e}}{m_{i}}A_{eix} & \frac{m_{e}}{m_{i}}A_{eiy} & \frac{m_{e}}{m_{i}}A_{eiz} & \frac{m_{e}}{m_{i}}C_{eix} & \frac{m_{e}}{m_{i}}C_{eiy} & \frac{m_{e}}{m_{i}}C_{eiz}\\
\frac{m_{e}}{m_{i}}C_{iex} & \frac{m_{e}}{m_{i}}C_{iey} & \frac{m_{e}}{m_{i}}C_{iez} & \frac{m_{e}}{m_{i}}A_{iex} & \frac{m_{e}}{m_{i}}A_{iey} & \frac{m_{e}}{m_{i}}A_{iez}
\end{array}\right.\\
\\
\\
\\
 & \left.\begin{array}{cccccccccc}
\omega_{e} & 0 & 0 & 0 & \frac{-V_{e0z}}{c}\omega_{e} & \frac{V_{e0y}}{c}\omega_{e} & 0 & 0 & \frac{m_{i}}{m_{e}}\frac{k_{B}T_{n}|\vec{k}|}{m_{e}\omega_{e}}\left(-ik_{x}-\gamma_{x}\right) & 0\\
0 & \omega_{e} & 0 & \frac{V_{e0z}}{c}\omega_{e} & 0 & \frac{-V_{e0x}}{c}\omega_{e} & 0 & 0 & \frac{m_{i}}{m_{e}}\frac{k_{B}T_{n}|\vec{k}|}{m_{e}\omega_{e}}\left(-ik_{y}-\gamma_{y}\right) & 0\\
0 & 0 & \omega_{e} & \frac{-V_{e0y}}{c}\omega_{e} & \frac{V_{e0x}}{c}\omega_{e} & 0 & 0 & 0 & \frac{m_{i}}{m_{e}}\frac{k_{B}T_{n}|\vec{k}|}{m_{e}\omega_{e}}\left(-ik_{z}-\gamma_{z}\right) & 0\\
-\omega_{i} & 0 & 0 & 0 & \frac{V_{i0z}}{c}\omega_{i} & \frac{-V_{i0y}}{c}\omega_{i} & 0 & 0 & 0 & \frac{k_{B}T_{n}|\vec{k}|}{m_{e}\omega_{i}}\left(-ik_{x}-\gamma_{x}\right)\\
0 & -\omega_{i} & 0 & \frac{-V_{i0z}}{c}\omega_{i} & 0 & \frac{V_{i0x}}{c}\omega_{i} & 0 & 0 & 0 & \frac{k_{B}T_{n}|\vec{k}|}{m_{e}\omega_{i}}\left(-ik_{y}-\gamma_{y}\right)\\
0 & 0 & -\omega_{i} & \frac{V_{i0y}}{c}\omega_{i} & \frac{-V_{i0x}}{c}\omega_{i} & 0 & 0 & 0 & 0 & \frac{k_{B}T_{n}|\vec{k}|}{m_{e}\omega_{i}}\left(-ik_{z}-\gamma_{z}\right)\\
0 & 0 & 0 & 0 & -cik_{z} & cik_{y} & -\frac{m_{i}}{m_{e}}|\vec{k}|V_{e0x} & \frac{m_{i}}{m_{e}}|\vec{k}|V_{i0x} & 0 & 0\\
0 & 0 & 0 & cik_{z} & 0 & -cik_{x} & -\frac{m_{i}}{m_{e}}|\vec{k}|V_{e0y} & \frac{m_{i}}{m_{e}}|\vec{k}|V_{i0y} & 0 & 0\\
0 & 0 & 0 & -cik_{y} & cik_{x} & 0 & -\frac{m_{i}}{m_{e}}|\vec{k}|V_{e0z} & \frac{m_{i}}{m_{e}}|\vec{k}|V_{i0z} & 0 & 0\\
0 & cik_{z} & -cik_{y} & 0 & 0 & 0 & 0 & 0 & 0 & 0\\
-cik_{z} & 0 & cik_{x} & 0 & 0 & 0 & 0 & 0 & 0 & 0\\
cik_{y} & -cik_{x} & 0 & 0 & 0 & 0 & 0 & 0 & 0 & 0\\
0 & 0 & 0 & 0 & 0 & 0 & \vec{\gamma}\cdot\vec{V}_{e0}-i\vec{k}\cdot\vec{V}_{e0}+\xi\bar{n}_{0} & \xi\bar{n}_{0} & 0 & 0\\
0 & 0 & 0 & 0 & 0 & 0 & \xi\bar{n}_{0} & \vec{\gamma}\cdot\vec{V}_{i0}-i\vec{k}\cdot\vec{V}_{i0}+\xi\bar{n}_{0} & 0 & 0\\
0 & 0 & 0 & 0 & 0 & 0 & D_{ei} & \frac{2m_{e}\nu_{ei}P_{io}}{(m_{e}+m_{i})k_{B}T_{n}\bar{n}_{0}} & L_{ei} & -2\frac{m_{e}\nu_{ei}}{m_{e}+m_{i}}\\
0 & 0 & 0 & 0 & 0 & 0 & \frac{2m_{i}\nu_{ie}P_{eo}}{(m_{e}+m_{i})k_{B}T_{n}\bar{n}_{0}} & D_{ie} & -2\frac{m_{i}\nu_{ie}}{m_{e}+m_{i}} & L_{ie}
\end{array}\right],\;\mathrm{where}
\end{align*}
\begin{align*}
A_{\alpha\beta}= & \frac{\vec{\gamma}}{|\vec{k}|}\frac{P_{\alpha0}\omega_{\alpha}}{k_{B}T_{n}\bar{n}_{0}}-\frac{5}{3}\frac{i\vec{k}}{|\vec{k}|}\frac{P_{\alpha0}\omega_{\alpha}}{k_{B}T_{n}\bar{n}_{0}}-\frac{4}{3}\frac{(\vec{V}_{\alpha0}-\vec{V}_{n})\frac{\omega_{\alpha}}{|\vec{k}|}}{k_{B}T_{n}/m_{\alpha}}\frac{m_{n}\nu_{\alpha n}}{m_{\alpha}+m_{n}}-\frac{4}{3}\frac{(\vec{V}_{\alpha0}-\vec{V}_{\beta0})\frac{\omega_{\alpha}}{|\vec{k}|}}{k_{B}T_{n}/m_{\alpha}}\frac{m_{\beta}\nu_{\alpha\beta}}{m_{\alpha}+m_{\beta}},\\
C_{\alpha\beta}= & \frac{4}{3}\frac{(\vec{V}_{\alpha0}-\vec{V}_{\beta0})\frac{\omega_{\beta}}{|\vec{k}|}}{k_{B}T_{n}/m_{\alpha}}\frac{m_{\beta}\nu_{\alpha\beta}}{m_{\alpha}+m_{\beta}},\\
D_{\alpha\beta}= & -2\frac{m_{\alpha}\nu_{\alpha n}}{m_{\alpha}+m_{n}}-2\frac{m_{\alpha}\nu_{\alpha\beta}P_{\beta0}}{(m_{\alpha}+m_{\beta})k_{B}T_{n}\bar{n}_{0}}-\frac{2}{3}\frac{m_{n}\nu_{\alpha n}}{m_{\alpha}+m_{n}}\frac{\left|\vec{V}_{\alpha0}-\vec{V}_{n}\right|^{2}}{k_{B}T_{n}/m_{\alpha}}-\frac{2}{3}\frac{m_{\beta}\nu_{\alpha\beta}}{m_{\alpha}+m_{\beta}}\frac{\left|\vec{V}_{\alpha0}-\vec{V}_{\beta0}\right|^{2}}{k_{B}T_{n}/m_{\alpha}},\\
L_{\alpha\beta}= & -i\vec{k}\cdot\vec{V}_{\alpha0}+\frac{5}{3}\left(\vec{\gamma}\cdot\vec{V}_{\alpha0}\right)+2\frac{m_{\alpha}\nu_{\alpha n}}{m_{\alpha}+m_{n}}+2\frac{m_{\alpha}\nu_{\alpha\beta}}{m_{\alpha}+m_{\beta}}.
\end{align*}
}\caption{Matrix $H_{5}$ containing the linearized and Fourier transformed
electromagnetic five-moment fluid equations. In this work $\vec{V}_{e0}$
and $\vec{V}_{i0}$ are set to zero, and $P_{e0}$ and $P_{i0}$ are
set to $k_{B}T_{n}\bar{n}_{0}$.\label{fig:H_Matrix}}
\end{sidewaysfigure}

As can be seen from the homogeneous solution~(\ref{eq:solutionHomogenious})
from Section(\ref{sec:Solution}), the eigenvalues of $H_{5}$ are
the complex frequencies $\omega_{j}$ of the supported waves, where
the real part is the frequency of oscillation, and the imaginary part
is the dissipation rate. The eigenvectors of $H_{5}$ are the polarizations
of the supported waves, that is, they determine the relative amplitudes
and phases for the dynamical variables in $\vec{X}$. 

This formulation is equivalent to deriving the complete linear dispersion
relation for the equation set~(\ref{eq:generalIono}), and solving
it for complex $\omega_{j}$ as a function of real $\vec{k}$. However,
although finding the exact analytical dispersion relation for~(\ref{eq:generalIono})
is possible with due patience, finding the frequencies that satisfy
the dispersion relation still requires finding the 16 complex roots
of a polynomial, and does not determine the polarization vectors.
In our matrix based approach all this work is done by standard matrix
analysis tools. We solve for the eigenvectors/eigenvalues numerically
using numpy.linalg.eig from the NumPy Python library.

In order to evaluate electrostatic waves, we will compare results
using the full equations~(\ref{eq:generalIono}) with results from
the same equations, except with the curl of the electric field assumed
to be zero. To find the reduced equations substitute $\vec{E}=-\vec{\nabla}\phi$
into the equations~(\ref{eq:generalIono}). With this substitution
Faraday's law provides that the magnetic perturbation $\left(\delta\vec{B}\right)$
is zero, and so the three equations of Faraday's law decouple from
the other equations. However, Ampere's law becomes an over-determined
set of three equations for the one variable~$\left(\phi\right)$.
To render these equations consistent, recall that a vector field can
be decomposed into a curl-free part and a divergence-free part, where
the curl-free part is uniquely determined by the divergence of the
vector field. Therefore, with the assumption $\vec{\nabla}\times\vec{E}=0$,
it is sufficient to replace Ampere's law with an equation for $\vec{\nabla}\cdot\vec{E}$,
which we can get by taking the divergence of Ampere's law. In fact,
taking the divergence of Ampere's law with $\vec{E}=-\vec{\nabla}\phi$
gives the time derivative of the Poisson equation, which is equivalent
to the Poisson equation, since the latter has no inhomogeneous term.
So in summary, the electrostatic version of the 5-moment equations
is obtained by dropping Faraday's law and replacing Ampere's law with
its divergence, which is equivalent to using the Poisson equation. 

It should also be noted that taking the curl of Ampere's law under
the assumption $\vec{E}=-\vec{\nabla}\phi$ results in the equation
$\vec{\nabla}\times\vec{J}=\vec{\nabla}\times\vec{\nabla}\times\delta\vec{B}/\mu_{0},$
where $\vec{J}$ is the current density. Generally, because the curl
of the current is not zero, this equation cannot be satisfied when
$\delta\vec{B}=0$, and so the electrostatic waves are not solutions
for the electromagnetic equations. It is generally not consistent
to assume that $\vec{\nabla}\times\vec{E}$ is exactly zero. Rather,
in using electrostatic waves, there is an assumption that $\vec{\nabla}\times\vec{E}$
is sufficiently small such that the electrostatic waves are good approximations
for the actual electromagnetic waves.

The matrix $H_{5ES}$ for the electrostatic version of equation~(\ref{eq:matrixEqnWithNonlin})
is given in Figure~S.1 of the Supplementary Information, where
\[
\vec{X}_{ES}=\left(V_{ex},V_{ey},V_{ez},V_{ix},V_{iy},V_{iz},\left|\vec{k}\right|\phi,\delta N_{e},\delta N_{i},\delta P_{e},\delta P_{i}\right).
\]
The electrostatic waves are obtained as the eigenvalues/eigenvectors
of $H_{5ES}$.

\section{Appendix B: Integral Approximations\label{sec:Appendix-C:-Integral_approximations}}

This appendix describes an additional approximation for the integrals
in the first sum of equation~(\ref{eq:laplaceIntegralFormSolution}),
for the case of modes that have a minimum operating frequency, such
as the Whistler wave, which becomes cutoff when the operating frequency
drops too low. This result goes along with the result~(\ref{eq:drivenSteadyStateSolution})
that applies for modes such as the Alfvén wave, which can propagate
all the way down to DC. We also make some comments on the integrals
for modes below the operating frequency. 

The integral approximation~(\ref{eq:integralWithLinearizations})
for the propagating modes was obtained by linearizing about $k_{z}=k_{0z}$
such that $\mathrm{real\left(\omega_{j}(0,k_{0y},k_{0z})\right)}=\omega_{0}$.
This is only possible for propagating modes, that is, for modes that
can achieve the frequency $\omega_{0}$ for some $k_{z}$. An important
class of modes that we need to include in our model are those which
may not always be able to go low enough in frequency. For example,
when the Whistler wave is cutoff, above the $E$ region, it is because
the frequency cannot go low enough. In this case the greatest contribution
to the integral comes at closest approach to $\omega_{0}$, which
occurs at $k_{z}=0$, and the saturation of frequency as $k_{z}\rightarrow0$
can be modeled by a Taylor series expansion about $k_{z}=0$,\renewcommand{\theequation}{B.\arabic{equation}}\setcounter{equation}{0}
\begin{equation}
\omega_{j}\left(k_{z}\right)\cong\frac{1}{2}\left.\frac{\partial^{2}\omega_{j}}{\partial k_{z}^{2}}\right|_{k_{z}=0}k_{z}^{2}+\omega_{jrmin}+\left.i\omega_{ji}\right|_{k_{z}=0},\label{eq:quadraticModelAtZero}
\end{equation}
where the linear term is omitted because there must be symmetry about
$k_{z}=0$, and we drop terms higher than second order. Numerical
examination of the dependence of $\omega_{j}$ on $k_{z}$ for the
Whistler wave confirms that this model is reasonable in the region
around $k_{z}=0$. In fact, this being the case, we note that for
large $k_{z}$ the curvature becomes negligible compared to the slope,
and we would expect this quadratic model to recover the solution~(\ref{eq:drivenSteadyStateSolution})
from the linear model in the event that there exists some large $k_{z}=k_{0z}$
where $\mathrm{real}\left(\omega_{j}(k_{0z})\right)=\omega_{0}$,
if the slope at $k_{z}=k_{0z}$ is matched. Therefore, for the higher
frequency modes that saturate their lower bound at $k_{z}=0$, we
can use the quadratic model for both propagating and non-propagating
cases, and thereby recover a continuous result across the propagating/non-propagating
divide. To do this we devise the rule of fitting the quadratic form
$ak_{z}^{2}+\zeta$ at the point of closest approach to $\omega_{0}$,
and if this occurs away from $k_{z}=0$, meaning that equality was
achieved and the mode propagates, we fit the slope at $k_{0z}$:
\begin{equation}
\omega_{j}\left(k_{z}\right)=\frac{1}{2k_{0z}}\left.\frac{\partial\omega_{j}}{\partial k_{z}}\right|_{k_{z}=k_{0z}}k_{z}^{2}+\omega_{0}+\left.i\omega_{ji}\right|_{k_{z}=k_{0z}}-\frac{k_{0z}}{2}\left.\frac{\partial\omega_{j}}{\partial k_{z}}\right|_{k_{z}=k_{0z}}.\label{eq:quadraticModelAtk0z}
\end{equation}
This form approaches $0/0$ as $k_{0z}\rightarrow0$, and so when
closest approach happens at $k_{z}=0$ we replace it with the form~(\ref{eq:quadraticModelAtZero}).
By the symmetry property of $\omega_{j}$ about $k_{z}=0$, the quadratic
approximation becomes exact as $k_{z}\rightarrow0$ (higher order
terms vanish), and so the forms~(\ref{eq:quadraticModelAtZero})~and~(\ref{eq:quadraticModelAtk0z})
are consistent, and there will be continuity around $k_{z}=0$. 

Substituting either the form~(\ref{eq:quadraticModelAtZero}) or
the form~(\ref{eq:quadraticModelAtk0z}) into the integral form~(\ref{eq:laplaceIntegralFormSolution})
and using $\vec{h}_{j}^{\prime}$ and $a_{j}^{\prime}$ from~(\ref{eq:integralWithLinearizations})
gives,
\begin{eqnarray*}
 &  & \varint_{-\infty}^{\infty}\mathrm{d}k_{z}\frac{\vec{h}_{j}^{\prime}(k_{z}-k_{0z})a_{j}^{\prime}(k_{z}-k_{0z})}{ak_{z}^{2}+c}\mathrm{e}^{ik_{z}z}\\
 & = & \varint_{-\infty}^{\infty}\mathrm{d}k_{z}\frac{\vec{h}_{j}^{\prime}(k_{z}-k_{0z})a_{j}^{\prime}(k_{z}-k_{0z})}{\left(k_{z}-\sqrt{-ac}/a\right)\left(k_{z}+\sqrt{-ac}/a\right)a}\mathrm{e}^{ik_{z}z},
\end{eqnarray*}
where either $a=\frac{-i}{2}\left.\frac{\partial^{2}\omega_{j}}{\partial k_{z}^{2}}\right|_{k_{z}=0}$
and $c=i\left(\omega_{0}-\omega_{jrmin}\right)+\left.\omega_{ji}\right|_{k_{z}=0}$,
or $a=\frac{-i}{2k_{0z}}\left.\frac{\partial\omega_{j}}{\partial k_{z}}\right|_{k_{z}=k_{0z}}$
and $c=\left.\omega_{ji}\right|_{k_{z}=k_{0z}}+\frac{ik_{0z}}{2}\left.\frac{\partial\omega_{j}}{\partial k_{z}}\right|_{k_{z}=k_{0z}}$.
There are two simple poles, and applying partial fraction expansion
allows application of the residue theorem. Omitting these standard
integration steps we obtain for the integral,
\begin{equation}
\vec{h}_{j}^{\prime}\left(\pm\sqrt{-ac}/a-k_{0z}\right)a_{j}^{\prime}\left(\pm\sqrt{-ac}/a-k_{0z}\right)\frac{i}{2\sqrt{-ac}}\mathrm{e}^{\pm i(\sqrt{-ac}/a)z},\label{eq:solutionFromQuadratic}
\end{equation}
where the $\pm$ option is chosen to give exponential decay away from
the source, depending on the sign of $z$, and the negative option
for $k_{0z}$ is used with the - sign. Testing finds that this relation
does indeed provide for continuity of the wavelength, and rapid reduction
of the Whistler mode dissipation scale length to about ten kilometers,
across the propagating/non-propagating altitude boundary. Also, it
can be shown that the solution~(\ref{eq:solutionFromQuadratic})
reduces to the solution~(\ref{eq:drivenSteadyStateSolution}) when
$k_{0z}$ is large.

The quadratic solution~(\ref{eq:solutionFromQuadratic}) is appropriate
for the Whistler wave, and probably also for the higher frequency
waves. For the Alfvén wave we should continue to use the driven steady-state
solution~(\ref{eq:drivenSteadyStateSolution}), because the quadratic
solution~(\ref{eq:solutionFromQuadratic}) is not appropriate. The
Alfvén wave has a different form of dispersion relation, that crosses
through zero-frequency with a non-zero $k_{z}$ derivative. For example,
the ideal Alfvén wave dispersion relation is simply $\omega=\pm v_{A}k_{z}$,
which is most naturally viewed as having two antisymmetric branches,
but which for purposes of comparing with the quadratic dispersion
relations~(\ref{eq:quadraticModelAtZero})~and~(\ref{eq:quadraticModelAtk0z}),
which are symmetric, can be viewed as having upward and downward V-shaped
branches, which are symmetric and linear except for having a singularity
at $k_{z}=0$. So for the Alfvén wave the linearized approximation
of equation~(\ref{eq:drivenSteadyStateSolution}) is highly appropriate.

The validations given in Section~\ref{sec:Solution} also apply to
the quadratic form~(\ref{eq:solutionFromQuadratic}), as can be seen
by writing out the expression for the residue location in the case~(\ref{eq:quadraticModelAtk0z}).
The residue location comes out to be $k_{0z}\sqrt{1+2i\omega_{ji}/(k_{0z}v_{gj})}$.
Expanding the square-root shows that the first two terms are exactly
the same as those found for the linear approximation of Section~\ref{sec:Solution},
and so the same validations apply.

It remains to develop approximations for the dispersion relations
of the Ion wave, Thermal wave, and of the lowest-frequency cutoff
mode. However, there is every indication that polynomial approximations
can be found as in the above. As long as the denominator is a polynomial
in $k_{z}$, we can solve the integral using the residue theorem,
even if roots must be found numerically. Whether these must be higher
order is not yet known. The biggest obstacle to a higher order approximation
is estimation of the higher orders of derivative of $\omega_{j}$.

\section*{Open Research}

This is a theoretical paper that does not rely on data. No special
software has been used in this research.

\section*{Acknowledgements}

This material is based upon work supported by the National Science
Foundation (NSF) under NSF grant AGS-1344300. 


\end{document}


\part*{Supplementary Information}

\setcounter{page}{1}\renewcommand{\thesection}{S.\arabic{section}}\setcounter{section}{0}\renewcommand{\thefigure}{S.\arabic{figure}}\setcounter{figure}{0}\renewcommand{\theequation}{S.\arabic{equation}}\setcounter{equation}{0}\thispagestyle{plain}This
file contains the Supplementary Information that is referenced in
the paper, Cosgrove, R. B. (2022), An Electromagnetic Calculation
of Ionospheric Conductance that seems to Override the Field Line Integrated
Conductivity. 

\lhead{}

\section*{Contents of this file}
\begin{enumerate}
\item The Matrix $H_{5ES}$
\item Parameter Choices
\item Additional Commentary on Energy Conservation
\item Analytic Dependence for Polarization Vectors
\item Validation Figures and Elaboration
\item Reviews of this and Predecessor Papers, with Replies
\end{enumerate}

\section{The Matrix $H_{5ES}$}

The matrix $H_{5ES}$ is shown in detail in Figure~\ref{fig:ES_Matrix}.

\begin{sidewaysfigure}
{\tiny{}
\begin{align*}
H_{5ES}= & i\left[\begin{array}{cccccc}
\nu_{en}+\nu_{ei}-i\vec{k}\cdot\vec{V}_{e0}+\gamma_{x}V_{e0x} & \Omega_{e0}\hat{B}_{z}+\gamma_{y}V_{e0x} & -\Omega_{e0}\hat{B}_{y}+\gamma_{z}V_{e0x} & -\nu_{ei}\sqrt{\frac{m_{e}}{m_{i}}} & 0 & 0\\
-\Omega_{e0}\hat{B}_{z}+\gamma_{x}V_{e0y} & \nu_{en}+\nu_{ei}-i\vec{k}\cdot\vec{V}_{e0}+\gamma_{y}V_{e0y} & \Omega_{e0}\hat{B}_{x}+\gamma_{z}V_{e0y} & 0 & -\nu_{ei}\sqrt{\frac{m_{e}}{m_{i}}} & 0\\
\Omega_{e0}\hat{B}_{y}+\gamma_{x}V_{e0z} & -\Omega_{e0}\hat{B}_{x}+\gamma_{y}V_{e0z} & \nu_{en}+\nu_{ei}-i\vec{k}\cdot\vec{V}_{e0}+\gamma_{z}V_{e0z} & 0 & 0 & -\nu_{ei}\sqrt{\frac{m_{e}}{m_{i}}}\\
-\nu_{ie}\sqrt{\frac{m_{i}}{m_{e}}} & 0 & 0 & \nu_{in}+\nu_{ie}-i\vec{k}\cdot\vec{V}_{i0}+\gamma_{x}V_{i0x} & -\Omega_{i0}\hat{B}_{z}+\gamma_{y}V_{i0x} & \Omega_{i0}\hat{B}_{y}+\gamma_{z}V_{i0x}\\
0 & -\nu_{ie}\sqrt{\frac{m_{i}}{m_{e}}} & 0 & \Omega_{i0}\hat{B}_{z}+\gamma_{x}V_{i0y} & \nu_{in}+\nu_{ie}-i\vec{k}\cdot\vec{V}_{i0}+\gamma_{y}V_{i0y} & -\Omega_{i0}\hat{B}_{x}+\gamma_{z}V_{i0y}\\
0 & 0 & -\nu_{ie}\sqrt{\frac{m_{i}}{m_{e}}} & -\Omega_{i0}\hat{B}_{y}+\gamma_{x}V_{i0z} & \Omega_{i0}\hat{B}_{x}+\gamma_{y}V_{i0z} & \nu_{in}+\nu_{ie}-i\vec{k}\cdot\vec{V}_{i0}+\gamma_{z}V_{i0z}\\
i\omega_{e}\hat{k}_{x} & i\omega_{e}\hat{k}_{y} & i\omega_{e}\hat{k}_{z} & -i\omega_{i}\hat{k}_{x} & -i\omega_{i}\hat{k}_{y} & -i\omega_{i}\hat{k}_{z}\\
\frac{m_{e}}{m_{i}}\frac{\omega_{e}}{|\vec{k}|}\left(\gamma_{x}-ik_{x}\right) & \frac{m_{e}}{m_{i}}\frac{\omega_{e}}{|\vec{k}|}\left(\gamma_{y}-ik_{y}\right) & \frac{m_{e}}{m_{i}}\frac{\omega_{e}}{|\vec{k}|}\left(\gamma_{z}-ik_{z}\right) & 0 & 0 & 0\\
0 & 0 & 0 & \frac{m_{e}}{m_{i}}\frac{\omega_{i}}{|\vec{k}|}\left(\gamma_{x}-ik_{x}\right) & \frac{m_{e}}{m_{i}}\frac{\omega_{i}}{|\vec{k}|}\left(\gamma_{y}-ik_{y}\right) & \frac{m_{e}}{m_{i}}\frac{\omega_{i}}{|\vec{k}|}\left(\gamma_{z}-ik_{z}\right)\\
\frac{m_{e}}{m_{i}}A_{eix} & \frac{m_{e}}{m_{i}}A_{eiy} & \frac{m_{e}}{m_{i}}A_{eiz} & \frac{m_{e}}{m_{i}}C_{eix} & \frac{m_{e}}{m_{i}}C_{eiy} & \frac{m_{e}}{m_{i}}C_{eiz}\\
\frac{m_{e}}{m_{i}}C_{iex} & \frac{m_{e}}{m_{i}}C_{iey} & \frac{m_{e}}{m_{i}}C_{iez} & \frac{m_{e}}{m_{i}}A_{iex} & \frac{m_{e}}{m_{i}}A_{iey} & \frac{m_{e}}{m_{i}}A_{iez}
\end{array}\right.\\
\\
\\
\\
 & \left.\begin{array}{ccccc}
i\hat{k}_{x}\omega_{e} & 0 & 0 & \frac{m_{i}}{m_{e}}\frac{k_{B}T_{n}|\vec{k}|}{m_{e}\omega_{e}}\left(-ik_{x}-\gamma_{x}\right) & 0\\
i\hat{k}_{y}\omega_{e} & 0 & 0 & \frac{m_{i}}{m_{e}}\frac{k_{B}T_{n}|\vec{k}|}{m_{e}\omega_{e}}\left(-ik_{y}-\gamma_{y}\right) & 0\\
i\hat{k}_{z}\omega_{e} & 0 & 0 & \frac{m_{i}}{m_{e}}\frac{k_{B}T_{n}|\vec{k}|}{m_{e}\omega_{e}}\left(-ik_{z}-\gamma_{z}\right) & 0\\
-i\hat{k}_{x}\omega_{i} & 0 & 0 & 0 & \frac{k_{B}T_{n}|\vec{k}|}{m_{e}\omega_{i}}\left(-ik_{x}-\gamma_{x}\right)\\
-i\hat{k}_{y}\omega_{i} & 0 & 0 & 0 & \frac{k_{B}T_{n}|\vec{k}|}{m_{e}\omega_{i}}\left(-ik_{y}-\gamma_{y}\right)\\
-i\hat{k}_{z}\omega_{i} & 0 & 0 & 0 & \frac{k_{B}T_{n}|\vec{k}|}{m_{e}\omega_{i}}\left(-ik_{z}-\gamma_{z}\right)\\
0 & i\frac{m_{i}}{m_{e}}\vec{k}\cdot\vec{V}_{e0} & -i\frac{m_{i}}{m_{e}}\vec{k}\cdot\vec{V}_{i0} & 0 & 0\\
0 & \vec{\gamma}\cdot\vec{V}_{e0}-i\vec{k}\cdot\vec{V}_{e0}+\xi\bar{n}_{0} & \xi\bar{n}_{0} & 0 & 0\\
0 & \xi\bar{n}_{0} & \vec{\gamma}\cdot\vec{V}_{i0}-i\vec{k}\cdot\vec{V}_{i0}+\xi\bar{n}_{0} & 0 & 0\\
0 & D_{ei} & \frac{2m_{e}\nu_{ei}P_{io}}{(m_{e}+m_{i})k_{B}T_{n}\bar{n}_{0}} & L_{ei} & -2\frac{m_{e}\nu_{ei}}{m_{e}+m_{i}}\\
0 & \frac{2m_{i}\nu_{ie}P_{eo}}{(m_{e}+m_{i})k_{B}T_{n}\bar{n}_{0}} & D_{ie} & -2\frac{m_{i}\nu_{ie}}{m_{e}+m_{i}} & L_{ie}
\end{array}\right],\;\mathrm{where}
\end{align*}
\begin{align*}
A_{\alpha\beta}= & \frac{\vec{\gamma}}{|\vec{k}|}\frac{P_{\alpha0}\omega_{\alpha}}{k_{B}T_{n}\bar{n}_{0}}-\frac{5}{3}\frac{i\vec{k}}{|\vec{k}|}\frac{P_{\alpha0}\omega_{\alpha}}{k_{B}T_{n}\bar{n}_{0}}-\frac{4}{3}\frac{(\vec{V}_{\alpha0}-\vec{V}_{n})\frac{\omega_{\alpha}}{|\vec{k}|}}{k_{B}T_{n}/m_{\alpha}}\frac{m_{n}\nu_{\alpha n}}{m_{\alpha}+m_{n}}-\frac{4}{3}\frac{(\vec{V}_{\alpha0}-\vec{V}_{\beta0})\frac{\omega_{\alpha}}{|\vec{k}|}}{k_{B}T_{n}/m_{\alpha}}\frac{m_{\beta}\nu_{\alpha\beta}}{m_{\alpha}+m_{\beta}},\\
C_{\alpha\beta}= & \frac{4}{3}\frac{(\vec{V}_{\alpha0}-\vec{V}_{\beta0})\frac{\omega_{\beta}}{|\vec{k}|}}{k_{B}T_{n}/m_{\alpha}}\frac{m_{\beta}\nu_{\alpha\beta}}{m_{\alpha}+m_{\beta}},\\
D_{\alpha\beta}= & -2\frac{m_{\alpha}\nu_{\alpha n}}{m_{\alpha}+m_{n}}-2\frac{m_{\alpha}\nu_{\alpha\beta}P_{\beta0}}{(m_{\alpha}+m_{\beta})k_{B}T_{n}\bar{n}_{0}}-\frac{2}{3}\frac{m_{n}\nu_{\alpha n}}{m_{\alpha}+m_{n}}\frac{\left|\vec{V}_{\alpha0}-\vec{V}_{n}\right|^{2}}{k_{B}T_{n}/m_{\alpha}}-\frac{2}{3}\frac{m_{\beta}\nu_{\alpha\beta}}{m_{\alpha}+m_{\beta}}\frac{\left|\vec{V}_{\alpha0}-\vec{V}_{\beta0}\right|^{2}}{k_{B}T_{n}/m_{\alpha}},\\
L_{\alpha\beta}= & -i\vec{k}\cdot\vec{V}_{\alpha0}+\frac{5}{3}\left(\vec{\gamma}\cdot\vec{V}_{\alpha0}\right)+2\frac{m_{\alpha}\nu_{\alpha n}}{m_{\alpha}+m_{n}}+2\frac{m_{\alpha}\nu_{\alpha\beta}}{m_{\alpha}+m_{\beta}}.\\
\hat{k}= & \frac{\vec{k}}{\left|\vec{k}\right|}
\end{align*}
}\caption{Matrix $H_{5ES}$ containing the linearized and Fourier transformed
electrostatic five-moment fluid equations. In this work $\vec{V}_{e0}$
and $\vec{V}_{i0}$ are set to zero, and $P_{e0}$ and $P_{i0}$ are
set to $k_{B}T_{n}\bar{n}_{0}$.\label{fig:ES_Matrix}}
\end{sidewaysfigure}

\section{Parameter Choices\label{subsec:Parameter-Choices}}

\begin{figure}[p]
\thisfloatpagestyle{plain}\includegraphics{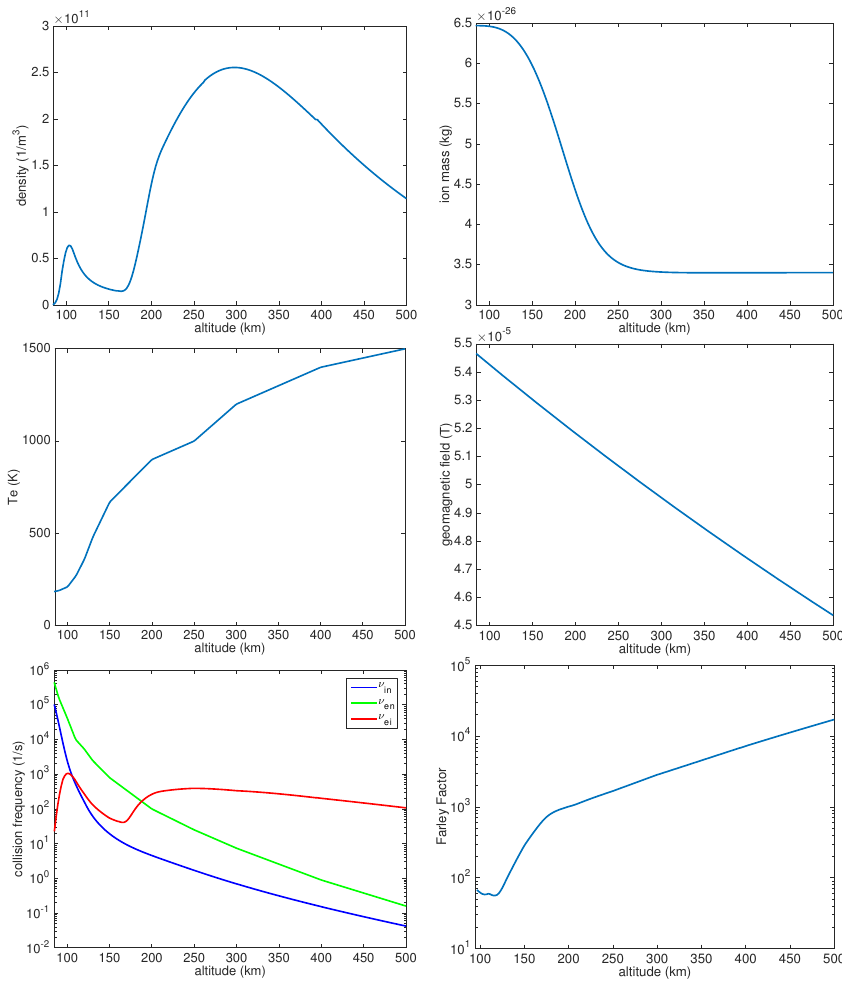}

\caption{Ion mass, electron temperature, geomagnetic field, and collision frequencies
used in all calculations, along with the associated Farley factor.
A density profile is also shown, but is used only for the green curve
of panel~l, Figure~2.
Otherwise the density is kept constant with altitude, with values
as identified in the text. This figure is reproduced from \citeauthor{cosgrove-2016}~{[}2016{]}.\label{fig:IonosphereParameters}}
\end{figure}

The ionospheric parameters used in the analysis are the same as used
by \citeauthor{cosgrove-2016}~{[}2016{]} (see their Appendix~B),
and are reproduced here for reference. An electron density profile
derived from Sondrestrom incoherent scatter radar data, with $E$
region arc and substantial $F$ region, is shown in the top left panel
of Figure~\ref{fig:IonosphereParameters}. However, this profile
is used only for the green curve of panel~l, Figure~2.
Otherwise the density is kept constant with altitude, with values
as identified in the text. A profile for the ion mass is shown in
the top right panel of Figure~\ref{fig:IonosphereParameters}. A
collision frequency profile, including $\nu_{in}$, $\nu_{en}$, and
$\nu_{ei}$, is shown in the bottom left panel of Figure~\ref{fig:IonosphereParameters}.
The collision frequency profiles up to 1000~km come from the textbook
by \citeauthor{gurevich-1978}~{[}1978{]}. However, $\nu_{ei}$ is
proportional to density, and so the $\nu_{ei}$ profile was rescaled
to reflect the density profile in Figure~\ref{fig:IonosphereParameters},
which is slightly different from that assumed by \citeauthor{gurevich-1978}~{[}1978{]}.
When used with the various constant vertical density profiles in the
text, the $\nu_{ei}$ profile is rescaled for each. The neutral collisions,
$\nu_{in}$ and $\nu_{en}$, are set to zero above 1000~km. The formula
\begin{equation}
\nu_{ei}=54.5\frac{n_{i}Z_{i}^{2}}{T_{e}^{3/2}},\label{eq:electron-ion-coll-freq}
\end{equation}
from \citeauthor{schunk+nagy-2000}~{[}2000{]} is used above 1000~km,
where $T_{e}$ is the electron temperature in Kelvin, $n_{i}$ is
the ion density in $\mathrm{cm^{-3}}$, and $Z_{i}$ is the ion charge
number. We assume $n_{i}=n$, $Z_{i}=1$, and use the electron temperature
profile shown in the middle left panel of Figure~\ref{fig:IonosphereParameters}.
Above 1000~km, the $T_{e}$ profile is motivated by \citeauthor{abe+etal-1997}~{[}1997{]}.
In the region below 1000~km, the $T_{e}$ profile is taken from Gurevich's
text {[}\citealp{gurevich-1978}, nighttime{]}, and is incorporated in
$\nu_{en}$ and $\nu_{ei}$. The temperature in the region below 1000~km
is used only to compare the collision frequency equation~(\ref{eq:electron-ion-coll-freq})
with the values from the text by \citeauthor{gurevich-1978}~{[}1978{]}.
The formula (\ref{eq:electron-ion-coll-freq}) is found to match exactly
the value given by \citeauthor{gurevich-1978}~{[}1978{]} at 1000~km,
although (\ref{eq:electron-ion-coll-freq}) gives negligibly larger
values around the $F$ region peak (323~s$^{-1}$ as compared with
300~s$^{-1}$, at 315~km). Another formula, $\nu_{ei}=\left[34+4.18\ln\left(T_{e}^{3}/n\right)\right]nT_{e}^{-3/2}$,
appears in {[}\citealp{kelley-1989}{]}, where it is attributed to \citeauthor{nicolet-1953}~{[}1953{]};
it gives nearly identical results to equation (\ref{eq:electron-ion-coll-freq}).

We have favored the \citeauthor{gurevich-1978}~{[}1978{]} results
only because that text seems to have a complete treatment that is
possible to follow without too much difficulty. \citeauthor{gurevich-1978}~{[}1978{]}
expands the electron distribution function in terms of Legendre polynomials
and solves the Boltzmann equation with the Boltzmann collision integral
and the Rutherford scattering cross section. \citeauthor{schunk+nagy-2000}~{[}2000{]}
explain that their collision frequency formula is derived from a 13-moment
velocity distribution, which arises from an expansion in Hermite polynomials
about a Gaussian. They reference Chapman and Cowling~{[}1970{]} for
the Chapman-Cowling collision integral, derived for a general inverse
power force law. We have compared the results from these two methods
and find that they are essentially identical, for our example.

\section{Additional Commentary on Energy Conservation\label{subsec:Additional-Commentary-on-energy_conservation}}

Because these are plasma waves, with degrees of freedom and modes
of energy storage beyond the electromagnetic, the landscape for energy
conservation is more complex than for the usual TL theory.\footnote{In this work we consider the case where the neutral wind is zero,
and so the ionosphere is not a source of energy. Wave amplitudes decrease
with time. But when the wind is nonzero it is possible for the ionosphere
to be a source of energy, which would manifest as wave amplitudes
that increase with time, that is, unstable waves. In this case there
is likely no steady state for the system, and so the steady state
treatment is not applicable, and neither is the concept of admittance
(or conductance). It is necessary to use a time-domain description,
and to limit the time horizon such that wave amplitudes remain small,
or otherwise to include their non-linear interactions.} Consider a source with an internal admittance that is positive-real,
attached to a TL section with characteristic admittance $Y_{0}$.
For the usual case where $Y_{0}$ is positive-real, it is easy to
understand that when there is no back reflected wave in the TL section,
the system absorbs energy by carrying it away, and the input admittance
$(Y_{in\mathrm{Matched}})$ may be matched to the source. And when
there is a back reflected wave the energy is returned, which is consistent
with the fact that the input admittance $(Y_{in\mathrm{OC}})$ becomes
imaginary, and so is no longer matched to the source. 

However, the situation is somewhat different for the ionosphere. For
the ionosphere $Y_{0}$ is imaginary, and so if there is no back-reflected
wave in the TL section the input admittance $(Y_{in\mathrm{Matched}})$
is imaginary, and so not well matched to the source. The mismatch
means that the energy is returned, but in this case it is not returned
from within the TL section. Rather, the energy never makes it into
the TL section at all, that is, the wave amplitudes in the TL section
are very small.

On the other hand, when $Y_{0}$ is imaginary and there \emph{is}
a back-reflected wave in the TL section, the input admittance $(Y_{in\mathrm{OC}})$
is real and so may be matched to the source. In this case the energy
may be absorbed even though there is a back-reflected wave within
the TL section. How is this possible without wave dissipation, which
is not needed for the admittance calculation? In this case the amplitude
of the reflected wave may be reduced by nonlinear interaction with
the incident wave, with associated heating. Since the heating terms
decrease with the square of the wave amplitudes, the fractional reduction
of the reflected wave decreases with the wave amplitudes, even as
energy conservation is maintained.

And finally we have the case where $Y_{0}$ is imaginary and there
is a back-reflected wave in the TL section, but the input admittance
$(Y_{in\mathrm{OC}})$ becomes negative-real, instead of positive-real.
In this case the source and TL section are severely mismatched, and
so the net Poynting flux may actually be into the source. Where is
this energy coming from? In this case we must appeal to energy conversion.
From the analysis surrounding Table~3 in \citeauthor{cosgrove-2016}~{[}2016{]},
it appears that kinetic energy may be converted into electromagnetic
energy {[}also see \citealp{vasyliunas+song-2005}{]}, and so this may
supply the energy that is needed to support a negative ionospheric
conductance. If so, this is a feature that distinguishes collisional
plasma waves, where the mass of the charge carriers may be significant. 

\section{Analytic Dependence for Polarization Vectors\label{subsec:Analytic-Dependence-Polarization-Vec}}

Here we provide details on the method for approximating the eigenvector
$k_{z}$-dependence by an analytic function. The eigenvectors, $\vec{h}_{j}$,
which make up the columns of the matrix $U$, are, by convention,
constrained to have a complex modulus of one. Hence, variations of
the eigenvectors occur through transformations under the 16-dimensional
special unitary group, $SU(16)$. It is well known that special unitary
transformations can be parameterized about the identity by $\vec{\theta}\in\mathbb{R}^{16^{2}-1}$
(in the 16-dimensional case), where the transformation is realized
by multiplication by the matrix $\mathrm{e}^{-i\vec{\theta}\cdot\mathbf{J}}$,
and $\mathbf{J}$ is a vector of matrices, which are the $16^{2}-1$
generators of $SU(16)$ {[}e.g., \citealp{mathews+walker_1970}; \citealp{Tung_1985}{]}.
Although we cannot be sure that the dependence of $\theta$ on $k_{z}$
is analytic, we can obtain an analytic function that approximates
the eigenvector transformation over the contributing part of the peaked
integrand by linearizing the dependence of $\vec{\theta}$ on $k_{z}$,
that is, $\vec{\theta}\cong\left(k_{z}-k_{0z}\right)\left.\frac{\partial\vec{\theta}}{\partial k_{z}}\right|_{k_{z}=k_{0z}}$.
Therefore, we will approximate the eigenvector evolution operator
by, 
\begin{equation}
\Theta\left(k_{z}-k_{0z}\right)=\mathrm{e}^{-i\left(k_{z}-k_{0z}\right)\left.\frac{\partial\vec{\theta}}{\partial k_{z}}\right|_{k_{z}=k_{0z}}\cdot\mathbf{J}}.\label{eq:su16}
\end{equation}

The only problem is that it is not completely clear if this exponential
function will allow the boundary terms associated with the residue
theorem to vanish. However, using again our assumption that the main
contribution to the integral can be localized around $k_{z}=k_{0z}$,
there is no harm in multiplying the integrand by an analytic function
such as $\mathrm{e}^{\pm i(k_{z}^{3}-k_{0z}^{3})/\sigma^{3}}$, and
choosing $\sigma$ to be sufficiently large that the integral is not
affected. Choosing the $\pm$ sign to give exponential decay, the
$k_{z}^{3}$ dependence will dwarf any exponential growth that might
possibly come from~(\ref{eq:su16}), regardless of the choice for
$\sigma$. Thus, we can assume $\sigma$ is sufficiently large that
the integral is not affected, apply the residue theorem, and the residue
will not be affected either. 

Note that introduction of the form~(\ref{eq:su16}) is really a formality,
so that we can apply the residue theorem and then later argue that
$\Theta\cong I$, based on numerical evaluation of the integrals.
Hence we need never actually specify the derivatives $\left.\frac{\partial\vec{\theta}}{\partial k_{z}}\right|_{k_{z}=k_{0z}}$. 

\section{Validation Figures and Elaboration\label{subsec:Validation-Figures}}

\begin{figure}[p]
\thisfloatpagestyle{plain}\includegraphics{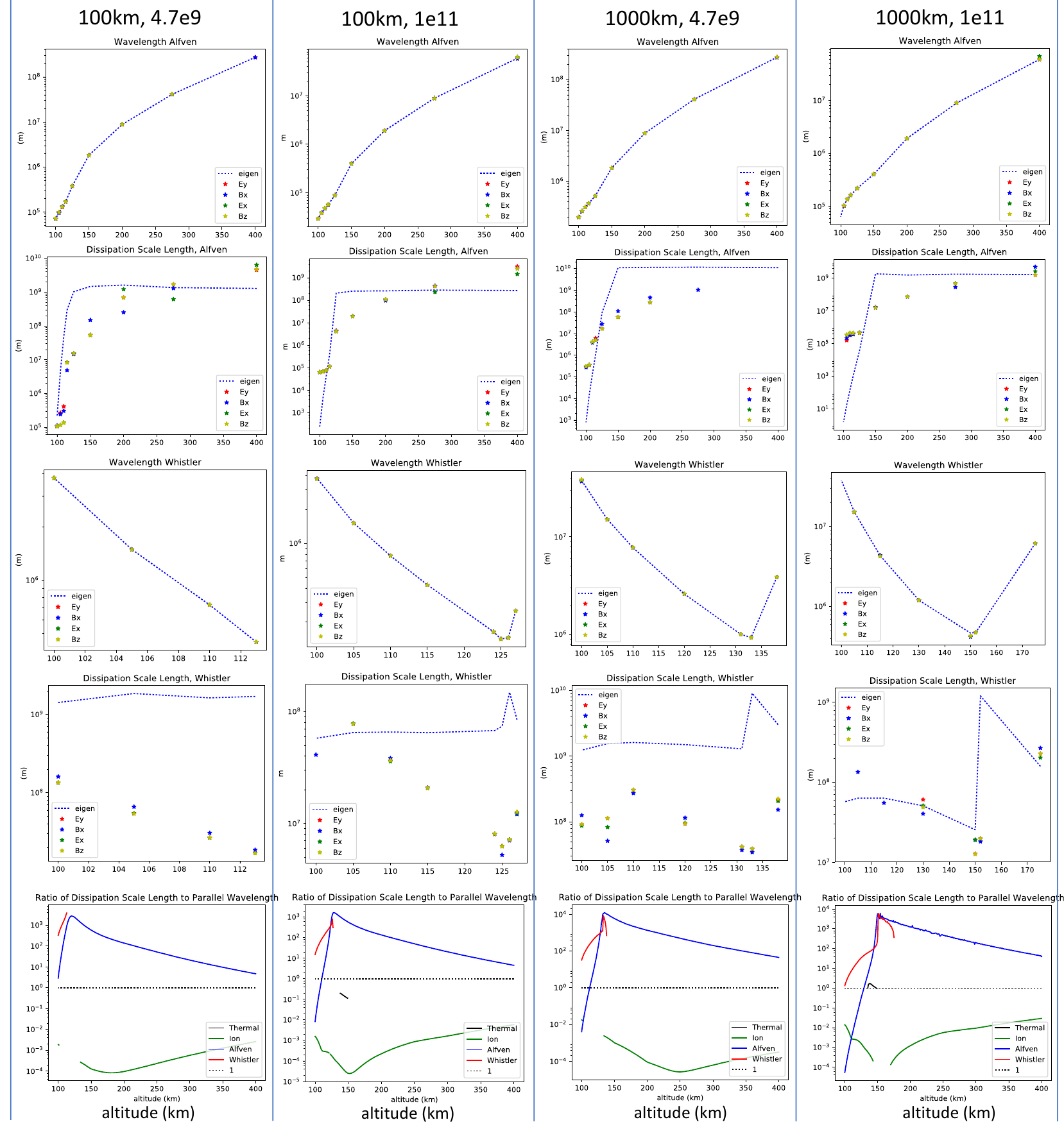}

\caption{Validation of parallel wavelength and dissipation scale length. The
transverse-wavelength and density are shown at the tops of the columns.
For reference, the ratio of dissipation scale length to parallel wavelength
under the wavepacket interpretation is shown in the bottom row, for
the four modes of Section~6.
\label{fig:Validation-of-wavelength-dissipScale}}
\end{figure}

\begin{figure}[p]
\vspace{-0.75in}
\thisfloatpagestyle{plain}\includegraphics[viewport=0.918in 2.104348in 546.975bp 792bp,clip]{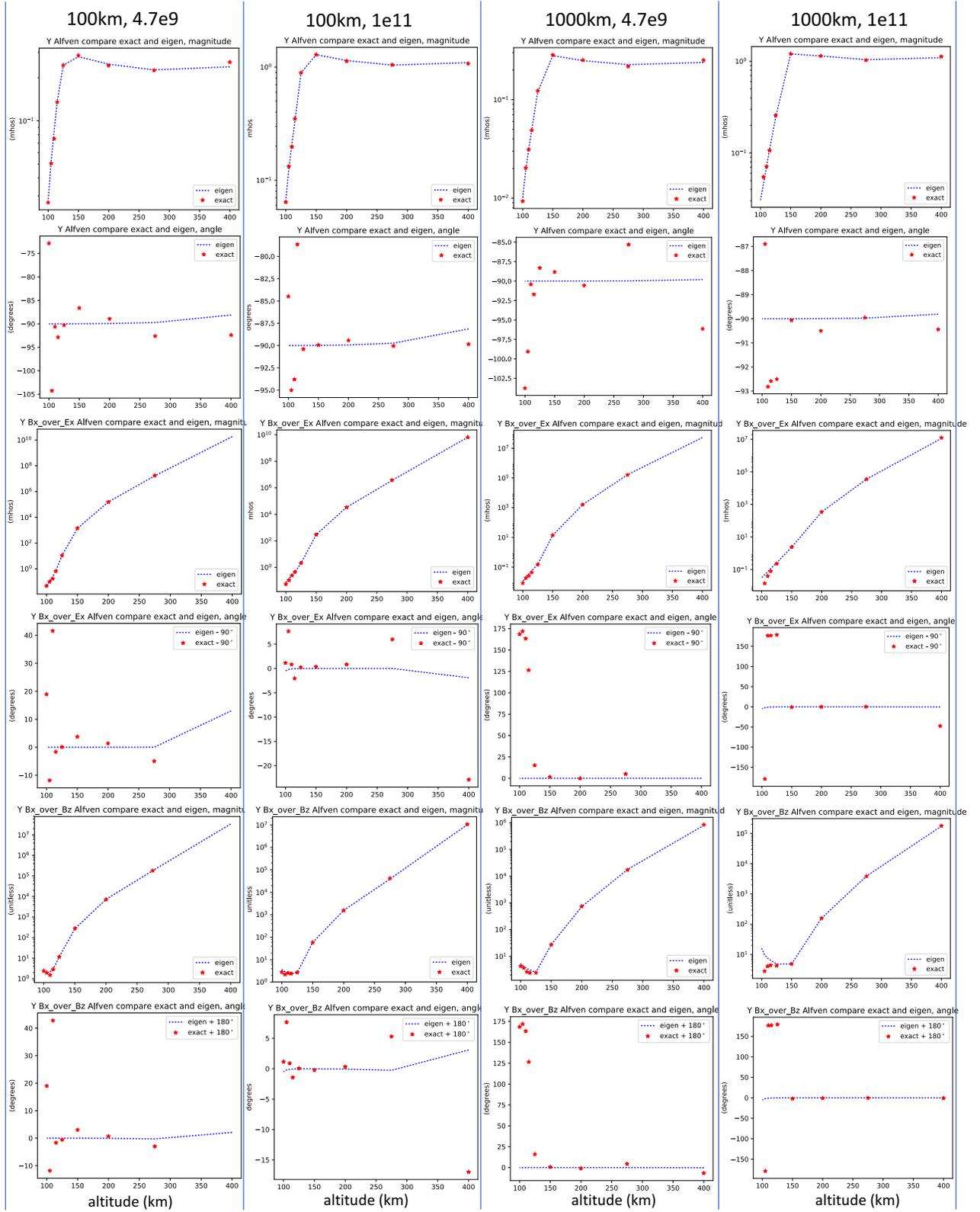}\caption{Validation of the Alfvén wave polarization vectors. The transverse-wavelength
and density are shown at the tops of the columns. \label{fig:Validation-of-Alfven_polarization}}
\end{figure}

\begin{figure}[p]
\vspace{-0.75in}
\thisfloatpagestyle{plain}\includegraphics[viewport=0.918in 2.104348in 546.975bp 792bp,clip]{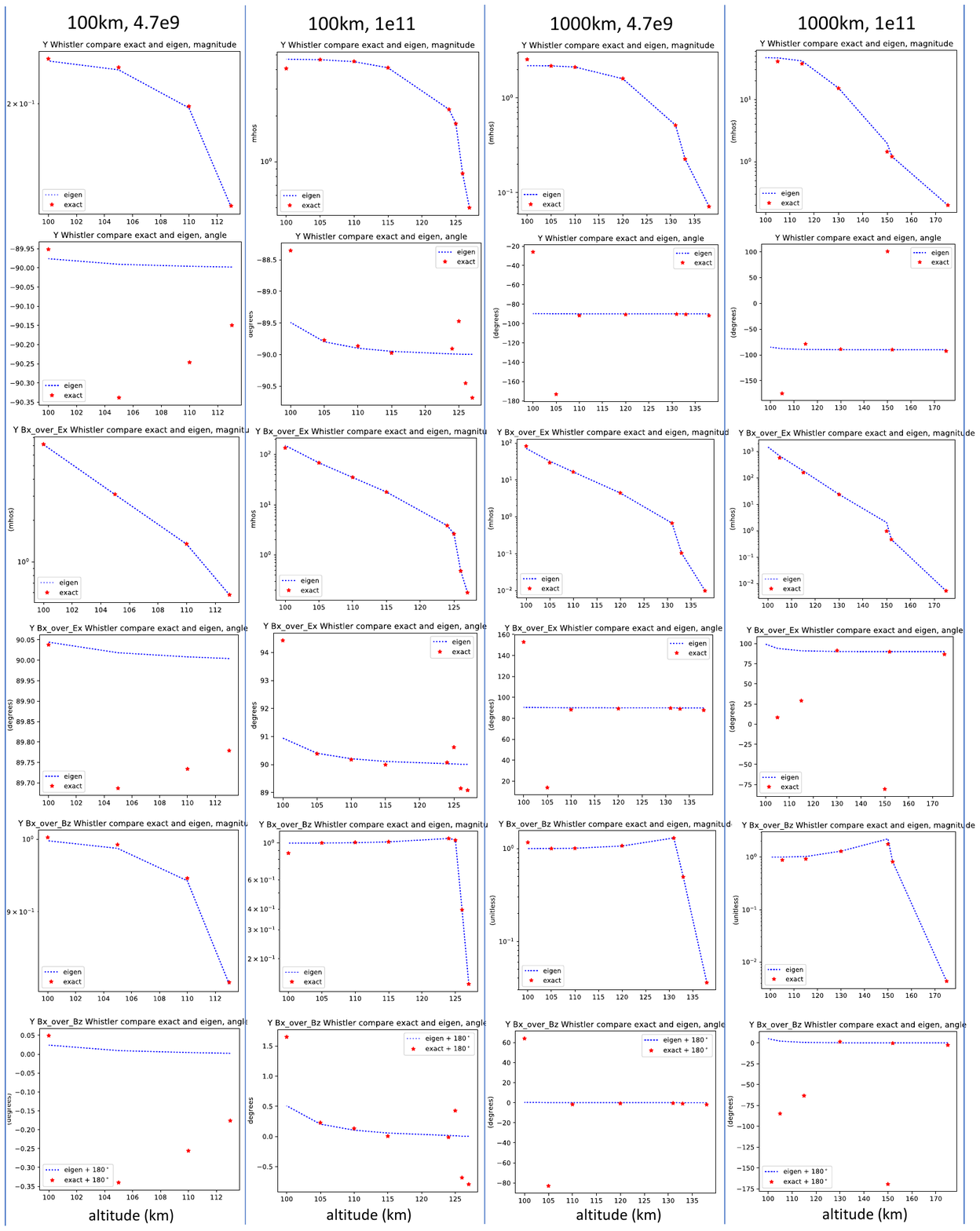}\caption{Validation of the Whistler wave polarization vectors. The transverse-wavelength
and density are shown at the tops of the columns. \label{fig:Validation-of-whistler_polarization}}
\end{figure}

Figures~\ref{fig:Validation-of-wavelength-dissipScale}-\ref{fig:Validation-of-whistler_polarization}
show comparisons of the exact (i.e., obtained by numerical integration)
and eigenmode results at the four corners of the modeling domain,
which domain is described in Section~7. While
there is some numerical scatter in the integration results, they nevertheless
provide a pretty strong indication of convergence to the simple wave-packet
interpretation, without corrections. However, there are some anomalies
that we discuss in this supplementary section. At this time it is
difficult to tell if these anomalies represent actual predictions,
numerical inadequacies such as aliasing or windowing, or the result
of a failure of our assumption that the source~(10)
possesses a far zone (which would be a failure of the validation method,
and not necessarily an invalidation). 

Results for the wavelength and dissipation scale length are shown
in Figure~\ref{fig:Validation-of-wavelength-dissipScale}, for both
modes. Results are shown separately for $E_{y}$, $B_{x}$, $E_{x}$,
and $B_{z}$, since these are all integrated separately. For reference,
the ratio of the dissipation scale to the wavelength is also shown
in the bottom row, as it obtains from the simple wavepacket interpretation.
With respect to the wavelength, it is evident that the results agree
almost exactly in all cases. The only caveat is that it was impossible
to obtain results at the lowest altitude (100~km) for one of the
four corners, specifically at the corner with transverse wavelength
1000~km and density $10^{11}\,\mathrm{m}^{-3}$. This combination
of altitude, transverse wavelength, and density produces the lowest
ratio of dissipation scale length to parallel wavelength (bottom right
panel of Figure~\ref{fig:Validation-of-wavelength-dissipScale}),
where in this case we mean the dissipation scale length derived from
the group velocity. The dissipation scale length for the Whistler
mode is about equal to its wavelength, while for the Alfvén mode it
is much less, under these conditions. This situation might be expected
to pose a challenge to our method of fitting a damped sinusoidal wave
to the output from the numerical integration, since the signal may
not look like a wave; it may not extend far enough to display oscillation,
and it may not extend past the near zone. 

It is interesting, however, that in less severe cases it is still
possible to fit the Alfvén mode, even though the dissipation length
predicted from the group velocity is quite a bit less than the wavelength.
In these cases it is still possible to see the Alfvén mode oscillating
like a wave, as though it had a much longer dissipation scale: the
fits succeed with a reasonable visual appeal, and the dissipation
length for the Alfvén mode comes out much longer than that predicted
from the group velocity, as can be seen in the low-altitude region
of the three right-most columns in Figure~\ref{fig:Validation-of-wavelength-dissipScale}.
It is entirely possible that this represents a real prediction, and
that we should be using a somewhat longer dissipation scale length
at the bottom of the $E$ region (for the Alfvén mode). The group
velocity is itself difficult for us to calculate, since it requires
a finite difference evaluation of the derivative of $\omega_{jr}$
(see jittery trace in the bottom right panel of Figure~\ref{fig:Validation-of-wavelength-dissipScale}).
The fairly-low frequencies that we are evaluating are associated with
$k_{z}$ very close to the value where $\omega_{jr}$ becomes zero
(for the Alfvén wave), and $\omega_{jr}$ is very sensitive to $k_{z}$
in this area. Concern about this close proximity was actually the
major reason we decided it necessary to perform the numerical integrations,
as a rigorous test. However, it is not clear which of the two results
is more reliable, because the bandwidth and sampling requirements
for the bottom of the $E$ region were difficult to realize for the
numerical precision (double precision) and number of processors (11)
that were available. Also, it may be that we are seeing only a near-zone
effect. 

Results for the polarization vectors are shown in Figures~\ref{fig:Validation-of-Alfven_polarization}~and~\ref{fig:Validation-of-whistler_polarization},
for the Alfvén and Whistler modes, respectively. Since the polarization
vector is arbitrary up to normalization, it is represented by plotting
the ratio $B_{x}/B_{z}$, along with the two admittances $B_{x}/(\mu_{0}E_{y})$
and $B_{x}/(\mu_{0}E_{x})$, where $\mu_{0}$ is the permeability
of free space. Both magnitude and angle are shown for these complex
numbers. 

For the case of magnitude the agreement is really quite good, with
the only exception being, again, the Alfvén mode in the lower \emph{E}
region when the transverse wavelength is 1000~km, and the density
$10^{11}\,\mathrm{m}^{-3}$. For the case of phase, however, the anomalies
in the lower $E$ region extend to the other three cases, and the
Whistler wave is also involved. For the most part these anomalies
have a scattered appearance, suggesting that they are due either to
the numerical limitations mentioned above, or to highly-variable near-zone
effects. The regions where this scatter is pronounced generally exhibit
evidence of either aliasing (in the case of the Whistler wave), or
of a situation where the near zone extends too far from the source,
such that the signal does not look completely wavelike (in the case
of the Alfvén mode).

Nevertheless, some of the anomalies appear sufficiently consistent
as to be worthy of comment. Looking at Figure~\ref{fig:Validation-of-Alfven_polarization},
in the lower \emph{E-}region, for the 1000~km wavelength cases, the
admittance $B_{x}/(\mu_{0}E_{x})$ and the ratio $B_{x}/B_{z}$ are
negated for the Alfvén wave. This is the area, also noted above, where
the dissipation scale length computed from the group velocity is quite
a bit less than the wavelength (bottom row of Figure~\ref{fig:Validation-of-wavelength-dissipScale}),
and also quite a bit less than that calculated by numerical integration.
Since the results seem to be consistent in the region below 115~km
in altitude it is possible that this represents a real correction
to the results obtained from the eigenmodes. However, we do not feel
comfortable to implement these corrections until it is possible to
verify them with a higher sampling density and wider bandwidth. Also,
it may again be that we are seeing a near-zone effect. 

Another anomaly concerns the Whistler mode in the altitude regime
where, as discussed in the paper, it is probably more analogous to
the fast magnetosonic wave than to the usual whistler wave. In this
regime the numerical integrations are producing negative phase velocities
and admittances that are conjugates of those found from the eigenvectors.
These conjugations have been reversed before plotting, since the effect
seemed quite clear, and so cannot be seen in the figures. We have
tested these possible corrections in the model calculation by applying
the conjugations and negating the wavevector over the appropriate
altitude regime, and found that the effect is almost unnoticeable.
Although we would like to test these results with a higher sampling
density and wider bandwidth, it appears that they are not of any consequence.

We have also tested the effect of lengthening the dissipation scale
length in the model calculation, since there was an altitude range
where the dissipation scale length computed from the group velocity
was short enough to have an effect on the Alfvén wave, in the lower
\emph{E} region. In this case we do find a modest effect for the case
when the tangent function signature is in evidence, which is the shortest
wavelength case. The tangent function signature is somewhat sharpened.
The wavelength for the Alfvén wave is very short in the lower \emph{E}-region,
and lengthening the dissipation scale allows this very-short-wavelength
mode to participate more in the dynamics. However, the effect at the
top of the ionosphere is quite modest, in our testing. And there was
no noticeable effect at any altitude for the other three cases, when
the tangent function signature was not in evidence. 

In the body of the paper we have explained that the anomalies seen
in the lower \emph{E} region may be simply near-zone effects that
are particularly associated with the $\delta(z$) in the source~(10),
and not relevant to the response of the plasma to radiation-zone waves
incident from above. To resolve this matter we would like perform
the calculations using a much larger number or processors, and with
quadruple precision arithmetic. This would allow us to verify that
the computed signal remains unchanged when the sampling density and
bandwidth are increased. However, if near-zone effects are actually
the culprit then numerical improvements are not going to solve the
problem, and it may be necessary to appeal to a time-domain approach
for validation and/or correction. But since such a time-domain simulation
has not, to our knowledge, been done before, it may be that there
are significant obstacles (some of which were mentioned in Section~2).
Hence, it may be quite difficult to fully validate (or correct) the
eigenvectors and dissipation scale lengths in the lower $E$ region,
and it may even be that circuit-theory concepts like conductivity
and conductance are not really applicable there, due to a high degree
of source-dependency. 

However, since the ionospheric conductance is used and will likely
continue to be used for quite some time, we have to do something.
Therefore, it seems appropriate that for the time being we should
content ourselves with the arguments given in Section~4,
that radiation-zone waves traveling into the $E$ region from above
will not excite the near-zone effects that the $\delta(z$) in the
source~(10) excites, and will instead excite other
radiation-zone waves, which may be characterized by the eigenvectors
and eigenvalues, and by the usual wave-packet interpretation.

In conclusion, although there are some possible corrections indicated
from our attempts at an exact analysis (i.e., the numerical integration
of equation~(12)), we both do
not find the corrections to be sufficiently reliable, and we do not
find that the corrections would have a large effect. The altitude
regime where the corrections would apply is below the transition region
that has been found to block electric field penetration, and this
is one likely reason that the effects do not appear very dramatic
for a signal entering from above. Another is that the model calculation
naturally excludes the Alfvén mode, and favors the Whistler mode,
at the bottom of the \emph{E} region. As long as this exclusion is
a correct prediction, the behavior should not be very sensitive to
the properties of the Alfvén mode at the bottom of the \emph{E} region.
And one thing seems almost certain, implementing corrections like
these is not going to enhance conformance with electrostatic theory
in any way. Thus we feel it most appropriate to implement the eigenmode-based
model in it's pure form, using the usual wave-packet interpretation.
This model is a baseline electromagnetic model that is the most natural
extension of the current electrostatic baseline.

\section{Reviews of this and Predecessor Papers, with Replies}

\noindent Reviewer/Referee = black

\noindent \textcolor{blue}{RBC = blue}

\subsection{Reviews of {[}\emph{Cosgrove}, 2016{]}, ``Does a localized plasma
disturbance in the ionosphere evolve to electrostatic equilibrium?
Evidence to the contrary,'' published in JGR.}

\subsubsection{Reviewer \#1:}

Reviewer \#1 Evaluations: \\
Recommendation: Return to author for major revisions \\
Significant: There are major errors or gaps in the paper but it could
still become significant with major changes, revisions, and/or additional
data. \\
Supported: Mostly yes, but some further information and/or data are
needed. \\
Referencing: Yes \\
Quality: Yes, it is well-written, logically organized, and the figures
and tables are appropriate. \\
Data: Yes \\
Accurate Key Points: Yes \medskip{}
\\
Reviewer \#1 (Formal Review for Authors (shown to authors)): This
work is worth consideration for publication if questions are addressed.
The implications of this work can affect nearly the whole field of
ionospheric plasma physics. Attached is my full review.\medskip{}
\\
Overall impression:
\begin{itemize}
\item Challenging the ubiquitous electrostatics in ionospheric plasma community,
this paper is exciting and engaging. Whether electromagnetic dynamics
actually settle to the electrostatic state is question unanswered
worth exploring. It can potentially affect most of ionospheric plasma
physics. However, I have some major questions, as well as minor suggestions.
\end{itemize}
\medskip{}
Major Questions:

\medskip{}

(1) Section 4 (Distribution of effects around the source). Comparison
to the Farley distance using plane-wave analysis may have a flaw.
Going back to the heart of the Farley factor, Farley points out that
by going into a coordinate system where the z-dimension is squished
by a Farley factor, one can work with Poisson\textquoteright s equation.
The problem is in the coordinate transformation; because the z dimension
is squished, everything along the z dimension is effectively brought
closer to each other, where one might need to consider near-field
effects. Like how ray-tracing cannot describe near-field effects,
the plane wave analysis cannot fully model near-field effects. Can
you rule out near-field-like effects providing the mechanism to reach
the electrostatic state? Granted you are in a regime that is traditionally
considered far field, and the Farley factor/coordinate-system is built
derived from electrostatics, but can you address this concern?\medskip{}

\textcolor{blue}{The short answer is that near-zone effects almost
certainly depend on the specific source that is driving the region
of interest, and, of course, also depend on how close the source is
to the region of interest. As discussed in Section~4
under the heading ``Validating the Steady-State Solution,'' the
concept of admittance arises in the context of a source-independent
characterization of the system. Admittance (and other circuit quantities)
are descriptions of those aspect of the system that are independent
of the specific source, and so involve an inherent assumption that
the source is sufficiently far away that near-zone effects can be
ignored. If near-zone effects were required to produce conformance
with electrostatic theory, that would be a major modification of the
current understanding of electrostatic theory, and would cast severe
doubt on its use when the source is in the magnetosphere.}

\textcolor{blue}{This referee was reviewing the 2016 article {[}\citealp{cosgrove-2016}{]},
where the source was in the E region. Their comment is not relevant
to the 2022 paper where it is specifically stated that we are deriving
the admittance of the ionosphere as seen from above. By studying the
admittance we are inherently omitting near-zone effects. And, this
seems physically appropriate when the source is in the magnetosphere.}

\textcolor{blue}{However, even with respect to the 2016 article where
the source was in the E region, it should be noted that near-zone
effects are generally a phenomena associated with sources localized
in three dimensions. These kinds of problems are normally treated
by expanding the solution to Maxwell's equations as a sum of spherical
harmonics, to describe the two angular dimensions. The spherical harmonics
are used because they are eigenfunctions of the Laplacian, which reduces
the equations to linear ones for the two angular dimensions. Substituting
this expansion results in a non-linear equation for a radial function
describing the variation in the third (radial) dimension, with solution
that is a sum of spherical and Bessel functions: the }\textcolor{blue}{\emph{near
zone}}\textcolor{blue}{{} is the region where the Bessel function dominates,
and the }\textcolor{blue}{\emph{far zone}}\textcolor{blue}{{} is the
region where the spherical function dominates. }

\textcolor{blue}{By contrast, in both the 2016 and 2022 papers I am
considering an infinite planar source, because I am deriving the Fourier
components of the response. This (different) geometry calls for expanding
the solution to Maxwell's equations as a sum of plane waves. Like
the spherical harmonics, the plane waves are eigenfunctions of the
Laplacian. Unlike the spherical harmonics, the plane waves are three-dimensional
functions (not 2D), and substituting the sum of plane waves into Maxwell's
equations results in a completely linear problem; there is no non-linear
``radial'' equation remaining to be solved. So actually the issue
of near-zone and far-zone does not arise, at least not with the usual
meaning. Even for the E region source in the 2016 paper, the effects
that might be described by the term ``near zone'' effects are just
that there is a possibility of the source coupling to non-propagating
modes, which might be able to transport energy over short distances.
This is actually the referee's next question, although as already
explained, it is not relevant to the 2022 paper.}

\medskip{}

(2) The discrepancy between Alfven wave dispersion and the electrostatic
state is certainly convincing. However, can one rule out the other
modes contributing to a superposition where the electrostatic state
is met? One can imagine energy being siphoned out of the transmission
line via coupling to modes perpendicular to B0. Examination of the
numerical model described by Dao {[}2013{]}, electric fields are not
strictly confined on field lines hooked up parallel to the dynamo.
It seems that the parallel-propagating Alfven waves couple to transverse
E/M waves, that couple into anti-parallel waves that send some Poynting
flux, outside of the transmission line, back towards the dynamo {[}Figure
7.1 in Dao Thesis{]}. By focusing strictly on Alfven waves that propagate
along B0, you may have thrown away the dynamics that lead to an electrostatic
state.\medskip{}

\textcolor{blue}{At least for the 2022 paper, there is no assumption
that any of the waves propagate ``along magnetic field lines'' in
any sense of the term. In equation~(12)
of the 2022 paper we have given the general (linearized) solution
for a homogeneous plasma, and in equations~(14)~and~(B.3)
we have given approximate solutions that omit near-zone effects, and
which apply to the two modes that were found capable of propagating
any significant distance (at the frequencies relevant to ionospheric
science). Hence, to the extent that electrostatic theory might be
produced in the ionosphere without invoking near-zone effects, the
solution in terms of these two waves should be able to model it. This
applies also to the solution for the vertically inhomogeneous ionosphere
found in Section~5, since the
only additional approximation is to model the ionosphere as a stack
of thin homogeneous slabs. I have already explained, above, that in
seeking to characterize the ionosphere by an admittance we have a
license to omit near-zone effects.}

\textcolor{blue}{As far as the effects seen in Figure 7.1 from the
Dao thesis, to the extent they are not near-zone effects, or effects
from coupling to non-propagating modes, or nonlinear effects, or artifacts
from some of the modeling limitations described in Section~2
of the 2022 paper, it should be possible to produce them with the
two-wave description of the 2022 paper. However, doing this would
require modeling the response to a source that is localized in the
transverse dimension, and this must be done by superposing a sufficiently-broad
transverse spectrum of waves. While doing this is a long-term goal,
that capability has not yet been developed. Also, if the effects come
from coupling to non-propagating modes then they are highly dissipative,
and so would tend to prevent the electric field from ``mapping''
through the ionosphere; so there does not seem to be any chance that
they could help to establish electrostatic equilibrium. The results
from Dao {[}2013{]} are very interesting and demand further investigation,
but they are not directly relevant to the paper under discussion here.}

\textcolor{blue}{By the way, since equation~(12)
is the general (linearized) solution for a homogeneous plasma, the
near-zone effects that are omitted in the model can in principle be
derived from a numerical evaluation of the integrals in equation~(12).
The results of an initial attempt at doing this have been described
in Section~\ref{subsec:Validation-Figures} of the Supplementary
Information. For the particular source that was employed, it was found
that near-zone effects are not in evidence, except in the lower E
region. And with the source above the ionosphere, the model finds
that energy does not make it down to these altitudes. Also, the source
that was employed was localized in a thin sheet, whereas in the real
world the source seems better thought of as a wave carrying energy
into the region, which would seem less subject to near-zone effects.}

\textcolor{blue}{In concluding this response, it should be noted that
the generally accepted idea is that Alfvén waves mediate the transition
to the electrostatic state. Some of the referees to follow do not
seem to accept this idea, but nevertheless it is the standard idea.
So it is valuable and important to show that it is not the case, even
if there are possibly other ways to reach electrostatic equilibrium,
that is, other ways that somehow defy the logical analysis attempted
above. }\medskip{}

(3) Figures 9-12: For most cases, the dissipation lengths were near
or exceeded the Farley length. Can one rule out coupling to other
modes as a means of unaccounted dissipation? However, supporting your
claim, it seems that mode coupling cannot explain how the Farley length
can exceed the dissipation length.\medskip{}

\textcolor{blue}{This comment is only relevant to one aspect of the
comparison to electrostatic theory, concerning the dissipation scale
length. And, as the referee explains, only applies to the specific
cases where the dissipation length was found to exceed the Farley
mapping distance, whereas sometimes the reverse is found. And regardless
of any of this, in the 2022 paper the dissipation scale length was
found sufficiently long that its specific value does not affect the
results.}\medskip{}

(4) Going back to the root of electrostatics. At an electromagnetic
steady state, do you expect nonzero dE/dt and dB/dt? There seems to
be a disconnect between an electromagnetic stead state and an electrostatic
state that is difficult to reconcile.\medskip{}

\textcolor{blue}{In the model from the 2022 paper, the solution is
the driven steady-state solution, which oscillates at the source frequency
$\omega_{0}$. In the mathematics it is possible for $\omega_{0}$
to be zero, and the Alfvén wave is capable of propagating at zero
frequency. However, perfectly frozen systems do not exist in reality,
and so the most idealized case appropriate for consideration is the
limit $\omega_{0}\rightarrow0$. To be fully rigorous, the answer
to the referees question lies in determining if this limit is well
behaved. However, since realistically the system is never perfectly
frozen, we avoid this question and evaluate the model at finite $\omega_{0}$,
which is chosen to be a realistic operating point for ionospheric
phenomena. This is done in the 2022 paper by assuming a transverse
drift (relative to the field lines) of 40~m/s, and then calculating
the frequency from the transverse wavelength chosen for analysis. }

\textcolor{blue}{In the 2016 paper we also use other transverse drift
velocities, and observe without proof that the limit $\omega_{0}\rightarrow0$
appears to be well behaved. However, $\omega_{0}$ is the real part
of the frequency, which also has an imaginary part that remains both
finite and sizable as $\omega_{0}\rightarrow0$. This imaginary part
is not included in any electrostatic description, which would seem
to say that electrostatic theory is different from the $\omega_{0}\rightarrow0$
limit.}

\medskip{}

(5) Do you expect a limit where the electrostatics is valid? By showing
there is some limit where electrostatics work will give credibility
to calculations as a sanity check.\medskip{}

\textcolor{blue}{In the context of the 2022 paper this issue is addressed
by showing that the model would reproduce electrostatic theory exactly
if the waves had the ``right'' properties (see column~a of Figure~3),
but that this requires artificial modification of the waves. We have
not found any combination of plasma conditions and wavelength that
cause the waves to have the needed properties. However, the electrostatic
input admittance (i.e., the field line integrated conductivity) is
reproduced very closely in some cases (see panels~a1~and~a2 of
Figure~4), even though the
electric field does not map in the expected way.}

\medskip{}

\noindent Minor Suggestions:\textbf{\medskip{}
}

Figure 3: Instead of component numbers (1-12), label the ticks with
actual parameters (\textquoteleft Vex\textquoteright , \textquoteleft Vey\textquoteright ,
\textquoteleft Vez\textquoteright , \dots ). Perhaps also label what
the modes are (Alfven, O, X, fast, etc.); this will enable the figure
to stand alone for ease of the reader. A line plot doesn\textquoteright t
seem appropriate, since linear interpolation (connecting the dots)
between the 12 components has no physical meaning. A stem plot may
keep components separate and clearly illustrate which components is
zero or non-zero.\medskip{}

\textcolor{blue}{All of these suggestion have been implemented. The
stem plot was a good idea.}\medskip{}

Figure 6. The stippled areas aren\textquoteright t physical, nor do
they have meaning. You may just want to \textquotedblleft white out\textquotedblright{}
or \textquotedblleft cross hatch\textquotedblright{} these areas to
make clear that stippled areas show nothing meaningful.\medskip{}

\textcolor{blue}{I could not figure out a good way to ``white out''
or ``cross hatch'' the stippled area, since it's boundary is a curve,
and thresholding could not work. So I decided to leave the stippled
area as is, since the description ``stippled'' does seem to work
to identify it.}\medskip{}

Figure 9-14: A 1 to 1 reference line where ratio of lengths are equal
will aid the reader determining the difference between the Farley
and dissipative lengths.\medskip{}

\textcolor{blue}{The reference lines have been added, as suggested.}\medskip{}

Line 36: \textquotedblleft Horizontal\textquotedblright{} is ambiguous.
Horizontal at the magnetic pole is different from horizontal at the
equator. Somehow make clear that B is vertical in your paper.\medskip{}

\textcolor{blue}{The paragraph has been reworded to make it clear
that $\vec{B}_{0}$ is perpendicular and vertical.}\medskip{}

Line 191: Typo? \textquotedblleft to be written\textquotedblright{}
to be followed by \textquotedblleft as\textquotedblright ?\medskip{}

\textcolor{blue}{The word ``as'' has been added, as suggested.}\medskip{}

(13) \textbf{Reviewer comment:} Line 376. \textquotedblleft In this
case\textquotedblright{} should precede a comma.\medskip{}

\textcolor{blue}{A comma has been added, as suggested.}\medskip{}

Line 407: Section \textquotedblleft 7 8\textquotedblright ?\medskip{}

\textcolor{blue}{The word ``and'' has been added between the 7 and
the 8.}\medskip{}

Line 630: \textquotedblleft be seen my comparing\textquotedblright ;
typo at \textquotedblleft my\textquotedblright ?\medskip{}

\textcolor{blue}{The word ``my'' has been changed to ``by'', as
it was supposed to be.}\medskip{}

Line 908: \textquotedblleft components in \textquoteleft X\textquoteright \textquotedblright .
What is X?\medskip{}

\textcolor{blue}{The reference to $X$ has been removed, as it wasn't
needed and apparently didn't help with clarity.}

\subsubsection{Reviewer \#2:}

\noindent Reviewer \#2 Evaluations: \\
Recommendation: Publish in present form \\
Significant: Yes, the science is at the forefront of the discipline.
\\
Supported: Yes \\
Referencing: Yes \\
Quality: Yes, it is well-written, logically organized, and the figures
and tables are appropriate.\\
Data: Yes \\
Accurate Key Points: Yes \medskip{}
\\
Reviewer \#2 (Formal Review for Authors (shown to authors)): \medskip{}
\\
Review of manuscript \#2015JA021672: Does a localized plasma disturbance
in the ionosphere evolve to electrostatic equilibrium? Evidence to
the contrary, by Russell B. Cosgrove\medskip{}

In reviewing this manuscript I find myself in a strange position.
Throughout my professional life I have considered the ionosphere (in
contrast to the solar chromosphere) under dc or electrostatic aspects,
whereas this paper introduces me to the ac aspects. Thus I was quite
fascinated by what I read and tried to absorb. However, for a really
critical evaluation I am lacking familiarity with subject and willingness
to invest the time I would need for checking the derivations and results
beyond plausibility. The scenario of the paper is rather special.
An E region dynamo, narrow in height extent, and the allowance for
short wavelengths and high winds require consideration of spatial
and temporal scales that are outside most of the constellations of
M-I coupling. The questions are the excitation and propagation of
electromagnetic perturbations in the Alfvén or other modes under inclusion
of collisions, the related energy transport, and whether electrostatic
equilibrium can be reached. The paper is of high quality. Every step
in deriving solutions of the full set of wave equations in a collisional
medium has been described in a transparent and physically understandable
way. The difference between the classical Alfvén wave admittance and
the real one in the ionosphere has been very clearly elaborated. Key
is the matching condition between the dynamo and the adjacent ionosphere.
The comparison of the real admittance, calculated from eigenvector
solutions, with the electrostatic values in Table 2 offers valuable
insights. The same is true for the quantitative solutions using wavelet
integrals (equation 21) laid down in Figures 9 to 14. Although the
more typical application of the magnetospheric/ionospheric physicist
is to consider a high-altitude generator and waves impinging from
above with time scales orders of magnitude longer than treated here
and to treat the ionospheric damping in electrostatic approximation,
I do see the high educational value of studying the submitted manuscript.
Furthermore, I realize that there are more directly applicable physical
situations like ionospheric modification experiments. In summary and
realizing my shortage of real competence I recommend the paper for
publication in JGRA in the submitted form.\bigskip{}
\\
\textcolor{blue}{Thank you very much for agreeing to review this paper!
By giving this work your scrutiny you have helped me to feel much
more confident that the results are correct. You say that you are
not an expert in the ionosphere, but you seem to have the expertise
necessary to understand this work: I am buoyed by the degree to which
you seemed to understand my main points and single them out for emphasis.
I think that wave-analysis is not so common for ionospheric specialists,
but is maybe much more common in other areas of space plasmas, so
maybe that is what was required.}

\subsection{Reviews of ``Failure of the Electrostatic Assumption via an Exact
Solution for the Linear Waves of the Electromagnetic 5-Moment Fluid
Equations, plus 150 km Echoes,'' submitted to JGR in 2018. This was
the first version of the 2022 paper.}

\noindent \textcolor{blue}{This is the first version of the paper
that eventually became the final, 2022 paper. This first version did
not derive the rigorous solutions given in later versions. The solution
for the eigenvectors/eigenvalues was described, but they were not
assembled into a full solution for homogenous plasma. Instead, the
wavepacket interpretation was described for each wave mode, as in
Cosgrove {[}2016{]}. The full solution was added to later versions
to address some of the difficulties of the reviewers, who were not
aware that the electrostatic solution should be derivable from the
``waves'' (which really just means working in the Fourier domain).
Also, this version included only an abbreviated description of how
the results could be extended to a model for the vertically inhomogeneous
ionosphere, which was described as future research. The main result
was found by integrating the wavevector over altitude, to show that
there was over $90^{\circ}$ of phase rotation even for a 100~km
transverse wavelength.}

\subsubsection{{\large{}Reviewer \#1}}

\noindent Reviewer \#1 (Formal Review for Authors (shown to authors)):

I do not recommend publication of this paper in its present form.
There are several misunderstandings of basic principles, and much
unnecessary polemic against a confusing concept the author calls \textquotedbl the
electrostatic assumption\textquotedbl . There is also a serious mistake
of physics. \medskip{}

\textcolor{blue}{The above text is a summary that does not identify
any specific criticisms, except that the reviewer had a hard timing
understanding my explanation of the topic. The reviewer also indicates
his/her basic unhappiness about the situation. In using the word}
\textquotedblleft polemic,\textquotedblright{} \textcolor{blue}{the
reviewer says that my language is too strong and is offensive to them.
In using strong language like} \textquotedblleft several misunderstandings
of basic principles\textquotedblright{} \textcolor{blue}{and} \textquotedblleft serious
mistake of physics\textquotedblright{} \textcolor{blue}{the reviewer
says that they are very unhappy about the results in the paper, and
really do not want them to be true. To the extent that the review
explains the reasons for using these terms, I refute them completely
below. }

\textcolor{blue}{\medskip{}
}

1. There are at least three different things that may be called \textquotedbl electrostatic
assumption\textquotedbl :\textcolor{blue}{\medskip{}
}

(a) The electrostatic assumption proper, as in \textquotedbl Electrostatics\textquotedbl{}
chapters of elementary textbooks: curl E =0, hence E given by electric
potential, governed by Poisson's equation; no magnetic field, current
J (if any) has curl J =0 (possible in conducting media only if conductivity
is isotropic and either uniform or depending only on electric potential).\textcolor{blue}{\medskip{}
}

(b) What are called \textquotedbl electrostatic waves\textquotedbl{}
in plasma physics: these are electromagnetic waves in which the magnetic
perturbation is sufficiently small to be neglected, delta B <\textcompwordmark <
delta E (Gaussian units), or phase velocity <\textcompwordmark <
c.\textcolor{blue}{\medskip{}
}

(c) The conventional treatment of large-scale ionospheric electrodynamics,
in which it is assumed curl E =0 but the electric potential is determined,
not from Poisson's equation but by relating E to J through the ionospheric
Ohm's law and requiring div J =0; magnetic disturbance field assumed
small compared to geomagnetic field (but not zero); time derivatives
in momentum equation neglected and plasma flow assumed given by E
x B drift.\medskip{}

\textcolor{blue}{The above concerns the Introduction section of the
paper. There is no criticism of the paper. The reviewer attempts to
give his or her own definitions for the topics in the paper (Case
(b) and Case (c) are fairly close, see more below). These provide
insight into the nature of the definition that the reviewer would
understand. I have used this knowledge to make the Introduction more
palatable to readers like the reviewer. \medskip{}
}

2. The main objective of the paper as well as of a published earlier
one by the author {[}Cosgrove 2016{]} is to criticize the \textquotedbl electrostatic
assumption\textquotedbl . The discussion is very confusing: the verbal
definition given by the author clearly corresponds to (a), but the
arguments seem to be directed against (b) or sometimes (c). To my
mind, the criticisms based on wave models are mostly irrelevant. Case
(a) does not have any waves (except possibly in some highly contrived
situations), and it never really applies in the ionosphere. Case (b)
is just an example of terminology not to be taken too literally (there
are others, e.g. \textquotedbl magnetosphere\textquotedbl{} is not
a sphere): anybody working in the field knows that \textquotedbl electrostatic
waves\textquotedbl{} are really electromagnetic, and that the validity
of neglecting delta B must be verified in every instance. Case (c)
invokes assumptions that limit its validity to time scales longer
than wave propagation times. \medskip{}

\textcolor{blue}{The above concerns the Introduction section of the
paper, where the topics are being defined. The reviewer says that
it was not clear I was discussing Case (c), as opposed to Case (a)
(using the reviewers categorization for the present). In fact, the
first sentence of the paper defines the main topic of the paper, which
is pretty close to what the reviewer calls Case (c). Later, the fifth
paragraph (Line 62) distinguishes Case (b) as a related but distinct
case. However, apparently, the reviewer could not follow the explanations,
and so I have rewritten them to be, hopefully, more clear. However,
it is very hard for me to understand how the reviewer could have so
dramatically misinterpreted the opening sentence as to think it addressed
Case (a). The opening sentence reads \textquotedblleft The assumption
of electrostatic equilibrium in ionospheric plasma physics {[}Cowling,
1945; Martyn, 1955; Dagg, 1957a, 1957b; Farley, 1959; Farley, 1960;
Spreiter and Briggs, 1961{]} begins with the assumption that setting
the time derivatives to zero in the combined Maxwell and fluid-momentum
equations (hereafter, Maxwell/momentum equations) determines a valid
approximation for the electric field, which is obtained by solving
the resulting elliptic equation for the electric potential.\textquotedblright{}
Note that my sentence reads \textquotedblleft in ionospheric physics\textquotedblright{}
and identifies the \textquotedblleft fluid-momentum equation,\textquotedblright{}
neither of which are relevant to Case (a). Case (a) has nothing to
do with fluid theory or the ionosphere; it involves only the Maxwell
equations. Also, the references given all discuss Case (c) and ionospheric
physics. }

\textcolor{blue}{In fact, I think my definition is quite precise.
Setting the time derivatives to zero in the Maxwell equations immediately
gives $\vec{\nabla}\times\vec{E}=0$ and $\vec{J}=\vec{\nabla}\times\vec{B}/\mu_{0}$,
so that $\vec{\nabla}\cdot\vec{J}=\vec{\nabla}\cdot\left(\vec{\nabla}\times\vec{B}\right)/\mu_{0}=0,$
which are two of the equations given by the reviewer. Setting the
time derivatives to zero in the momentum equations gives ``steady
state'' expressions for the ion and electron velocities, which, using
a well known procedure, are the equations commonly solved to obtain
what the reviewer calls the ionospheric Ohm's law, $\vec{J}=\overleftrightarrow{\sigma}\vec{E}$.
This is the third equation mentioned by the reviewer. Hence, it should
be easy to see that the definition given in the first paragraph of
the paper corresponds with what the reviewer called Case (c) (although
the reviewer does add one extraneous provision that isn't normally
used, which is that the ``plasma'' $\vec{E}\times\vec{B}$ drifts).
Also, it is well known that combining these three equations gives
an elliptic equation for the electric potential, as stated in the
paper. }

\textcolor{blue}{With respect to Case (b), Line 62 of the original
submission reads, ``A related but distinct concept to that of electrostatic
equilibrium is that of electrostatic plasma waves {[}e.g., Bernstein,
1958, Dougherty and Farley, 1963, Farley, 1963{]}, where the time
derivative is set to zero only in Faraday\textquoteright s law (i.e.,
$\vec{\nabla}\times\vec{E}$ set to zero), but the other time derivatives
are retained.'' Setting the time derivative $\left(\partial\delta\vec{B}/\partial t\right)$
to zero in Faraday's law essentially requires that $\delta\vec{B}$
is zero (or, sufficiently small), as indicated by the reviewer for
Case (b). Line 66 refers the reader to Section 2 for more detail.
Beginning on Line 200 of Section 2 there is quite a bit of additional
detail, including specific mention of the requirement that $\delta\vec{B}\cong0$.
I do not, however, give the particular inequality given by the reviewer,
because I don't feel that it can be proven to be a sufficient condition.
Rather, the validity of ``electrostatic waves'' (as Case (b) is
termed both in the article, and by the reviewer) is examined using
the full machinery derived in the paper.}

\textcolor{blue}{In writing a paper it is appropriate that I give
my own definition for the topic, and there is no need for it to correspond
with a definition given by a reviewer, which may not be accurate to
what I want to discuss. A solution obtained by solving the identified
equations with the identified time derivatives set to zero, etc.,
is the definition of my topic. The reviewer does not say what part
of this they do not understand. The reviewer does not identify any
error in the paper.}

\textcolor{blue}{The above also states} \textquotedblleft \dots criticisms
based on wave models are mostly irrelevant.\textquotedblright{} \textcolor{blue}{I
will refute this below, where the point is raised again. But let me
first mention that the reviewer themselves identified the Case (b)
as one of waves, and also said ``}Case (c) invokes assumptions that
limit its validity to time scales longer than wave propagation times.''
\textcolor{blue}{So the reviewers position against ``wave models''
seems to be somewhat inconsistent. \medskip{}
}

3. The discussion of sources and boundary conditions (lines 25-47)
contains a misunderstanding. It is true that steady-state solutions
of Maxwell's equations require the presence of sources somewhere,
but if one is solving the equations within a limited volume of space
only, with boundary conditions specified, it is NOT true that the
sources must be WITHIN the volume - they may be outside, indicated
by appropriate boundary conditions (e.g., I have a reading lamp that
runs on DC from a battery; if I put the lamp inside a large metal
box and connect it to the battery outside the box, I can certainly
describe the field inside the box by electrostatic equations.)\medskip{}

\textcolor{blue}{The above concerns the Introduction section, and
only the Introduction section. The purpose of the criticized material
was to address a possible objection to the goals of the paper, where
some might argue that since the electrostatic solution does solve
certain related equations, it must be the physical solution. The material
seeks to establish that this argument is not air tight, and thus that
it is appropriate to check if the electrostatic solution is the physical
one by the more rigorous means employed in the paper. What the reviewer
has written is nearly correct, but does not contradict the essential
point made in those lines (or anywhere else) of the paper. The key
word is ``appropriate,'' in that the enforced boundary conditions
must match very closely to those that would be produced by a real
and narrow source that lies almost completely outside the boundary.
When boundary conditions are chosen arbitrarily or incompletely, they
may not match to any real source, and so a non-physical solution may
be obtained. This is the main point that was made, and it is not contradicted
by what the reviewer has written.}

\textcolor{blue}{I said the reviewer is ``nearly'' correct because
there does remain a minor technical disagreement. In the material
referred to by the reviewer, the paper is asserting that the electrostatic
solution is (contained in) a solution of the source-free equations
of motion, if it is physical (i.e., if it does not have a singularity).
This point may be proven simply by noting that any valid solution
must be an analytic function, and so a solution of the source free
equations (inside some region) can be analytically continued to a
solution over all of space. This implies that the solution should
decay toward thermal equilibrium (since no sources), and so if it
is unchanging it must actually be thermal equilibrium. The only way
around this is if the solution is singular, which is therefore the
conclusion. The reviewer has not made any argument against this reasoning,
and may have read more into the point than was intended.}

\textcolor{blue}{However, I have decided that this point is rather
technical, and does not really say very much about whether electrostatic
theory might be useful as an approximation. So I have omitted it in
later versions of the paper. The response to the next reviewer question
includes a much more compelling reason for testing electrostatic theory.}

\textcolor{blue}{\medskip{}
}

4. Another misunderstanding is the statement (lines 74-75) that \textquotedbl the
wave equations are obtained by performing a Fourier transform in space
and a Laplace transform in time\textquotedbl . The wave equation
is one with the operator d\textasciicircum 2/dt\textasciicircum 2
-V\textasciicircum 2 d\textasciicircum 2/dx\textasciicircum 2;
Fourier and Laplace transforms are simply mathematical techniques
for solving equations, useful under certain restricted conditions.
Which one to use in which dimension is a question of convenience and
has nothing to do with physics. The convention of complex frequency
and real wave-number is convenient when solving an initial value problem:
given a periodic spatial disturbance at t=0, how does the wave evolve
in time? The opposite convention of real frequency and complex wave-number
is equally convenient when solving a boundary value problem: given
a periodic temporal disturbance at a boundary surface ( e.g., in Farley
1959 or in the author's figure 3), how does the wave propagate into
the interior? There is nothing unphysical or otherwise wrong with
(lines 81-82) \textquotedbl a Fourier transform in time and two dimensions
of space, and a Laplace transform in the remaining spatial dimension\textquotedbl{}
if it helps to solve the equation (e.g., Laplace transform in one
spatial dimension and Fourier transform in the other two is standard
technique for solving Laplace/Poisson equation in three-dimensional
planar geometry.) \textcolor{blue}{\medskip{}
}

\textcolor{blue}{The above again concerns the Introduction section
of the paper. Again, the purpose of the criticized material was to
address a possible objection to the goals of the paper, where some
might argue that since the electrostatic solution does solve certain
related equations, it must be the physical solution. The material
seeks to establish that this argument is not air tight, and thus that
it is necessary (or reasonable) to check if the electrostatic solution
is the physical one by the more direct means employed in the paper.
The point in the paper is, again, that an arbitrarily or incompletely
specified boundary condition may not correspond to a real source,
and so it is possible for electrostatic theory to give an inappropriate
answer. This motivates the analysis in the paper where we actually
test this question.}

\textcolor{blue}{In this case there is a much more compelling argument
for doubting electrostatic theory, than in the answer to the previous
question. The reviewer says, ``}There is nothing unphysical or otherwise
wrong with (lines 81-82) 'a Fourier transform in time and two dimensions
of space, and a Laplace transform in the remaining spatial dimension'
if it helps to solve the equation.\textcolor{blue}{'' I am in the
difficult position of having to say that this approach may sometimes
give inappropriate results. I have in fact already said this in \citeauthor{cosgrove-2016},
and this may have been a difficult thing for the reviewer to hear.
The reviewer says that it is a ``}question of convenience,''\textcolor{blue}{{}
whether we solve the equations of motion for the initial value problem,
or for a boundary value problem, that it ``}has nothing to do with
physics\textcolor{blue}{.'' This is in contradiction with what was
asserted in the paper, and in \citeauthor{cosgrove-2016} (Section~7).}

\textcolor{blue}{To elaborate on the arguments given in the paper,
the laws of physics are framed within the causative structure of spacetime,
which allows for formulation of the initial value problem: the allowed
instantaneous physical states of the system are parameterized by configuration
variables, and the equations of motion define how the configuration
changes with time. Because the time evolution is unique, the configuration
space at any instant of time maps one-to-one to the physical solution
space, meaning that any configuration can be chosen, and there corresponds
a single, physical time-evolving solution. Here, ``instant of time,''
is more precisely a space-like surface in spacetime, and specifying
the configuration means specifying an analytic function for the distribution
of events on that surface. That this is the proper framing of the
physical problem is ignored in solving the boundary value problem
that comes from making the electrostatic assumption. }

\textcolor{blue}{In making the electrostatic assumption there arises
a boundary value problem, meaning that it is necessary to specify
the events on a time-like surface. But unlike the events on a space-like
surface, the set of events on a time-like surface do not map one-to-one
to the solution space, because only certain temporal sequences of
events are allowed; the sequence of events must be as predicted by
the (source-free) equations of motion, which places a constraint on
the allowed boundary conditions. In using the electrostatic assumption
we ignore this constraint, and specify the boundary condition (which
is a sequence of events) without checking to see if the constraint
is satisfied. The only way around taking this risk (that we are specifying
an unphysical boundary condition) is to actually solve the equations
of motion.}

\textcolor{blue}{In fact it can be seen that the electrostatic boundary
condition is not really correct. The boundary condition in electrostatic
theory is normally specified by taking the Fourier transform in time,
and then solving for a Fourier component. If the frequency is non-zero
the boundary condition is oscillation over infinite time, but more
commonly the frequency is taken to be zero, so that the boundary condition
is no-change over infinite time. It is immediately apparent that this
does not match to the boundary condition that arises from the physical
initial value problem, where there is initially an undisturbed plasma,
until the source turns on, and then the source runs for a finite amount
of time, and (we assume) approaches a steady state.}

\textcolor{blue}{That this difference may be very important can be
seen by looking at the general expression for the conductivity, which
for example can be found in equation~A9 of {[}Cosgrove 2016{]}. The
conductivity depends on the frequency $\omega$, but it does not depend
on the wavevector $\vec{k}$. Since the Fourier transform has been
used $\omega$ is a real number, whereas if the Laplace transform
had been used as for the initial value problem, $\omega$ would be
a complex number. This is a serious issue because Figure~16 of {[}Cosgrove
2016{]} (right hand column) shows that including the imaginary part
of $\omega$ dramatically changes the conductivity. The reason for
this is exemplified in Figure~8 of {[}Cosgrove 2016{]} (left hand
column, for the Alfvén wave), which shows that the imaginary part
of $\omega$ remains large as the real part goes to zero. This occurs
because the imaginary part of $\omega$ determines the dissipation-rate
of the wave, and the dissipation is expected to remain as the real
part of frequency goes to zero.}

\textcolor{blue}{So this shows that there is a very concrete difference
between the solutions found through the Laplace-transform/initial-value
approach and the Fourier-transform/boundary-value approach. The conductivities
will be different, and this means that the conditions satisfied on
the boundary may be significantly different not just in the distant
past before the source turned on, but at all times.}

\textcolor{blue}{The difference between the two solutions may be viewed
as the difference between two different operating points on the dispersion
surface. I have to be careful with terminology because the term ``dispersion
relation'' is generally applied to specific wave modes, whereas by
``dispersion surface'' I mean all the solutions of the complex-plane-extension
of the equation that arises in solving the eigenvalue problem for
the linearized equations of motion, which is $\det\left(\omega I-H_{5}(\vec{k})\right)=0$,
where $\det$ indicates the determinant of the matrix, and $H_{5}$
is the matrix defined in the paper. In solving the initial value problem
$\omega$ is complex and $\vec{k}$ real, because the Laplace transform
is used for time and the Fourier transform for space. The 16 solutions
found in this case give the causal wave modes that enter into the
exact initial-value or driven-steady-state solutions of the equations
of motion. When sources are absent these modes decay with time according
to the imaginary part of $\omega$, and so we say that they are causal.}

\textcolor{blue}{But if we extend the domain of the determinant equation
to include complex values of $\vec{k}$ additional solutions become
possible. For example, there are solutions with $\omega$ entirely
real and $k_{z}$ complex, which are associated with solving the equations
of motion when the Laplace transform has been applied to the $z$-axis
and the Fourier transform to time and the two remaining spatial axes.
In Cosgrove {[}2016{]} it was explained that when $\omega=0$ only
one of these solutions has all finite components in the eigenvector,
and that this is the ``usual'' electrostatic solution. Specifically
this solution was given in equation~(19) of Cosgrove {[}2016{]}.
It decays with the Farley mapping distance and has the ``usual,''
zero-frequency conductivities found by setting $\omega=0$ in the
conductivity equations (e.g., equations~(A9) of \em{Cosgrove} {[}2016{]}).
As already noted, these conductivities are different from those found
when solving the initial value problem, where $\omega$ has a substantial
imaginary part that must be retained. In addition it seems almost
certain that the real part of $k_{z}$ (wavelength) will be different,
and that the eigenvectors will be different, and so the plasma configuration
should be quite different. These are different solutions that come
from very different operating points on the dispersion surface, with
the electrostatic solution having $\omega$ real and $k_{z}$ complex,
and the initial-value (and driven-steady-state) solution having $\omega$
complex and $k_{z}$ real.}

\textcolor{blue}{Since the solutions are different, they cannot both
be the physical solution. By virtue of the arguments based in causality,
it is clear that the initial-value and driven-steady-state solutions
are the ones guaranteed to be physical. So we must prefer the driven-steady-state
solution over the electrostatic solution, and expect that these two
may be quite different in some cases. In the event that they are in
fact quite different, the reason for the failure of the electrostatic
solution can be found in the choice of boundary condition, which did
not actually match to the condition that arises from causal evolution.
Hence we have a very good reason for testing electrostatic theory,
which we do by comparing it to an electromagnetic solution that should
contain it, if the electrostatic solution is accurate.}

\textcolor{blue}{The above is an elaboration on the arguments put
forth in the paper, which are also presented in \citeauthor{cosgrove-2016}.
}\textcolor{red}{The reviewer does not counter these arguments in
any way. The reviewer does not mention the discussion of causality
and the initial value problem that are given in both papers.}\textcolor{blue}{{}
While the reviewer is correct that} \textquotedblleft The opposite
convention of real frequency and complex wave-number is equally convenient
when solving a boundary value problem,\textquotedblright{} \textcolor{blue}{this
is irrelevant to the assertion made in the paper, which is that the
electrostatic solution is not guaranteed to be physical, and that
therefore it is appropriate to undertake the investigation in the
paper. In order to keep other readers from getting side-tracked in
this way, in later versions I have significantly reduced the discussion
of why electrostatic theory needs to be tested. In the final 2022
version I have essentially eliminated this discussion, and referred
to \citeauthor{cosgrove-2016} for the heavy lifting on this apparently
touchy subject. If the electromagnetic answer comes out different
from the electrostatic answer than that in-and-of-itself is an important
result, and afterward maybe the community will be more interested
in hearing how the electrostatic theory could have gone wrong.}

\textcolor{blue}{The reviewer also makes a semantical objection to
the use of the term ``wave equations,'' which I think I agree with.
To address this objection I have changed my terminology generally
to refer to ``equations of motion,'' which also helps to make my
point. \medskip{}
}

5. The serious mistake of physics, which is also present in the previously
published paper {[}Cosgrove 2016{]} and affects the conclusions of
both, is the author's assumption (lines 224-287 and elsewhere) that
an ionospheric wave can couple to a source (e.g., neutral-wind dynamo)
ONLY if its wavelength and its transverse phase velocity match the
source velocity and wavelength, respectively. The paper does not explain
clearly what is \textquotedbl transverse phase velocity\textquotedbl ;
it is V\_phase k /k\_x where k\_x is the wave-number in the direction
of source motion (horizontal in figure 3), but a casual reader might
misinterpret it as V\_phase k\_x /k. The assumption is that coupling
is possible only for wave modes for which (figure 3 geometry, theta
= angle from vertical)\textcolor{blue}{\medskip{}
}

V\_phase = V\_source sin theta \textcolor{blue}{\medskip{}
}where theta is the propagation direction. For Alfven waves, V\_phase
= V\_A cos theta and the coupling condition is\textcolor{blue}{\medskip{}
}

tan theta = V\_A /V\_source\textcolor{blue}{\medskip{}
}which can always be satisfied by choosing theta. For sound waves
(a simpler example than the whistler mode waves invoked by the author),
V\_phase = V\_s independent of theta, and the coupling condition is\textcolor{blue}{\medskip{}
} 

sin theta = V\_s /V\_source\textcolor{blue}{\medskip{}
}

which can be satisfied only if V\_s < V\_source. The author's assumption
would imply that an obstacle moving through a fluid can couple to
the fluid if it is moving supersonically, but not if it is moving
subsonically -{}-{}- which is obviously absurd. \medskip{}

\textcolor{blue}{In the linear approximation it can be proven that
``}an ionospheric wave can couple to a source (e.g., neutral-wind
dynamo) ONLY if its wavelength and its transverse phase velocity match
the source velocity and wavelength\textcolor{blue}{,'' and so what
the reviewer has characterized as a ``}serious mistake of physics\textcolor{blue}{''
is not any such thing. To make this clear an exact solution to the
(linearized) equations of motion has been included in all later versions
of the paper. A source is defined that oscillates with frequency $\omega_{0}$,
and which is a pure sinusoid in the transverse spatial dimension (transverse
to the geomagnetic field). The exact solution then finds that the
steady-state response oscillates exactly at frequency $\omega_{0}$,
and is exactly a (matching) pure sinusoid in the transverse spatial
dimension. (The source is equation~(10), and the solution
is equation~(12), in the final
2022 paper.) This is all that there is to the selection criterion
that is used in the paper. So there is no assumption like what the
reviewer is suggesting, it is proven.}

\textcolor{blue}{The best way I can understand what the reviewer is
saying is that they may be thinking of non-linear effects. But in
this paper we analyze the linear case, and why linearization is appropriate
has been explained thoroughly in the paper, and has also been explained
in the responses to other reviewer questions. So I will not include
that material here.}

\textcolor{blue}{\medskip{}
}

The fallacy lies in the implicit assumption that coupling must be
to a single simple wave, overlooking the possibility that coupling
can also occur by modification of the entire flow field when the waves
propagate sufficiently fast relative to motion of the source. All
the conclusions about non-coupling modes, both in this paper and in
Cosgrove 2016, are thereby invalidated.\textcolor{blue}{\medskip{}
}

\textcolor{blue}{The paper does not assume} \textquotedblleft that
coupling must be to a single simple wave\textquotedblright . \textcolor{blue}{All
the waves are considered, and four waves are found capable of operating
in the frequency/wavelength range relevant to ionospheric phenomena.
Of these four only two can transmit energy more than a few kilometers,
and these two are included in the analysis using a best-case assumption
for energy transfer through the ionosphere (and are both included
in the model that is completed in later versions of the paper). Three
other radio frequency waves are found to be much too high in frequency,
and anyway they do not follow geomagnetic field lines and so could
not facilitate the mapping of electric field along geomagnetic field
lines. Since there are 16 equations in the electromagnetic five-moment
fluid equations, there are 8 potential waves (each wave corresponds
to two eigenvectors, one for forward propagation, and one for reverse
propagation). This leaves one other possible wave, which was found
in the paper to be evanescent, and therefore not capable of transmitting
energy. Hence, all the modes have been included in the analysis, and
all those capable of transmitting energy are included in the calculations.}

\textcolor{blue}{Also, the reviewer asserts} \textquotedblleft \dots overlooking
the possibility that coupling can also occur by modification of the
entire flow field when the waves propagate sufficiently fast relative
to motion of the source.\textquotedblright{} \textcolor{blue}{I am
not able to relate this to any kind of physical reasoning. No physical
principle is invoked, and no equation is mentioned. The idea of ``}modification
of the entire flow field\textcolor{blue}{'' sounds non-causal. So
the only way I can interpret this statement is as an appeal to non-linearity,
in which case it does not deny the results in the paper. The paper
analyzes the linear problem and that has been well justified, both
in the paper and in these responses.}

\textcolor{blue}{The paper absolutely allows for wave propagation
that is faster (or much faster) than the source velocity. Note that
the source has infinite size in two dimensions and does not change
its location, so there is no stacking up of waves as in the shock-wave
case that the reviewer mentioned; that is a non-linear phenomenon
and so is not a suitable analogy. The selection criteria used in the
paper has been proven by an exact solution (see previous response),
and is also highly intuitive.}

\textcolor{blue}{The intuition for coupling into the plasma is described
in the beginning of Section~4. Basically, think
of a source that tickles the edge of a large plasma ``pool.'' For
energy to be transported any significant distance away from the source,
propagating modes have to be excited, or otherwise there will be only
a near-zone effect. The modes that will be excited are those that
match well to the tickling motions, whereas modes that do not match
well will be confounded. Expressing the equations of motion in the
Fourier domain we see that different Fourier components are coupled
only by non-linear terms, and so for a very slight tickling where
the source is a pure sinusoid, only the (transverse) wavelength of
the source should be excited. The excitations should also match the
frequency of the source. If there is more than one propagating mode
with this transverse wavelength and frequency, they may both be excited,
and these may have different wavelengths in the remaining direction
(i.e., the direction away from the source and into the plasma). This
is the intuition that applies to the linear domain, that is for small
excitations. And this way of thinking is well supported by well accepted
physical examples, such as transmission by an antenna and excitation
of resonances in a structure or circuit.}

\textcolor{blue}{The next step is to deal with a smaller ``pool''
of plasma where the waves may reach the other edge and reflect back.
The transmission line theory invoked in the paper allows for analysis
of the steady state reached after the waves bounce back and forth
as many times as is needed (to reach steady state). Under the traditional
MI coupling model, this is the process that is supposed to lead to
electrostatic equilibrium, and the reviewer even seems to acknowledge
this. So the analysis in the paper tests whether this process really
will lead to electrostatic equilibrium, or whether the steady state
may instead be markedly different. How much non-linearities can modify
this process is a whole other question, which should not apply for
sufficiently small-amplitude disturbances.}

\medskip{}

6. The only valid results that could be shaped into a publishable
paper are the descriptions of various wave properties and modes. The
author should delete completely all polemics about electrostatic approaches
(anything necessary to be said about them has been said already, at
more than sufficient length, in Cosgrove 2016), which includes changing
the title and completely rewriting the abstract. \medskip{}

\textcolor{blue}{It is odd to use the terms }\textcolor{black}{``polemic''}\textcolor{blue}{{}
and }\textcolor{black}{``criticism,''}\textcolor{blue}{{} instead
of the more even-handed term ``analysis.'' It is odd to refer to
my earlier paper with }``anything necessary to be said about them
has been said already, at more than sufficient length.''\textcolor{blue}{{}
It sounds as though the reviewer is not denying the paper is correct,
but never-the-less feels it should not be said.}

\textcolor{blue}{Almost all the objections were focused on material
in the Introduction, which material served only to provide a reason
for the analysis (and these objections have all been refuted). Otherwise,
the reviewer has stated }``All the conclusions about non-coupling
modes, both in this paper and in Cosgrove 2016, are thereby invalidated.''\textcolor{blue}{{}
But the selection criteria have been proven by an exact solution (see
two responses previous). The idea that an antenna transmits waves
according to its size and transmitter frequency is well established,
and the reviewer cannot criticize it unless they wish to appeal to
nonlinear phenomena.}

\textcolor{blue}{At the beginning of the review, the reviewer states}
\textquotedblleft \dots the criticisms based on wave models are mostly
irrelevant.\textquotedblright{} \textcolor{blue}{He/she seems to believe
that the ``wave model'' is somehow incomplete, like there is a separate
DC response that is not included. Actually, I have found this to be
a common misconception. So I have added material to the paper that
demonstrates a complete and general solution to the linearized electromagnetic
5-moment fluid equations. This is the solution that was used above
to prove the selection criteria (coupling criteria). Hence, the paper
can be understood as an analysis of this exact solution, and deriving
from it an approximate calculation for the vertically inhomogeneous
ionosphere. The approximations are all carefully documented, so that
future researchers may continue to evaluate them.}

\textcolor{blue}{So the most valid way to interpret the reviewers
statements is as an appeal to non-linear phenomena to save electrostatic
theory. This is the only way to interpret the otherwise meaningless
statement ``}\ldots coupling can also occur by modification of the
entire flow field when the waves propagate sufficiently fast relative
to motion of the source.''\textcolor{blue}{{} But even if it were the
case that nonlinear phenomena create electrostatic equilibrium (highly
unlikely!), it would not make either paper wrong; rather, the papers
would have shown that linear phenomena cannot produce electrostatic
equilibrium, a very interesting and unexpected result, especially
considering that electrostatic theory is itself linearized in the
same way!}

\textcolor{blue}{There is something causing a level of emotion in
the reviewer, and I think it has to be acknowledged that there is
a probable conflict of interest. The reviewer was unable to identify
any technical flaws in the paper, but they really cannot accept that
the results might be true. So they have attempted to block publication
through grandiose condemnations without technical merit.}

\medskip{}

The following should also be noted:\medskip{}

(a) The terminology for various wave modes now used by the author
is totally confusing. It is true that there is not complete unanimity
among various subfields on terminology, but this does not give the
author a license to redefine established terms according to his whims.\medskip{}

\textcolor{blue}{I did my best to use the most common terminology
that seemed applicable. The naming of waves is well known to result
in a zoo of terminology. Sometimes waves are named for the sources
that commonly produce the waves. Sometimes waves are named for the
simplifications that go into deriving their dispersion relations.
Sometimes waves are named for observational characteristics. Sometimes
waves are named for people. Sometimes the same waves are obtained
in different approximations, and given different names. Sometimes
waves just have different names in different fields. The reviewer
did not say which of the waves he/she thought was ill named, nor suggest
any names that should be used. In revisiting my naming scheme, I can
find nothing to change. Again, the reviewer has used strong language
(}``totally confusing\ldots does not give the author a license to
redefine established terms according to his whims''\textcolor{blue}{)}
\textcolor{blue}{without sufficient explanation, and thus has revealed
an emotional reaction to this paper. }

\medskip{}

(b) The basic equations (between lines 166 and 167) (credited to Schunk
and Nagy 2000, but I cannot find them in their book) are missing some
terms: (1) gravity in the momentum equations (can be neglected only
if all motions are horizontal, which manifestly is not the case for
several of the wave modes discussed), (2) the term Q (v\_i - v\_n)
in the momentum equations (momentum loading from photoionization).

\textcolor{blue}{I did forget to mention that I omitted gravity and
photoionization. I have corrected this in the revision. Of course,
no one would argue that gravity or photoionization are required to
achieve electrostatic equilibrium.}

\medskip{}

It should be pointed out that all effects of waves on the neutrals
are neglected -{}- neutral equations are not included. 

\textcolor{blue}{I have added this to the revision, although it's
not very relevant. No one would argue that neutral dynamics is needed
to achieve electrostatic equilibrium.}

\medskip{}

I would not consider the thermal equilibrium state of the ionosphere
\textquotedbl trivial\textquotedbl{} (line 174) -{}- not with gravity
included.

\textcolor{blue}{The reviewer is correct that including gravity makes
the equilibrium non-trivial. But I didn't include gravity, and again,
no one would argue that gravity is needed to achieve electrostatic
equilibrium.}

\medskip{}

{[}end of referee's report{]}

\subsubsection{{\large{}Reviewer \#2}}

\noindent This paper investigated a fundamental assumption used extensively
in ionospheric physics related to the electrostatic \textquotedblleft mapping\textquotedblright{}
of electric field between E and F region. Using the five- moment equations
for ionospheric plasma species together with the full Maxwell\textquoteright s
equations, the author also tried to test a hypothesis of using electrostatic
waves as approximations for electromagnetic waves in ionospheric dynamics.
Based on the linearized five-moment equations plus the Maxwell equations,
the wave modes are studied using the eigenvectors of the system. Based
on the reviewer\textquoteright s understanding, this theoretical study
found that 1) for transverse scale larger than 0.1 km, the (linear)
electrostatic modes differ significantly from the electromagnetic
modes; 2) electric field structures with transverse scale size < 100
km do not map from 400 km to 100 km in altitude; 3) the effective
ionospheric admittance (or conductance which is used more commonly
in M-I coupling) is a function of transverse scale size. 1) and 2)
are not so surprising given the fact of not using an electrostatic
assumption, and 3) is quite interesting and may have significant impact
on future studies of M-I coupling.\medskip{}

\textcolor{blue}{Ionospheric models are almost exclusively electrostatic,
and this includes models that operate over the transverse scale range
from 1~km to 100~km: so item (1) will (unfortunately) come as an
unpleasant surprise to the people who make and use these models. Also,
under electrostatic theory, 100~km transverse scale electric fields
easily map through the ionosphere, and 10~km transverse scales also
map: so item (2) differs from the results under the electrostatic
assumption by at least a factor of 10. So I think all three of these
results will come as a surprise to those who believe electrostatic
theory is reasonably accurate, and there are many such people in the
ionospheric community. }

\noindent \medskip{}

\noindent The reviewer found this kind of study intriguing, and theoretical
studies focusing on basic ionospheric physics should always be encouraged
in general. However, there are some major concerns about the clarity
of the methodology and the applicability of the theoretical conclusions
to complicated ionospheric configurations. In my own opinion, the
title, abstract and conclusion should also be improved or adjusted
based on proper physical interpretation of the results. Therefore
the reviewer does not think the manuscript ready for publish in JGR
yet, before a significant amount of improvement/clarification is done.
\medskip{}

\textcolor{blue}{This study develops methodology that could be applied
to more complicated configurations, but we have to walk before we
can run. The purpose of this study is to show that electrostatic and
electromagnetic theories can sometimes differ in important ways, in
order to motivate the much bigger job of executing and testing the
model that is proposed (which has been completed in later additions
of the paper). If the two theories disagree for the simplest of plasma
configurations (a homogeneous plasma), well that has never been shown
before, and it gives reason to work more on the topic. }

\textcolor{blue}{I don't actually find any reference to the conclusion
in the reviewers comments below, so I will comment here. The concluding
section has been reworked and hopefully is less antagonizing now.
But to be clear, I cannot find any problem with the physical interpretation
given in the original paper. I hope that the new Section~2 will provide
the needed additional rigor to convince the reviewer that this material
should go out to the broader community. It is difficult material with
lots of subtlety, and so I can only be so successful in making it
easy to understand. But there are now many more equations to back
things up, and so the correctness of the physical interpretation should
now be connected to the correctness of the mathematical development,
which should allow for a more concrete discussion.}

\noindent \medskip{}

\noindent \begin{flushleft}
\textbf{Major concerns }\medskip{}
\par\end{flushleft}

\noindent The reviewer\textquoteright s major concerns are a) in the
abstract, b) in the description related to equation (1) and c) in
the derived eigenvectors shown in Figure 4. \medskip{}

\noindent a) In the Abstract:\medskip{}

\noindent In the abstract, the author said \textquotedblleft the electrostatic
assumption has not been validated, either observationally or in theory.\textquotedblright{}
This statement is not quit true since observations have already shown
the mapping of large-scale magnetospheric electric field to ionospheric
reference altitudes (e.g., Figure 2 of Gonzales et al. 1980). The
validity of the electrostatic assumption and electric field mapping
actually depends on the spatial and temporal scale of the problem
of interest. Therefore it is not accurate to say that the electrostatic
mapping of electric field has not been seen in datasets, it is more
about the limitation of the mapping assumptions. \medskip{}

\textcolor{blue}{I have changed the abstract to read ``...and observational
evidence is weak'' in the revision. The Gonzales 1980 paper used
an incoherent scatter radar (ISR), and these measure electric field
in the F region, well above the E region altitudes where I am finding
that mapping can fail. ISR can only be used for E region electric
fields through the assumption that there }\textcolor{blue}{\emph{is}}\textcolor{blue}{{}
mapping. Also, mapping from the magnetosphere to the F region }\textcolor{blue}{\emph{is}}\textcolor{blue}{{}
predicted by the electromagnetic theory; it is predicted because of
the very long wavelength of the Alfvén wave. So the Gonzalez 1980
paper, while important for other reasons, does not actually address
a case where different results are expected. }

\textcolor{blue}{I believe this paper is the first time significant
deviations between the electrostatic and electromagnetic theories
have been found (for the ionosphere). So what is needed now is an
observational/experimental validation that addresses cases where the
two theories give different results, and this could not have been
done before in any definitive way, because the opposing electromagnetic
theory (the one from this paper) did not exist before. So this is
really what I meant, although I agree that my language could have
been misinterpreted, and so the revision is an improvement. }

\noindent \medskip{}

\noindent In the abstract \textquotedblleft Using an exact solution
for the Maxwell..\textquotedblright{} -- the results presented in
the manuscript is not an exact solution of the full equation set (are
exact solutions to the coupled fluid/Maxwell equations ever exist?),
rather, it\textquoteright s only possible to derive the Eigen system
of the linearized equations, which is not an exact \textquotedblleft solution\textquotedblright .\medskip{}

\textcolor{blue}{I have added the word ``linearized'' to the sentence
in question (line 20). (The word ``linear'' was in the title, by
the way, and still is.) I did obtain an exact solution to the linearized
equations, and the revised paper has been modified to make this clear
(Section~2). (Of course, it is not anything amazing to have an exact
solution to a set of linear equations.) }

\noindent \medskip{}

\noindent In the abstract \textquotedblleft using either electrostatic
waves or the electrostatic assumption can lead to results that are
not just quantitatively wrong, but also qualitatively wrong----such
as failure to predict instability\textquotedblright . This statement
is likely too strong especially given that no observational evidence
is provided in this study. While it is fine to present results which
contradict previous studies, the author needs to provide more quantitative
evidence to convince potential readers about why all these assumptions
are wrong, how much they are off compared to the results presented
in this study, etc. \medskip{}

\textcolor{blue}{I have removed that sentence from the abstract. However,
at the risk of being argumentative, I do want to say that I don't
think the observations are all that relevant at this time. From the
well accepted laws of fundamental physics, when there is an electromagnetic
solution that appears air-tight, it has to be considered above any
electrostatic solution, unless there actually exist observations that
definitively indicate otherwise. It is very unlikely that there could
have been a paper making such contrary observations, since the results
I'm presenting here are new. There are some clear distinctions in
the predictions of the two theories. One I will mention is the phase
rotation of over $90^{\circ}$, in going from the F region to the
E region, for any scale size less than 100~km. (The phase rotation
under electrostatic theory is zero for all scale sizes.) Another has
to do with the relationship between field-aligned current and transverse
electric field in a homogeneous plasma excited by a dynamo on one
edge, there is a $90^{\circ}$ difference in the predicted phase relationships
between these two quantities. These two are both major deviations
from electrostatic theory; they are important wave effects, and they
are quantitative predictions. And then there are also the differences
in conductivity, and in dissipation scale length, which are also quantitative
differences that are not minor.}

\textcolor{blue}{Researchers should try to test these predictions,
and if they ever find a case where the electrostatic theory works
better than the electromagnetic theory there will be quite a conundrum.
But until that happens, I feel it is irresponsible to continue using
the electrostatic theory in cases where it gives a different prediction,
without at least citing this work. Of course, I understand that this
is because I have confidence in my analysis, whereas the rest of the
world may not be ready to believe it yet. It is for this reason that
the abstract should be somewhat bold. I don't want this paper to be
ignored as just another finding that the two theories agree except
for very minor differences; there appear to be major differences;
so I think the abstract needs to make people take notice and deal
with it. My 2016 paper has been ignored so far, even though it shows
important differences also.}

\noindent \medskip{}

\noindent b) In Equation (1) \medskip{}

\noindent Equation (1) is a relative comprehensive set of fluid equations
for the dynamic evolution of partially ionized plasmas (without coupling
back to neutrals). The author needs to justify why so many terms are
needed given that they are different by orders of magnitude when applying
to ionospheric plasma. For example, for most ionospheric plasma phenomena
of interest, the time derivative terms in the fluid equations are
much smaller than the other terms. This is simply seen in the ion
velocity equation (the fourth equation), the time derivative term
has a magnitude of $1/\tau$, where $\tau$ is the response time to
a new set of forces, or a time scale of interest. The Lorentz term
has a magnitude of $\Omega_{i}$, which is the ion gyro frequency.
Thus for most ionospheric phenomena of interest, $1/\tau\ll\Omega_{i}$.
A similar analysis can be done for the friction term as well since
$1/\tau\ll\nu_{in}$. Thus the acceleration term can be neglected.
The convective term is also small according to the same rationale,
and none of these terms are scale dependent in equation (1). Therefore
although the time derivative terms may introduce interesting wave
dynamic, are they really dominate the whole solution to equation (1)?
Also the electron-ion collision frequency is at least one order of
magnitude lower than the electron-neutral and ion neutral frequency
across all altitude range in the E and lower F ionosphere, so terms
associated with $\nu_{ie}$ ($\nu_{ei}$) are also negligible in equation
(1). \medskip{}

\textcolor{blue}{The reviewer brings up some very interesting questions!
However, they are really relevant to different questions than the
one I am addressing in this paper, which is whether electrostatic
theory really leads to a reasonable approximation for all the cases
that we currently think it does. To answer this question, it is certainly
permissible and in fact desirable to not make approximations of any
kind, to the extent possible. Any approximation I make opens-up a
possible objection. The electrostatic theory exists because it is
not so easy to refute directly the arguments that go into it. However,
if I can show differences using an approach that does not make approximations,
then it means that somehow the arguments for the electrostatic theory
are subtly wrong. This is a publishable result, whether or not I am
able to convincingly refute the usual arguments for electrostatic
theory through intellectual discourse. With this goal in mind there
is value in using equations that are more accurate than necessary,
if it does not prevent their solution. The paper is already very long,
and there simply isn't room for such things, even if I were able to
word them convincingly.}

\textcolor{blue}{I also want to mention that the time derivative terms
do not only describe the acceleration and convection, they also describe
the decay of the disturbance as energy is dissipated. So there is
possibly more risk than the reviewer is considering, in setting a
time derivative to zero. This is just about being cautious.}

\textcolor{blue}{A more direct answer that addresses one specific
part of the question is as follows: If an equation is linearly coupled
to other equations, then before setting a time derivative to zero,
it is really necessary to check the sensitivity in the other equations.
So to do it properly we should first diagonalize the equations (I'm
assuming we are talking about the Fourier transformed and linearized
equation set). To do this, let $U$ be the matrix with the eigenvectors
of $H_{5}$ as columns. Then the matrix $D=-iU^{-1}H_{5}U$ is diagonal,
and the equation of motion for the $j$th component becomes,
\begin{equation}
\frac{\partial X_{j}^{\prime}}{\partial t}+D_{jj}X_{j}^{\prime}=0,
\end{equation}
where $\vec{X}^{\prime}=U^{-1}\vec{X}$, and there is one of these
scalar equations for each eigenvector (See equation~(2) in the revised
paper). So $X_{j}^{\prime}$ is the amplitude of the $j$'th mode,
and these are just the same waves discussed in the paper. Now the
evolution is completely decoupled, and we can clearly see the effect
of setting a time derivative to zero. If we set the $j$th time derivative
to zero, it means that $X_{j}^{\prime}$ must be zero, and so there
can be no energy in that wave mode. So this allows us to clearly understand
the effect of setting a time derivative to zero, and it amounts to
discarding the associated wave mode. But if we set a time derivative
to zero before diagonalizing, the effect ripples through all the equations,
and all the wave modes can be affected. To verify that this is a good
approximation it is necessary to evaluate the sensitivity of each
important wave mode, because it is possible that a very small change
in the terms of the equation where the time derivative was set to
zero adds up to a significant change in a wave mode that depends on
all the equations coupled together.}

\textcolor{blue}{The point is that debunking the usual arguments for
the electrostatic assumption is likely possible, but that it requires
long and subtle arguments that are not likely to convince people.
But if by not making approximations I find discrepancies, I effectively
prove that such arguments exist. This serves for the purposes of this
paper. And since this paper is already very long, I feel that examination
of what approximations are actually OK must be left to another paper.
Otherwise the bar would be set too high, and it would not be possible
to publish these important results.}

\noindent \medskip{}

\noindent The author claims that \textquotedblleft Linearization is
justified by the idea that we are exploring the dynamical evolution
of small perturbations (i.e., second order terms negligible) about
some equilibrium state, and that these represent characteristic motions
of the system about the equilibrium.\textquotedblright{} This is somewhat
inconsistent/confusing with the motivation of this study. For example
the equilibrium states are derived by setting all the time derivatives
to be zero, which is basically a version of the electrostatic assumption.
If such \textquotedblleft equilibrium\textquotedblright{} states are
somewhat independent and cannot be described by the linearized equation
sets used here, then the theory presented in this manuscript relate
is only a modification to the \textquotedblleft given\textquotedblright{}
equilibrium state, which is based on the electrostatic assumption.
\medskip{}

\textcolor{blue}{The reviewer is correct that the equilibrium state
being linearized about is a special case of electrostatic equilibrium,
known as thermal equilibrium. However, the electrostatic equilibrium
solutions (other than thermal equilibrium) are also derived by linearizing
the equations about thermal equilibrium. It is true that other background
states are sometimes linearized-about to derive electrostatic equilibrium,
but this is a heuristic extension, and the paper uses this same heuristic
extension to develop the electromagnetic model for an inhomogeneous
plasma, in Section~7. But the canonical manifestations for both electrostatic
equilibrium and the solutions developed in Section~2 are as linearizations
with respect to thermal equilibrium. Both reduce to thermal equilibrium
when the excitation is zero, and both apply only to small excitations
about thermal equilibrium. So the solutions developed in Section~2
are appropriate for testing electrostatic equilibrium in canonical
form, and the model proposed in Section~7 is appropriate for testing
the heuristic extension of electrostatic equilibrium to an inhomogeneous
background.}

\textcolor{blue}{Normally authors simply state that non-linear terms
are dropped without mentioning the role of the background state, and
it is easily forgotten even that there is one. You might say that
I have gone above and beyond the call of duty by including so much
discussion of the linearization process, and it may have been a source
of confusion, especially since I failed to explain that electrostatic
equilibrium is linearized in the same way. So the reviewers comment
is useful and discussion has been added to the paper.}

\noindent \medskip{}

\noindent The author may also need to justify the choice of ignoring
important terms such as the neutral wind and density gradient quantitatively,
given that the ionosphere is not a localized plasma system but is
largely driven by solar radiation and the neutral thermosphere. In
other words, when talking about scales above several hundred kilometers,
the ionosphere is not only stratified vertically, but also exhibits
longitude/latitude variations especially in the equatorial region
where interesting phenomena such as plasma instabilities occur.\medskip{}

\textcolor{blue}{The reviewer is correct that inhomogeneities and
internal sources (like wind) are very important. However, this paper
is an early-stages investigation of the theoretical limits of the
electrostatic assumption. For this purpose, I feel it is appropriate
to consider simple gedanken experiments, in order to understand the
breakdown of the assumption in simplest form. We have to correct the
theory before we can correct the models. Section 6 of the original
submission (Section~7 in the revision), begins the development of
a more realistic model, which could be used to analyze a vertically
stratified ionosphere. But it is simply not practical to complete
and evaluate this model in the present paper (update, the model has
been implemented in later versions of the paper, at the expense of
making them very long). There needs to be a motivation to pursue a
more complex model, and the purpose of the present paper is to provide
that motivation. Analyzing the homogeneous case does allow comparison
of the electrostatic and electromagnetic theories, when we think they
should both apply, and so serves the purpose. The paper does also
contain an analysis of electric field mapping through the inhomogeneous
ionosphere using a sort of best-case analysis based on the idea of
the model in Section~7. This is how the $90^{\circ}$ ``electrical
thickness'' was calculated. But it is simply not possible to go further
at this time. (Update, I did go much farther in the later editions,
and made the whole model, since this paper was not accepted.)}

\noindent \medskip{}

\noindent In the linearization process, $v_{i}=\delta v_{i}$ and
$v_{e}=\delta v_{e}$, but in the matrix form shown in Figure 1, additional
variables such as $V_{i0x}$, $V_{e0x}$ occurs in $H_{5}$, which
are not defined anywhere in the manuscript, and are not consistent
with what\textquoteright s described in the manuscript. Thus it is
unclear how to get the linearized version of equation (1) shown in
Figure 1.

\textcolor{blue}{Those variables are the zeroth order velocities,
which are zero in this application. I noticed I also forgot to define
the zeroth order pressures, and $\hat{B}$. Sentences have been added
to define all these variables and explain how they are set (Appendix
II). Thank you for actually looking at the messy matrix!\medskip{}
}

\noindent In line 180, the author states that \textquotedblleft After
the Fourier and Laplace transforms these equations are linear and
homogenous\textquotedblright , equation (2) is basically derived from
dropping all the non-linear and zeroth-order terms. However, in line
186, the author says \textquotedblleft To be clear, $H_{5}$ is formed
without dropping any terms or making any approximations\textquotedblright .
These two statements are inconsistent. 

\textcolor{blue}{Actually, $H_{5}$ (a matrix) was defined as the
linear part, and so the statement is technically correct. Although
I have retained that sentence, the paragraph is now altered so that
I think the reader will not have your question anymore. Also, the
zeroth order equations are satisfied, they are not dropped; thermal
equilibrium satisfies the zeroth order equations, and then the perturbation
terms are determined by the higher order equations (where we drop
second order and above).}

\noindent \medskip{}

\noindent Specific Clarifications for equation (1): \medskip{}

\noindent Variables undefined: $Q$, $\xi$, $\epsilon_{0}$, $\mu_{0}$,
$K_{B}$. 

\textcolor{blue}{These variables are the generation rate, the coefficient
for the recombination rate, the permittivity of free space, the permeability
of free space, and the Boltzmann constant, respectively. I have added
their definitions to the text, and mentioned that $Q$ is chosen to
balance recombination in zeroth order. }

\noindent \medskip{}

\noindent Coordinates system for generating Matrix $H_{5}$ undefined:
e.g., define the $\vec{k}$ vector and $\vec{B}_{0}$ vector. 

\textcolor{blue}{Sorry there was a typo and I had both $\vec{B}_{0}$
and $\vec{B}_{g}$ for the geomagnetic field. I have also now indicated
specifically that $\vec{k}$ is the wavevector for the Fourier transform.
The matrix $H_{5}$ actually does not assume a coordinate system,
but in plotting the eigenvectors I place $\vec{B}_{0}$ along $\hat{z}$
and take $k_{x}=0$, which I did forget to mention (now said in lines
591-592). These things are corrected in the revision, and I have also
added to the figure captions for the eigenvectors, to support the
casual reader.}

\noindent \medskip{}

\noindent c) The Eigen Vectors \medskip{}

\noindent First of all, the x-axis labeled in Figure 4 is not consistent
with the definition of state vector $X$ used in equation (2). 

\textcolor{blue}{I assume the reviewer is referring to the fact that
the $\delta$s are missing for some of the variables, which was a
mistake. However, adding the $\delta$s requires shrinking the font
more than is desirable, so I have decided to solve the problem by
omitting the $\delta$s from the definitions of those variables, making
the figure correct. Possibly, the reviewer is instead referring to
the fact that the figure axis contains all upper case variables. If
so, then the reviewer may have missed where some of the lower-case
variables were rescaled, with the rescaled variables denoted by upper
case. This was done right after equation~(2), in line 182 of the
original submission, and for the revision it is in Appendix II (lines
1520-1521). }

\noindent \medskip{}

\noindent In Figure 4, the author shows examples of four Eigen modes
calculated using specific parameters, however, the detailed parameters
used for computing the eigenvectors are not provided except for density,
altitude and transverse scale. These three alone does not provide
enough information about the average state of the ionosphere considered
as an example and how the matrix $H_{5}$ is generated. The author
may want to consider providing a detailed table/list of all the parameters
used in matrix $H_{5}$ to derive the normalized eigenvectors. The
author should also consider provide reasonable validations in order
to show that the Eigen system is calculated correctly, if possible.
\medskip{}

\textcolor{blue}{The figures in the appendix provide the remaining
variables as a function of altitude. So providing the altitude does
actually provide all the needed information. I have added sentences
in a few places referring the reader to the appendix (Appendix I in
the revision).}

\textcolor{blue}{Regarding the correctness of the eigensystem, in
Cosgrove {[}2016{]} I compared the eigenvalues with the standard,
non-collisional dispersion relations for the X-mode, O-mode, Z-mode,
whistler mode, fast Alfvén mode, and shear Alfvén mode. The results
were very good, with differences arising only in ways that make sense
for collisional effects, and considering the relationship between
the whistler and fast-mode waves. The ``standard'' dispersion relations
were taken from {[}\citealp{stix-1992}{]}. In the current paper I have
added the continuity and momentum equations, but I have checked the
eigenvalues against those from the Cosgrove {[}2016{]} version and
everything looks good!}

\textcolor{blue}{I have also checked the numerical accuracy by multiplying
$H_{5}$ by the eigenvectors and comparing to multiplication by the
eigenvalues. The accuracy is generally very close to 15 decimal places.}

\noindent \medskip{}

\noindent The eigenvectors shown in Figure 4 are also confusing. For
example in the Alfven mode, there is a non-zero $\delta B_{z}$ component
in the solution. What does a finite $\delta B_{z}$ component mean
in the Alfven mode physically (maybe the reviewer is mis-interpreting
the results since no information about the background B field is given.
Is it in the z direction?). Why all the electric field components
are zero in all the modes shown in Figure 4? What does it mean physically
that all the E-field components are zero in the Eigen vector? Also
the Ion and Thermal modes are not coupled to any electromagnetic fields,
does that mean they should be eliminated in the equations before the
calculations in order to simplify the analysis? \medskip{}

\textcolor{blue}{With respect to the electric fields, they are not
actually zero, for any of the modes; it's just that they are sufficiently
small that they cannot be seen with the linear axes. Using plots with
logarithmic axes allows, in some ways, a better comparison, but since
these are signed quantities I decided that was not a good solution.
Also, it should be understood that all the variables can be rescaled,
which changes the relative amplitudes within a single eigenvector;
so the electric fields could, arbitrarily, be made the largest components.
It was actually quite difficult to choose good relative scalings for
the variables, ones that did not seem too arbitrary, and which balanced
the components relatively well. The units for electric field are related
to those for magnetic field by a factor with units of velocity. The
decision was to set this factor equal to the speed of light, which
means the plots actually contain magnetic field in Gaussian units.
As we can seen by referring to Figure~3 of the Cosgrove {[}2016{]}
paper, this choice makes the electric field dominate for the radio
frequency modes, but for the lower phase velocity modes the magnetic
field takes over. This is consistent with the idea that the ``displacement
current'' is negligible in the ionosphere, where there is an assumption
that the radio frequency waves are not relevant. However, I think
your point is well taken in the sense that since I am not showing
the radio frequency modes, I should have used a different normalization
for this plot that allows electric field to be seen.}

\textcolor{blue}{By the way, the rescaling freedom does not affect
the eigenvalues, or any of the physical quantities such as conductivity;
it is a sort of gauge freedom. The purpose of these figures is to
compare one mode to another, for example, to see that the Whistler
wave has a larger magnetic perturbation than the Alfvén wave. This
purpose remains valid even with the rescaling ambiguity. I have added
a sentence explaining about this rescaling ambiguity, and I'm sorry
it did not appear in the original submission. The figure caption has
also been modified. I think maybe this is one of the things the reviewer
meant about the physical interpretation not being clear, which I can
understand. That was an important point, where my explanation was
insufficient, but nothing actually wrong. Due to the difficulty in
explaining so much material I have decided to omit these kinds of
plots in the final 2022 version of the paper.}

\textcolor{blue}{With respect to $\delta B_{z}$ for the Alfvén wave,
the prototype Alfvén wave is derived in a collisionless low-frequency
approximation that finds three modes, the shear mode, fast mode, and
slow mode. The fast mode wave transitions into the whistler wave,
when moving to the high frequency regime, and so they correspond to
the same eigenvector {[}Cosgrove 2016{]}. The shear mode is the one
normally just called the Alfvén wave by ionospheric physicists. The
slow mode is the one normally called the ion-acoustic wave by ionospheric
physicists. The fast mode normally has a $\delta B_{z}$ component,
but the shear mode does not. So the existence of a $\delta B_{z}$
component for the collisional (shear) Alfvén wave does not seem so
surprising, it just means that full inclusion of collisions mixes
somewhat the properties of the collisionless fast and shear modes.
Beyond this it's not clear what kind of answer is desired. One could
attempt to simplify the equations and to figure out precisely which
terms control $\delta B_{z}$ in the collisional Alfvén mode, but
that is a very big job that is not very relevant to this study. If
the question is actually whether there could be a mistake in the matrix
$H_{5}$, that is why I have shown that matrix in all its gory detail.
The $\delta B_{z}$ component also shows in the Cosgrove {[}2016{]}
paper, which constitutes a completely separate derivation of the (very
similar) equations and also implementation in Matlab instead of Python.
So while I hope there will some day be an independent derivation,
I think the matrix $H_{5}$ is correct.}

\textcolor{blue}{Also, the Ion (i.e., ion-acoustic) and Thermal waves
are electromagnetic waves that are well approximated by electrostatic
waves. I did not show that they were completely decoupled, only that
they can't transmit energy very far. As far as whether there should
be simplifications that somehow eliminate them, the purpose of this
work is best satisfied by avoiding approximations except when necessary.
Making approximations opens-up the possibility that the solution could
be wrong, and I want to squash that possibility. Actually, I think
the best way to remove the Ion and Thermal waves (when desired) is
to first solve for the eigenvectors/eigenvalues exactly, and then
just drop the associated eigenvectors/eigenvalues.}

\noindent \medskip{}

\noindent Minor concerns/Clarifications \medskip{}

\noindent The key points 1 and 2 are too broad, key point 3 does not
provide useful information as well. The key points should convey the
main points and conclusions of the article. 

\textcolor{blue}{Agreed, I have made the key points more informative.}

\noindent \medskip{}

\noindent The author may also consider make the title more focused
and specific. \textquotedblleft plus 150km echoes\textquotedblright{}
seems less related to the first part of the title. 

\textcolor{blue}{Agreed, I have changed the title for the revision,
and removed the subject of 150~km echoes entirely. It will have to
go in another paper.}

\noindent \medskip{}

\noindent Line 25, what kind of sources? Please specify. Line 27-31,
the reviewer is not quite clear about what these descriptions mean
without examples or qualitative analysis. \medskip{}

\noindent \textcolor{blue}{I have added a couple of sentences to the
Introduction defining the term ``source'' (lines 102-107). But since
the Introduction is now changed quite a bit, the reviewer may have
to give me additional feedback on the current state of clarity. The
other reviewer seemed to be triggered by that material, and so I wanted
to de-emphasize it, rather than go into it more. This relates to my
point above about the subtlety involved in debunking the usual arguments
for the electrostatic assumption; they perhaps read as technical objections.
(Although, I think the point about the imaginary part of frequency
being omitted is quite clear and convincing.)}\medskip{}

\noindent Line 112: \textquotedblleft the usual dynamo theory {[}e.g.,
Richmond, 1973; Kelley, 2009{]} needs to be reworked, in order to
correctly account for the loading effects of the surrounding plasma.\textquotedblright{}
The dynamo theory applies to a magnetized plasma system driven by
the neutral thermosphere, which is also determined by the spatial
variations (e.g., gradient) of ionospheric plasma. None of these factors
are included in the analysis, so how does the linear theory presented
here apply to the re-work of the usual dynamo theory? Can the author
provide more details? For example, does the pre-reversal enhancement
phenomenon need to be revised?\medskip{}

\textcolor{blue}{Actually, the effects from back-reflection associated
with ``}spatial variations'' \textcolor{blue}{are fully included
in the model for a vertically inhomogeneous ionosphere, which was
described in the section ``Modeling the Ionospheric Admittance.''
And this model has been implemented in later versions of the paper
and is now the main focus. Also, the effects from back-reflection
were discussed in Section~6 of {[}Cosgrove 2016{]}, where it was
shown that they do produce a more normal, real-valued conductance.
So the reviewer is correct that these effects do allow for producing
a more normal dynamo loading effect, but is not correct that they
are not included in the analysis. The original version of the paper
began by discussing a homogeneous plasma, and then only at the end
discussed the more general case with back reflection. But that discussion
comes much earlier in the final, 2022 version.}

\textcolor{blue}{Note that the reviewer is commenting on the discussion
of the material from my previous 2016 paper; line~112 is in the section
(of the original submission) titled ``Previous Work'' and provides
a summary of {[}Cosgrove 2016{]}. I agree that the summary was inadequate,
and in all later versions of the paper is added a section called ``gedanken
experiment'' that describes the relationship between cases with and
without back reflection. I think that this section may better highlight
the importance of back-reflection, compared with what was actually
said in {[}Cosgrove 2016{]}, although the ``wave-Pedersen conductivity''
was originally developed there.}

\textcolor{blue}{The pre-reversal enhancement is definitely something
I would like to look at, since the interplay between the E and F regions
may be important there, although I'm not sure about scale.}

\textcolor{blue}{By the way, dynamos are normally treated using electrostatic
theory, which is also derived from the linearized equations. So using
the linearized equations does not limit the ability to test electrostatic
theory, for dynamos or otherwise.}

\noindent \medskip{}

\noindent Line 138: please define the physical meaning of the term
\textquotedblleft electric thickness.\textquotedblright{} Line 150:
please define the physical meaning of the term \textquotedblleft electric
length\textquotedblright{} \medskip{}

\textcolor{blue}{The electrical thickness of a region, with respect
to a particular wave, is the amount of phase rotation suffered by
the wave in traversing the region. In this case the region is the
ionosphere. So the two terms are related in that the ``electrical
thickness'' of the ionosphere is the ``electrical length'' from
the ``top'' of the ionosphere to the ``bottom,'' where 400~km
and 100~km are the examples used for top and bottom, respectively,
in the paper. Sentences have been added to later versions.}

\noindent \medskip{}

\noindent \medskip{}

\noindent Line 229: \textquotedblleft the dynamo layer will move relative
to the geomagnetic field lines at approximately the wind velocity\textquotedblright{}
This statement is generally true for the F-region dynamo but not for
the E- region dynamo.\medskip{}

\textcolor{blue}{I have modified this by adding ``ignoring polarization
electric fields, which will depend on the detailed configuration''
within the sentence mentioned. I believe this makes the statement
true, and here is why. Ignoring diffusion, the ion and electron velocities
derived from the steady-state momentum equations may be written,
\begin{eqnarray*}
\vec{v}_{i} & = & \frac{\rho_{i}}{1+\rho_{i}^{2}}\left(\frac{\vec{E}_{\perp}}{B}+\vec{u}\times\hat{b}\right)+\frac{\vec{E}_{\parallel}}{\rho_{i}B}+\frac{1}{1+\rho_{i}^{2}}\left(\frac{\vec{E}_{\perp}}{B}+\vec{u}\times\hat{b}\right)\times\hat{b}+\vec{u},\\
\vec{v}_{e} & = & \frac{-\rho_{e}}{1+\rho_{e}^{2}}\left(\frac{\vec{E}_{\perp}}{B}+\vec{u}\times\hat{b}\right)-\frac{\vec{E}_{\parallel}}{\rho_{e}B}+\frac{1}{1+\rho_{e}^{2}}\left(\frac{\vec{E}_{\perp}}{B}+\vec{u}\times\hat{b}\right)\times\hat{b}+\vec{u},
\end{eqnarray*}
where $\rho_{i}$ is the ratio of the ion-neutral collision frequency
to the iono-cyclotron frequency, $\rho_{e}$ is the ratio of the electron-neutral
collision frequency to the electron-cyclotron frequency, $\vec{u}$
is the neutral wind velocity, $\vec{B}=B\hat{b}$ is the geomagnetic
field, $\vec{E}_{\perp}$ is the electric field perpendicular to $\vec{B},$
and $\vec{E}_{\parallel}$ is the electric field parallel to $\vec{B}$
{[}e.g., Cosgrove and Tsunoda, 2003, https://doi.org/10.1029/2002JA009728{]}.
The perpendicular electric field ($\vec{E}_{\perp})$ is zero in the
frame of reference moving with the geomagnetic field lines (because
field lines move with the $\vec{E}\times\vec{B}$ velocity). Also,
in the lower E region, $\rho_{i}\gg1$ and $\rho_{e}\ll1.$ Using
these results the velocities are approximated by,
\begin{eqnarray*}
\vec{v}_{i} & \cong & \frac{\vec{E}_{\parallel}}{\rho_{i}B}+\vec{u},\\
\vec{v}_{e} & \cong & -\frac{\vec{E}_{\parallel}}{\rho_{e}B}+\vec{u}\times\hat{b}\times\hat{b}+\vec{u}=-\frac{\vec{E}_{\parallel}}{\rho_{e}B}+u_{\parallel}\hat{b},
\end{eqnarray*}
where $u_{\parallel}$ is the component of the wind along the geomagnetic
field. Since $\vec{E}_{\parallel}$ is very small, the ion velocity
is equal to $\vec{u}$, and so this is the layer velocity as claimed
in the paper. However, things are not this simple, because the electrons
have a different velocity, meaning that there is a current. We can't
really discuss the parallel component unless diffusion is included,
but maybe we can agree to ignore the parallel phenomena (ambipolar
diffusion and the ambipolar electric field). So the question is, will
the ion current cause charge buildup and an opposing electric field
that stops the ion motion, or will the ion current find a way to close,
preventing the formation of the opposing field and allowing the motion
to continue? Previous research {[}e.g., Shalimov et al., JGR, 1998,
103, 11617; Cosgrove and Tsunoda, GRL, 2001, 28(8), 1455{]} has shown
that the answer to this question is highly geometry dependent, depending,
for example, on coupling to the F region, which is a potential current
closure path. }

\textcolor{blue}{For a long thin ``dynamo'' with the wind blowing
across it there can be an electrojet-like effect where the polarization
field builds up and drives a large Cowling channel current, and when
this works really well the polarization field could essentially stop
the ion motion, as the reviewer may be suggesting. But this is a relatively
specialized effect and there are many ways it can be defeated. For
example, if the dynamo is not long enough the strong Cowling channel
current will not be able to close completely, and a polarization field
will build up opposing it, which will drive a Hall current that closes
or partially closes the original wind-driven ion current---so then
the ion motion can continue or partially continue. Or, if the dynamo
is not thin enough, the initial polarization field may map to the
F region and likely find a lot of extra conductivity up there, such
that there is closure through the F region and again, the polarization
field doesn't get very large and the ion motion can continue. So I
would argue that the electrojet case is special and rare, and that
the more generic configuration does not allow for a fully formed polarization
electric field, and therefore the layer will typically move at or
somewhat below the wind velocity.}

\textcolor{blue}{I think this is sufficient for the purposes of the
present paper, where we are just seeking an estimate of the typical
layer velocity, in order to characterize the frequencies associated
with ionospheric phenomena. All of this is only to justify the 40~m/s
layer velocity used in the paper, which could also be justified by
observations, for example, ISR.}

\subsection{Reviews of the Revision of ``Failure of the Electrostatic Assumption
via an Exact Solution for the Linear Waves of the Electromagnetic
5-Moment Fluid Equations, plus 150 km Echoes,'' submitted to JGR
in 2018. This was the second version of the 2022 paper.}

\textcolor{blue}{This first revision of the paper was treated as a
resubmission by JGR. This revision still did not include implementation
of the model for a vertically inhomogeneous ionosphere, although the
equations for that model were given with almost the same level of
detail as in the final, 2022 paper. Completing the implementation
was described as future work. Otherwise, this revision added most
of the elements of the final, 2022 paper, including writing down the
exact solution, writing down the approximate solution found by complex
plane integration, and addition of the gedanken experiment, which
was added to better explain the work in {[}Cosgrove 2016{]}.}

\subsubsection{Reviewer \#3:}

Reviewer \#3 Evaluations: \\
Recommendation (Required): Reject \\
Significant (Required): There are major errors or gaps in the paper
but it could still become significant with major changes, revisions,
and/or additional data. \\
Supported (Required): No \\
Referencing (Required): Yes \\
Quality (Required): The organization of the manuscript and presentation
of the data and results need some improvement. \\
Data (Required): Yes \\
Accurate Key Points (Required): No\\

\noindent Reviewer \#3 (Formal Review for Authors (shown to authors)):\medskip{}

I accepted to review the paper out of curiosity. Once I looked at
the paper, I wondered why I had accepted. In the end I am not sure
that the paper adds much to the sum of our knowledge. A complicated
argument is made about what seems a highly idealized model. Moreover,
it seems neither of the first two referees sees a strong reason to
publish. It has taken a while for me to understand what the other
referees and I find so troubling about the paper. It is in its negative
style and rather iconoclastic ethos. 

\medskip{}

\textcolor{blue}{The review does not mention that the paper finds
a greater than $90^{\circ}$ phase rotation from the top to the bottom
of the ionosphere, for a 100~km transverse wavelength. So the electric
field does not map! This is absolutely a new and important result!!
It is only by omitting this result that the reviewer can pretend that
``}I am not sure that the paper adds much to the sum of our knowledge.\textcolor{blue}{''}

\textcolor{blue}{Otherwise, the only reason given} \textcolor{blue}{for
rejecting the paper is the style of writing; the paper is said to
be ``}\textcolor{black}{iconoclastic.}\textcolor{blue}{'' Nothing
is suggested to be wrong with the methodology or with the results,
anywhere in what the reviewer has written.}

\textcolor{blue}{As far as the analysis being idealized, the paper
analyzes a homogeneous plasma with boundary, and explains how the
analysis could be extended to a vertically inhomogeneous ionosphere
in future work (which has now been completed). Analyzing a homogeneous
plasma (with boundary) allows for deriving a very conclusive comparison
to electrostatic theory, and since the major conclusion involves the
parallel wavelength being short, it is clear that it also applies
for inhomogeneous plasma. Especially given that the issue seems to
be very contentious, taking a careful step-by-step approach would
seem to be the appropriate way to proceed.}

\textcolor{blue}{Anyway the model for the vertically inhomogeneous
ionosphere has now been completed.}

\textcolor{blue}{Finally, the reviewer says a number of things that
seem unfitting. First, the reviewer says, ``}Once I looked at the
paper, I wondered why I had accepted.\textcolor{blue}{'' In this
sentence the reviewer seems to be saying that they don't want to do
the review, and that they aren't approaching the task with a good
attitude. Then, the reviewer says, ``}It has taken a while for me
to understand what the other referees and I find so troubling about
the paper. It is in its negative style and rather iconoclastic ethos.\textcolor{blue}{''
In this sentence the reviewer seems to be saying that their review
is not based on the scientific content of the paper.}

\medskip{}

While one learns that the author finds the electrostatic approach
to describing ionospheric equilibrium states limited, one does not
see how it should be replaced. 

\medskip{}

\textcolor{blue}{Apparently the reviewer did not read Section~6,
which was titled ``Modeling the Ionospheric Admittance,'' and was
about how a practical model could be developed in the future. For
example in the first paragraph of Section~6 is the sentence, ``We
now describe a transmission line analysis methodology that could be
used to develop a more quantitative model for the ionospheric input
admittance as a function of transverse wavelength and frequency, and
which we intend to develop in future work.'' This model has now been
completed and the results were described in two unpublished version
of the paper before the final 2022 version.}

\bigskip{}

\noindent I found the authors first principles hard to grasp initially.
For example, I could not understand why if curl E is unavoidably not
zero, it means a steady state is impossible. I thought it might mean
the system was like chaos for example. However, that cannot be so
if one can demonstrate the problem with a linearised system. Accordingly,
I think that what the author has in mind is that in any final equilibrium
state achieved by setting up a perturbation to a previously static
system there would be a departure in the magnetic field from its original
state as well as a time-stationary electric field. I don't find this
unduly controversial but nor does it seem revolutionary. \medskip{}

\textcolor{blue}{Above is the reviewers main commentary on the nature
and importance of the results. It seems that the reviewer did not
read the section titled ``Summary of Results,'' because in the first
paragraph of that section the paper states,}
\begin{quote}
\textcolor{blue}{``\ldots it is found that electric fields with transverse
scale sizes of 100 km and below will not map from 400 km to 100 km
in altitude, and there is a profound effect on the ionospheric input-admittance.
The actual input admittance depends on the transverse scale size,
and will be very different from the field-line-integrated ionospheric
conductivity when the transverse scale size is less than 100 km.''}
\end{quote}
\textcolor{blue}{The reviewer did not mention this result. Since the
reviewer seems to have some knowledge of ionospheric physics, the
reviewer would have known that this is a new and unexpected result,
if they had read the section ``Summary of Results.''}

\textcolor{blue}{The reviewer has also said ``}I could not understand
why if curl E is unavoidably not zero, it means a steady state is
impossible.\textcolor{blue}{'' But the paper does not say any such
thing, and the reviewer does not refer to a line number or even to
a specific section. So I am unable to understand how the reviewer
developed such a misunderstanding. What the reviewer wrote after ``}I
think that what the author has in mind\textcolor{blue}{\ldots ''
is also not at all correct. In the first sentence the reviewer wrote
``}I found the authors first principles hard to grasp initially.\textcolor{blue}{''
So it would appear that the reviewer did not understand the paper.
This would be important if the reviewer had given an indication of
where they became confused, or challenged any of the reasoning. But
since they did not do any of these things, and apparently did not
read the section ``Summary of Results,'' it would appear that the
issue is actually that the reviewer did not try very hard to understand
the paper.}

\medskip{}

\noindent The obvious example of an initial ionospheric field introduced
by a local dynamo effect certainly is very likely to launch an Alfvén
wave during an intermediate stage. However, what happens to that signal
far away determines how the magnetic disturbance of the wave will
evolve. For example, if it propagates through the magnetosphere and
ends up in the opposite hemisphere it can bounce back and forth until
a steady system of field aligned current is formed. In the final steady
state one would then envisage a global structure where there is a
global (i.e. common electrostatic field in both hemispheres) and a
field aligned current system between hemispheres to even out stress.
There would be a magnetostatic field associated with the overall current
system as the magnetic field has to shear or twist to provide the
field aligned current flow. 

\medskip{}

\textcolor{blue}{The reviewer does not seem to be aware that the paper
is denying that there is always a ``}\textcolor{black}{common}\textcolor{blue}{{}
}electrostatic field in both hemispheres\textcolor{blue}{'' (where
I am speaking of the E regions). Again, it would seem that the reviewer
has not read enough of the paper even to know what conclusions are
claimed.}

\textcolor{blue}{The conclusion that was stated, for example, in the
section ``Summary of Results'' (Section~8), was developed in the
analysis of Section~5 titled ``Modified View of Electric Field Mapping.''
In Section~5, the wavelengths of the Alfvén and Whistler waves were
used to calculate the phase rotation for a signal traveling from the
F region at 400~km to the E-region at 100~km, for a transverse wavelength
of 100~km and density of $10^{11}\,\mathrm{m}^{-3}$. It was found
that in this case there is over $90^{\circ}$ of phase rotation just
in one direction of travel. With the wavelength being this short it
is obvious that there will not be a ``}\textcolor{black}{common}\textcolor{blue}{{}
}electrostatic field in both hemispheres\textcolor{blue}{'' for the
100~km altitude. In this case, there is no need to think about the
back-reflected wave to make the conclusion. The back-reflected wave
will have the same wavelength, and superposing these waves cannot
increase the wavelength.}

\textcolor{blue}{However, a detailed discussion of the importance
of the back reflected wave was given in Section~6, where some standard
results from transmission line theory were described. It was explained
that a $90^{\circ}$ phase rotation for one-way travel would result
a short circuit input admittance seen looking down from 400~km, if
the bottom of the ionosphere is regarded as an open circuit. It was
explained that this happens because the back reflected wave returns
with $180^{\circ}$ of phase rotation, and thus (nearly) cancels the
electric field of the incident wave. It was explained how the formulas
from transmission line theory allow for treating more complex problems.
It would appear that the reviewer did not read any of this material.}

\textcolor{blue}{The reviewers comments imply that they did not read
either Section~5 (Modified View of Electric Field Mapping), Section~6
(Gedanken Experiment), or Section~8 (Summary of Results).}

\medskip{}

\noindent In reading again the responses to referees and rebuttals
I found that the comment that I resonated with most was the query
about orders of magnitude and the implicit issue of what physical
process linked to particular scale lengths where breakdown occurred.

\medskip{}

\textcolor{blue}{I do not know what ``query'' the reviewer is referring
to. The best I can do is to imagine what question I might ask that
would involve similar words. Hence, the ``breakdown'' referred to
in the paper is the breakdown of electrostatic theory that occurs
when the parallel wavelength becomes comparable to the thickness of
the ionosphere. The ``physical process'' that causes the unexpected
shortening of the wavelength is the high rate of collisions, which
are not included in the dispersion relations for the MHD waves. Another
issue is that dispersion relations are often evaluated with $\omega$
purely real and $\vec{k}$ complex, and this may produce an artificially
long wavelength.} \medskip{}

\noindent I wrote the preceding paragraph in an attempt to picture
how I would have liked to see what the author is talking about.

\medskip{}

\textcolor{blue}{It seems that the reviewer does not realize that
the main results involve the response of the ionosphere to a source
placed above it. So the field line above the source, which may (or
may not) connect to the other hemisphere, is not part of the system
being analyzed. If one likes, one can think of the wave produced by
the source as coming from the other hemisphere, but that is not necessary;
one can also think of it as coming from the magnetosphere, or wherever.
So the reviewers discussion of hemisphere-to-hemisphere propagation
is not relevant.}

\textcolor{blue}{I might have taken this to mean that the discussion
was inadequate, if I thought that the reviewer had read the paper.
But instead I think that what happened is that the reviewer scanned
the figures and noticed one of the figures from the section ``Gedanken
Experiment,'' where the cases with and without a back-reflected wave
were compared. Since the reviewer did not read that section they misinterpreted
the purpose of the figure.}

\medskip{}

\noindent The author adopts a bracing, even iconoclastic, style. I
think that that is alienating to the reader. The issues are not entirely
original. Indeed, I found the author's 2016 paper more amenable in
style. However, even better, I would firmly recommend reading the
Vasyliunas (2012) paper that is referenced. Vasyliunas is also raising
very similar criticisms of traditional theory. The major difference
is in the style of presentation and in Vasyliunas' clear interest
in explaining the challenges to electrostatic theory in terms of physical
behaviour, even if approximate. Most significant to me is Vasyliunas'
modest but incisive approach to expressing what shortcomings there
are in the traditional approximate theories that have been used for
half a century or so. The ultimate issue of publication hinges on
what new knowledge there is in the paper. I am not clear that I have
learnt much but there may be a publication here.\medskip{}

\textcolor{blue}{Except for the last two lines, the reviewer is criticizing
the writing style, which they feel is not sufficiently humble. Regarding
the last two lines, it is apparent that the reviewer did not read
the ``Summary of Results'' section, because the reviewer did not
mention the results claimed in the following line:}
\begin{quotation}
\textcolor{blue}{``\ldots it is found that electric fields with transverse
scale sizes of 100 km and below will not map from 400 km to 100 km
in altitude, and there is a profound effect on the ionospheric input-admittance.
The actual input admittance depends on the transverse scale size,
and will be very different from the field-line-integrated ionospheric
conductivity when the transverse scale size is less than 100 km.''}
\end{quotation}
\textcolor{blue}{Assuming that the reviewer has some knowledge of
ionospheric physics, if they had read these lines they could not have
said that the paper does not contain ``new knowledge,'' without
pointing to problems with the methodology. The reviewer did not indicate
that they thought anything was wrong with the methodology.}

\textcolor{blue}{\medskip{}
}

\noindent However, a much more direct less argumentative style needs
to be adopted so one may see where icons may be allowed to stand and
where they need replacement.\textcolor{blue}{\medskip{}
}

\textcolor{blue}{It appears that the reviewer became angry at the
``}\textcolor{black}{iconoclastic style}\textcolor{blue}{'' in the
Introduction, and decided to review the paper without reading very
much of it. The review is almost solely concerned with directing me
to be more tactful. The review does not mention the main results,
or the methods, not even to challenge them.\medskip{}
}

\noindent Notes by line: Line 8 \textquotedbl contraindicate\textquotedbl{}
is usually a medical term. I could not see what special instance led
to its use here. Line 29 and 34 Possible to give physical basis for
scale? Line 42 \textquotedbl miss-using\textquotedbl{} is wrong:
\textquotedbl misusing\textquotedbl{} Line 64 \textquotedbl I-or
we, to involve the reader..\textquotedbl{} This can grate. I'd use
\textquotedbl we\textquotedbl{} (unapologetically) or the passive
voice. Line 148 If there is a neutral atmosphere driven dynamo present
or magnetospheric motion imposed from above, do the Matthews and Walker
constraints on electrostatic solutions apply. I think not. Rhetorical
questions like \textquotedbl So why should we expect to derive an
approximation in this way?\textquotedbl{} are polemical in many people's
view. It is more sensible to not ask rhetorical questions especially
if there is a risk of an emotional response in the reader. 

Line 184 Repetitious?

Line 1441 \textquotedbl Thinking in terms of \textquotedbl Pedersen
and Hall currents\textquotedbl{} and \textquotedbl polarization electric
fields\textquotedbl{} and \textquotedbl efficient mapping along geomagnetic
field lines\textquotedbl{} allows a form of intuition for ionospheric
physics based in electrostatic theory. But how reliable is this intuition?\textquotedbl{}
I couldn't really find the answer to this question. Although rhetorical
questions may work well in lecturing they do not really suit this
sort of work.

\medskip{}

\textcolor{blue}{It is useful for the reviewer to point out that the
rhetorical questions were annoying. Otherwise, the above portion concerns
minor details and I will not bother to respond.}

\subsubsection{Reviewer \#4:}

Reviewer \#4 Evaluations: \\
Recommendation (Required): Return to author for minor revisions \\
Significant (Required): Yes, the paper is a significant contribution
and worthy of prompt publication. \\
Supported (Required): Yes \\
Referencing (Required): Yes \\
Quality (Required): Yes, it is well-written, logically organized,
and the figures and tables are appropriate. \\
Data (Required): Yes \\
Accurate Key Points (Required): Yes

\medskip{}

\paragraph*{Basic Comments by A. M. Hamza}

It is only legitimate to question some of the conventional assumptions
used in the space plasma literature. To challenge what one could describe
as the \textquoteright Electrostatics\textquoteright{} Kuhnian paradigm
may not be acceptable to many, but if we are to move forward we need
to transcend any kind of \textquoteright mob psychology\textquoteright{}
as Paul Feyerabend would have it and test the grounds beyond the \textquoteright Electrostatics
paradigm.\textquoteright{} I believe the paper ought to be published
with some corrections as described below.

\paragraph*{Corrections}
\begin{itemize}
\item The actual length of the paper is too long, and the author ought to
be able to shorten the paper given the fact that his 2016 paper is
already long and published. I will leave it up to the author\textquoteright s
judgement to see a reduction in size. The case against the validity
of the electrostatic approximation should not be difficult to make
concisely.
\item The ratio used in the paper to quantify the electrostatic approximation
needs to be slightly modified. An analytical treatment can be pushed
to check the validity of the numerical integrations for all four cases
treated in the paper.
\begin{equation}
r=\frac{\left|\nabla\times\mathbf{E}\right|}{\left|\nabla\cdot\mathbf{E}\right|}\label{eq:curlOverDiv}
\end{equation}
In fact this ought to be a comparison of the perpendicular component
of the electric field to the parallel one. If we write the electric
field in terms of its perpendicular and parallel components to the
wave vector k, respectively, we can easily show that the ratio r ought
to be replaced by:
\begin{eqnarray*}
\mathbf{E} & = & E_{\parallel}\hat{\mathbf{k}}+\mathbf{E}_{\perp}=E_{\parallel}\hat{\mathbf{k}}-\hat{\mathbf{k}}\times\left(\hat{\mathbf{k}}\times\mathbf{E}\right)\\
r & = & \frac{\left|\mathbf{E}_{\perp}\right|}{\left|E_{\parallel}\right|}
\end{eqnarray*}
This is easy to establish for a plane wave with $\mathrm{e}^{i\left(\boldsymbol{k}\cdot\boldsymbol{x}-\omega t\right)}$
since the curl becomes a cross product and the divergence a dot-product.
\begin{equation}
r=\frac{\left|\hat{\mathbf{k}}\times\mathbf{E}\right|}{\left|\hat{\mathbf{k}}\cdot\mathbf{E}\right|}=\frac{\left|\mathbf{E}_{\perp}\right|}{\left|E_{\parallel}\right|}\label{eq:curlOverDiv_withKvector}
\end{equation}
Notice that this ratio is nothing but the tangent of the angle between
the electric field and the wave vector, $\theta_{kE}$. A naive suggestion
would have $\left|\tan\theta_{kE}\right|\ll1$, which in turn suggest
small angles, but how small before violation of the electrostatic
approximation.\\
\medskip{}
The next step consists of evaluating the ratio r by determining the
dispersion relation, which is done in books, and I will cite the classical
book on Plasma Waves by Stix (no need for reference since this is
a classic and any student of plasma physics ought to be familiar with
its content). \\
\medskip{}
The dispersion relation is obtained from:
\begin{equation}
\nabla\times(\nabla\times\mathbf{E})+\frac{1}{c^{2}}\frac{\partial^{2}\mathbf{E}}{\partial t^{2}}=-\frac{4\pi}{c^{2}}\frac{\partial\mathbf{j}}{\partial t}
\end{equation}
Which can be written as:
\begin{equation}
\underline{\underline{\mathbf{D}}}\cdot\mathbf{E}=0\label{eq:dispersion_Hamza}
\end{equation}
With
\begin{equation}
D_{ij}=c^{2}k_{i}k_{j}+(\omega^{2}-k^{2}c^{2})\delta_{ij}+4\pi i\omega\sigma_{ij}
\end{equation}
Where the induced current has been written in terms of the conductivity
$\sigma$
\begin{equation}
\mathbf{j}=\underline{\underline{\sigma}}(\omega,\mathbf{k})\cdot\mathbf{E}
\end{equation}
The electrostatic approximation consists of assuming that the perpendicular
component of the electric is much much smaller than the parallel component,
i.e. setting the perpendicular component to zero, to zeroth order,
gives us the electrostatic dispersion relation, which then needs to
verified for consistency by calculating the perpendicular component
and showing that it is indeed much smaller than the parallel component
as assumed. \\
\medskip{}
Let us set $\mathbf{E}_{\perp}=0$ in the dispersion relation above
\begin{equation}
\underline{\underline{\mathbf{D}}}\cdot E_{\parallel}\hat{\mathbf{k}}=0\label{eq:electrostatic_dispersion_Hamza}
\end{equation}
this leads to
\begin{equation}
\omega^{2}\hat{\mathbf{k}}+4\pi i\omega\underline{\underline{\sigma}}\cdot\hat{\mathbf{k}}=0
\end{equation}
which in turn leads to the electrostatic dispersion relation:
\begin{equation}
\omega(\omega+4\pi i\hat{\mathbf{k}}\cdot\underline{\underline{\sigma}}\cdot\hat{\mathbf{k}})=0
\end{equation}
We now evaluate the first order correction to the perpendicular component
of the electric field.
\begin{equation}
[\omega^{2}\hat{\mathbf{k}}+4\pi i\omega\underline{\underline{\sigma}}\cdot\hat{\mathbf{k}}]E_{\parallel}+(\omega^{2}-k^{2}c^{2})\mathbf{E}_{\perp}+4\pi i\omega\underline{\underline{\sigma}}\cdot\mathbf{E}_{\perp}=0
\end{equation}
This can be further simplified using the electrostatic dispersion
relation derived above.
\begin{equation}
4\pi i\omega[\underline{\underline{\sigma}}\cdot\hat{\mathbf{k}}-\hat{\mathbf{k}}\cdot\underline{\underline{\sigma}}\hat{\mathbf{k}}\hat{\mathbf{k}}]+(\omega^{2}-k^{2}c^{2})\mathbf{E}_{\perp}+4\pi i\omega\underline{\underline{\sigma}}\cdot\mathbf{E}_{\perp}=0
\end{equation}
We now use the following identity
\begin{equation}
\underline{\underline{\sigma}}\cdot\hat{\mathbf{k}}-\hat{\mathbf{k}}\underline{\underline{\sigma}}\cdot\hat{\mathbf{k}}\hat{\mathbf{k}}=\hat{\mathbf{k}}\times(\underline{\underline{\sigma}}\cdot\hat{\mathbf{k}})
\end{equation}
The equation for the perpendicular component of the electric field
becomes
\begin{equation}
4\pi i\omega\hat{\mathbf{k}}\times(\underline{\underline{\sigma}}\cdot\hat{\mathbf{k}})E_{\parallel}+(\omega^{2}-k^{2}c^{2})\mathbf{E}_{\perp}+4\pi i\omega\underline{\underline{\sigma}}\cdot\mathbf{E}_{\perp}=0
\end{equation}
Let us compare the ratio of the last two terms, with the perpendicular
component of the electric field, in the electrostatic approximation,
i.e.,
\begin{equation}
\left.\frac{\left|4\pi i\omega\underline{\underline{\sigma}}\cdot\mathbf{E}_{\perp}\right|}{\left|(\omega^{2}-k^{2}c^{2})\mathbf{E}_{\perp}\right|}\right|_{\omega=-4\pi i\hat{\mathbf{k}}\cdot\underline{\underline{\sigma}}\cdot\hat{\mathbf{k}}}\ll1\label{eq:omegaFirstAppears}
\end{equation}
This allows us to calculate $\mathbf{E}_{\perp}$ when the electrostatic
dispersion relation is satisfied
\begin{equation}
\mathbf{E}_{\perp}=-i\left.\frac{4\pi\omega\hat{\mathbf{k}}\times(\underline{\underline{\sigma}}\cdot\hat{\mathbf{k}})E_{\parallel}}{(\omega^{2}-k^{2}c^{2})}\right|_{\omega=-4\pi i\hat{\mathbf{k}}\cdot\underline{\underline{\sigma}}\cdot\hat{\mathbf{k}}}
\end{equation}
Which finally allows us to set the condition for the validity of the
electrostatic approximation:
\begin{equation}
\frac{\left|\mathbf{E}_{\perp}\right|}{\left|E_{\parallel}\right|}=\left|\frac{4\pi\omega}{\omega^{2}-k^{2}c^{2}}\hat{\mathbf{k}}\times(\underline{\underline{\sigma}}\cdot\hat{\mathbf{k}})\right|_{\omega=-4\pi i\hat{\mathbf{k}}\cdot\underline{\underline{\sigma}}\cdot\hat{\mathbf{k}}}\ll1\label{eq:condition_for_electrostat_Hamza}
\end{equation}
To move forward we need the expressions for the components of the
conductivity tensor. \\
\medskip{}
In a reference frame with the geomagnetic field pointing in the z-direction,
the conductivity tensor is given by:
\begin{eqnarray*}
\sigma_{P} & = & \sigma_{xx}=\sigma_{yy}=\frac{\omega_{pi}^{2}}{\Omega_{i}^{2}+\nu_{in}^{2}}\nu_{in}+\frac{\omega_{pe}^{2}}{\Omega_{e}^{2}+\nu_{en}^{2}}\nu_{en}\\
\sigma_{H} & = & -\sigma_{xy}=\sigma_{yx}=-\frac{\omega_{pi}^{2}}{\Omega_{i}^{2}+\nu_{in}^{2}}\Omega_{i}+\frac{\omega_{pe}^{2}}{\Omega_{e}^{2}+\nu_{en}^{2}}\Omega_{e}\\
\sigma_{\parallel} & = & \sigma_{zz}=\frac{\omega_{pi}^{2}}{\nu_{in}}+\frac{\omega_{pe}^{2}}{\nu_{en}}\\
\sigma_{xz} & = & \sigma_{zx}=\sigma_{yz}=\sigma_{zy}=0
\end{eqnarray*}
we next choose the wave vector to be in the $(x,z)$ plane $\mathbf{k}=k(\cos\theta,0,\sin\theta)$.\\
\medskip{}
The validity condition can be written as:
\begin{equation}
\left|(\hat{\mathbf{k}}\cdot\underline{\underline{\sigma}}\cdot\hat{\mathbf{k}})[\hat{\mathbf{k}}\times(\underline{\underline{\sigma}}\cdot\hat{\mathbf{k}})]\right|\ll\left|(\hat{\mathbf{k}}\cdot\underline{\underline{\sigma}}\cdot\hat{\mathbf{k}})^{2}-\frac{k^{2}c^{2}}{16\pi^{2}}\right|
\end{equation}
Using the Petersen and Hall conductivities, we can explicitly write
the condition for validity of the electrostatic approximation:
\begin{eqnarray*}
\hat{\mathbf{k}}\cdot\underline{\underline{\sigma}}\cdot\hat{\mathbf{k}} & = & \sigma_{P}\cos^{2}\theta+\sigma_{\parallel}\sin^{2}\theta\\
\left|\hat{\mathbf{k}}\times(\underline{\underline{\sigma}}\cdot\hat{\mathbf{k}})\right|^{2} & = & \sigma_{H}^{2}\cos^{2}\theta+(\sigma_{P}-\sigma_{\parallel})^{2}\sin^{2}\theta\cos^{2}\theta
\end{eqnarray*}
For wave propagation along the background geomagnetic field, cos \textgreek{j}
\ensuremath{\approx} 0, the condition is always satisfied. For small
angles, on the other hand, we obtain the following condition:
\begin{eqnarray*}
\hat{\mathbf{k}}\cdot\underline{\underline{\sigma}}\cdot\hat{\mathbf{k}} & \approx & \sigma_{P}\\
\left|\hat{\mathbf{k}}\times(\underline{\underline{\sigma}}\cdot\hat{\mathbf{k}})\right| & \approx & \sigma_{H}
\end{eqnarray*}
Which leads to the condition:
\begin{equation}
\frac{\sigma_{H}}{\sigma_{P}}\ll\left|1-\left(\frac{kc}{4\pi\sigma_{P}}\right)^{2}\right|\label{eq:electrostatic_condition_final_Hamza}
\end{equation}
Similar results may be obtained and can be verified for all four cases
addressed in the paper.
\end{itemize}

\paragraph*{Conclusion}

I recommend the publication of the paper provided the author verifies
that the results are consistent with the argument made in this review.

\medskip{}

\begin{figure}[p]
\thisfloatpagestyle{plain}\includegraphics{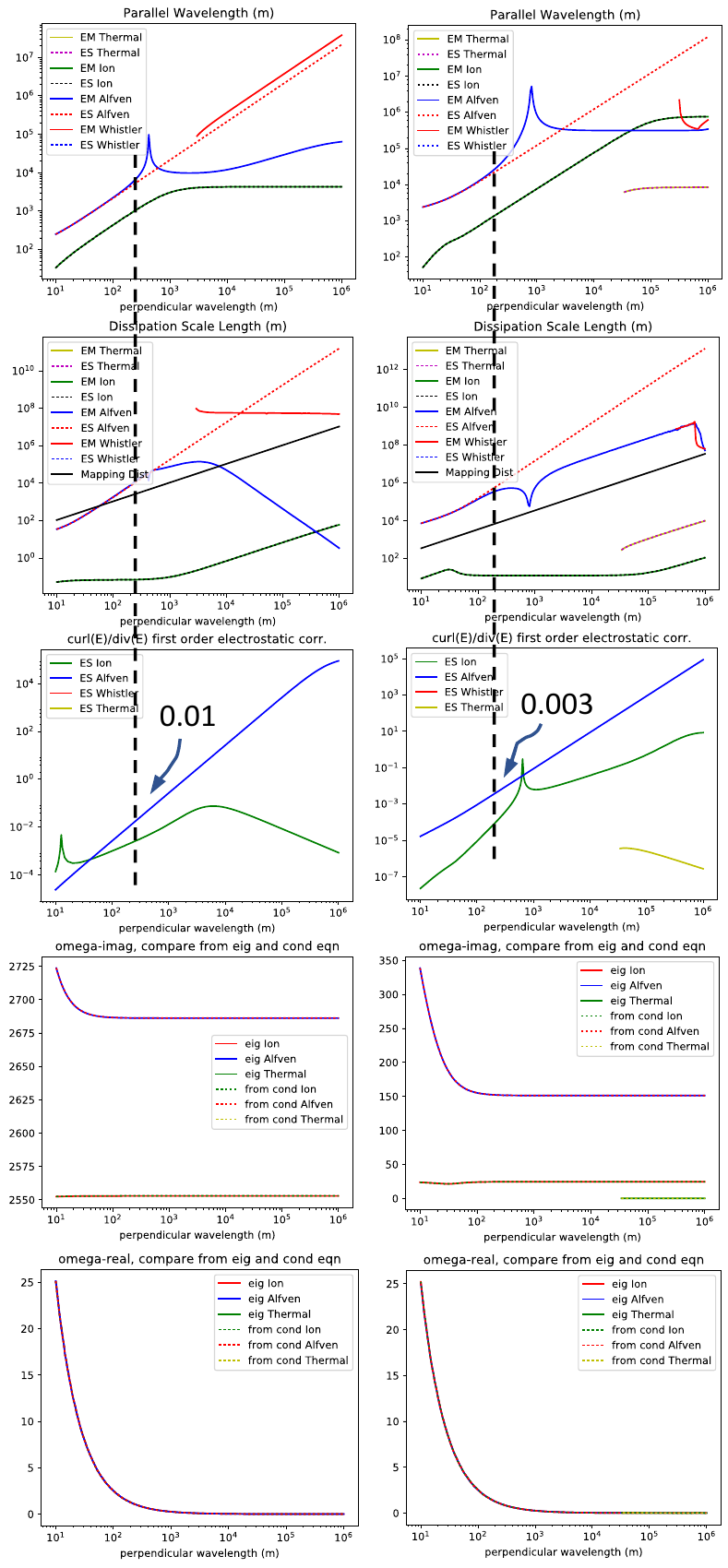}

\caption{\textcolor{blue}{Comparison of different proposed criteria for validity
of electrostatic theory, as a function of the wavelength in the direction
perpendicular to the geomagnetic field. The left column is for an
altitude of 100~km and the right column is for an altitude of 145~km.
Top row: comparison of the wavelength in the direction parallel to
the geomagnetic field, for electrostatic waves (dashed lines) and
electromagnetic waves (solid lines). Second row: comparison of the
dissipation scale lengths for electrostatic waves (dashed lines) and
electromagnetic waves (solid lines). Third row: ratio $\left|\hat{\mathbf{k}}\times\mathbf{E}\right|/\left|\hat{\mathbf{k}}\cdot\mathbf{E}\right|$
computed from a first order correction to electrostatic theory (equation~(\ref{eq:condition_for_electrostat_Hamza})).
Fourth and fifth rows: comparison of the imaginary and real parts,
respectively, of the eigenvalues of the matrix $H_{5ES}$ with the
solutions of $\omega=-4\pi i\hat{\mathbf{k}}\cdot\underline{\underline{\sigma}}(\omega)\cdot\hat{\mathbf{k}}$.
\label{fig:Comparison-of-different-electrostatic-criteria}}}

\end{figure}

\textcolor{blue}{This reviewer is very positive and asks for only
a minor revision. They request one minor change and require only that
the results should come out consistently. The reviewer writes ``}\textcolor{black}{The
ratio used in the paper to quantify the electrostatic approximation
needs to be slightly modified.}\textcolor{blue}{'' And then they
demonstrate the alternative quantity they would prefer to see, and
ask me to verify that it comes out consistently. I have done this
now and the results are shown in Figure~\ref{fig:Comparison-of-different-electrostatic-criteria}.
They are quite consistent and I thank the reviewer for suggesting
this additional validation.}

\textcolor{blue}{To elaborate, in the paper I used the electromagnetic
waves to calculate $\left|\nabla\times\mathbf{E}\right|/\left|\nabla\cdot\mathbf{E}\right|=\left|\hat{\mathbf{k}}\times\mathbf{E}\right|/\left|\hat{\mathbf{k}}\cdot\mathbf{E}\right|$,
which is a quantity that should be small if the electrostatic waves
are to agree with the electromagnetic ones. The reviewer instead shows
how this quantity can be estimated in first order from the electrostatic
waves themselves. It is this quantity that they would prefer to see,
which they give in their equation~(\ref{eq:condition_for_electrostat_Hamza}).
This review is uploaded by permission, and the reviewer represents
that their derivation either comes from or is equivalent to a derivation
from the classical textbook by Stix {[}\citealp{stix-1992}{]}. }

\textcolor{blue}{Since I did not include either of the quantities
in the body of the final 2022 paper, I have plotted the reviewer's
quantity (equation~(\ref{eq:condition_for_electrostat_Hamza})) in
the third row of Figure~\ref{fig:Comparison-of-different-electrostatic-criteria}.
The top two rows show the method of comparing the electrostatic and
electromagnetic waves that is used in the final 2022 paper. The top
row compares the wavelengths parallel to the geomagnetic field and
shows that the electrostatic and electromagnetic results diverge as
the wavelength perpendicular to the geomagnetic field becomes longer.
The second row shows similar results for the dissipation scale length.
The left column shows results at 100~km in altitude, and the right
column shows results at 145~km in altitude. Note that these are necessary
conditions, but they are not sufficient conditions as there exist
other important wave properties (e.g., characteristic admittance,
etc.).}

\textcolor{blue}{I believe that the comparisons in the top two rows
are more conclusive, because they show that these waves would give
different results when used in the TL theory calculation from the
paper. Especially in the case of the parallel wavelength, the finding
that it can be short enough to produce a resonance is very important,
and it appears from Figure~\ref{fig:Comparison-of-different-electrostatic-criteria}
that the electrostatic waves miss this result. For the ratio~(\ref{eq:condition_for_electrostat_Hamza}),
on the other hand, it is not clear exactly how small the ratio has
to be for the electrostatic waves to give the same results as the
electromagnetic waves.}

\textcolor{blue}{To compare the different criteria shown in Figure~\ref{fig:Comparison-of-different-electrostatic-criteria},
let us take it one mode at a time. For the Alfvén mode the ratio~(\ref{eq:condition_for_electrostat_Hamza})
appears to be consistent with the criteria in the top two rows if
we take the largest allowed value to be between 0.01 and 0.005, which
seems quite reasonable. For the Ion wave, the criteria in the top
two rows do not show any difference between the electrostatic and
electromagnetic waves. However, these criteria are only necessary
criteria, they are not sufficient criteria. Thus it is not inconsistent
that for the 145~km altitude (right column), the ratio~(\ref{eq:condition_for_electrostat_Hamza})
is seen to surpass 1 when the transverse wavelength exceeds 150~km;
there may be other wave properties that do not agree so well as the
ones shown in the top two rows. And at the 100~km altitude (left
column) the ratio~(\ref{eq:condition_for_electrostat_Hamza}) is
everywhere less than 0.08 for the Ion wave. For the Thermal wave,
the criteria in the top two rows do not show any difference between
the electrostatic and electromagnetic waves, and the ratio~(\ref{eq:condition_for_electrostat_Hamza})
is everywhere less than $5\times10^{-6}$. There is no electrostatic
counterpart for the Whistler wave and so it does not appear. Hence
these results are all consistent.}

\textcolor{blue}{In addition, in doing this I have also compared the
reviewers expression for $\omega$ in equations~(\ref{eq:omegaFirstAppears})-(\ref{eq:condition_for_electrostat_Hamza})
with the eigenvalues of the matrix $H_{5ES}$, and found that they
agree. The agreement is shown in the fourth and fifth rows of Figure~\ref{fig:Comparison-of-different-electrostatic-criteria},
where the fourth row shows the imaginary part and the fifth row shows
the real part. The real part is just the frequency derived from the
(assumed) 40~m/s transverse velocity and the transverse wavelength,
and so is the same for all the modes.}

\textcolor{blue}{This comparison helps to validate both calculations.
The expression $\omega=-4\pi i\hat{\mathbf{k}}\cdot\underline{\underline{\sigma}}(\omega)\cdot\hat{\mathbf{k}}$
given by the reviewer (in Gaussian units) is a nonlinear equation
in $\omega$, and clearly $\omega$ has a large imaginary part, since
we expect $\underline{\underline{\sigma}}(\omega)$ to be mostly real.
I am able to solve this equation by finding the eigenvalues and eigenvectors
of the matrix $H_{5ES}$, where there are multiple solutions corresponding
to the multiple modes of propagation. The conductivity tensor $\underline{\underline{\sigma}}(\omega)$
is calculated from the eigenvectors, and \citeauthor{cosgrove-2016}
showed that this gives the same results as direct evaluation of the
conductivity equations with the imaginary part of $\omega$ included.
The result is very different from the conductivity calculated without
including the imaginary part of $\omega$, which is evident from the
fact that there are multiple solutions, and we note that the imaginary
part is largest for the Alfvén mode. Plots of conductivity that directly
demonstrate the importance of the imaginary part of $\omega$ may
be found in Figure~16 of \citeauthor{cosgrove-2016} (errata: in the
figure caption ``equation~(1)'' should be ``equation~A9'').
This provides a nice additional validation and I thank the reviewer
for bringing the analytical results to my attention. }

\subsubsection{Decision Letter from JGR}

From: jgr-spacephysics@agu.org \\
Subject: 2018JA026274 (Editor - XXXXX): Decision Letter \\
Date: XXXX XX, XXXX at 9:44:30 AM PST \\
To: russell.cosgrove@me.com \\
Cc: russell.cosgrove@sri.com \\
Reply-To: jgr-spacephysics@agu.org\medskip{}

\noindent Dear Dr. Cosgrove,\medskip{}

\noindent Thank you for submitting your manuscript \textquotedbl Breakdown
of Electrostatic Theory via Exact Linearization and Solution of the
Electromagnetic 5-Moment Fluid Equations\textquotedbl{} to the Journal
of Geophysical Research - Space Physics.\medskip{}

\noindent At long last I have received two reviews of your manuscript
(see below/attached). One reviewer was particularly tardy despite
many reminders from AGU staff and myself. I apologise for the long
delay.\medskip{}

\noindent Both reviewers raise major concerns with the paper despite
very different summary evaluations. The key issues are essentially
not enough new science, and far too long. As a result I am declining
your manuscript for publication in Journal of Geophysical Research
- Space Physics.\medskip{}

\noindent I am enclosing the reviews, which you may find helpful if
you decide to revise your manuscript. I am sorry that I cannot be
more encouraging at this time.\medskip{}

\noindent Thank you for your continued support of JGR-Space Physics.\medskip{}

\noindent Yours sincerely,\medskip{}

\noindent XXXXX \\
Editor \\
JGR-Space Physics\medskip{}

\textcolor{blue}{Reviewer \#2 said the following about the three results
they found in the paper: }\textcolor{black}{``1) and 2) are not so
surprising given the fact of not using an electrostatic assumption,
and 3) is quite interesting and may have significant impact on future
studies of M-I coupling.}\textcolor{blue}{''}

\textcolor{blue}{Reviewer \#4 said ``}\textcolor{black}{To challenge
what one could describe as the \textquoteright Electrostatics\textquoteright{}
Kuhnian paradigm may not be acceptable to many, but if we are to move
forward we need to transcend any kind of \textquoteright mob psychology\textquoteright{}
as Paul Feyerabend would have it and test the grounds beyond the \textquoteright Electrostatics
paradigm.\textquoteright{} I believe the paper ought to be published
with some corrections as described below.}\textcolor{blue}{''}

\textcolor{blue}{So clearly two of the four reviewers did not concur
with the editors conclusion of ``}\textcolor{black}{The key issues
are essentially not enough new science, and far too long.}\textcolor{blue}{''
A likely reason for the editors conclusion is that the two negative
reviews did not report the results found in the paper.}

\textcolor{blue}{What is obvious is that the paper was highly controversial,
with strong opinions on both sides. This suggests that the paper could
be very important, but that we won't know until it has been debated
by the wider community.}

\subsection{Reviews of ``Breakdown of Electrostatic Theory via Exact Linearization
and Solution of the Electromagnetic 5-Moment Fluid Equations,'' submitted
to JASTP in 2019. This was the third version of the 2022 paper.}

\textcolor{blue}{In this version, the Introduction section was dramatically
shortened from the previous version, since that material seemed to
be antagonizing the reviewers, who were not really reviewing the principle
material in the paper. Otherwise the material was essentially the
same, although some language was revised for clarity.}

\subsubsection{Reviewer \#1 (the only one)}

This manuscript extends the previous work of Cosgrove {[}2016{]} and
provides some improvements on that previous work, but this manuscript
also has a number of problems. The notable improvements on the previous
work are:
\begin{enumerate}
\item This work adds the continuity and energy equations to the equation
set, and includes the effects of additional wave modes that result
from these extensions (Ion and Thermal waves).
\item Section 5 demonstrates how limiting the calculations to a finite slab
and including reflections results in convergence to the electrostatic
result. This resolves one of the most serious defi- ciencies with
the Cosgrove {[}2016{]} results. That paper considered an infinite
uniform plasma with no reflections, which is a scenario that never
exists in nature.
\end{enumerate}
\textcolor{blue}{The reviewer does not report the main results found
in the paper (that the electric field does not map through the ionosphere,
and that the ionospheric conductance is not the field line integrated
conductivity, etc). For example, the reviewer does not report the
following lines from the first paragraph of the section ``Summary
of Results and Discussion,''}
\begin{quote}
\textcolor{blue}{``\ldots it is found that electric fields with transverse
scale sizes of 100 km and below will likely not map from 400 km to
100 km in altitude, with profound effect on the ionospheric input-admittance.
The actual input admittance in this case will depend on the transverse
scale size, and should be very different from the field-line-integrated
ionospheric conductivity\ldots ''}
\end{quote}
\textcolor{blue}{A proper review reports all the important results
that are claimed, and then if they think there are problems with the
methodology, they report on those. Instead, this reviewer ignores
the ``Summary of Results and Discussion'' section and incorrectly
states that ``}no results are presented\textcolor{blue}{,'' and
then says that ``}This modeling approach could be useful in the future\textcolor{blue}{''
(see below for these two remarks).}

\textcolor{blue}{In addition, this reviewer is not accurate in the
results that they do report: the paper does not find that ``}limiting
the calculations to a finite slab and including reflections results
in convergence to the electrostatic result\textcolor{blue}{''. The
Abstract does say ``\ldots electrostatic predictions can be conditionally
recovered\ldots '', but the reviewer has ignored the word ``conditionally,''
which changes the meaning.}

\textcolor{blue}{In addition, the reviewer misrepresents the results
from {[}Cosgrove 2016{]}: item~(2) of the ``}notable improvements\textcolor{blue}{''
refers (inaccurately) to a summary of the results }\textcolor{blue}{\emph{from}}\textcolor{blue}{{}
{[}Cosgrove 2016{]}, and so that material certainly does not correct
any ``}serious deficiencies\textcolor{blue}{'' in that} \textcolor{blue}{paper,
since it came from that paper. Section~6 of {[}Cosgrove 2016{]} considered
the back-reflected wave and derived the same ``wave-Pedersen'' conductivity
that accounts for the back-reflected wave (equation~(14) of {[}Cosgrove
2016{]}, which is equivalent to equation~(17) in this paper). The
reviewer should have known this because the sentence right before
equation~(17) reads,}
\begin{quote}
\textcolor{blue}{``Making this same comparison, Cosgrove {[}2016{]}
suggested defining a wave-Pedersen conductivity (their equation (14))\ldots ''}
\end{quote}
\textcolor{blue}{Also, it should be noted that the case without a
back-reflected wave is relevant to small transverse scales, where
the wave may dissipate before meeting with any significant changes
in plasma density.}

\textcolor{blue}{And importantly, although it is of course true that
the natural plasma is never entirely uniform, that does not mean it
is not very informative to compare the electrostatic and electromagnetic
results for a uniform plasma. We expect electrostatic theory to work
for a uniform plasma, and so if it does not we have learned that there
is a major gap in our understanding.}\\
\textcolor{blue}{\medskip{}
}

Section 6 of this manuscript goes on to describe a new model formulation
for the ionosphere based on multiple transmission line segments, but
this model has not yet been implemented and no results are presented.
This modeling approach could be useful in the future, but it seems
premature to publish any information on it here since it has not been
realized or tested yet.\\
\medskip{}

\noindent \textcolor{blue}{In saying ``}This modeling approach could
be useful in the future\ldots\textcolor{blue}{'' the reviewer is
acknowledging that the methodology seems to be correct. But the reviewer
tries to avoid this by saying ``}no results are presented\textcolor{blue}{,''
which is simply not true, as already documented by the lines extracted
from the paper that appear just above. The results that the reviewer
did not report were those that could be derived from this methodology
without a full implementation. The following text from the abstract
specifically addresses this matter,}
\begin{quote}
\textcolor{blue}{``The model will be fully executed in future work,
but an approximate version is used to estimate the phase rotation
for electrical signals traversing the ionosphere: our example calculations
imply that this \textquotedblleft electrical thickness\textquotedblright{}
generally exceeds 90\textopenbullet{} for transverse scale-sizes of
100 km and below\ldots ''}
\end{quote}
\noindent \medskip{}

\noindent The remainder of this review critques the other portions
of the manuscript for which results have been presented. There are
several outstanding issues requiring additional examination. This
manuscript is also very poorly organized and would be much easier
to read with substantial reorganization and removal of superfluous
material.\medskip{}

\noindent \textcolor{blue}{The reviewer is trying to say that the
only results presented are those that he frames below, involving an
E region source. This is simply not true.} \textcolor{blue}{The reviewer
did not report the main results, which are clearly stated in both
the Abstract and in the first paragraph of the section ``Summary
of Results and Discussion'' (and also in the Conclusions section,
in a more general way). To prove this I have extracted text from both
these sections, which appears above.}\medskip{}

\noindent Major Scientific Issues 1.

\medskip{}

\begin{enumerate}
\item The treatment of the E-region wind forcing as a boundary condition
at a single surface is not justified. This manuscript repeatly discusses
the E-region wind dynamo as a boundary condition of the form of Equation
6, and refers to \textquotedblleft the surface of the dynamo\textquotedblright{}
as an inifinitely thin surface. In reality the neutral winds are distributed
throughout the E-region. The neutral wind dynamo is being forced by
winds distributed throughout the 100 to 130 km region, not by a point
layer at 100 km. The wave properties computed in this paper change
dramatically over the 100 to 130 km region, so treating the neutral
wind source as an inifitely thin layer at one altitude is unjustifiable.\medskip{}
\\
\textcolor{blue}{This is not the geometry that the paper addresses.
For example see Figure~9, where the main result is derived. The main
results were derived for a source outside the ionosphere, such as
a source in the magnetosphere, or in the conjugate hemisphere. And
in this case it makes sense to begin with the simple case where there
are no winds in the ionosphere. What the reviewer is referring to
is that it was also explained how a similar analysis could be applied
to a localized source in the E region. While this ``flipped'' geometry
is not as realistic, it still provides for a representative calculation
where we can assess whether electrostatic theory is working as we
expect. It is a very well recognized technique of physics to study
relatively simple configurations, in order to understand the essential
physics, which then helps us to understand more complex situations.
Section~5 is titled ``Gedanken Experiment and Breakdown of Electrostatic
Theory,'' in reference to the technique that Einstein famously applied
to discover relativity.}\medskip{}
\item The meaning of a \textquotedblleft breakdown in electrostatic theory\textquotedblright{}
is poorly defined. There are two different \textquotedblleft electrostatic
theories,\textquotedblright{} and a failure of one does not imply
a failure of both. The first is the theory of electrostatic waves
predicted by the Vlasov-Poisson system of equations (or fluid limits).
This is a dynamical theory describing the time evolution of plasma
parameters at small scales. This theory is extensively used for the
description of 1 small-scale plasma irregularities involved in scintillation,
for the Farley-Buneman and other E-region instabilities, and for the
derivation of incoherent scatter radar theory. All of these applications
are at scales less than 100 m, so the results of this paper showing
the electrostatic wave theory is valid at small scales are in agreement
with the existing literature. The second is the theory of electrostatic
equilibrium, which is a boundary value problem describing steady state
solutions with no reference to how the plasma evolved to those solutions.
The core point of Vasyliunas {[}2012{]} is to describe how electrostatic
equilibrium arises even though the proper description of the transient
evolution must be electromagnetic. Is the objective of this paper
to describe when the electrostatic theory of waves is not valid, or
is it to describe when electrostatic equilibrium will not be reached
from a given initial condition?\medskip{}
\\
\textcolor{blue}{As in previous editions, the paper absolutely explains
that there are two versions (or degrees) of electrostatic theory,
and that they have to be examined separately. It was explained that
the terms ``electrostatic theory'' and ``electrostatic waves''
would be used to differentiate between them. These two are separately
analyzed in the paper! That the reviewer does not seem to know this
suggests that they didn't read much of the paper. \medskip{}
}\\
\textcolor{blue}{For example in the second to last paragraph of the
Introduction section the paper reads,\medskip{}
}\\
\textcolor{blue}{\quad{}''Electrostatic equilibrium is obtained
by solving the equations of motion as though they determine the solution
to a boundary value problem. By \textquotedblleft electrostatic theory,\textquotedblright{}
or \textquotedblleft full electrostatic theory,\textquotedblright{}
for emphasis, we refer to the practice of assuming that this solution
is an approximation for the steady state solution found through the
Laplace transform. A weaker form of the electrostatic assumption involves
the substitution of electrostatic waves for electromagnetic waves.
The validity of this assumption is a distinct question from that of
electrostatic theory, which does not include wavelike effects. We
also analyze this weaker assumption, by substituting the Poisson equation
for the Maxwell equations.''}\bigskip{}
\\
\textcolor{blue}{I am afraid that the reviewers statement ``}\textcolor{black}{\ldots so
the results of this paper showing the electrostatic wave theory is
valid at small scales are in agreement with the existing literature.}\textcolor{blue}{''
is not correct. I do not like to incriminate people when it is not
necessary, but it seems that the reviewer is not familiar with all
the literature. And without publishing this paper it seems likely
that the mistake would be made in the future, even though it was not
already made. The reviewer did not give a reference to where others
have identified the maximum scale. I am aware of results for collisionless
waves, but for collisional waves the closest result of which I am
aware was actually given by one of the earlier (very positive) reviewers,
although they did not indicate a publication. Maybe they are having
the same problem I am having. }\medskip{}
\textcolor{blue}{}\\
\textcolor{blue}{Finally, I do not agree that the ``core point''
of the Vasyliunas {[}2012{]} paper was to show how electrostatic equilibrium
arises; I don't believe he made a conclusion on whether it arises.
I understood the ``core point'' of {[}Vasyliunas, 2012{]} to be
that we should employ a causal analysis, in order to test electrostatic
theory. And that is exactly what I am doing.}
\item The manuscript, and the previous Cosgrove {[}2016{]} work, do make
an intriguing point about how the existence of a steady state solution
does not guarantee that steady state solution is physically reachable
from any particular initial condition.\medskip{}
\\
\textcolor{blue}{It is nice that the reviewer has found something
positive to say.}
\item The manuscript does not clearly explain how electrostatic waves are
associated with corresponding electromagnetic waves. The discussion
of an \textquotedblleft electrostatic Alfven wave\textquotedblright{}
is particularly confusing since Alfven waves are electromagnetic by
definition. The matrix H5 has 16 eigenvectors, whereas the matrix
H5ES has 11. How is the author mapping between the two different sets
of eigenvectors, especially since no one-to-one mapping can exist
for sets of different sizes?\medskip{}
\\
\textcolor{blue}{The third paragraph of Section~4.1 reads, \medskip{}
}\\
\textcolor{blue}{''Where the electromagnetic and electrostatic waves
agree the dashed lines should lie on top of the solid lines. Examining
the figures, the Ion and Thermal waves appear to be well described
by the corresponding electrostatic waves. However, the Alfvén wave
appears to be electrostatic only up to about 100 m in transverse wavelength,
and then radically diverges, with the associated electrostatic wave
generally having a much longer parallel wavelength. And the Whistler
wave appears to have no electrostatic counterpart at all; we could
find no comparable electrostatic wave.''\medskip{}
}\\
\textcolor{blue}{The correspondence between electrostatic and electromagnetic
waves is found using the basic technique of the paper. First, the
eigenvalues/eigenvectors of the matrices $H_{5}$ and $H_{5ES}$ are
sorted and searched so that all waves capable of operating at the
requisite frequency/wavelength are found. Then, the correspondence
of the electrostatic and electromagnetic waves is found simply by
observing that the wavelengths and dissipation scale lengths are exactly
the same, although in the case of the Alfvén wave they are observed
to converge to be the same only below about 100~m in transverse scale.
This can be clearly seen in the figures (Figures 5, and 6, for parallel
wavelength and dissipation scale length), and seems a quite reasonable
way to associate the waves. Also, the usual Alfvén wave treatment
does not include collisions, and anyway I believe that even for collisionless
Alfvén waves (shear mode) it is known that they become electrostatic
below some scale. No electrostatic wave was found to correspond to
the Whistler wave, and also there is the matter of the radio frequency
waves (i.e., X mode, O mode, and Z mode). So the matter of the different
number of equations does not suggest of a problem.}
\item The manuscript makes no attempt to discuss the relevant observational
literature. There has been substantial recent interest in investigating
the Alfven wave dynamics in the ionosphere. This includes C/NOFS observations
demonstrating that interhemispheric coupling in MSTIDs is electromagnetic
{[}Burke et al., 2016a, b{]}, and many observations of Alfven wave
dynamics at high-latitudes from Swarm {[}e.g. Park et al., 2017; Pakhotin
et al., 2018; Miles et al., 2018{]}. All of these publications are
already discussing all transport of elec- tromagnetic energy along
field lines as being carried by Alfven waves, not by a \textquotedblleft mapping\textquotedblright{}
process, exactly as recommended by Vasyliunas {[}2012{]}. The manuscript
claims their theory predicts different altitudes for the turning between
the ion drift direction and the E \texttimes{} B direction than electrostatic
theory (Line 975). The electrostatic prediction of the location of
this turning has been throughly experimentally verified, most notably
during the Joule II sounding rocket campaign where independent measurements
of electric field, ion drift, and neutral wind were all performed
simultaneously {[}Sangalli et al., 2009{]}.\medskip{}
\\
\textcolor{blue}{The paper was already plenty long and included plenty
of material. An observational study would be a whole other study.
Since the electromagnetic theory is more general than the electrostatic
theory, the former can be used to critique the latter without recourse
to data. As far as the turning point, it would depend on scale size
and, of course, would depend very strongly on the neutral density
profile. Without critiquing the Sangalli et al. study, I note that
the reviewer did not say that the neutral density profile was measured.
At any rate, the calculation presented here was not tested, because
it did not exist.}
\end{enumerate}
\medskip{}

\noindent Minor Scientific Issues\medskip{}

\begin{enumerate}
\item The quoted conductivities have the wrong units for conductivities.
Figures 5, 6, and 13 present conductivities in milli-mho (mS) where
the correct unit of a conductivity should be S/m, not S or mS. The
unit of S (mho) should only be used for conductances (integrated conductivities)
or admittances. The wave conductivities defined in Equation 17 have
units of admittance times wavenumber, which are S/m is the admittance
has units of S. 2.\medskip{}
\\
\textcolor{blue}{The reviewer is correct, those panels were labeled
with the wrong units.}
\item The matrix in Figure B2 is missing a row from the top half. This matrix
should be 11 by 11, but the top half of the figure only has 10 rows.\medskip{}
\\
\textcolor{blue}{The reviewer is NOT correct, all 11 rows of the matrix
were present. The reviewer must not have counted very carefully.}
\end{enumerate}
Organizational and Presentation Issues
\begin{enumerate}
\item Section 2 spends a lot of time on tangential problems that distract
from the main thread and confuse the reader. Section 2 could be made
much shorter and much more clear by removing these superfluous tangents.
The main thread is from Equations 1, 6, 7, 8, 10, 11, 12, and Appendix
C. Equation 1 is the prime equation of interest, Equation 6 defines
the driven boundary conditions, Equation 7 is the Laplace transform
of 1, Equation 8 is the fully rigorous solution, and then 10, 11,
12, and appendix C all contain approximate forms of 8. All of the
discussion around equations 3, 4, 5, and 9 are solutions to different
problems from the driven boundary value problem of primary interest
that are only tangentially relevant to the rest of the paper. Equation
3 is the transient solution to the initial value problem with no driving,
which is a different problem all together. The derivation of 4 and
5 is described by the manuscript as \textquotedblleft somewhat heuristic,\textquotedblright{}
and they do not add anything that is not already covered by the fully
rigorous soltuion to Equation 8. Equation 9 is the solution to the
driven boundary value problem when the driver abruptly turns off at
t0, which is a different problem from what Equation 8 solves.\medskip{}
\\
\textcolor{blue}{The reviewer has studied the equations and not found
anything wrong with them. So the reviewer should not have withheld
the results from the editor. I find the additional information helpful
for developing an intuitive understanding of the solution, which I
think is very important.}
\item Many critical details are buried in appendices such that the reader
needs to constantly flip back and forth. Appendix B needs to be read
nearly in entirety before attempting to under- stand Section 2. The
definition of $\Omega$ needed to understand Equation 10 is buried
in Appendix C.\medskip{}
\\
\textcolor{blue}{I don't understand why this is the case. More details
were needed from the reviewer. With respect to $\Omega$, just after
equation~(10) the paper reads ``\ldots where $\Omega(k_{z}-K_{0z})$
is a parameterization of the 16-dimensional unitary transformations
about the identity, with parametric dependence linearized (described
more fully in Appendix C).'' Except in the case of a very intrepid
reader, this seems enough information, especially since $\Omega$
was taken to be the identity shortly thereafter.}
\item Figures 3 and 4 use a normalization that results in several salient
features of the eigenvectors not being visible in the plots. It is
impossible to tell which values are small non-zero numbers. The text
describes multiple features that cannot be seen in these plots as
they are currently presented.\medskip{}
\\
\textcolor{blue}{I agree with this comment. Earlier reviewers have
also criticized the normalization and I see now that I could have
done better. I had used the normalizations from {[}Cosgrove 2016{]},
but they were not appropriate for the plots in this paper since in
this case the X, O, and Z mode waves were not included in the plots.
The electric field is very large for these waves, and accommodating
it swamps the electric field in the other modes. Anyway I have dropped
these kinds of plots entirely from the final 2022 paper. Similar stem
plots appeared in Cosgrove {[}2016{]}, and they seemed to be creating
a distraction.}
\item The definition of wave conductivity, Equation 17, does not appear
until page 16, which is several pages after the first discussion of
wave conductivities in Figures 5 and 6 starting on page 13 (line 494).\medskip{}
\\
\textcolor{blue}{The wave conductivity was derived in {[}Cosgrove
2016{]}, and so was not a new thing. Anyway I have moved this section
earlier; it now appears right after the Introduction.}
\end{enumerate}

\subsubsection{Decision Letter from JASTP}

From: Journal of Atmospheric and Solar-Terrestrial Physics <em@editorialmanager.com>
\\
Subject: Decision on submission to Journal of Atmospheric and Solar-Terrestrial
Physics \\
Date: XXXX XX, XXXX at 12:10:51 PM PST \\
To: \textquotedbl Russell B. Cosgrove\textquotedbl{} <russell.cosgrove@me.com>
\\
Reply-To: Journal of Atmospheric and Solar-Terrestrial Physics <atp-eo@elsevier.com>\\
\medskip{}
Manuscript Number: JASTP-D-19-00027 \, 

\noindent Breakdown of Electrostatic Theory via Exact Linearization
and Solution of the Electromagnetic 5-Moment Fluid Equations\medskip{}
 

\noindent Dear Dr. Cosgrove,

Thank you for submitting your manuscript to Journal of Atmospheric
and Solar-Terrestrial Physics.  I regret to inform you that following
review,  your manuscript has been rejected for publication in JASTP.
  I took time to read the manuscript in detail myself, and my own
views are in accord with the reviewer.  My most specific comment is
the lack of discussion and detailed description/justification of how
electrostatic waves are associated with corresponding electromagnetic
waves.  I view this as a major weakness in the study as it stands.
 We appreciate you submitting your manuscript to Journal of Atmospheric
and Solar-Terrestrial Physics and thank you for giving us the opportunity
to consider your work. \medskip{}
\, 

\noindent Kind regards, \, \,  XXXXXX\,  Editor \,

\noindent Journal of Atmospheric and Solar-Terrestrial Physics\medskip{}
 \,

\noindent Editor and Reviewer comments: \, \,  \\
Reviewer \#1: See attached PDF.\medskip{}

\textcolor{blue}{The reviewer did not tell the editor about the results
claimed in the paper, which are the same as those in the previous
JGR submission and can be read in my responses to the reviewer above.
Basically, the wavelength of the waves is too short to be consistent
with electrostatic equilibrium on the 100~km transverse scale. The
reviewer did not mention this result, even though they did not indicate
any problems with the methodology that was used to derive it.}

\textcolor{blue}{Instead, the reviewer reports certain other results
that are not in the paper, and criticizes a geometry that is not relevant
to the main results from the paper. It does not seem possible that
the problem is in the paper's presentation, because the results that
the reviewer did not report are stated clearly both in the Abstract,
and in the section ``Summary of Results and Discussion.'' These
results were derived in Figure~9, which shows clearly that the reviewers
geometry is not the one that was used. Please see the specific responses
above.}

\textcolor{blue}{It is very odd that I should have to write such a
thing. I do not like doing it, but the reviewer has left me no choice.}

\textcolor{blue}{About the correspondence of the waves, I discussed
it above, and I repeat that discussion here for reference: First,
the eigenvalues/eigenvectors of the matrices $H_{5}$ and $H_{5ES}$
are sorted and searched so that all waves capable of operating at
the requisite frequency/wavelength are found. Then, the correspondence
of the electrostatic and electromagnetic waves was found simply by
observing that the wavelengths and dissipation scale lengths were
exactly the same, although in the case of the Alfvén wave they were
observed to converge to be the same only below about 100~m in transverse
scale. This can be clearly seen in the figures (Figures 5, and 6,
for parallel wavelength and dissipation scale length), and seems a
quite reasonable way to associate the waves. Also, I believe that
even for collisionless Alfvén waves it is known that they become electrostatic
below some scale. No electrostatic wave was found to correspond to
the Whistler wave, and also there is the matter of the radio frequency
waves (i.e., X mode, O mode, and Z mode). So the matter of the different
number of equations does not suggest of a problem. It would not have
been hard to revise the description in the paper.}

\subsection{Reviews of ``The Electromagnetic Ionosphere: a Model,'' submitted
to Phys. Rev E in 2020. This was the fourth version of the 2022 paper.}

\textcolor{blue}{The implementation of the model for a vertically
inhomogeneous ionosphere was added to this version of the paper, and
initial results were given. In the previous version the model had
been described and the equations given, but it had not been completely
implemented. To make room the discussion of the properties of the
wave modes was substantially streamlined. The sections were also substantially
reorganized, to address criticism from the previous reviewer. The
Introduction was modified to address the additional content. With
the implementation of the full model came the additional findings
for longer wavelengths, which might be considered even more surprising.
Hence there was also substantial modification of the conclusions.\medskip{}
}

\noindent \textcolor{blue}{(note, this journal/editor uses the term
``referee'' instead of ``reviewer,'' and so I will try to adapt
to this)}

\subsubsection{Referee \#1}

\noindent -{}-{}-{}-{}-{}-{}-{}-{}-{}-{}-{}-{}-{}-{}-{}-{}-{}-{}-{}-{}-{}-{}-{}-{}-{}-{}-{}-{}-{}-{}-{}-{}-{}-{}-{}-{}-{}-{}-{}-{}-{}-{}-{}-{}-{}-{}-{}-{}-{}-{}-{}-{}-{}-{}-{}-{}-{}-{}-{}-{}-{}-{}-{}-{}-{}-{}-{}-{}-{}-

\noindent Report of the First Referee -{}- EU12156/Cosgrove

\noindent -{}-{}-{}-{}-{}-{}-{}-{}-{}-{}-{}-{}-{}-{}-{}-{}-{}-{}-{}-{}-{}-{}-{}-{}-{}-{}-{}-{}-{}-{}-{}-{}-{}-{}-{}-{}-{}-{}-{}-{}-{}-{}-{}-{}-{}-{}-{}-{}-{}-{}-{}-{}-{}-{}-{}-{}-{}-{}-{}-{}-{}-{}-{}-{}-{}-{}-{}-{}-{}-

This is a long and confusing paper on a one-dimensional steady-state
model of ionospheric coupling using an electromagnetic transmission
line model. While it is generally an improvement to use the full Maxwell's
equations rather than the electrostatic equations, the calculations
presented here are not well described or motivated and produce rather
confusing results. \medskip{}

\textcolor{blue}{In keeping with the pattern established by referees
of earlier versions, this summary does not acknowledge the central
results of the paper. Only the single word ``confusing'' is provided.
The referees additional comments below also make no mention of the
results, or of the important Figures~12-15 and their discussion in
the Model Results section of the paper. A few of the earlier figures
are cited as ``confusing,'' but these are not the figures that contain
the main results, and anyway the confusion is easily cleared up. Although
the results are summarized in the Abstract, it is still usual for
referees to summarize them for the editor, say something about their
importance, and confirm that the body of the paper contains them.
It is usual for this material to appear in a review regardless of
what the referee has to say about the methodology.}

\textcolor{blue}{And also the referee has not indicated any errors
in the methodology. The referee has indicated some confusion, and
there is one specific area indicated below. However, the referee's
confusion appears to be associated with not having read the relevant
paragraph in the paper (and there was also a reference in the Abstract).
That paragraph is quoted in the specific response below. Together
with the referee's apparent failure to read the Model Results section,
there is a strong indication that the referee has not read enough
of the paper to write the review they have written. Other indications
of this may be found below.}

\textcolor{blue}{After failing to report the results, the referee
says that they are ``confusing.'' So it seems that what the referee
actually means is that they did not expect these results, so much
so that they will not repeat them. The referee also says that the
calculation is not well ``motivated.'' The unexpected results are
the motivation for the calculation. But the reviewer simply refuses
to engage with the material.}

\textcolor{blue}{It is not accurate to characterize this new calculation
as ``one dimensional.'' Actually, the model calculates the Fourier
components of a three-dimensional signal propagating in a vertically
stratified ionosphere. After compiling the Fourier components we can
do the inverse Fourier transform to the spatial domain, and describe
a three-dimensional signal. Describing such a three-dimensional signal
with a finite-difference model would require a super computer, whereas
this model can be run on an ordinary personal computer.}\medskip{}

In addition, time-dependent models of the ionosphere exist that are
more complete than this model. The transmission line approach is similar
to that employed by Mallinckrodt and Carlson in 1978 (ref 7), but
has been largely replaced by finite difference modeling of the equations
(e.g., refs 27, 29, 30, 32, 37-40). Thus, it is hard to think of this
work as an advance in the field. \medskip{}

\textcolor{blue}{The referee has compared the model with certain ``time-dependent''
and ``finite difference'' models, although the model presented in
the paper is a frequency-domain model that does not require finite-differencing.
Time domain models are not appropriate for the purposes of the paper.
It is well known that admittance is defined in the frequency domain,
and so a frequency-domain calculation is far more appropriate. The
Introduction and Background sections of the final, 2022 paper explain
why this is the case, and so I will not duplicate that explanation
here.}

\textcolor{blue}{But I do have to point out the basic fallacy in arguing
that since we have one kind of calculation, we don't want to know
anything about other kinds of calculations. In scientific research,
providing an alternative kind of calculation is always an ``advance''
of the field. And in this case the proof is in the pudding, since
this work contains (for example) the unexpected result that the ionospheric
conductance can be very different from the field line integrated conductivity.
No time-domain model has ever even tested this question, and this
serves as evidence that they are not appropriate for the purpose.
But for some reason the referee has not told the editor about the
results in the paper.}\medskip{}

In particular, it is confusing to be considering various wave modes
in a model without frequency dependence. The Alfven wave and the whistler
mode wave occupy very different regimes in frequency, which does not
seem to be considered in this model. \medskip{}

\textcolor{blue}{This is not a correct account of what is in the paper.
The model has frequency dependence, and the frequencies are chosen
in accordance with typical ionospheric scale sizes and drift velocities.
This is mentioned in the abstract as ``based on characteristic ranges
of spatial and temporal scale,'' and then described fully in Section~VA,
where for example is the sentence,}
\begin{quote}
\textcolor{blue}{``Given knowledge of reasonable wind velocities
and a choice for the transverse scale-sizes of interest within the
dynamo (i.e., in the direction along the boundary), the frequencies
that must be matched by the wave modes are thus determined.''}
\end{quote}
\textcolor{blue}{\quad{}Concerning the referee's statement about
the Whistler and Alfvén waves occupying different frequency regimes,
there is no existing research that allows the referee to judge this
matter. The existing dispersion relations for waves are not valid
in the E region ionosphere, because they do not fully include the
collision terms in the 5-moment equations. This is the case for all
the published dispersion relations of which I am aware (excepting
{[}Cosgrove 2016{]}), and the referee has not supplied a reference.}\medskip{}

The results are also rather confusing. For example, in figures 8-10,
the plots seem to show parallel wavelengths in excess of 1000 km at
altitudes of 100-150 km, even though the ionosphere is strongly inhomogeneous
on such scales.\medskip{}

\textcolor{blue}{First, the ``results'' in figures 8-10 are not
the main results of the paper, they are just some intermediate plots
and secondary results.}

\textcolor{blue}{Anyway, there is nothing confusing or strange about
this. Inhomogeneity does not constrain the local wavelength. The wavelength
at a given altitude depends only on the local ionospheric parameters
at that altitude, like the density and collision-frequency and etc.
The model handles the vertical inhomogeneity by piecing together many
short pieces of waves, where the properties of the waves, including
wavelength, change with altitude. This is how transmission line theory
works.}

\textcolor{blue}{And by the way, over most of that altitude range
the wavelength for both modes is less than 1000~km, and generally
much less. Perhaps the referee should look at the figures again.}\medskip{}

It is difficult to reconcile the results of the model to any observable
quantities. For example, I would have expected plots of the calculated
electric and magnetic field perturbations as a function of altitude,
but such results are not presented. \medskip{}

\textcolor{blue}{This statement is not correct. It appears that the
referee has not looked at the figures in the section ``Model Results,''
which appears right before the concluding section. Figure~14e shows
plots of electric and magnetic field versus altitude. But the main
observable studied in the paper is the ionospheric input admittance,
the real part of which is the ionospheric conductance. Figures~13-15
all contain plots of the ionospheric conductance, comparing it to
the field line integrated conductivity. The ionospheric conductance
is an important parameter determining the inner boundary condition
for global MHD models of the magnetosphere. So while I admit that
this parameter is difficult to measure directly, it is heavily relied
on for global modeling, where it is calculated electrostatically,
as the field line integrated conductivity. This paper shows that this
electrostatic calculation is unreliable.}\medskip{}

As one final point, \textquotedbl Poisson\textquotedbl{} is misspelled
as \textquotedbl Poison\textquotedbl{} through the paper.\medskip{}

\textcolor{blue}{This statement is not correct. In searching the paper
I find one instance of ``Poison'' and five instances of ``Poisson.''
It does not seem that the reviewer has read much of the paper. This
is a specious claim, and one wonders what is motivating the referee
to be so hostile. Does the referee have a conflict of interest?}

\subsubsection{Referee \#2}

\noindent -{}-{}-{}-{}-{}-{}-{}-{}-{}-{}-{}-{}-{}-{}-{}-{}-{}-{}-{}-{}-{}-{}-{}-{}-{}-{}-{}-{}-{}-{}-{}-{}-{}-{}-{}-{}-{}-{}-{}-{}-{}-{}-{}-{}-{}-{}-{}-{}-{}-{}-{}-{}-{}-{}-{}-{}-{}-{}-{}-{}-{}-{}-{}-{}-{}-{}-{}-{}-{}-

\noindent Report of the Second Referee -{}- EU12156 Cosgrove

\noindent -{}-{}-{}-{}-{}-{}-{}-{}-{}-{}-{}-{}-{}-{}-{}-{}-{}-{}-{}-{}-{}-{}-{}-{}-{}-{}-{}-{}-{}-{}-{}-{}-{}-{}-{}-{}-{}-{}-{}-{}-{}-{}-{}-{}-{}-{}-{}-{}-{}-{}-{}-{}-{}-{}-{}-{}-{}-{}-{}-{}-{}-{}-{}-{}-{}-{}-{}-{}-{}-

This manuscript presents the development of a 1D, electromagnetic
ionosphere model using the 5-moment, two fluid equations and based
on the concept of stacked transmission lines driven by a source outside
the ionosphere. The model itself is interesting; however, the presentation
in the manuscript is both hard to follow and steeped with unnecessary
re-derivations of well known results, and it is unclear why the model,
and the development of the model, is not rooted in observational data.\textcolor{blue}{\medskip{}
}

\textcolor{blue}{The referee does not report the results to the editor.
There is an insinuation that the results are ``well known,'' but
the results discussed in this regard by the referee (below) are not
the central results of the paper; and anyway the referee does not
provide a reference, and so appears to be referring to results that
are not valid in the E region ionosphere. The results claimed in the
Abstract, which the referee does not acknowledge, are contained in
Figures~12-15 of the Model Results section. It is usual for this
material to appear in a review regardless of what the referee has
to say about the methodology.}

\textcolor{blue}{And with this referee it is quite clear that they
are not suggesting any errors in the methodology (which is summarized
in the Abstract). The referee does say that the methodology is ``hard
to follow,'' but they do not mention any specific areas where they
had difficulty. Instead, they say that new methodology should not
be introduced, that the results can be derived in a different way.
But even if the results could be derived in a different way, this
is not an appropriate criticism of the paper; the referee should be
judging whether the methodology is correct, and if the methodology
is new that is a good thing.}

\textcolor{blue}{The referee suggests that the paper contains ``re-derivations.''
However, I do not think this is the case, and the referee has not
supplied a reference. The ionospheric conductance is mostly in the
E region of the ionosphere, including the lower E region. Analyzing
this region requires keeping all the collision terms in the 5-moment
equations, and I do not know of any reference that contains the needed
results. Since the referee does not mention the issue of the E region
collision frequencies anywhere in the review, I suspect they have
not considered it, and that there is in fact no reference available.
It is hard to imagine how the ionospheric community could have persisted
in relying completely on electrostatic theory if such a reference
existed, since that reference would have to show the importance of
electromagnetic waves in the ionosphere. }

\textcolor{blue}{Also, the derivation given in the paper includes
quite a bit of analysis of potential sources of error relevant to
the model, and it seems unlikely that this material would have been
covered by other research that did not having the specific goal of
developing the model. The referee has complained about the length
of the derivation, but the alternative is a model that is less rigorous
and therefore less defendable. Although I would like to believe that
the referee's failure to acknowledge the main results was not intentional,
the unexpected nature of these results require that the paper be rigorous
and complete. Including detail allows the diligent reader to understand
the derivation.}

\textcolor{blue}{With respect to the comment about observational data,
my reply follows just below.\medskip{}
}

Below are detailed several major points that, if well addressed, would
significantly improve the manuscript. They are presented in no particular
order.

1) The ionosphere is a well observed system, so why not use the observations
to both constrain and test the model?\textcolor{blue}{\medskip{}
}

\textcolor{blue}{When a calculation employs simplifications or approximations
to an accepted physical theory, the simplifications or approximations
can be tested theoretically by performing a more complete calculation
based on the same accepted physical theory. This kind of a test provides
insight beyond that provided by an experimental test, because the
reason for any discrepancy can be identified as the simplification
or approximation that was removed. In addition, what is possible in
the ionosphere is observational testing, not experimental testing,
where the difference lies in the ability to rigorously analyze the
errors. Because observational testing does not allow for rigorous
error analysis, it is subject to a confirmation bias. Hence, theoretical
testing of the approximations should always be done first, when possible,
with observational testing coming afterward.}

\textcolor{blue}{In particular, the purpose of the paper is to test
whether the simplifications that lead to electrostatic theory are
appropriate for calculating the ionospheric conductance, given vertical
profiles for the local ionospheric parameters. (The ionospheric conductance
is the real part of the input admittance seen from the top of the
ionosphere.) It is possible that we need to use a more complete form
of our accepted physical theory, which is electromagnetics. The model
is meant to supply this more complete form, for testing, and to form
the basis for a replacement if one is found necessary (which it is). }

\textcolor{blue}{It is possible that the referee was confused by the
use of the term ``model,'' given that there are such things as ``assimilative
models'' and ``empirical models,'' which involve data. Assimilative
models are formed by combining the best possible theoretical model
with data, and should never be based on inapplicable theory. Getting
the theory as correct as possible is the first step, and that is the
goal of the paper.}

\textcolor{blue}{Finally, it is naive to suggest that the ionosphere
is a ``well observed system.'' The E region ionosphere, where is
most of the ionospheric conductivity, is quite inaccessible. The ionospheric
community works very hard to collect what data it can, but the challenges
are great. Satellites can survive only for a very short time in the
E region, and so do not go there except to die. Incoherent scatter
radar cannot measure electric field in the E region, except by assuming
perfect mapping from the F region, which is exactly the assumption
being tested in this work. Lidar has so far been altitude limited
such that it can say little about the E region. Green line imagers
provide intriguing E-region observation that are not resolved in altitude.
TEC measurements are dominated by the topside of the F region. Imaging
coherent scatter radar sees backscatter only when irregularities are
present. Ionosondes are mostly limited to vertical density information
(although there exist advanced systems that are able to obtain certain
horizontal information). Rockets can include a nice suite of instruments,
but do not stay aloft very long, and the instruments are very hard
to reuse. The preceding should not be read as a criticism of the ionospheric
community! The point is that the E region ionosphere is very difficult
to access and definitive experiments are not, at present, possible.}

\textcolor{blue}{See further comments below on whether ``electrostatic
theory'' is the best term to use.\medskip{}
}

A significant portion of the manuscript is spent deriving the wave
modes of the 5-moment mode and determining whether the various modes
can propagate in the ionosphere, only to re-discover that only the
Alfven and whistler modes can propagate. Why not leverage known research
that these are the two relevant modes and simply focus on them from
the beginning?\textcolor{blue}{\medskip{}
}

\textcolor{blue}{This has already been answered. I am not aware of
any results for waves in the 5-moment equations that do not make approximations
with respect to the collision frequency terms, and which are therefore
not valid in the E region. Since the E region and collisions in general
are critical to the ionospheric conductance, approximations with respect
to the collision terms would not be appropriate for this study. The
referee has suggested there is ``known research,'' but has not supplied
a reference. Such a reference would imply electromagnetic effects
in the ionosphere, and so the existence of such a reference would
be inconsistent with the fact that electrostatic theory is well-accepted
in the ionosphere.}

\textcolor{blue}{Also, the paper does not find that ``}only the Alfvén
and whistler modes can propagate\textcolor{blue}{'' in the ionosphere.
Two additional propagating modes are found. These modes are omitted
from the model because they have very short dissipation scale lengths
and minimal coupling to the Alfvén and whistler modes. If included,
they would only serve to reduce the agreement with electrostatic theory.
This is explained in Section~VC. However, one of these modes has
characteristics showing that it may be involved in certain anomalous
ionospheric observations from the equatorial region. So the methodology
described in the paper opens up new opportunities for studying collisional
waves in the ionosphere.\medskip{}
}

Also, it is stated in the manuscript that \textquotedblleft It is
difficult to know where the model stands with respect to observational
studies, since it is really necessary to evaluate each observational
campaign in light of the model expectations for the particular event
or statistical sampling of events.\textquotedblright{} Of course it
is true that this model or any realistic model will produce different
results for each event; that is the entire point of producing an accurate
model, after all. Therefore, a particular event should be chosen as
a baseline for comparison. As it stands, this paper makes no direct
contact with any observation, making it impossible to judge the accuracy
of the model. The end goal of this model is to be incorporated into
the CCMC and InGEO models for public use, but the zeroth order test
for inclusion in these global models is that the ionosphere model
reproduce, at least, basic observations.\textcolor{blue}{\medskip{}
}

\textcolor{blue}{First, the ``end goal'' was meant as a long-term
goal after additional development, which is why it was only mentioned
in the second to last paragraph of the paper. It is not appropriate
to criticize a paper for failure to achieve }\textcolor{blue}{\emph{future}}\textcolor{blue}{{}
goals. The immediate goal is to show that an electromagnetic calculation
of the ionospheric conductance gives, in some cases, very different
results from the approximate electrostatic calculation that is used
in current practice. This is a purely theoretical result that is important,
because it strongly suggests that the current theoretical approach
is not reliable. It does this by showing that a more complete theory
(electromagnetic versus electrostatic) gives very different results.
Observations are not required to derive this kind of result, which
is lucky because contrary to the referee's assertion, the needed observations
do not exist. Satellites are in re-entry at E-region altitudes. There
is no way to obtain the simultaneous vertical and horizontal resolution
that is needed, and in fact, electrostatic theory is often used to
help the process along with regularization {[}e.g., Nicolls, Cosgrove,
and Bahcivan, JGR, 2014{]}. We have been relying on the electrostatic
calculation without good observational }\textcolor{blue}{\emph{or}}\textcolor{blue}{{}
theoretical support. But the theoretical test is possible, and is
what this paper provides for the first time.}

\textcolor{blue}{Second, I don't agree with the referee's statement
that there has to be observational validation before a model can be
included in CCMC, and definitely not for InGeo. If there is a theoretical
demonstration that the electrostatic calculation of conductance is
problematic, researchers need access to the alternative theory. Otherwise,
observational testing is not even possible, except by those who work
directly with me. In the interest of advancing science, models of
all kinds should be shared, not kept private.}

\textcolor{blue}{Third, the referee's comment that the ``paper makes
no direct contact with any observation'' is misleading.} \textcolor{blue}{The
model calculates many observables. The most important one is the ionospheric
conductance, which can (in principle) be measured by a satellite constellation
at the top of the ionosphere {[}e.g., \citealp{burke+etal-2017}{]},
and determines the inner boundary condition needed for global MHD
modeling. But the model also calculates altitude profiles for electric
field, magnetic field, ion velocity, electron velocity, ion density,
electron density, ion temperature, and electron temperature. The referee
only means that applications to specific observed events are left
to future research.\medskip{}
}

2) Wave modes of the 5-moment system, naming conventions, and descriptions.
As noted above, a considerable portion of the manuscript is spent
deriving the wave modes of the 5-moment, two fluid system; however,
this is a well studied system of equations, whose wave modes are also
well known. So, why spend so much time re-deriving them?\textcolor{blue}{\medskip{}
}

\textcolor{blue}{This has already been answered. I do not know of
any such results and the referee has not supplied a reference. The
issue is a lack of full inclusion of the collision frequency terms
in the 5-moment equations. To my knowledge, the results I have given
here are the only ones that fully include these collision terms (excepting
{[}Cosgrove 2016{]}). And this full inclusion is necessary to treat
the E region, which is essential to this work. \medskip{}
}

Importantly, why after re-deriving known results are different naming
conventions employed, i.e., why is contact not made with existing
literature for ease of comparison and establishing a common language?

Specifically, why \textquotedblleft create\textquotedblright{} two
modes, thermal and ion (acoustic mode/short wavelength limit of the
MHD slow mode), rather than refer to them by their conventional names?
Also, the Alfven mode seems to be incorrectly labelled, as detailed
below.\textcolor{blue}{\medskip{}
}

\textcolor{blue}{This is about what are the best names to use for
the waves found in the analysis. To my knowledge ``ion acoustic mode''
is a standard name, which I shortened to ``ion mode,'' as stated
on page~13, and which is standard practice in the ionospheric community.
Is the referee wishing that I retain the word ``acoustic'' throughout
the paper? I can certainly do that, but was hoping to avoid using
a two-word name.}

\textcolor{blue}{Also, it is not clear what name the referee is suggesting
I use for what I called the ``thermal mode.'' In my understanding
there are three MHD modes plus a non-propagating entropy mode. Perhaps
the latter is the thermal mode, which can propagate when collisions
are included? The referee has not supplied any references and so their
assertion of waves being ``well known'' with ``conventional''
names is not convincing. What is well known is that there is a literal
zoo of names for wave modes, and in many cases the names are associated
with the simplifications that were used to derive them. Since I do
not use these simplifications the associations are not necessarily
clear (more below).}

\textcolor{blue}{The names I have chosen are ones I think will be
comfortable for the ionospheric community, but may not be so comfortable
for researchers from other communities. There does not seem to be
any way to avoid this compromise.}

\textcolor{blue}{In the comment about the Alfvén mode, the referee
did not mention any specific figures or text, but to the extent the
comment is understandable it is addressed below. Also, I have additional
reason for not using the MHD naming scheme that I will save for below.\medskip{}
}

The wave modes derived seem incorrect, although this is difficult
to evaluate because the dispersion relations of the modes are not
presented in a manner that is easy to interpret. The whistler wave
lies on the fast magnetosonic dispersion surface and transitions from
the fast mode at large, MHD, scales to the whistler mode for scales
k\_perp d\_i and/or k\_parallel d\_i >\textasciitilde{} 1, where d\_i
is the ion inertial length. Similarly, in the 5-moment system, the
large-scale, shear Alfven dispersion surface transitions to the Alfven
ion cyclotron mode for k\_parallel d\_i >\textasciitilde{} 1 but k\_perp
d\_i \textasciitilde < 1, or the kinetic Alfven wave for k\_perp
d\_i >\textasciitilde{} 1, which can then couple to the first ion
Bernstein mode if k\_perp d\_i >1 and k\_parallel d\_i > 1, and eventually
becomes electrostatic for scales d\_e \textasciitilde{} 1. If you
are in the limit that the Alfven speed is larger than the electron
thermal speed, i.e., electron beta is less than the mass ratio, then
rather than becoming a kinetic Alfven wave, the large-scale Alfven
wave transition to an inertial Alfven wave. It is also possible, for
beta < 1 but not mass ratio small, to have combined aspects of the
kinetic Alfven and inertial Alfven wave. Given these facts, it is
entirely unclear to me why the whistler wave and shear Alfven wave
are discussed as being simultaneously relevant to the system, since
if the whistler wave is relevant, the scale range of interest must
lie in the ion cyclotron, kinetic Alfven wave, or inertial Alfven
wave range, none of which have simple linear dispersion relations.\textcolor{blue}{\medskip{}
}

\textcolor{blue}{First, I note that the referee has given a very complex
recipe for sorting the modes and their properties. All of this complexity
is eliminated by the methodology adopted in this paper.}

\textcolor{blue}{And second, there is still the same issue as before.
To my knowledge, the results I have given here are the only ones that
fully include the collision frequency terms in the 5-moment equations;
and these terms are critical in the E-region ionosphere, where the
neutral collision frequencies are very high. So if the modes seem
different that is not relevant. The explanation is that fully including
the collision frequency terms makes a big difference, at least in
the E region. As far as the linear/non-linear comment is concerned,
the dispersion relations found in the paper are not limited to linear
ones, as is evident from Figure~8 (for example). The paper also explains
that the 5-moment equations are fluid equations and so do not include
kinetic effects, which is sufficient to ensure that the treatment
is more general than electrostatic theory, since the latter doesn't
include kinetic effects either.}

\textcolor{blue}{There is also a more fundamental issue here about
what is meant by ``dispersion relation'' and ``wave.'' Formally,
a system of 16 equations-of-motion supports 16 eigenmodes, where a
``wave'' involves two of these modes, because the direction of propagation
can be reversed. There is a single dispersion relation corresponding
to these 16 equations, which will be a polynomial having 16 roots,
one for each mode. (The polynomial comes from the eigenvalue problem
for the equations of motion, as described in Section III of the paper.)
We can imagine solving the single polynomial dispersion relation for
the roots where, in analogy with the quadratic equation, we will obtain
a solution involving square and higher roots with $\pm$ signs. There
will be 16 possible permutations for the $\pm$ signs, and each permutation
can be viewed as producing a sub-dispersion relation for a single
mode. Contrast this picture with what the referee is describing: when
the referee refers to a ``dispersion relation'' for a ``wave,''
they are actually referring to a limited portion of one of these sub-dispersion
relations, what might be called a ``dispersion branch,'' as now
explained. }

\textcolor{blue}{To my knowledge, no one has ever written-out the
full analytical solution for the (single) dispersion relation of the
(collisional) electromagnetic 5-moment equations, with all the square
and higher roots and $\pm$ signs. What is normally done instead is
to employ various physical assumptions to approximate the sub-dispersion
relations over limited regions, resulting in what I call dispersion
branches; these are waves as categorized by physical type. The different
dispersion branches are one way to define different kinds of waves,
and result in the well known ``zoo of waves'' with a complex relationship
among them (hence the referee's description above). Under this definition,
a term like ``whistler wave'' may only be applicable over a certain
range of altitude (for example). This approach is useful in that it
facilitates a physical understanding of wavelike disturbances that
may be detected in data. But in a case where it is desired to form
a continuous wave-description over a range of physical conditions,
such as over a range of collision frequencies (range of altitudes,
for the ionosphere), the approximations will breakdown in certain
areas. So we cannot use these (approximate) dispersion branches to
form a model that is reliable and accurate over the full range of
ionospheric altitudes. }

\textcolor{blue}{Instead, the present paper uses standard matrix analysis
tools to solve for all 16 roots exactly (numerically), so that all
the terms in the 5-moment equations can be retained, including all
collision terms over all altitudes. What are normally considered different
wave branches may be fused together over altitude, because the approximations
that separate them are not used. This is described in some detail
in the paper. The referee is apparently objecting to this and requiring
instead that I use the many approximate wave expressions (dispersion
branches). Attempting to follow the referee's suggestion would not
result in a satisfactory model, because the approximations could never
be defended over the full height of the ionosphere, owing to the wide
range of collision frequencies that exist; trying to piece together
the many approximate dispersion branches would be a nightmare; there
is no one-to-one map between the physically-defined waves and the
more formally-defined waves in the paper. And in fact, I do not believe
that there exist any of the approximate results that are applicable
to the high-collision-frequencies of the E region ionosphere; there
are probably new ``collisional'' dispersion branches that have not
been previously studied, but which are included in the model of the
present paper.\medskip{}
}

The exploration of which of the electromagnetic modes has electrostatic
counterparts in various limits is also strange, and I do not understand
why it is done; however, this is again well explored territory in
various wave textbooks and published papers. For instance, it is stated
that \textquotedblleft We find that electrostatic waves are good approximations
for electromagnetic waves when the transverse wavelength is less than
about 100 m.\textquotedblright{} 100 m is approximately the electron
inertial length, based on a density of \textasciitilde 10\textasciicircum 9
m\textasciicircum -3 used in the manuscript. Scales below the electron
inertial length for the Alfven dispersion branch within the 5-moment
model are electrostatic; therefore, the statement in the manuscript
and, very lengthy, derivation of this result were obvious from the
outset based on the physics contained in the model. Note that for
a more realistic density of 10\textasciicircum 11, this scale becomes
\textasciitilde 10m, significantly altering the conclusions in the
manuscript, which could be easily ameliorated by using dimensionless
units, i.e., the electrostatic transition scale for the Alfven branch
is always d\_e in the 5-moment model. Note also that the fact that
the fast/whistler branch does not have an electrostatic counterpart
is well known. The fast mode branch, when followed from large to small
scales at fixed propagation angle, transitions from fast to whistler
to upper hybrid to R wave, all of which are fully electromagnetic.\textcolor{blue}{\medskip{}
}

\textcolor{blue}{The referee asserts that it is well known that the
waves are electromagnetic, at least above a few hundred meters in
scale. How can we reconcile this with the fact that electrostatic
theory is considered completely reliable in the ionosphere? Either
ionospheric scientists are completely ignorant of the results the
referee asserts, or the referee is in error. So this is yet again
the same issue: given that the referee has not supplied a reference,
and does not mention collision frequency, the most likely explanation
is that the referee is referring to results that do not fully include
the collision frequency terms. The collisional effects are very important
in the E region, and results may be quite different if they aren't
fully included. }

\textcolor{blue}{The calculations in this work }\textcolor{blue}{\emph{do}}\textcolor{blue}{{}
fully include the collision frequency terms, but require a numerical
computation of the eigenvalues/eigenvectors, and so it is difficult
to evaluate if the electron inertial length is always the transition
scale. Given the ionospheric communities total reliance on electrostatic
theory, this community apparently believes that collisional effects
greatly increase the scale at which electromagnetic effects become
important.}

\textcolor{blue}{The referee has discussed the transition from the
fast mode to the whistler mode as associated with changes in scale.
But in the paper is found an apparently similar mode transition associated
with changes in }\textcolor{blue}{\emph{collision frequency}}\textcolor{blue}{,
with transverse scale held constant. So the collisional effects are
very important, and given that they have been used as part of the
justification for electrostatic theory in the ionosphere, it is necessary
to re-examine the electromagnetic nature of the waves with collisions
fully included. Ionospheric models are almost exclusively based on
either electrostatic waves or full electrostatic theory, and so apparently
have assumed that collisional effects eliminate the electromagnetic
nature that the referee is referring to. We need to know if these
models are reliable. The referee's statement that the waves are electromagnetic
implies that they believe these models are not reliable, which reinforces
the need for the present paper.}

\textcolor{blue}{Also, the referee's comment concerns only secondary
results of the paper that come essentially for free, from the formalism
developed for the central question of the ionospheric conductance.
So the comment concerning a ``very lengthy derivation'' is inapplicable.
When secondary results are easily obtained, they are presented, but
there is no room to pursue them further. Also, the length of the derivation
should not be an issue on which the paper is evaluated. What matters
is the correctness of the derivation, and proving correctness requires
detail.\medskip{}
}

The above points could be significantly clarified if rather than the
lengthy derivations given in the paper, a table supplying and summarizing
which modes exist or are evanescent at each altitude, together with
their properties, e.g., wavelength range, frequency, and the parameters
of the ionosphere, e.g., plasma beta, density, ion and electron temperature,
etc, at that altitude. Importantly, the wave properties should be
given in normalized units, e.g., k\_perp d\_i and omega / Omega\_ci,
where Omega\_ci is the ion cyclotron frequency, because the normalized
quantities determine the universal physics of the waves; whereas,
an Alfven wave with wavelength of, for instance, 1000 km has vastly
different meaning in the ionosphere vs a magnetar magnetosphere or
the solar wind.\textcolor{blue}{\medskip{}
}

\textcolor{blue}{Note that the referees list of parameters does not
include collision frequency. So it's the same issue: I do not know
of any derivations in the literature that fully include collision
frequency, and the way I manage to include it involves a numerical
calculation of the eigenvectors/eigenvalues. In this case, it is not
clear what the natural scalings should be. One likely change is the
normalization of frequency, which should probably include a collision
frequency contribution (and not just Omega\_ci, as the referee suggests).
I know this from the analytical derivation of a collisional dispersion
relation that I gave in my 2016 paper. Trying to figure out a universal
normalization within this semi-numerical scheme is not practical,
at least not at this time. And it is not needed to derive the central
results of the paper (which the referee has omitted from their report). }

\textcolor{blue}{The idea that the model could be based on the existing
approximate dispersion branches (for the zoo of waves) is not tenable,
as already argued above. Instead, the paper solves the equations of
motion in a rigorous manner, and derives the model from the solution.
Using a rigorous solution makes the model more reliable, which is
extremely important given that the results are apparently so unexpected
that both referees have omitted them from their reports.\medskip{}
}

3) It is very difficult to follow what assumption are employed in
different portions of the manuscript. For instance, density profiles
and other parameter profiles are given in figure A.1, but then an
array of various constant densities are stated throughout the manuscript
for computing quantities and producing figures, but the physical behavior
in the figures associated with the constant densities appears to be
due to the density profile shown in figure A.1.\textcolor{blue}{\medskip{}
}

\textcolor{blue}{The referee states ``}the physical behavior in the
figures associated with the constant densities appears to be due to
the density profile shown in figure A.1\textcolor{blue}{.'' The vertical
character the referee is seeing is caused by the collision frequency
profiles, not density, which again leads me to believe that the referee
has in mind results that do not incorporate collision frequency. The
referee is not familiar with the effects of the neutral collisions
on waves, because these are new results. A constant vertical profile
has been used for plasma density so that these new results will not
be obscured by plasma density effects.}

\textcolor{blue}{The density is given in the caption of every figure,
and it is stated repeatedly that they are constant vertical profiles.
The only exception is one trace in Figure~12C (green trace), as indicated
in the legend and caption and text. The referee does not say which
figures they are referring to, and so it is hard to know where additional
clarification is needed. Given that the density is stated in every
figure caption, there is no general inadequacy that I can discern.
However, I have added notations such as ``at all altitudes,'' so
that the density call-outs in the captions will be even more clear.}

\textcolor{blue}{\medskip{}
}

Related to figure A.1, some of the data in the figure is stated to
be \textquotedblleft derived from Sondrestrom incoherent scatter data.\textquotedblright{}
Is it previously published data? Is it averaged over some time period?
What is the date and time of the observation(s)? Also, why use textbook
data from 1978, reference 51, for estimating the collision frequency
in the same figure? There are surely newer, more accurate measurements
of ionospheric density and temperature from which you could calculate
the collision frequencies.\textcolor{blue}{\medskip{}
}

\textcolor{blue}{The Sondrestrom data is only used in one trace of
one figure (Figure~12C, the green trace), for purposes of comparison
to something more realistic than the constant profiles used everywhere
else (for the reasons just explained). So this data is essentially
irrelevant to the results and does not merit such a detailed pedigree.
Actually, the neutral collision frequencies do not depend on the plasma
density, but they do depend critically on the neutral density and
composition, which varies quite a bit and is also }\textcolor{blue}{\emph{very}}\textcolor{blue}{{}
difficult to measure in the ionosphere. For this reason, representative
neutral collision-frequency profiles are all that is really possible.
However, they are also all that is needed to demonstrate the behavior
of the model; we are not trying to model a specific event at this
time. Also, the paper describes the reasons for the choice and why
one other alternative was not used, while showing that it would give
similar results. Also, the referee does not reference any newer results
that would be superior for the purpose, and merely says ``surely''
there is one.\medskip{}
}

4) As noted in point 1 above, one of the goals of this model is to
be implemented in global modeling codes such as the CCMC and InGeo,
but I do not understand how the model as given in the current manuscript
could be implemented. Generally, global models need to compute the
parallel current and perpendicular electric field at the ionosphere
boundary; however, the ionosphere model in the manuscript is based
on waves propagating in the ionosphere driven by a harmonic source.
The source seems disconnected from what magnetosphere models would
supply to the ionosphere model. Additionally, magnetosphere models
rarely go to altitudes lower than \textasciitilde 1-2 R\_Earth, which
is an order of magnitude above what is considered in the current manuscript.\textcolor{blue}{\medskip{}
}

\textcolor{blue}{The paper presents results for the ionospheric input
admittance, which is the ratio of the parallel current to the divergence
of perpendicular electric field at the ionospheric boundary, and is
the quantity used by the global models. This quantity is plotted three
different ways in the concluding figure (Figure15, real part), and
is also plotted in Figure~14 and Figure~13. The definition of the
input admittance was given in equation~(1).}

\textcolor{blue}{The way that the input admittance can be calculated
from a wave theory is thoroughly covered in the paper, with the main
concepts being presented in Section~II. For example, at the bottom
of page~3 there are the sentences,}
\begin{quotation}
\textcolor{blue}{``The input admittance associated with a single
wave propagating away is commonly referred to as the characteristic
admittance of the wave, $Y_{0}$. It is this quantity that enters
into the well-known TL expression for the input admittance to a transmission
line terminated by a load admittance $Y_{L}$'' (the equation follows)}
\end{quotation}
\textcolor{blue}{I would suggest that the referee reread Section~II.
And also the referee should reread the Introduction, where it is explained
that transmission line (TL) theory is specifically appropriate for
calculating input admittance, and other circuit quantities. }

\textcolor{blue}{As far as applying the conductance model to global
magnetospheric modeling, there is a dependence on wavelength that
would have to be accounted for. The paper derives the Fourier components
of the conductance, and these would have to be combined to model a
particular disturbance. (Although, each Fourier component can be viewed
as representing the behavior at a particular scale.) This is discussed
in the concluding section. However, as mentioned above, deployment
to CCMC or InGeo is an ``end goal'' and we are not there yet. The
purpose of mentioning this goal was to show the potential for replacing
electrostatic theory with this model, as opposed to just showing the
limitations of electrostatic theory. }

\textcolor{blue}{As far as the altitude of lower boundary for magnetospheric
models, this problem is well known and the accommodations all involve
using in some way the ionospheric conductance (real part of the input
admittance), which is what is provided by the model. (Actually, the
model provides both the real and imaginary parts, but it appears the
imaginary parts are generally small, so far.)\medskip{}
}

The ionosphere model as currently described also has severe limitations,
100-1000km, on the transverse wavelengths of modes considered. A magnetosphere
model to which this ionosphere model would be coupled would inject
a broad spectrum of modes, far exceeding the limited range the current
ionosphere model can handle. Thus, at best, the ionosphere model as
described needs significant improvement before it can be employed
as part of a global model.\textcolor{blue}{\medskip{}
}

\textcolor{blue}{This is why the main purpose of the paper is to show
that an electromagnetic calculation of the ionospheric conductance
gives, in some cases, very different results from the approximate
electrostatic calculation that is used in current practice. This provides
the motivation to further develop the model.\medskip{}
}

5) There are many other, lesser issues with the manuscript that I
am reluctant to fully enumerate in light of significant criticisms
noted above, e.g., proper historical references for the 5-moment model;
multiple definitions of what it means to be electrostatic, some of
which are incorrect (formally, it only means partial B /partial t
= 0 or curl E = 0, not a fully steady-state electromagnetic system
as the first sentence of the manuscript states); the first reference
of section III is missing; red versus blue color coding in figure
6, etc\dots \textcolor{blue}{\medskip{}
}

\textcolor{blue}{The first reference of Section III is the reference
for the 5-moment equations; I am sorry that it was broken. It was
meant to be the textbook by Schunk and Nagy. Will that be sufficient?}

\textcolor{blue}{The ionospheric conductivity is calculated by setting
all time derivatives to zero. The conductance is calculated (electrostatically)
by integrating this conductivity along the geomagnetic field. This
can be discovered, for example, in the textbook by Kelley (1989) that
is referenced in the paper. (It is possible to have a frequency dependent
version, but in practice the frequency is always set to zero, which
is equivalent to setting the time derivatives to zero.) The Farley,
1959, reference describes the underlying methodology, which ionospheric
physicists call electrostatic theory, and which proceeds by setting
all time derivatives to zero. Also, looking at the classic electromagnetics
textbook by Jackson, I find as the first line of the first electrostatics
chapter, ``We begin our discussion of electrodynamics with the subject
of electrostatics---phenomena involving time-independent distributions
of charge and fields.'' So the standard usage of the term ``electrostatic
theory'' is for a theory of time-independent systems, and in physics
they must arise as the steady state of a time-evolving system. This
all supports the definition I have given for ``electrostatic theory.'' }

\textcolor{blue}{In the last paragraph of the Introduction I explained
the distinction between this and the methodology I think the referee
is referring to, which is commonly called ``electrostatic waves''
by ionospheric physicists. Apparently the referee did not read this
paragraph. This comment by the referee causes me to wonder if they
understood the purpose of the paper, which is to test the former ``electrostatic
theory,'' the one that is used to justify using the field line integrated
conductivity for the ionospheric input conductance. If the referee
did not understand this than I can see why they would have trouble
understanding the results in the paper. This is something I could
clarify in the Introduction, except that the referee has not offered
enough detail to enable me to understand where they became confused.
I thought that I had covered this quite well, for example in the second
paragraph of the paper there appears,}
\begin{quotation}
\textcolor{blue}{``To narrow the problem to one that is tractable,
we focus on modeling the \textquotedblleft ionospheric conductance\textquotedblright{}
{[}e.g., 10, 11{]} for a vertically stratified ionosphere\ldots{} which
is meant to represent the input admittance to the ionosphere as seen
from the magnetosphere \ldots . Under electrostatic theory it is derived
by integrating the (zero frequency) ionospheric conductivity along
the geomagnetic field. Dropping the assumption that it is purely real,
and dropping the assumption that it can be derived from electrostatic
theory, we arrive at an important problem that is well suited to TL
modeling, that of deriving the ionospheric input admittance.''}
\end{quotation}

\subsubsection{Referee \#3}

-{}-{}-{}-{}-{}-{}-{}-{}-{}-{}-{}-{}-{}-{}-{}-{}-{}-{}-{}-{}-{}-{}-{}-{}-{}-{}-{}-{}-{}-{}-{}-{}-{}-{}-{}-{}-{}-{}-{}-{}-{}-{}-{}-{}-{}-{}-{}-{}-{}-{}-{}-{}-{}-{}-{}-{}-{}-{}-{}-{}-{}-{}-{}-{}-{}-{}-{}-{}-{}-
\\
Report of the Third Referee -{}- EU12156/Cosgrove \\
-{}-{}-{}-{}-{}-{}-{}-{}-{}-{}-{}-{}-{}-{}-{}-{}-{}-{}-{}-{}-{}-{}-{}-{}-{}-{}-{}-{}-{}-{}-{}-{}-{}-{}-{}-{}-{}-{}-{}-{}-{}-{}-{}-{}-{}-{}-{}-{}-{}-{}-{}-{}-{}-{}-{}-{}-{}-{}-{}-{}-{}-{}-{}-{}-{}-{}-{}-{}-{}-

In this manuscript, the author develops a model of `ionospheric conductance'
via novel combination of Transmission Line theory and electromagnetic
5-moment fluid equations. The ionosphere is modeled as a stacked layer
of transmission lines, and the transmission lines are taken to support
waves allowed within the framework of the 5-moment fluid equations.
Development of the methodology is, to the referee's knowledge, novel,
and thus a major strength of the work. In additional to the methodological
advancement, the work has applicational importance, with the potential
to improve numerical simulations which employ models of ionospheric
conductance based on electrostatic theory. The work is interesting,
but the referee also thinks the presentation can be streamlined, and
clarified; below are several suggestions/questions toward this end:
\begin{enumerate}
\item It is the referee's opinion that clarity will improve if Section III
is distilled (to the extent possible) to focus on the steps in going
from Eq. 7 to Eqs. 12--14. For example, discussion of the homogeneous
solution before focusing on solution to the full driven problem may
provide physical insight, but it also adds, perhaps, inessential discussion
with respect to the results that are needed in Section IV. Clarity
of Section IV would benefit from a similar streamlining (i.e., shortening
to focus on the essential steps in the derivation needed to establish
the model).\medskip{}
\\
\textcolor{blue}{Even though other referees have made similar comments,
I do not wish to remove the material that provides physical insight.
And given that the correctness of the work is apparently being doubted
in indirect and unstated ways, I feel the need to retain as much rigor
as possible also. So in the final 2022 paper I have attempted to organize
the material by providing the reader with a bold-face labeling of
blocs of text, so that they can better choose what they wish to read
and what they may wish to skip. Hence, the main derivation begins
with an intuitive approach under the header ``The Intuitive Derivation
of the Steady-State Solution,'' and then after this continues with
the rigorous derivation under the header ``Validating the Steady-State
Solution.'' The reader who wants a user-level understanding of the
model does not need to read the second of these, which would be primarily
for the reader wanting to advance or further validate the model.}
\item The author mentions problems evaluating the model for transverse wavelengths
outside the range 100---1000 km; certainly numerical issues can creep
into solvers before an error is thrown, so how robust, with respect
to numerical precision, are the results within this range?\medskip{}
\\
\textcolor{blue}{The question of numerical accuracy is discussed at
the end of Section~7.4 of the final 2022 paper. With respect to whether
the resonance is real, the issue is in the accuracy of the parallel
wavelength. All indications are that the parallel wavelength is very
accurate, and certainly sufficient for prediction of the resonance.}\\
\textcolor{blue}{\quad{}With respect to the reflection that is keeping
the signal from penetrating, the issue is whether sufficiently small
changes can be resolved in the polarization vectors, so that the reflection
is accurately calculated. Since we see this effect also when only
the Alfvén wave is used in the model, it is informative to first consider
this simpler case. We find that plotting the components of the polarization
vector versus altitude reveals a smooth change without noise, and
so it appears that the changes in the polarization vectors are sufficiently
resolved. This should also validate the absolute-accuracy of the polarization
vectors, to the extent that there is no bias in the eigenvalue/eigenvector
calculation (which seems unlikely since it is basically numerical
root finding). Also, the eigenvalues/eigenvectors are checked by multiplying
the eigenvector by the matrix ($H_{5}$), and comparing the result
to the eigenvector multiplied to the eigenvalue. The agreement is
typically to 15 decimal places.}\\
\textcolor{blue}{\quad{}The accuracy is equally good for both modes.
However, when both modes are present and they become nearly degenerate
(which happens), there is an additional question of whether the modes
are sufficiently differentiated, so that the mode-mixing is accurate.
But this only affects the two mode case, and then only when they become
nearly degenerate. So this doesn't affect the basic conclusions.}\\
\textcolor{blue}{\quad{}Please see Section~7.4 for additional discussion,
including of why the range has been limited to 100-1000~km transverse
wavelength.}
\item I am curious about the sharp features in Fig. 15 a, around altitude
125 km (the feature is especially pronounced for the dark yellow curve
for density 8e10 m\textasciicircum -3). According to Fig. 10 c, the
parameters are evidently within the range of validity for the various
linear approximations of the theory, so is the feature numerical in
nature, or a physical prediction of the model?\medskip{}
\\
\textcolor{blue}{The final 2022 paper includes considerable discussion
of the electric field cutoff (Section~7.3), and of the spiky or kinky
features (Section~7.4). It appears that they are a real prediction,
although they are probably not being fully resolved. The explanation
for these features appears to be that the matrix $H_{5}$ is degenerate
or nearly degenerate at some E region altitude, with the Alfvén and
Whistler waves becoming nearly parallel to one another. This is definitely
an area where further research should be performed. However, the kinky
features go away when only the Alfvén wave is used in the model, while
all the other unexpected features remain (such as the resonance and
the greatly reduced conductance). So it seems clear that the kinky
features are not related to the main results.}\\
\textcolor{blue}{\quad{}It should be noted that this is the kind
of feature that may not be contained in the finite-difference, time-domain
models that are also sometimes used for the ionosphere, because of
the MHD simplifications they employ (to eliminate the radio-frequency
modes), and also because their altitude resolution is limited.}
\item For transverse wavelengths greater than 100 km, what does the equivalent
version of Fig. 10 c look like? Since the model is explored for transverse
wavelengths in the range 100---1000 km, demonstrating validity of
linear approximations over this full range will improve robustness
of presented results.\medskip{}
\\
\textcolor{blue}{Since validating the integral approximations seemed
to be very important, I have now developed a more rigorous approach
to validation. As a result, figures like Figure~10c are not used
in the final 2022 paper. Instead, I am now doing numerical integrations
for selected conditions and extracting the critical components for
comparison to the analytical solution. The results are reported in
Section~4 of the final 2022 paper, and elaborated
on in Section~\ref{subsec:Validation-Figures} of this Supplementary
Information. The results have given me a very high degree of confidence
in the critical parallel-wavelength parameter. They have also given
me high confidence in the polarization vectors, although there are
some inconclusive results near 100~km in altitude. The latter is
not likely to have an effect, because the model is not very sensitive
to changes in that area. The lack of sensitivity is probably because
the energy is not getting down that low in altitude. If the source
were instead placed in the lower E-region (to evaluate other physical
questions), the issue may become more important. Hence, this issue
is deserving of further research.}
\item Is it possible that the approximations in Fig. 10 c are an overestimate,
given that modes other than Whistler and Alfvén are assumed to not
dissipate energy? If this is an overestimate, does it have any meaningful
impact on conductance computed via the model?\medskip{}
\\
\textcolor{blue}{I'm not sure I understand the question. It is possible
that the answer to the previous question addresses this one also.
On the other hand, if the issue is the effect of omitting the Ion
and Thermal modes, then I think the referee is correct that omitting
them will tend to overestimate the conductance. The signal would tend
to be dissipated by coupling into those modes, and so would not penetrate
as deeply. The conductivity at lower altitudes would not make as much
of a contribution to the conductance. However, this is based on intuition
and needs testing.}
\item Can a limit be taken in the model to recover exactly (to machine precision)
known results from the electrostatic theory? Even if the electrostatic
results are provably wrong (assuming the presented model is a better
one), this agreement may help to reduce any doubt with respect to
numerical precision within the asserted range of numerical validity,
i.e. transverse wavelengths 100---1000 km.\medskip{}
\\
\textcolor{blue}{It does not appear that the waves have the right
properties to reproduce electrostatic theory exactly under any conditions
I have been able to uncover. However, for purposes of validating the
other aspects of the model (other than the absolute-accuracy of the
eigenvectors/eigenvalues), I have included a section in the final
2022 paper where the waves are artificially modified to have the properties
that were identified in Section~3. Doing
this results in an exact reproduction of electrostatic theory (column~a
of Figure~3). This shows
that the model is working as intended. It also validates the conditions
for electrostatic theory that were derived in Section~3
(and also the additional conditions when two modes are present, from
equation~21).}\medskip{}
\textcolor{blue}{The issues of resolution and numerical precision
are discussed in Section~7.4 of the final 2022 paper.}
\item For transparency, reproducibility, and for the benefit of other researchers
who might like to use such a model, the referee encourages the author
to make the implementation open-source and publicly available (if
at all possible).\medskip{}
\\
\textcolor{blue}{I absolutely agree, that is the plan! Although since
funding is lacking, if anybody thinks they might like to help with
this please feel free to contact me.}
\end{enumerate}

\subsubsection{Referee \#4}

-{}-{}-{}-{}-{}-{}-{}-{}-{}-{}-{}-{}-{}-{}-{}-{}-{}-{}-{}-{}-{}-{}-{}-{}-{}-{}-{}-{}-{}-{}-{}-{}-{}-{}-{}-{}-{}-{}-{}-{}-{}-{}-{}-{}-{}-{}-{}-{}-{}-{}-{}-{}-{}-{}-{}-{}-{}-{}-{}-{}-{}-{}-{}-{}-{}-{}-{}-{}-{}-
\\
Report of the Fourth Referee -{}- EU12156/Cosgrove \\
-{}-{}-{}-{}-{}-{}-{}-{}-{}-{}-{}-{}-{}-{}-{}-{}-{}-{}-{}-{}-{}-{}-{}-{}-{}-{}-{}-{}-{}-{}-{}-{}-{}-{}-{}-{}-{}-{}-{}-{}-{}-{}-{}-{}-{}-{}-{}-{}-{}-{}-{}-{}-{}-{}-{}-{}-{}-{}-{}-{}-{}-{}-{}-{}-{}-{}-{}-{}-{}-

I have struggled with this paper. And some of my assessments are given
as \textquotedbl maybe\textquotedbl{} because of my lack of familiarity
with the scope of Physical Review E (PRE) articles. Looking at the
Journal website I would think that the article is out of scope for
PRE - but I consider that to ultimately be an Editor decision. I am
somewhat surprised that the paper has not been submitted to journals
that are more geophysics/space physics in scope as the subject matter
is firmly in the area of space plasma physics. The topic of the paper
is wave propagation through the collisional ionosphere, taking into
account the variation of collision frequencies and plasma parameters
as a function of height.\medskip{}

\textcolor{blue}{I was hoping this journal might be able to find referees
who would not be subject to the groupthink that seems to be dominating
the reviews of this article.}\medskip{}

I consider the paper to be correct in the governing equations used
to generate the wave-propagation solutions. But I do not consider
the work to be particularly new, at least in terms of the physics.
What may be new is the methodology used to find solutions of the 16-variable
coupled equations. But the results are not particularly new either.
The author appears to be unaware of work that has been done in this
area over the years. I concur with the author that only recently have
models that use a thick, i.e., height resolved, ionosphere been developed.
But at the same time it is not at all clear that the thin ionosphere
models are in serious error for low enough frequency and long enough
vertical wavelength. In essence this is similar to asking when a WKB
approach should be used for wave propagation instead of a Snell's
Law approach.\medskip{}

\textcolor{blue}{The referee says that they consider the methodology
to be correct! But the referee does not tell the editor about the
main results claimed in the paper. The results are simply omitted,
and some of the subsidiary results are discussed in their stead (below).}

\textcolor{blue}{In fact, the referee makes the statement ``}\ldots it
is not at all clear that the thin ionosphere models are in serious
error\textcolor{blue}{\ldots ,'' which is in direct contradiction
to the results in the paper (the results that they did not report
to the editor). It is not possible that the referee could have missed
these results because the first sentence of the Abstract finishes
``computing the total conductance for a vertically stratified ionosphere,
which we compare to the field line integrated conductivity, finding
significant differences.'' And the referee's inclusion of the qualification
``}for low enough frequency and long enough vertical wavelength''\textcolor{blue}{{}
does not excuse them because (a) the sentence still contradicts the
results in the paper; (b) the Abstract specifically references ``effects
from short parallel wavelengths''; and (c) even were it not for (a)
and (b) the sentence would still be highly misleading.}

\textcolor{blue}{The referee began their review by writing ``}I have
struggled with this paper.\textcolor{blue}{'' I can only guess that
the referee is having a very hard time dealing with results that seem
to contradict what they have believed for years, but where they can
find no way to criticize the methodology. Rather than thinking that
they are being deliberately misleading, I would prefer to believe
that the referee is experiencing a short-circuiting of the brain that
is causing them to be uncharacteristically nonsensical. Their language
certainly does not seem hostile, so this seems likely.}\medskip{}

This could be considered a minor issue, but the author does not appear
to be knowledgeable about the low frequency wave modes in a magnetized
plasma. In space plasma physics these have been called the fast, shear,
and slow modes. The fast mode is the same as the mode identified by
the author as the whistler wave, whereas the shear mode is called
the Alfvén wave in the paper. This mode is sometimes called the Alfvén
mode. The slow mode is the mode that corresponds to sound waves, and
this mode may be related to one of the modes identified by the author
for waves below the ionospheric density peak.\medskip{}

\textcolor{blue}{I am certainly familiar with the MHD wave modes!
I have been working in this field for over 20 years and have read
many articles and books. What the referee is reacting to is the choice
of names for the modes, which are well in accord with the terminology
used in ionospheric science, but which differ from the terminology
used in magnetospheric science. In ionospheric science the term ``ion
acoustic wave'' or ``ion wave'' is used to denote the mode associated
with incoherent scatter radar; the term ``whistler wave'' is used
for very low frequency waves from the ionosphere and atmosphere that
are often detected on the ground; and the term ``Alfvén wave'' is
used to denote the waves associated with the MI coupling interaction.
There is a rough correspondence to the MHD ``slow mode,'' ``fast
mode,'' and ``shear mode,'' but I would not want to mislead the
reader into thinking they are the same.}

\textcolor{blue}{One well known difference is that the usual whistler
wave is derived using a high frequency approximation, whereas the
MHD fast mode is derived in a low frequency approximation {[}e.g.,
\citeauthor{stix-1992}, 1992{]}. In particular, I chose the name
``Whistler'' instead of ``fast mode'' because in Cosgrove {[}2016{]}
it was found that in the E region where these waves are able to propagate,
they agreed better with the standard dispersion relation for the whistler
wave than with the standard dispersion relation for the fast mode
wave, using the dispersion relations from \citeauthor{stix-1992},
{[}1992{]}. In general I prefer the ionospheric terminology for this
paper because using it implies that collisions are included in the
calculations, whereas MHD waves are generally non-collisional, which
makes them invalid in the ionosphere. The identification of waves
is well known to result in a zoo of names and it is not possible to
make everyone happy.}

\textcolor{blue}{Finally, this is all a distraction. The purpose of
the article is to compute the ionospheric conductance and electric
field mapping. The purpose is not to study the relationship of the
waves in the ionosphere with the MHD waves. While it seems likely
that the MHD results would mirror results in the ionosphere in many
ways, these results do not include collisions and so cannot be applied
in this work. While it is interesting to make these comparisons, such
comparisons are not the topic of this article, and are certainly not
as important as the actual topic (which the referee did not tell the
editor about).}\medskip{}

I have provided detailed comments in the files appended below. Not
knowing the author's preferences, I have included a version of the
comments where the files are embedded in the PDF as \textquotedbl pop-ups\textquotedbl{}
(EU12156\_with\_embedded\_comments.pdf). The second version lists
the comments separately (Summary of Comments on EU12156.pdf). The
author will notice that I have not provided detailed comments on some
of the more technical sections (specifically the discussion on methods
to solve the set of equations). I must confess that I did not find
those sections particularly useful, but I am willing to accept that
that could be a failing on my part.\medskip{}

\textcolor{blue}{Replies to the detailed comments are below. I appreciate
that this referee is more polite than some of the previous ones. But
the fact remains that the referee has not told the editor about the
results in the paper.}\medskip{}

But I also question the usefulness of the model for inclusion in other
models, such as models that couple the magnetosphere to the ionosphere
and neutral atmosphere. The author includes comments in the concluding
remarks discussing the incorporation of the model in other coupled
models as a future objective.\medskip{}

\textcolor{blue}{The ionospheric conductance is used by (almost?)
all global MHD models. This paper finds that the current method of
calculation is highly inaccurate, and provides a more rigorous replacement.}\medskip{}

The model is essentially 1-dimensional (height variation). It was
not clear how the model could be modified for 2-D or 3-D implementation.\medskip{}

\textcolor{blue}{This is a mischaracterization, the model calculates
the Fourier components of a 3D signal in the ionosphere. It is therefore
a 3D solution for small perturbations of an }\textcolor{blue}{\emph{initially}}\textcolor{blue}{{}
one-dimensional ionosphere. While it is true that extending the background
ionosphere to higher dimensionality would be difficult, the normal
way that conductance is calculated, as the field line integrated conductivity,
also assumes a one-dimensional background ionosphere.}\medskip{}

Why wouldn't a simple finite difference implementation on a 2-D or
3-D grid be sufficient? What makes the author's methodology for solving
the coupled equations particularly useful?\\
\medskip{}

\textcolor{blue}{``Simple'' and ``Sufficient'' are odd choices
of words, since those models are much more difficult to implement,
require much more computing power, and involve much more numerical
error. The reasons the method is more appropriate for the goals of
the paper have been described in the Introduction and Background sections
of the latest 2022 paper, and I will not repeat them here. However,
the fact that ``finite difference'' methods have not been used to
evaluate the ionospheric conductance (in any reference I can find,
after many comments by many referees) is a pretty strong indicator
that they are not appropriate for the problem.}\medskip{}

Why is it superior to other methods?

\medskip{}

\textcolor{blue}{The method is not ``superior,'' it is different,
with different strengths and weaknesses. It is a more analytical method
that is especially useful for steady-state quantities like the ionospheric
conductance. It provides more physical understanding with more clarity
in the approximations, but is not as flexible as the numerical simulations
with respect to geometry. Finding agreement between these two approaches
is the way that the field can be advanced. And again, the reasons
the method is more appropriate for the goals of the paper have been
described in the Introduction and Background sections of the latest
2022 paper, and I will not repeat all of them here.}\medskip{}

Detailed Comments from the Markup:\medskip{}

\noindent In my understanding the electrostatic approximation does
not mean steady state. It means that the magnetic field is non time-varying,
or varying so slowly that it can be ignored, and consequently only
Gauss's law and charge conservation are required to investigate the
variability of the electric field. The electric field can itself be
time-varying. A simple example is plasma oscillation. This mode is
purely electrostatic, but not time-stationary.\medskip{}

\textcolor{blue}{By ``steady state'' I do not mean no oscillation,
I mean the condition after the source has turned on and been on for
a long time. More specifically, I use the term ``steady state''
in the context of a solution found using the Laplace transform, where
there is a transient response that can be discarded. What remains
is commonly termed the ``steady-state solution,'' although it may
be an oscillatory solution, if the source that turned on was a harmonic
source. This is the solution used in transmission line theory, such
as the solution found in equation~(12),
which oscillates purely at frequency $\omega_{0}$. But this does
not mean that the magnetic field can be ignored.}\medskip{}

I draw the author's attention to Tu et al. (2014), Inductive-dynamic
magnetosphere-ionosphere coupling via MHD waves, J. Geophys. Res.
Space Physics, 119, 530--547, doi:10.1002/2013JA018982 who present
a fully electromagnetic model of wave propagation through the height-resolved
ionosphere. The results in that paper should be compared with those
presented here.\medskip{}

\textcolor{blue}{I had been aware of that reference, but did not consider
it very important since there was no field-aligned current in that
model. What I did miss was their later works, but anyway their later
works also did not include calculation of the ionospheric conductance,
and so were not any more relevant than the earlier time-domain simulations
that I did reference. As discussed in Sections~1~and~2,
there are a number of issues with time-domain simulations that make
them not so useful for the ionospheric conductance. So there are no
comparisons to be made with any of the time domain simulations of
which I am aware, so far.}\medskip{}

Again, I think a clarification about what the author means by electrostatic
might be in order. It is really a long wavelength approximation, so
that the ionosphere appears to be a thin shell, and the reflection
and transmission properties of the ionosphere are determined using
height-integrated conductivities.\medskip{}

\textcolor{blue}{The first paragraph of the Introduction contains
the words ``\ldots testing electrostatic theory which, when applicable,
arises from the long-wavelength limit of TL theory.'' Also, Section~2
(Section~3 in the final 2022 version)
was all about developing this idea of the long wavelength limit. Also,
the idea of the long-wavelength limit was developed in {[}Cosgrove
2016{]}, in defining the wave-Pedersen conductivity. So maybe the
referee simply forgot to remove their comment after finishing this
Section~2?}

\textcolor{blue}{Also, the referees comment is odd because it seems
to suggest that the parallel wavelength is long, without mentioning
that the paper specifically finds that the parallel wavelength is
NOT long. And it is not that the referee missed this result in the
paper, because in their last comment they specifically object to this
finding, based on a hand waving argument.}

\textcolor{blue}{Also, what was meant by the term ``electrostatic''
was explained in the last paragraph of the Introduction. In the latest
2022 paper that discussion is expanded and appears in the second and
third paragraphs of the Introduction. In both versions there was/is
substantial effort made to distinguish between two cases, and it was
explained that the terms ``electrostatic theory'' and ``electrostatic
waves'' would be used to differentiate between the two.}\medskip{}

I find the author's use of Alfvén and Whistler wave to be inconsistent
with the names assigned to the low frequency waves. In the MHD regime
(i.e., for waves with frequencies below the lowest ion cylcotron frequency,
assuming a multi ion species plasma) there are three wave modes for
a warm plasma. These are known as the fast, shear, and slow modes.
See, for example, Russell et al. (2016), Space Physics, Cambridge
University Press. The slow mode corresponds to sound waves. The shear
mode is sometimes referred to as the Alfvén mode, and this corresponds
to bending (or shearing) of magnetic field lines. Assuming that the
waves are first order this mode is non-compressive. On the other hand,
the fast mode is compressive with the magnetic pressure and plasma
pressure perturbations in phase. In a single ion species plasma the
fast mode does couple to the whistler mode, although the whistler
mode is usually thought of as an electron mode. In a multi ion species
plasma the fast mode only couples to the whistler mode for parallel
propagation. The presence of multiple ion gyro-resonances and ion-ion
hybrid frequencies means that the fast mode does not couple to the
whistler mode for non-parallel propagation. See, for example, Figure
13.7 in Russell et al. (2016).\medskip{}

\textcolor{blue}{The referee would like me to use the MHD terminology
for the waves. Please see my response to the referees earlier remark
along these lines. Basically, I do not want to use the MHD terms because
MHD waves are generally non-collisional, and a major point of this
article is that the collisions change the nature of the waves, and
in so doing give rise to conductivity. If I used the usual MHD waves
there would be no conductivity!}

\textcolor{blue}{Another point that was not mentioned yet in responding
to this referee (although was thoroughly covered in response to an
earlier referee), is that the ``waves'' in this article are formally
defined as eigenvectors, and are equated over the full altitude extent
of the ionosphere. The referee is instead thinking about waves defined
by making physical approximations, such as the high frequency approximation
that ions are immobile, or the low-frequency collisionless MHD approximations.
The latter approach may result in many different ``kinds'' of waves,
far more than there are eigenvectors. There is no one-to-one mapping
between these two different approaches to defining waves. In the approach
of this article, it arises that the physical nature of a wave may
change with altitude (changing collision frequency); but if they both
correspond to the same eigenvector then I use only one name. An example
is the relationship between the whistler and fast mode waves, which
were found in {[}Cosgrove 2016{]} to correspond to the same eigenvector
(using a smaller equation-set, but it also holds for the 5-moment
equations). So it is really not appropriate to use the MHD terminology
for the ``waves'' in this article.}\medskip{}

I consider that the \textquotedbl wave Pedersen conductivity\textquotedbl{}
is half the \textquotedbl DC\textquotedbl{} Pedersen conductivity
is in fact in agreement with our understanding of wave incidence and
reflection, even when using a \textquotedbl thin\textquotedbl{} ionosphere
model. In particular, if the characteristic parameter mu\_0{*}Sigma\_p{*}V\_Alfvén
is large, then the incident electric field is almost entirely reflected.
{[}mu\_0 is the permeability of free space, Sigma\_p is the height
integrated Pedersen conductivity, and V\_Alfvén is the Alfvén speed
above the conducting layer.{]} Furthermore, for this regime, the Incident
magnetic field is half the magnetic field just above the conducting
layer. The magnetic field of the reflected wave adds to the incident
wave field.This appears to be consistent with the results in Figure
4.\medskip{}

\textcolor{blue}{The wave-Pedersen conductivity is derived by superposing
the incident and reflected waves, that is exactly what it is all about,
and it was originally defined in {[}Cosgrove 2016{]}. A few paragraphs
before where the referee is commenting (on Section~2, which is Section~3
in the final 2022 version), there is the sentence ``The arrows in
the figure illustrate how the formula is derived, as the superposition
of two waves traveling in opposite directions, related by the reflection
coefficient at the load, $(Y_{0}-Y_{L})/(Y_{0}+Y_{L})$.'' The formula
referred is the (slightly generalized) standard transmission line
formula, and reading the section one finds that it was used to derive
the wave-Pedersen conductivity. In Cosgrove {[}2016{]} the wave-Pedersen
conductivity was derived by explicitly combining incident and reflected
waves, but in this article I wanted to emphasize that these are old
and well established ideas.}

\textcolor{blue}{In the final 2022 version of the article, artificial
waves are constructed where the wave-Pedersen conductivity is modified
to match the (regular, zero-frequency) Pedersen conductivity. It is
shown that the model then reproduces electrostatic theory exactly
(if the parallel wavelength and dissipation scale length are also
made long). This also shows that the interpretation given in the paper
for the ``wave-Pedersen conductivity'' is correct.\medskip{}
}

If the author is not going to include the non-linear term then why
is it discussed in the next paragraph? Indeed once equation (7) is
given I see little of value in the rest of this section. It strikes
me as a rather pedantic discussion that essentially leads to the conclusion
that the author will use linear perturbation theory when discussing
the eigenmodes of the system described by equation (7).\medskip{}

\textcolor{blue}{It is important for the article to give a rigorous
account of the model derivation, so that there cannot be broadly-stated
objections as were made by other referees. Reading some of the other
reviews, there appears to be a lack of understanding of certain key
points, for example that the DC response is fully included in this
approach. In addition, I feel it is important to develop an intuitive
understanding of the physics, and so I believe all the development
is very worthwhile. The patient reader will be rewarded if they read
this material carefully. I agree that this is known material that
could be considered pedantic, but there is a reason we learned all
this stuff in school. It is very powerful!}\medskip{}

Figure 6 has some value, but I would strongly encourage a different
normalization scheme, given the already existing knowledge of the
wave modes expected for the ionosphere (the fast, shear, and slow
modes as mentioned earlier). In particular, the electron velocities
should be normalized to the Alfvén speed. This is essentially the
normalization used for the ion velocities, the normalized ion velocity
= (c/V\_Alfven){*}ion velocity{*}B, where B is the ambient magnetic
field strength. The reason for doing thid is so that the reader can
see if there is any difference in the ion and electron velocities,
as it is that difference that gives the current density in the plasma.
The electric fields should also be renormalized by c/V\_Alfven). For
the shear mode this would ensure that the electric and magnetic fields
are plotted on a scale that makes them comparable. I further note
that the Whistler mode appears to have the properties I would expect
for the fast mode. The density, pressure, and parallel component of
the magnetic field perturbation are all in phase. Finally, in Figure
6 the thermal mode appears to have the characteristics of what I might
expect fot the slow mode, being mainly a pressure perturbation. I
do acknowledge that this could also be the ion mode. Perhaps the high
collision frequencies have separated the ion and electron motion.
This is may be a point worth making, especially since the stem plot
is drawn for 120 km altitude, i.e. the bottomside ionosphere.\medskip{}

\textcolor{blue}{These comments by the referee are potentially valuable,
as finding a good normalization for the waves was a challenge. The
referee is recommending normalization by the ideal Alfvén wave properties,
which for the Alfvén wave would allow for assessing how close to ideal
it is. This might be a good choice, although testing would be required,
and I note that the same normalization must be applied to all the
modes (if the figure is to serve its purpose of comparing the different
modes).}

\textcolor{blue}{However, the referee does not seem to appreciate
how much collisions will affect the modes in the E region, and continues
to insist that the modes be thought of simply as the ideal, collisionless
MHD modes. The relationship between whistler mode, fast mode, and
the single eigenvector they both correspond to was investigated in
Cosgrove {[}2016{]}. It was found that the fully-collisional eigenvector
corresponds better to the whistler wave, which is why that name was
chosen. Also, the relationship of the collisionless, shear Alfvén
mode to the fully-collisional eigenvector was investigated in Table~2
of Cosgrove {[}2016{]}, and profound differences were found.}

\textcolor{blue}{Regardless, the referee is going down the path of
trying to understand the relationship of the fully-collisional eigenvectors
to the physically defined, collisionless MHD waves. This is no doubt
interesting to them, but is not the topic of the article! I want to
discourage distractions like this. The point of the article is to
give a rigorous electromagnetic derivation of the ionospheric conductance,
and associated electric field ``mapping.'' Therefore I have removed
Figure~6 from the final 2022 version.}\medskip{}

I strongly recommend that the author refer to the \textquotedbl slow\textquotedbl{}
Alfvén wave as the shear mode. In classical MHD wave theory the slow
mode is the sound wave for a plasma where the thermal pressure is
much less than the magnetic pressure (i.e., sound speed <\textcompwordmark <
Alfvén speed).\medskip{}

\textcolor{blue}{I did make a mistake in the sentence commented on,
the word ``slow'' should have been ``shear.'' The slow mode is
the one that corresponds to the ion-acoustic wave, and the shear mode
to the Alfvén wave, in the ionospheric terminology. This is the sentence
where it was explained that what I called the Whistler wave could
also have been called the fast-Alfvén wave (referee calls it just
``fast mode''), and the latter would make more sense at higher altitudes.
Figure~7 was included specifically to make this point, although its
caption also reads ``slow'' where it should read ``shear'' (not
sure what happened with that). But since this eigenvector primarily
propagates at lower altitudes, where it better matches the whistler
wave {[}Cosgrove 2016{]}, the name ``Whistler'' was adopted.}\\
\medskip{}

Figure 8 and subsequent figures are very difficult to read, with the
colors blue, black and green hard to distinguish . It is also not
clear to me why the author spends so much time comparing the electrostatic
solutions to the electromagnetic. If the author is solving the electromagnetic
equations, then the stem plots would indicate how electrostatic the
waves are. Again, based on classical MHD wave theory I would expect
the fast (Whistler) mode or the shear (Alfvén) mode to generally not
have an electrostatic counterpart. The shear mode could become electrostatic
for short wavelengths (high wave number) if the wave frequency approaches
the ion gyro-frequency.\medskip{}

\textcolor{blue}{I am sorry about the lines being hard to distinguish.
These are PDF figures that are fully zoomable, but if the referee
is color blind that may not help. Although there are many possibilities,
I chose to compare the electrostatic and electromagnetic waves through
their parallel wavelengths, because if those are different and the
electromagnetic one is short (which can be the case!), the electrostatic
wave will give a wrong characterization of the ionosphere. A comparison
based on the stem plots would say little about the errors that could
be produced. I also note that an earlier referee said that the shear
mode becomes electrostatic at ``}Scales below the electron inertial
length\textcolor{blue}{\ldots ,'' which seems to me different from
a criteria based on the ion gyro-frequency. So it seems this material
is not all that well known, even in the collisionless case.}\medskip{}

This result is also expected from classical MHD wave theory. At sufficiently
high altitudes, where collisions are less important, we can use the
standard dispersion relations for the fast and shear mode. For the
fast mode the dispersion relation is given by omega\textasciicircum 2
= k\textasciicircum 2{*}V\_Alfvén\textasciicircum 2 assuming a cold
plasma. k is the wave number. If we split the wave number into parallel
and perpendicualr components then omega\textasciicircum 2 = (k\_perp\textasciicircum 2+k\_parl\textasciicircum 2){*}V\_Alfvén\textasciicircum 2
where k\_perp is the perpendicular wave number and k\_parl is the
parallel wave number. We note that if k\_perp is too large, such that
k\_perp\textasciicircum 2{*}V\_Alfvén\textasciicircum 2 > omega\textasciicircum 2,
then k\_parl must be imaginary, corresponding to an evanescent wave.
This condition does not apply to the shear mode, which has the dispersion
relation omega\textasciicircum 2 = k\_parl\textasciicircum 2{*}V\_Alfvén\textasciicircum 2.
This dispersion relation is independent of k\_perp, and so we expect
to have a propagating shear-mode wave regardless of the perpendicualr
wave number.\medskip{}

\textcolor{blue}{It's true that many of the results for the collisional
ionospheric waves mirror the results for the collisionless MHD waves.
But making a study of this relationship is another topic that is outside
the scope of this work. It was explained near the end of Section~III
(Section~4 in the final 2022 version) that the
Whistler wave requires a quadratic integral approximation, which results
in a quadratic dispersion relation, and is associated with a low frequency
cutoff. This is the point the referee is making and the existing discussion
is deemed sufficient.}\medskip{}

I again want to point out the conflation of two different uses of
the adjective electrostatic. In the context of wave theory electrostatic
does indeed mean ignoring curl\_E. But in the context of Pedersen
and Hall conductivities the time variation is ignored in the ion and
electron momentum equation. Clearly waves are the mean by which information
is transmitted from region of the magnetosphere to the other, and
indeed Alfvén (i.e. MHD) waves are the primary means by which the
electric fields and currents adjust. I do not disagree that a fully
electromagnetic wave theory is required. But these have been existance
for several years (again, I refer to Tu et al. (2014) as an example.\medskip{}

\textcolor{blue}{It appears that the referee did not read the last
paragraph of the Introduction section (Section~I), where this distinction
was clearly made and it was stated that the terms ``electrostatic
theory'' and ``electrostatic waves'' would be used to make the
distinction in the remainder. For example that paragraph contains
the line ``The validity of this electrostatic-wave theory is a distinct
question from that of the full electrostatic theory, which does not
include wavelike effects.'' So there is no conflation of terms, it
is just that the referee did not read all of Section~I (Introduction).}

\textcolor{blue}{I am glad that the referee now agrees that a fully
electromagnetic treatment is required, because their earlier comments
seemed to be in denial, and they did not tell the editor about the
main results in the paper.}

\textcolor{blue}{The fact that other kinds of electromagnetic models
exist is irrelevant, because different kinds of models have different
strengths and weaknesses.}\medskip{}

Because the stem plot in Figure 6 does not show the electric field
I am not sure that I concur with this conclusion. Faraday's law suggests
that if the vertical scale is 100 km and the electric field is 10
mV/m, then the magnetic field changes by \textasciitilde{} 600 nT/s
for a wave of 100 nT/s if the 100 km is an evanescence scale.\medskip{}

\textcolor{blue}{This comment was applied to the last line of the
paper, which read ``The electrical thickness is the amount of phase
rotation for a signal traversing the ionosphere, and we find that
it often exceeds $90^{\circ}$, meaning a complete failure of electric
field mapping, even for transverse wavelengths as large as 100 km.''
Figure~6 is irrelevant to this finding; it has nothing to do with
wavelength. And if the reviewer thinks they can use it to evaluate
whether the waves are correct, they cannot, because they do not have
any results for collisional waves that could be compared. The salient
figure is Figure~12 (Figure~2
in the final 2022 version), where was demonstrated the integration
of the wavevector over altitude, to get the phase rotation.}

\textcolor{blue}{The referee should have reported the findings in
the article to the editor, such as the contents of this last line.
This is especially true given that the referee did not find any problems
with the methodology. Instead, it appears the referee has relied on
a hand-waving argument to convince themselves that they didn't even
need to tell the editor what results the article claimed. This is
a typical feature of group think, where highly simplistic arguments
are deemed sufficient to uphold the existing paradigm, even over arguments
that are much more rigorous.}

\subsubsection{Decision Letter from Phys. Rev. E}

From: pre@aps.org \\
Subject: Your\_manuscript EU12156 Cosgrove \\
Date: XXXX XX, XXXX at 6:35:46 PM PST \\
To: russell.cosgrove@me.com \\
Reply-To: pre@aps.org\medskip{}

\noindent Re: EU12156    

Electromagnetic ionosphere: A model

by Russell B. Cosgrove\medskip{}

\noindent Dear Dr. Cosgrove,\medskip{}

\noindent The above manuscript has been reviewed by two of our referees.
Comments from the reports appear below. The Physical Review editors
wish to accept only papers that, in addition to being scientifically
sound, are important to the field and contain significant new results
in physics. Given the appended comments, we judge that these acceptance
criteria are not met.\medskip{}

\noindent We regret that consequently we cannot accept the paper for
publication in the Physical Review.\medskip{}

\noindent Yours sincerely,\medskip{}

\noindent XXXXXX \\
Associate Editor Physical Review E \\
Email: pre@aps.org \\
https://journals.aps.org/pre/\medskip{}
\\

\noindent P.S.  We regret the delay in obtaining these reports.\medskip{}

\textcolor{blue}{Three out of the four referees failed to tell the
editor about the results in the paper. None of the referees told the
editor that anything was wrong with the methodology. In fact one of
them specifically said ``}\textcolor{black}{I consider the paper
to be correct in the governing equations used to generate the wave-propagation
solutions. But I do not consider the work to be particularly new,
at least in terms of the physics.}\textcolor{blue}{'', and then did
not tell the editor what results the article claimed. Referees are
supposed to tell the editor what results a paper claims, and then
separately evaluate the methodology. The referees have not done their
job, and have instead acted as gate keepers for a potentially wrong
electrostatic paradigm.}

\subsection{Reviews of ``An Electromagnetic Calculation of Ionospheric Conductance
that seems to Override the Field Line Integrated Conductivity,''
submitted to JGR in 2022. This was the fifth version of the final
2022 paper.}

\textcolor{blue}{This version was very similar to the version submitted
to Phys. Rev. E. The main difference was that this version added modeling
results using artificially modified wave modes, which showed that
the model would exactly reproduce electrostatic theory if the identified
criteria were satisfied. These criteria were identified in the section
on the gedanken experiment. Also, the description of what was meant
by ``electrostatic theory,'' and the differentiation from ``electrostatic
waves'' was moved to the second paragraph in the Introduction. It
had previously been in the last paragraph of the Introduction, and
some of the reviewers had failed to discover it there.}

\subsubsection{Reviewer \#1}

{*}{*}{*}{*}{*}{*}{*}{*}{*}{*}{*}{*}{*}{*}{*}{*}{*}{*}{*}{*}{*}{*}{*}{*}{*}{*}{*}
\\
Reviewer \#1 Evaluations: \\
Recommendation (Required): Reject \\
Significant: No, the paper is not a significant advance or contribution.
\\
Supported: No \\
Referencing: Mostly yes, but some additions are necessary. \\
Quality: The organization of the manuscript and presentation of the
data and results need some improvement. \\
Data: No \\
Accurate Key Points: No \medskip{}
\\
Reviewer \#1 (Formal Review for Authors (shown to authors)): \medskip{}

This manuscript is to calculate ionospheric conductance using electromagnetic
solutions for collisional plasma. The transmission line theory is
applied to extend the calculation to stratified ionosphere. The major
conclusion is that an electromagnetic wave description is required,
as already advocated by a number of previous studies from prospects
of physical processes and sophisticated numerical simulations. This
study, in my opinion, did not significantly contribute to advance
our understanding of the ionosphere-thermosphere dynamics for the
reasons stated below. Therefore, I do not recommend its publication
in JGR Space physics.\medskip{}

\textcolor{blue}{The reviewer does not report the main results of
the paper to the editor, which are that the ionospheric conductance
is different from the field line integrated conductivity, and the
electric field does not map through the ionosphere. The referee suggests
that certain ``}\textcolor{black}{prospects of physical processes
and sophisticated numerical simulations}\textcolor{blue}{'' have
already found the results in the article, whereas in fact there are
no comparable results anywhere I can find. The reviewer does not give
any references to back up their claim.}\medskip{}

1. Transmission line (TL) theory is widely used for modeling signal
transmission along the specified path and often helpful to provide
physical insights to the process of interest. However, research in
ionospheric physics have long passed the stage that our understanding
of the system relies on simple concept of input-admittance. One serious
drawback of the TL theory is that the signal can only propagate along
the \textquotedbl line\textquotedbl , even the transverse scale
may be taken into consideration. In other words, oblique transmission
is not allowed. In fact, the wave transmission in the ionosphere is
3-D. The author cited Tu et al, 2014 to indicate that the electromagnetic
modeling of the ionosphere is one-dimensional without considering
transverse-scale dependence. This is not true, because later studies
are 2-D electromagnetic simulations (Tu and Song, A two-dimensional
global simulation study of inductive- dynamic magnetosphere ionosphere
coupling, JGR Space Physics, 121, doi:10.1002/2016JA023393, 2016;
Tu and Song, On the momentum transfer from polar to equatorial ionosphere,
Space Physics, 124, doi:10.1029/2019JA026760, 2019). There are also
other 2-D even 3-D electromagnetic studies (e.g., Lysak, et al, Modeling
of the ionospheric Alfvén resonator in dipolar geometry, JGR Space
Physics, 118, 1514-1528, doi:10.1002/jgra.50090, and references therein),
although these studies do not explicitly include plasma and neutral
dynamics.\medskip{}

\textcolor{blue}{The reviewer is not correct about the limitations
of the analysis in the article. The signal is not limited to propagation
``}\textcolor{black}{along a line}\textcolor{blue}{,'' and ``}\textcolor{black}{oblique
transmission}\textcolor{blue}{'' is fully supported. The article
in fact describes the propagation of three-dimensional perturbations
of a vertically stratified ionosphere, with no limitations as to the
direction of the wavevector or the Poynting vector. Also what was
said about {[}Tu et al., 2014{]} in the paper is correct, that article
contains a completely one-dimensional analysis that does not include
any transverse scale-size dependence. It is true that the article
failed to reference the later Tu and Song studies that were higher
dimensional, but it did reference other similar two- and three-dimensional
electromagnetic simulation studies that were done earlier, including
the Lysak study specifically mentioned by the reviewer. However, NONE
of these studies reported results for the ionospheric conductance,
and the reviewer has not referenced any such results. Before receiving
so many reviews of the various versions of this article I kept thinking
maybe I was missing something in the earlier studies, but that now
seems highly unlikely.}\\

\medskip{}
2. Our studies of the ionosphere-thermosphere now are using system
approach and focus on self-consistency. Both the plasma and neutrals
dynamics must be considered to provide a faithful description of the
coupled system. The TL theory used in the manuscript does not include
neutral dynamics, and plasma dynamics is not considered explicitly
either.\medskip{}

\textcolor{blue}{What the referee said about ``}plasma dynamics is
not considered explicitly\textcolor{blue}{'' is not correct. Other
than kinetic effects, the plasma dynamics is fully included in the
analysis, with the exception of nonlinear terms, which terms are also
not included in electrostatic theory. And the neutral dynamics is
also not included in electrostatic theory. So testing electrostatic
theory does not require inclusion of either neutral dynamics or nonlinear
terms. The physical justification for omitting neutral dynamics is
that the electrostatic state should arise over much shorter time scales
than is possible for any neutral response. The physical justification
for omitting the nonlinear terms is that they are negligible for sufficiently
small-amplitude signals, and it is useful to have an amplitude independent
characterization of the system; this is the basic recipe for circuit
theory and the definition of admittance depends upon it.}\medskip{}

3. In electromagnetic modeling of the magnetosphere-ionosphere-thermosphere
system, conductance is actually not necessary and is a by-product.
Conductance is needed in electrostatic theory to connect electrical
current density and electric field. Then from divergence-free of the
current density, the electric potential equation (Poisson equation)
is obtained and used to calculate electric field. In the electromagnetic
approach, Maxwell's equations are solved to determined electric field
and perturbation magnetic field. Plasma and neutral parameters are
governed by plasma and neutral transport equations. The current density
is actually not necessary either, which is merely a quantity to describe
difference between the electron velocity and ion bulk velocity.\medskip{}

\textcolor{blue}{The referee implies that the ionospheric conductance
is not important (and a similar statement was made in the first paragraph).
This assertion is something of a joke, since in addition to being
the standard quantity used to determine the inner boundary condition
for global MHD modeling, the ionospheric conductance is ubiquitous
in conversations of space physics. The ionospheric conductance is
a fundamental concept in MI coupling. But the calculation in the article
does not only produce the ionospheric conductance, it also gives all
the other field quantities as well, resolved in three dimensions.
More specifically, the Fourier components are derived, and these can
be assembled to describe an arbitrary three-dimensional disturbance.
At present, when only a few have yet been calculated, they can be
used to describe the scale-size dependence of the ionospheric response. }

\subsubsection{Decision Letter from JGR}

From: jgr-spacephysics@agu.org \\
Subject: 2022JA030862 (Editor - XXXX): Decision Letter \\
Date: XXXX XX, XXXX at 12:36PM \\
To: russell.cosgrove@me.com \\
Cc: russell.cosgrove@ucf.edu \medskip{}
\\
Dear Dr. Cosgrove: \medskip{}
\\
Thank you for submitting your manuscript to Journal of Geophysical
Research - Space Physics. as you suggested, I sent the manuscript
to two leading experts on the topic. one of them declined to review
and the other recommended to reject. I also take a careful reading
of the manuscript and agree with the referee. in magnetosphere-ionosphere
coupling, the electric engineering approach has already run its course.
now people agree that the momentum transfer between plasma and neutrals
are crucial to understand many phenomena. in your derivation, many
weak points can be found, mostly due to invalid approximations. you
may argue they are not wrong, but they are clearly not state-of-the-art
as required by the JGR-Space Physics. I am declining your manuscript
for publication in Journal of Geophysical Research - Space Physics.
\medskip{}
\\
I am enclosing the reviews, which you may find helpful if you decide
to revise your manuscript and submit to another journal. I am sorry
that I cannot be more encouraging at this time. \medskip{}
\\
Thank you for your interest in Space Physics. \medskip{}
\\
Sincerely, \medskip{}
\\
XXXX \\
Editor \\
Journal of Geophysical Research - Space Physics\medskip{}

\textcolor{blue}{The editor does not say anything specific in their
appraisal. There are general expressions of negativity, but there
is nothing that gives any insight into anything that might be wrong.}

\textcolor{blue}{In science, we use whatever methods give results,
just so long as the method is valid. A method has not ``}\textcolor{black}{run
its course}\textcolor{blue}{'' when it is still producing new and
important results. There are essentially no other papers that give
electromagnetic results for the ionospheric conductance. Although
there is a lot more discussion in Sections~1~and~2 of the final
2022 paper, this alone is proof that the methods in the paper are
``state of the art,'' to the extent that is a valid metric. And
having a variety of different methods available is a good thing.}

\subsection{Response to Reviews Concluded}

\textcolor{blue}{Beginning with the 2016 article where the concepts
were first introduced there have been 12 reviews of this work, with
5 being essentially positive. Of the reviews that were against, not
a single one told the editor about the results found in the paper,
and not a single one identified any specific problem with the methodology.
No doubt a few of these reviewers were simply not in a position to
understand the point of the article, and for some reason did not realize
that. But it is difficult to explain the whole phenomenon in this
way. Thus, it seems necessary to conclude that groupthink is preventing
the article from being published. This a situation where self publication
is appropriate. This is a thoroughly peer-reviewed article where the
reader may evaluate the reviews for themselves.}

\textcolor{blue}{The article presents what might be described as the
canonical electromagnetic calculation of the ionospheric conductance,
and of electric field mapping, etc. The methods are standard physics
methods, which have not yet been applied to ionospheric science. I
can find only one other article that compares an electromagnetic calculation
of conductance to the field line integrated conductivity, and that
article employs a boundary value calculation, the unreliability of
which was one of the major points of the Cosgrove {[}2016{]} article.
When a standard physics calculation provides clear results, that fact
needs to be known by the research community. This is especially true
when the results contradict a long-standing paradigm, where that paradigm
is not well supported by the published mathematical analysis.}
